\documentclass[iop,revtex4,numberedappendix,appendixfloats,twocolappendix]{emulateapj}
\usepackage{epsfig}
\usepackage{amsmath}
\usepackage{natbib}

\newcommand*{\bfrac}[2]{\genfrac{}{}{0pt}{}{#1}{#2}}
\newcommand{\lsun}{$\mathcal{L}_\Sun$}
\newcommand{\msun}{$\mathcal{M}_\Sun$}
\newcommand{\rsun}{$\mathcal{R}_\Sun$}
\newcommand{\lbol}{$\mathcal{L}_\mathrm{bol}$}
\newcommand{\teff}{$T_\mathrm{eff}$}
\newcommand{\logg}{log$(g)$}

\newcommand{\rjup}{$\mathcal{R}_\mathrm{Jup}$}
\newcommand{\mjup}{$\mathcal{M}_\mathrm{Jup}$}

\begin{document}

\title{Magellan Adaptive Optics first-light observations of the exoplanet $\beta$ Pic \lowercase{b}. 
II. 3--5~\lowercase{$\mu$m} direct imaging with M\lowercase{ag}AO+C\lowercase{lio}, and the empirical bolometric luminosity of a self-luminous giant planet
}

\author{Katie M. Morzinski$^{1,\star}$, Jared R. Males$^{1,\star}$, Andy J. Skemer$^{1,2,\dagger}$, Laird M. Close$^{1}$, Phil M. Hinz$^{1}$, T. J. Rodigas$^{3,\dagger}$,
Alfio Puglisi$^{4}$, Simone Esposito$^{4}$, Armando Riccardi$^{4}$, Enrico Pinna$^{4}$, Marco Xompero$^{4}$, Runa Briguglio$^{4}$,
Vanessa P. Bailey$^{5}$,
Katherine B. Follette$^{5}$,
Derek Kopon$^{6}$,
Alycia J. Weinberger$^{3}$, and Ya-Lin Wu$^{1}$
}

\affiliation{
	$^{1}$Steward Observatory, 933 N.\ Cherry Ave., University of Arizona, Tucson, AZ 85721, USA; \\
	$^{2}$Department of Astronomy and Astrophysics, 1156 High St., University of California, Santa Cruz, CA 95064, USA; \\
	$^{3}$Department of Terrestrial Magnetism, Carnegie Institute of Washington, 5241 Broad Branch Rd.\ NW, Washington, DC 20015, USA; \\
	$^{4}$INAF-Osservatorio Astrofisico di Arcetri, Largo E. Fermi 5, I-50125 Firenze, Italy; \\
	$^{5}$Kavli Institute of Particle Astrophysics and Cosmology, Stanford University, 382 Via Pueblo Mall, Stanford, CA 94305, USA; \\
	$^{6}$Harvard Smithsonian Center for Astrophysics, 60 Garden St., Cambridge, MA 02138, USA; \\
	$^{\star}$NASA Sagan Fellow; \\
	$^{\dagger}$NASA Hubble Fellow
	}

\date{\today}

\begin{abstract}
Young giant exoplanets are a unique laboratory for understanding cool, low-gravity atmospheres.
A quintessential example is the massive extrasolar planet $\beta$ Pic b, which is 9~AU from and embedded in the debris disk of the young nearby A6V star \object{$\beta$ Pictoris}.
We observed the system with first light of the Magellan Adaptive Optics (MagAO) system.  In Paper I \citep{males2014} we presented the first CCD detection of this planet with MagAO+VisAO.  Here we present four MagAO+Clio images of $\beta$ Pic b at 3.1~$\mu$m, 3.3~$\mu$m, $L^\prime$, and $M^\prime$, including the first observation in the fundamental CH$_4$ band.
To remove systematic errors from the spectral energy distribution (SED), we re-calibrate the literature photometry and combine it with our own data, for a total of 22 independent measurements at 16 passbands from 0.99--4.8~$\mu$m.
Atmosphere models demonstrate the planet is cloudy but are degenerate in effective temperature and radius.
The measured SED now covers $>$80\% of the planet's energy,
so we approach the bolometric luminosity empirically.
We calculate the luminosity by extending the measured SED with a blackbody and integrating to find log(\lbol/\lsun)~$= -3.78\pm0.03$.
From our bolometric luminosity and an age of 23$\pm$3~Myr, hot-start evolutionary tracks give a mass of 12.7$\pm$0.3~\mjup, radius of 1.45$\pm$0.02~\rjup, and \teff\ of 1708$\pm$23~K (model-dependent errors not included).
Our empirically-determined luminosity is in agreement with values from atmospheric models (typically $-3.8$ dex), but brighter than values from the field-dwarf bolometric correction (typically $-3.9$ dex), illustrating the limitations in comparing young exoplanets to old brown dwarfs.
\keywords{Planets and satellites: individual (Beta Pictoris b); Stars: individual (Beta Pictoris); Stars: planetary systems; Instrumentation: adaptive optics}
\end{abstract}

\section{Introduction}

Direct imaging offers the best opportunity to study self-luminous exoplanets as they cool from the heat of formation \citep{marley2007,fortney2008}.
The technique targets extrasolar planets on Solar-System-scale orbits,
rather than on short-period orbits where planets' energy budgets are dominated by heat from stellar insolation \citep{burrows2005,cowanagol2011}.
Development of the direct imaging technique is a pathway to characterizing true Solar System analogs.

In this work we study the giant ($\sim$10~\mjup) exoplanet $\beta$ Pictoris b, which orbits an A-type star on a Saturn-sized (9~AU) orbit.
$\beta$ Pic b is one of only a handful of directly-imaged exoplanets \citetext{\textit{e.g.}, \citealt{chauvin2005,hr8799,lagrange2009,lafreniere2010,carson2013,kuzuhara2013,rameau2013a,bailey2014,naud2014,macintosh2015}},
which are being characterized with optical--infrared spectrophotometry.
Being a massive exoplanet on a wider, non-highly-irradiated orbit, it presents an ideal opportunity to study the atmosphere and energy budget of a young Jovian-class object.

$\beta$ Pictoris is a nearby, young A6V \citep{gray2006} star, at 19.44$\pm$0.05~pc \citep{vanleeuwen2007} and 23$\pm$3~Myrs \citep{binks2014,mamajek2014}.
The star hosts an $\sim$1000-AU-wide gas-rich debris disk \citep{smithterrile1984} presented nearly edge-on.
Asymmetries in the disk and variability in high-resolution spectra indicative of dynamic comets \citep{kiefer2014} suggested the presence of one or more planets long before the discovery of $\beta$ Pic b itself \citep{lagrange1988,beust1990,betapicb_planetdisk,levison1994,betapicb_suspicion,betapicb_prediction,beust2007}.
Finally, the giant exoplanet $\beta$ Pic b was imaged \citep{lagrange2009} and confirmed \citep{lagrange2010} using adaptive optics (AO) at VLT/NaCo.

Observations to date indicate that the planet is embedded in and sculpts the debris disk \citep{kiefer2014,dent2014,apai2015,millarblanchaer2015,nesvold2015}.
The planet's semimajor axis is 9.1$\bfrac{+0.7}{-0.2}$~AU
\citetext{weighted mean of \citealt{macintosh2014}, \citealt{nielsen2014}, and \citealt{millarblanchaer2015}}
with an eccentricity $<$0.15--0.26 \citep{macintosh2014,nielsen2014,millarblanchaer2015}.
The combination of evolutionary models plus upper-limits based on radial-velocity non-detection indicate the planet's mass is between $\sim$7--15~\mjup\ \citep{lagrange2012rv,bonnefoy2013,bodenheimer2013,borgniet2014}.

\citet{chilcote2015} present the first H-band spectrum of $\beta$ Pic b, obtained with the Gemini Planet Imager (GPI) in Dec.\ 2013.
The H-band spectrum is best fit by \teff=1600--1700~K models, and has a distinctive triangular shape peaking at 1.68 $\mu$m, indicative of low surface gravity (\logg$\sim$3.5--4.5).
The spectral type of a young substellar object is affected by its gravity, effective temperature, and atmospheric properties, rather than being strictly a temperature sequence as it is for stellar objects and field brown dwarfs \citep{bowler2013,liu2013}.
Given the close match of the H-band spectral shape to both the 1000-K 2MASS 1207 b \citep{barman2011,skemer2011} and the $\sim$2000-K ROXs 42 b \citep{currie2014roxs}, both temperature and gravity are important in modulating its spectral shape, as also seen for young brown dwarfs \citep{rice2011,allers2013,dupuy2014,faherty2014}.

\citet{bonnefoy2014} present the J-band spectrum of $\beta$ Pic b obtained during commissioning of GPI.  The 1.11--1.35 $\mu$m spectrum shows water absorption, a rising continuum as seen in other late M and early L dwarfs, and FeH absorption.  The authors fit the spectrum to various sets of young low-mass and low-gravity objects, as well as model atmospheres, finding a best-fit SpT of L1$\bfrac{+1}{-1.5}$ and a best-fit \teff\ of 1650$\pm$150 K, ruling out ``cold-start'' formation.  Finally, they use radial velocity observations to set a constraint on the mass of $<$20~\mjup, similar to the $<$12~\mjup\ constraint for a 9-AU orbit found by \citet{lagrange2012rv}.

We observed $\beta$ Pic b at first-light of the Magellan adaptive optics system, ``MagAO'' \citep{close2012spie,morzinski2014spie} in December 2012.
In \citet{males2014} (hereafter Paper I) we analyze the short-wavelength observations with MagAO's visible-light camera ``VisAO'' and Gemini/NICI near-IR data.  
When the near-IR colors of $\beta$ Pic b are compared to $>$500 field BDs in $Y_s$, $J$, $H$, and $K_s$, we find that its colors are consistent with early- to mid-L dwarfs, as well as early-T dwarfs, while its absolute magnitude is consistent with early L dwarfs.
The best-fit spectral type of L2.5$\pm$1.5 and best-fit \teff\ of 1643$\pm$32~K show that $\beta$ Pic b matches the spectral types of early L field dwarfs but is cooler; thus, for $\beta$ Pic b as for many other young low-mass objects, the spectral sequence is not a temperature sequence but a luminosity sequence \citetext{\citealt{bowler2013}, \citealt{liu2013}, Paper I}.

This work (Paper II) describes the infrared observations (3--5 $\mu$m) with MagAO's IR camera ``Clio2'' (hereafter ``Clio''), and combines all of the available photometry to obtain a complete view of $\beta$ Pic b's optical-infrared spectral energy distribution (SED).
We present MagAO+Clio 3.1-$\mu$m, 3.3-$\mu$m, $L^\prime$, and $M^\prime$ images.
Next, we combine our data with spectrophotometry from the literature, determining the SED on a uniform photometric system.
Finally, we use the optical--infrared (O/IR) SED to measure the empirical bolometric luminosity of this young self-luminous planet.

\section{Observations and Data}

\subsection{Magellan Adaptive Optics}
MagAO (P.I.\ Laird Close) is a new facility instrument at the Magellan ``Clay'' telescope at Las Campanas Observatory, Chile \citep{close2012spie,close2013,morzinski2014spie}.  The 6.5-m primary mirror was fabricated at the Richard F.\ Caris Mirror Lab, Steward Observatory, Tucson, Arizona.
The 85-cm MagAO adaptive secondary mirror (ASM) glass shell was fabricated in Arizona, integrated with the electronics and mechanics by MicroGate, Bolzano, and A.D.S.\ International, Lecco, Italy, and integrated and tested with the wavefront sensor (WFS) with our partners at INAF-Arcetri Observatory, Florence, Italy.
The AO system updates at up to 1000 Hz and corrects turbulence via the 585-actuator ASM.  The WFS is a modulated pyramid \citep{ragazzoni1999,tozzi2008,esposito2010spie}.  On bright stars such as $\beta$ Pic A, MagAO flattens the wavefront to $\sim$100--150 nm rms, depending on seeing and wind.

There are two science cameras behind MagAO: VisAO and Clio.  VisAO (P.I.\ Jared Males) is a new diffraction-limited visible-light camera, taking full advantage of the pyramid WFS.  Clio (P.I.\ Phil Hinz) has been relocated to Magellan from its initial home at the MMT at Mt.\ Hopkins, Arizona, where it was developed for imaging low-mass companions at 3--5~$\mu$m \citep{heinze2003,clio2004,clio2006,hr8799mmt}.  Clio uses a prototype Hawaii-I HgCdTe array (one of the first MBE arrays) sensitive out to 5~$\mu$m, with the following filters as of first light at MagAO in 2012: $J$, $H$, $K_s$, [3.1], [3.3], [3.4], $L^\prime$, [3.9], Barr $M$, and $M^\prime$.  Clio has two pixel scales: the Narrow camera has 15.9 mas pixels and a 16 x 8'' field of view while the Wide camera has 28 mas pixels and a 28 x 14'' FoV (see Appendix~\ref{sec:app_astrom} for the plate scale calculation).  Further information is given at our ``Information for Observers'' website\footnote{\url{http://magao.as.arizona.edu/observers/}}.

Light is collected at the 6.5-m primary mirror of the Magellan Clay telescope, continues to the f/16 secondary, and is folded by the tertiary onto the Nasmyth platform.  A mounting ring encloses the compact optical table where both VisAO and the WFS are mounted (the ``W-unit''), while Clio is mounted to the Nasmyth ring.  The two science cameras are illuminated simultaneously by way of the dichroic window of Clio, for simultaneous observations in the optical and infrared.  MagAO was commissioned from Nov.\ 2012--Apr.\ 2013.

\subsection{Observations}
We observed $\beta$ Pic on the nights of UT 2012 December 1, 2, 4, and 7.  Table~\ref{tab:obs} lists the observing conditions and parameters, including precipital water vapor (PWV) from the forecaster at nearby La Silla Observatory\footnote{ \url{http://www.eso.org/gen-fac/pubs/astclim/forecast/meteo/ ERASMUS/las\_fore.txt}}.  Seeing ranged from 0\farcs5--1\farcs5.  We corrected 120--250 modes at a frame rate of 990 Hz, and residual wavefront error was $\sim$150 nm rms according to the VisAO focal plane measurement on Dec.\ 4.
Since first light, higher-order corrections at 100--130 nm are typical on a bright star when correcting 300 modes.

\begin{table}[htb]
	\caption{Observing conditions and AO parameters.}
	\label{tab:obs}
	\begin{center}
		\begin{tabular}{llllll} \hline \hline
			UT Date	&	Seeing	&	PWV	&	No.\ of			&	Freq. 		&	Filter	\\
			&			&	/mm	&	modes		&	/Hz		& 	\\
			\hline
			2012 Dec 01 & 0\farcs5--0\farcs7 &	3	& 250	&	990		& $[3.3]$ \\
			2012 Dec 02 & 0\farcs5--1\arcsec & $^\dagger$	&	250		&	990		& $L^\prime$	\\ 
			2012 Dec 04 &  0\farcs5--1\arcsec &	8.5	&	200	&	990		& $Y_s$, $M^\prime$	\\
			2012 Dec 07 &  1\arcsec--1\farcs5 &	4.6--5.7	&	200		&	990	& $[3.1]$	\\
		\hline
		\hline
		\multicolumn{6}{l}{}	\\ 
		\multicolumn{6}{p{0.95\linewidth}}{$^\dagger$Not recorded.}	\\
	\end{tabular}
	\end{center}
\end{table}

VisAO observations were limited by commissioning of the ADC (atmospheric dispersion compensator), seeing, and other issues to one good night of Dec.\ 4.  With Clio, we observed $\beta$ Pic in a different filter each night, always with the rotator off so that we were observing in angular differential imaging (ADI) mode \citep{marois2006adi}.

Table~\ref{tab:observations} summarizes the exposure time and field rotation obtained for each filter.  Only the deep (usually where the star is saturated) integrations are tabulated for Clio, as these are the frames that are combined via ADI processing to reveal the planet.  Shallow (unsaturated) integrations of $\beta$ Pic A are used to calibrate the photometry.

\begin{table}[htb]
	\caption{$\beta$ Pic observations obtained, this work.}
	\label{tab:observations}
	\begin{center}
		\begin{tabular}{lllll} \hline \hline
			Filter				& Date	 	& Instrument	& Total		& Field		\\
			& UT			& \&			& deep int.		& rot.		\\
			&			& camera		& /min.			& /deg.	\\
			\hline
			$[3.1]$			& 2012 Dec 07	& Clio Wide			& 98			& 88	\\
			$[3.3]$			& 2012 Dec 01	& Clio Narrow	& 70			& 95	\\
			$L^\prime$		& 2012 Dec 02	& Clio Narrow	& 80			& 80	\\
			$M^\prime$		& 2012 Dec 04	& Clio Narrow	& 107		& 109	\\
		\end{tabular}
	\end{center}
\end{table}

\begin{figure*}
	\begin{flushright}
		\includegraphics[height=0.33\linewidth,angle=90,trim=2.4cm 0.5cm 1.1cm 3.2cm,clip=true]{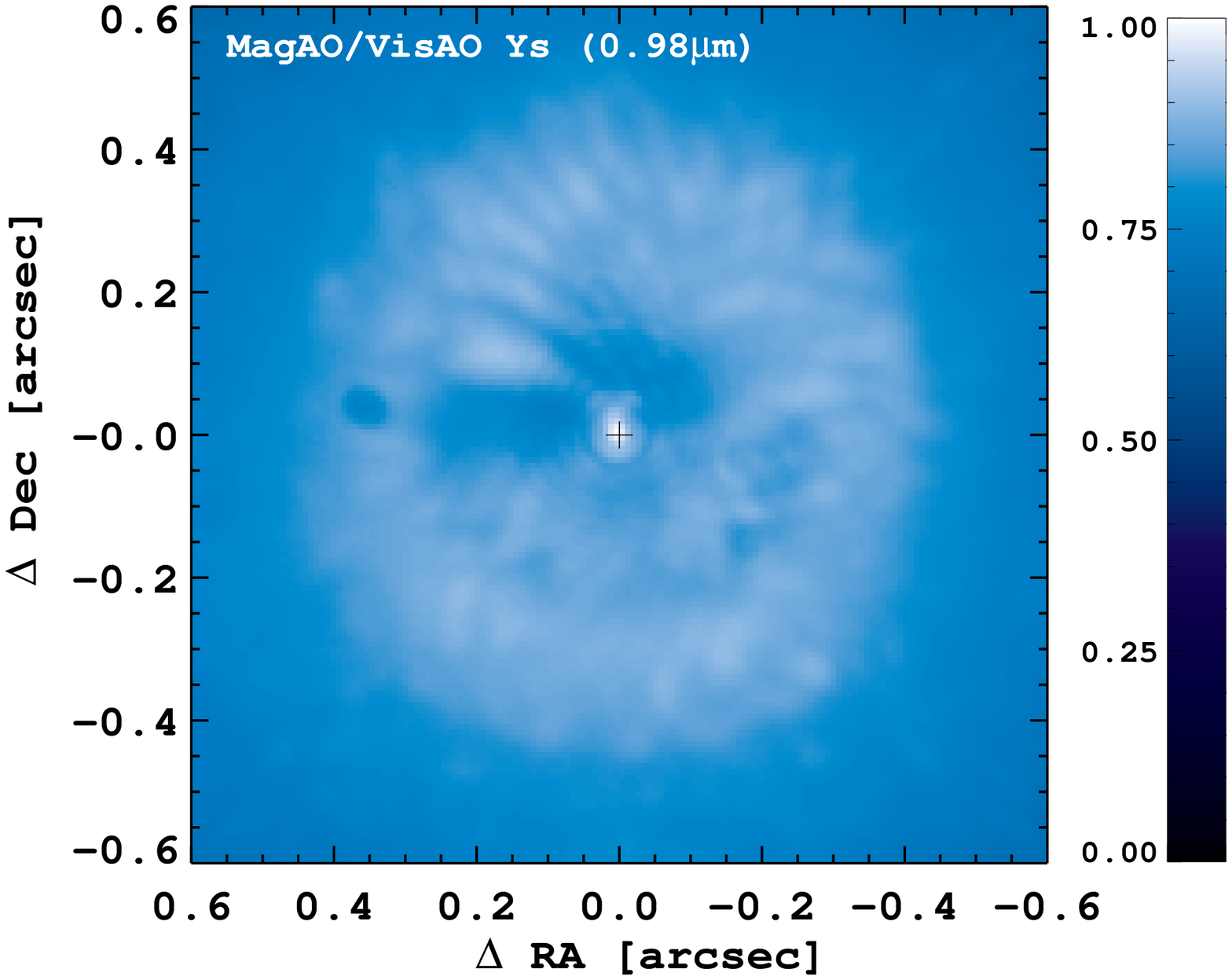}
		\includegraphics[height=0.33\linewidth,angle=90,trim=2.4cm 0.5cm 1.1cm 3.2cm,clip=true]{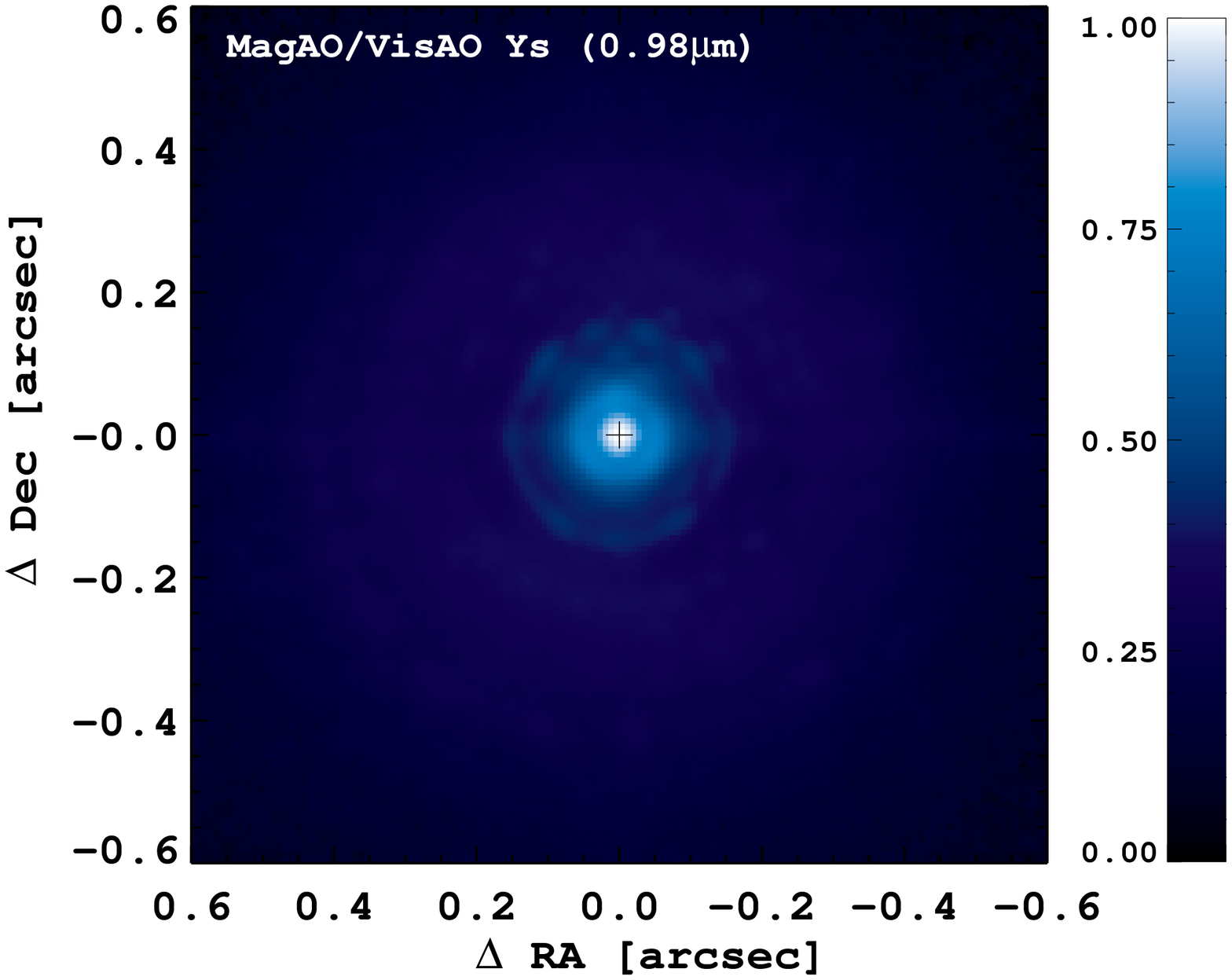}
		\includegraphics[height=0.33\linewidth,angle=90,trim=2.4cm 0.5cm 1.1cm 3.2cm,clip=true]{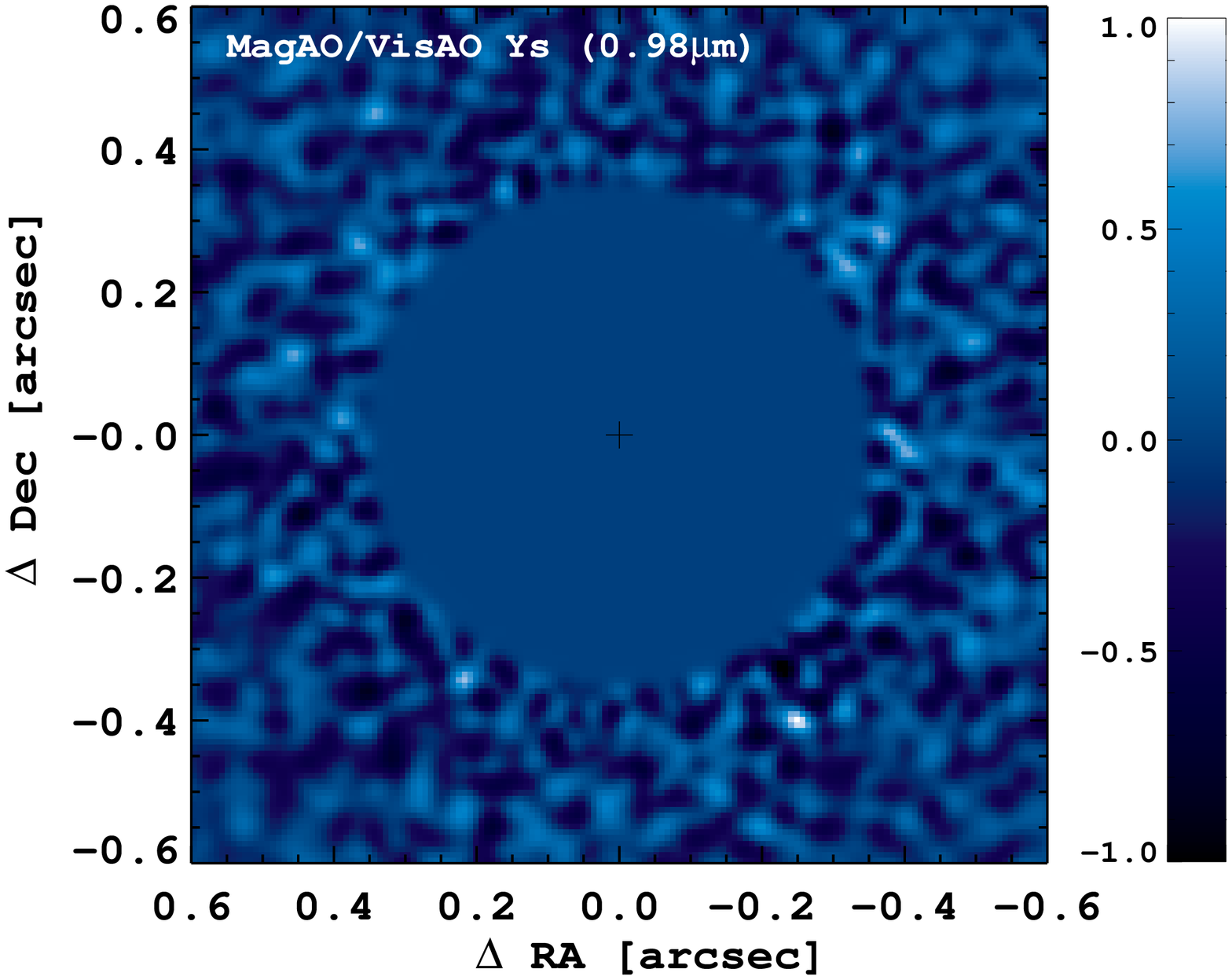}
		\includegraphics[height=0.33\linewidth,angle=90,trim=2.4cm 0.5cm 1.1cm 3.2cm,clip=true]{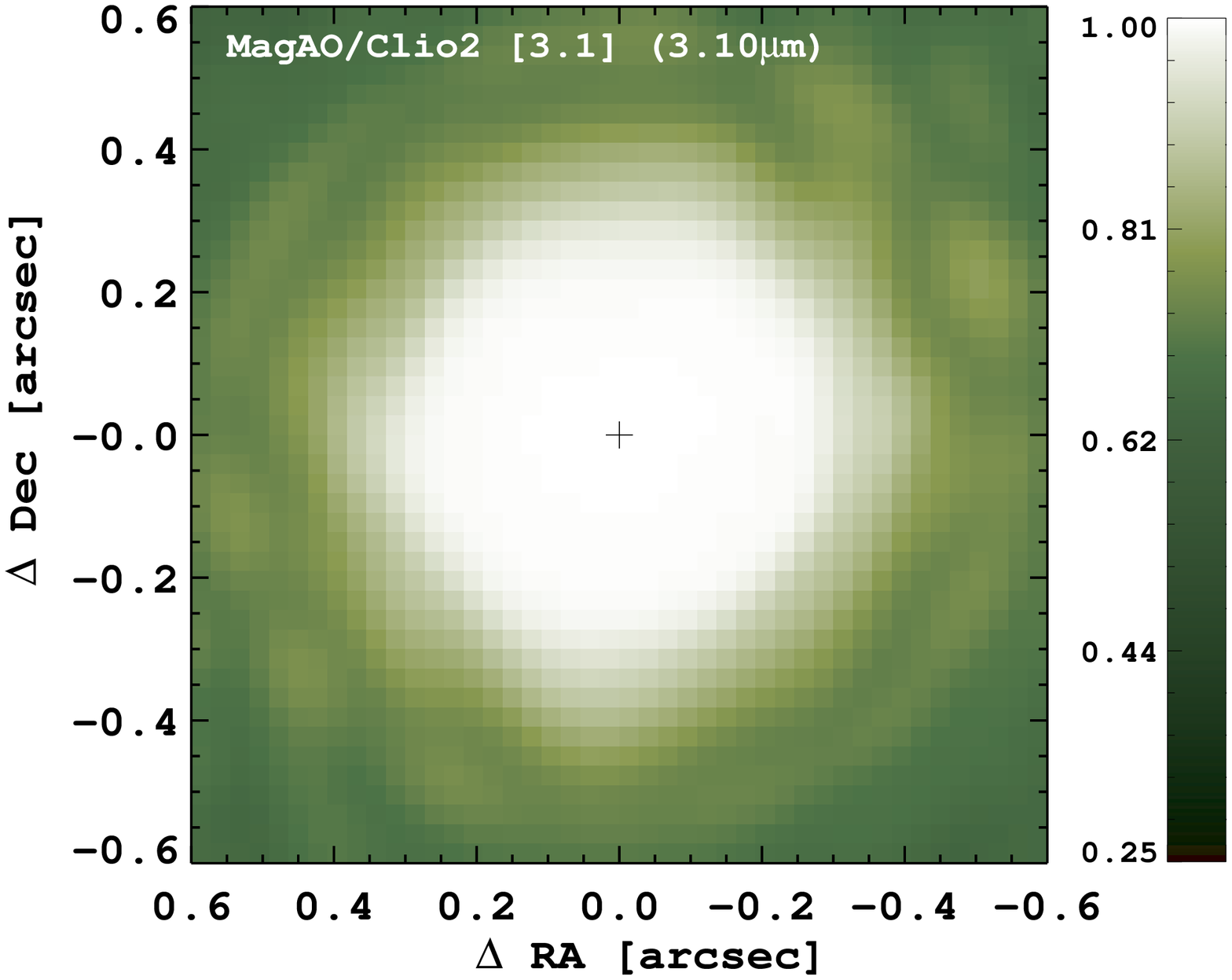}
		\includegraphics[height=0.33\linewidth,angle=90,trim=2.4cm 0.5cm 1.1cm 3.2cm,clip=true]{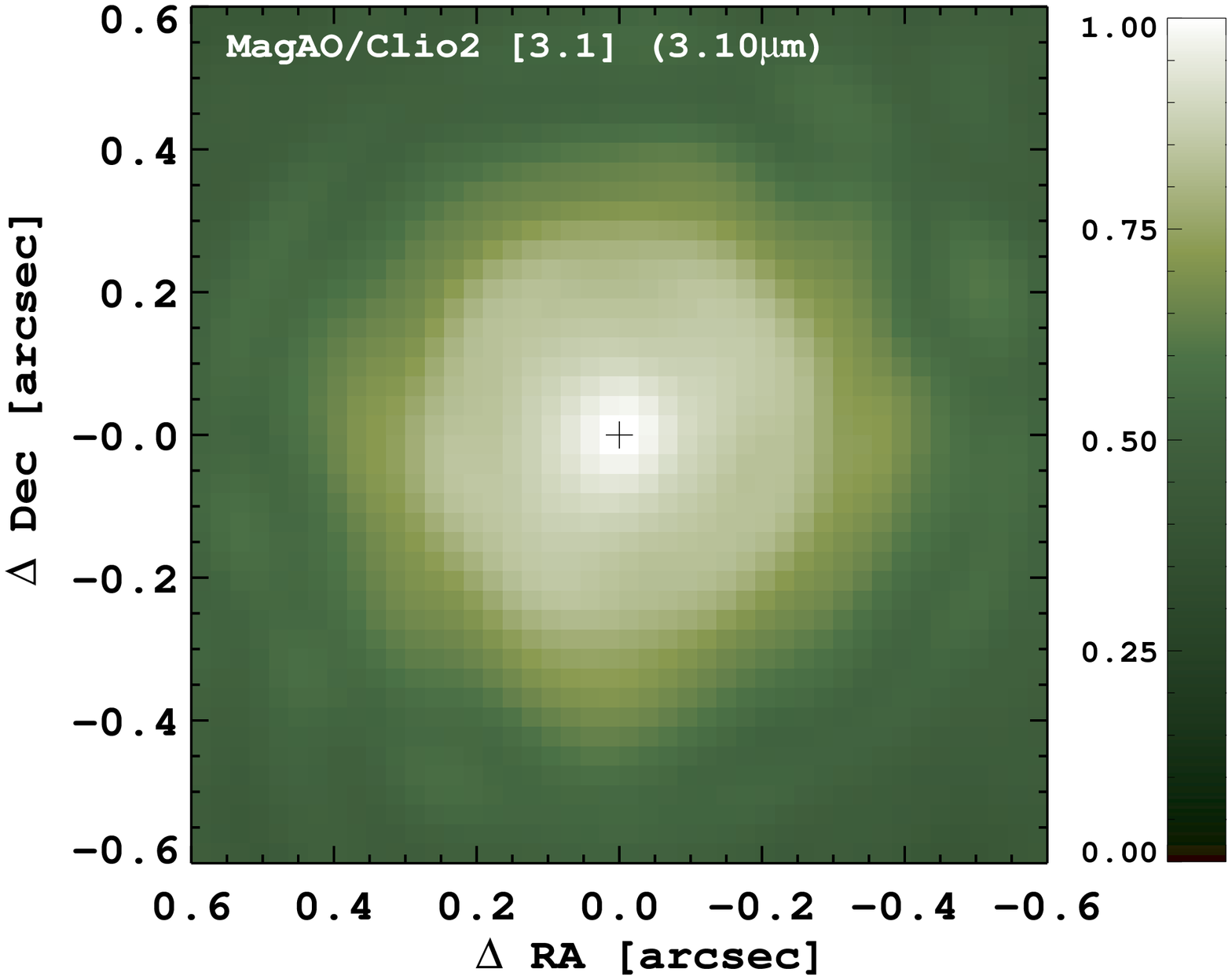}
		\includegraphics[height=0.33\linewidth,angle=90,trim=2.4cm 0.5cm 1.1cm 3.2cm,clip=true]{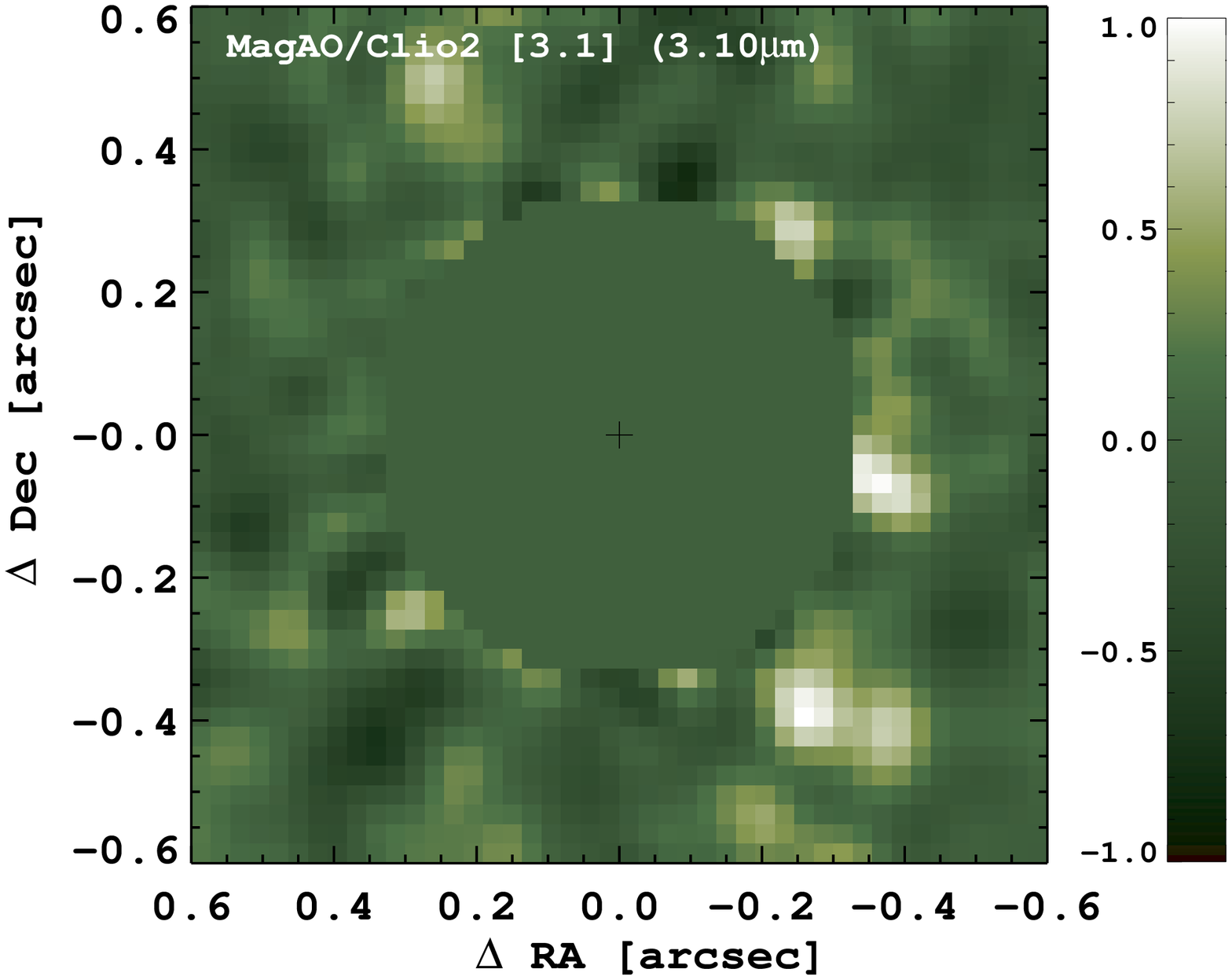}
		\includegraphics[height=0.33\linewidth,angle=90,trim=2.4cm 0.5cm 1.1cm 3.2cm,clip=true]{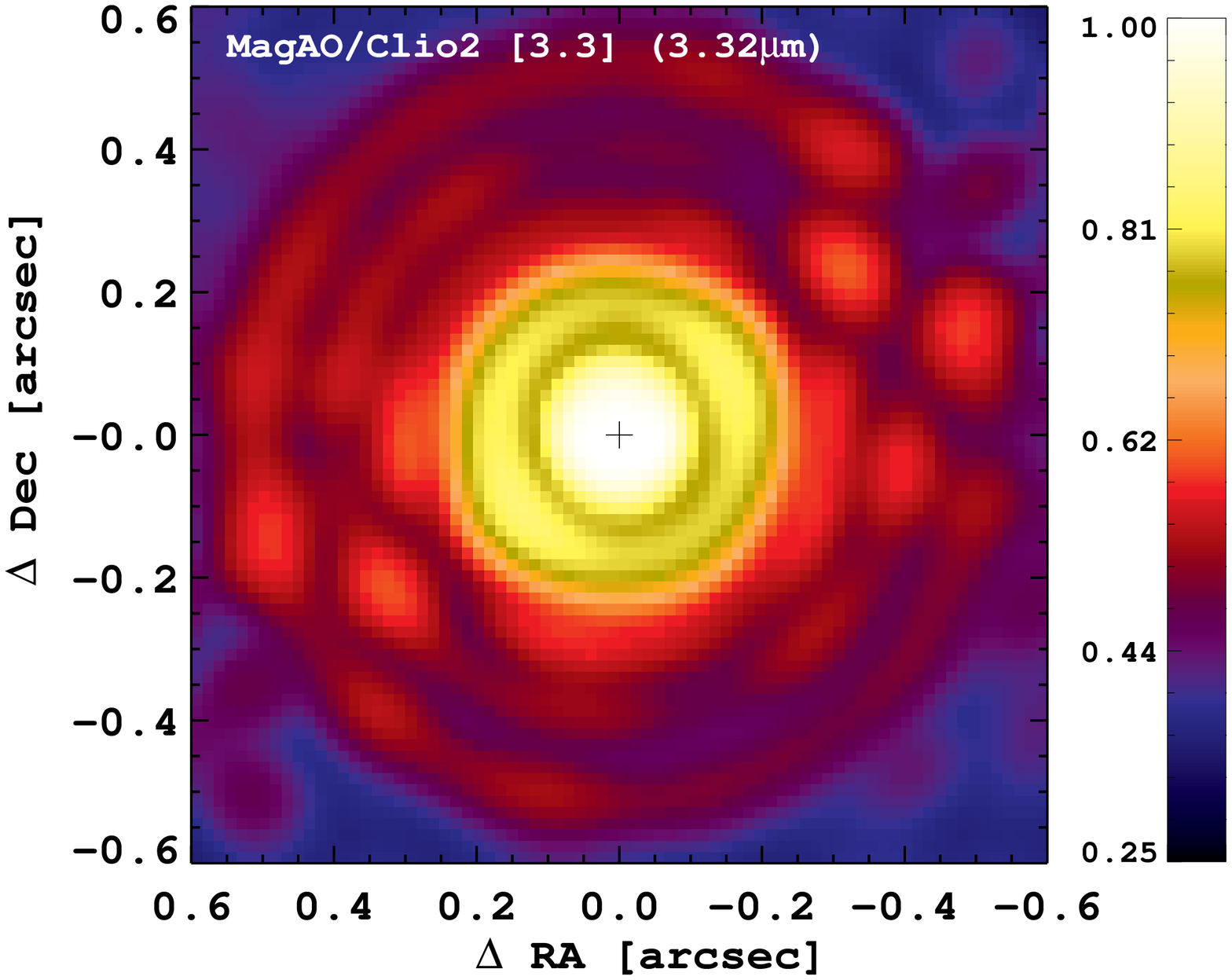}
		\includegraphics[height=0.33\linewidth,angle=90,trim=2.4cm 0.5cm 1.1cm 3.2cm,clip=true]{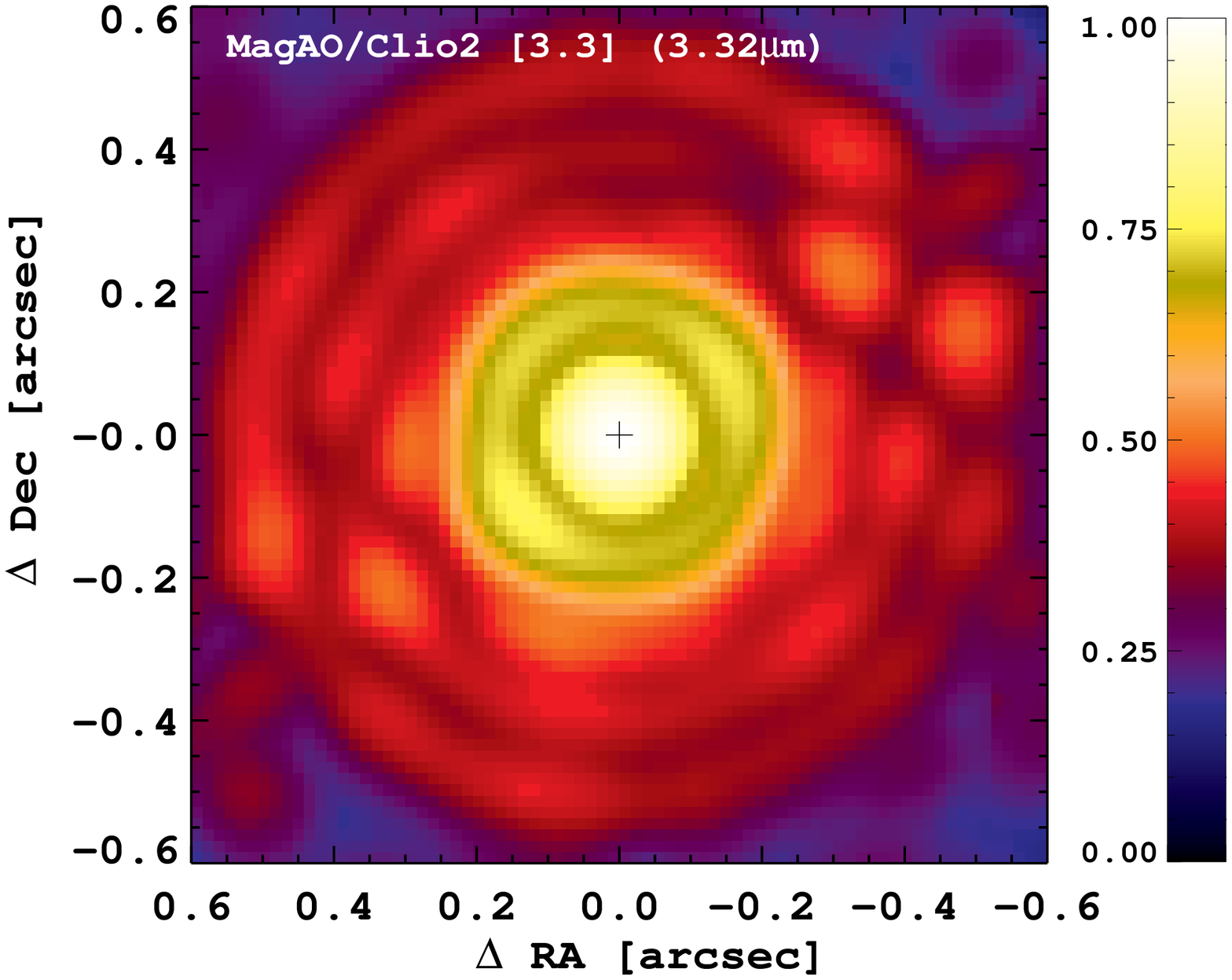}
		\includegraphics[height=0.33\linewidth,angle=90,trim=2.4cm 0.5cm 1.1cm 3.2cm,clip=true]{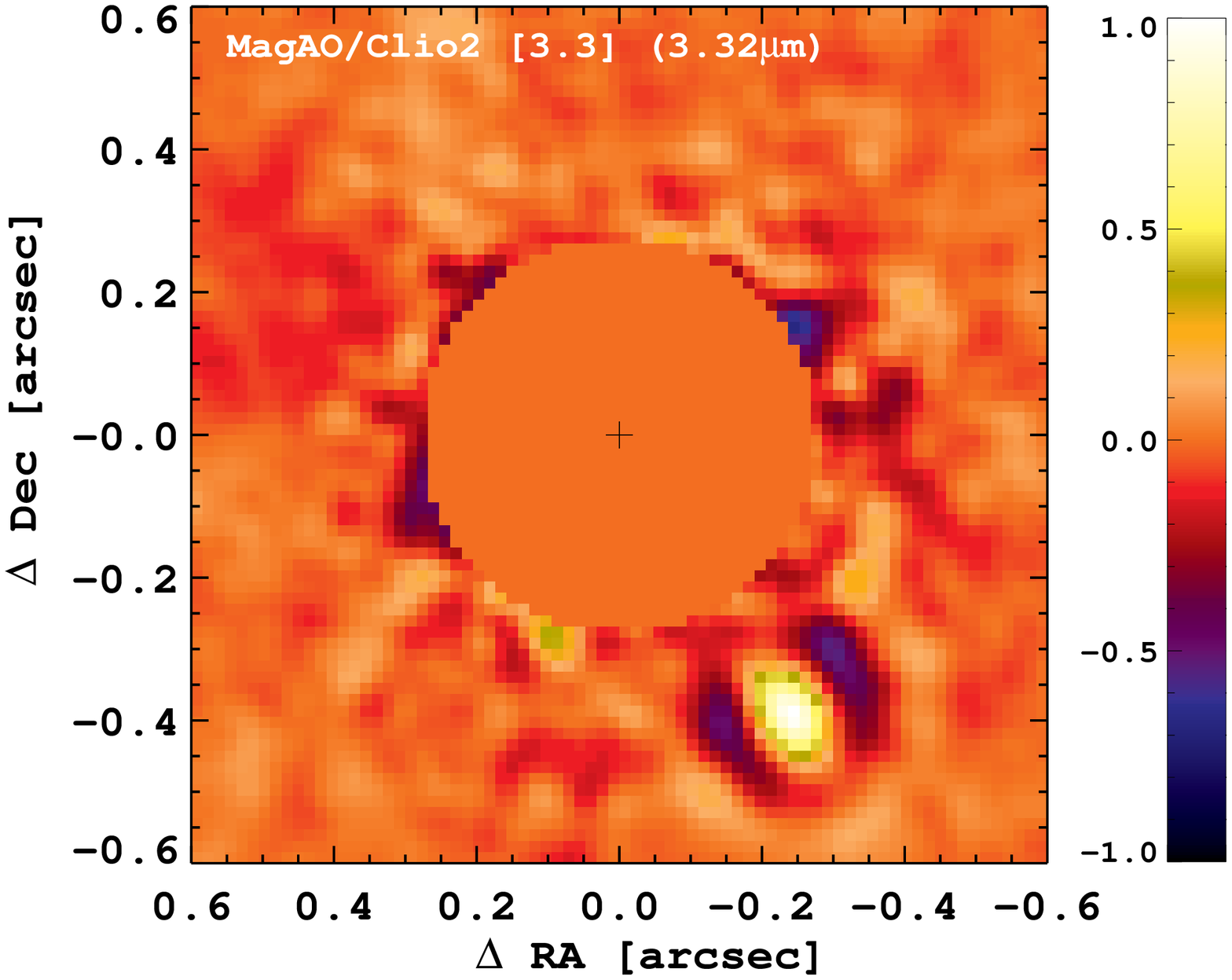}
		\includegraphics[height=0.33\linewidth,angle=90,trim=2.4cm 0.5cm 1.1cm 3.2cm,clip=true]{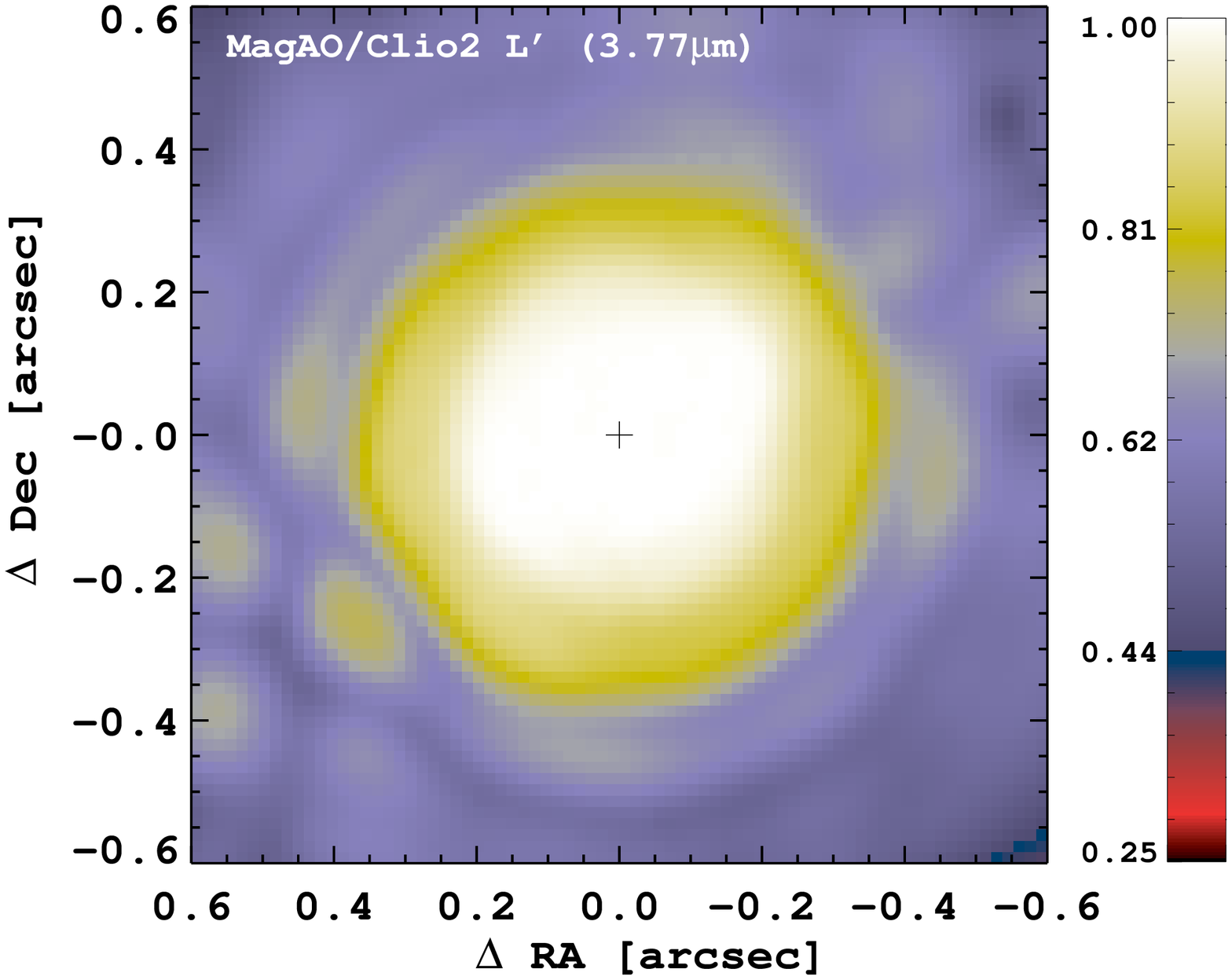}
		\includegraphics[height=0.33\linewidth,angle=90,trim=2.4cm 0.5cm 1.1cm 3.2cm,clip=true]{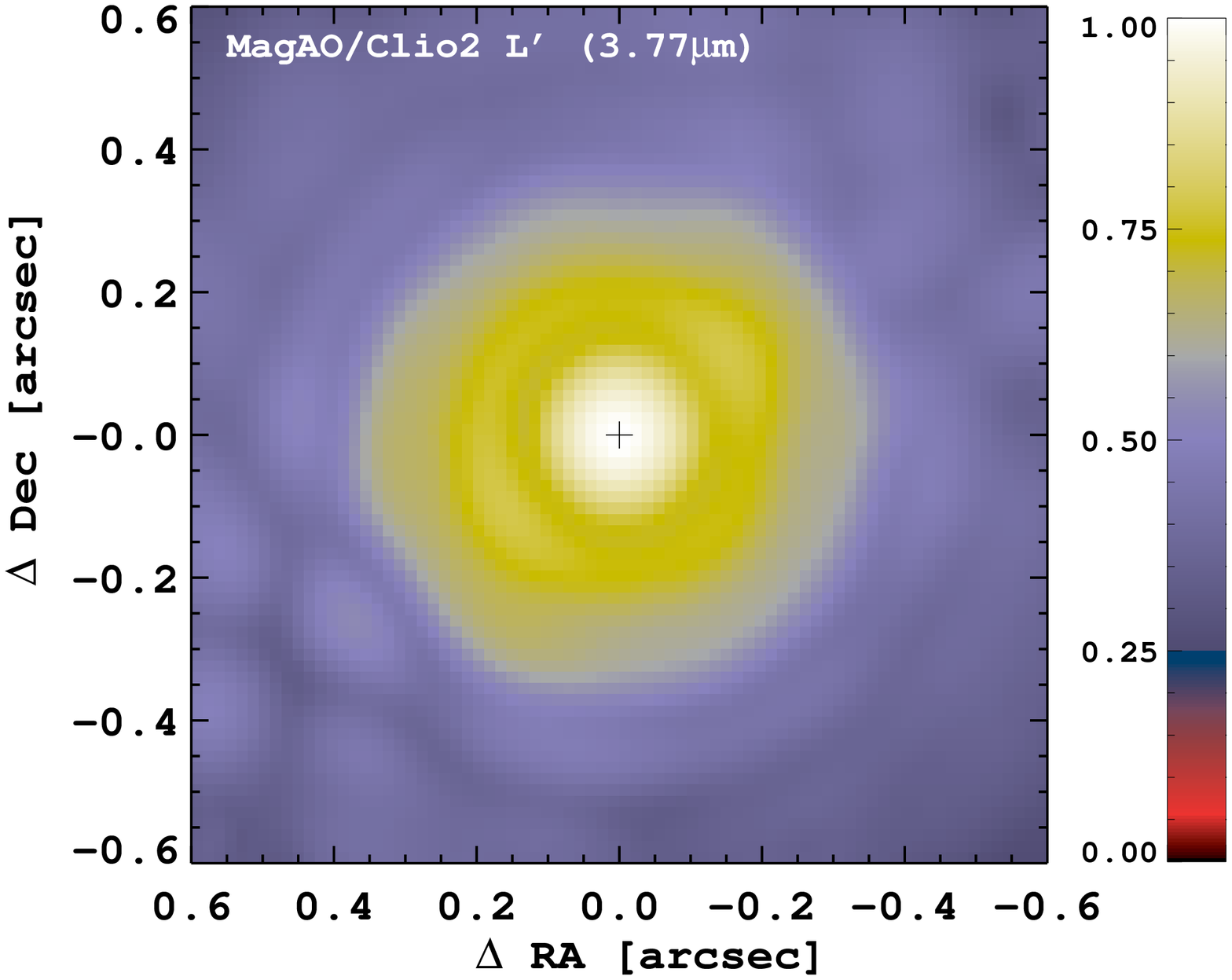}
		\includegraphics[height=0.33\linewidth,angle=90,trim=2.4cm 0.5cm 1.1cm 3.2cm,clip=true]{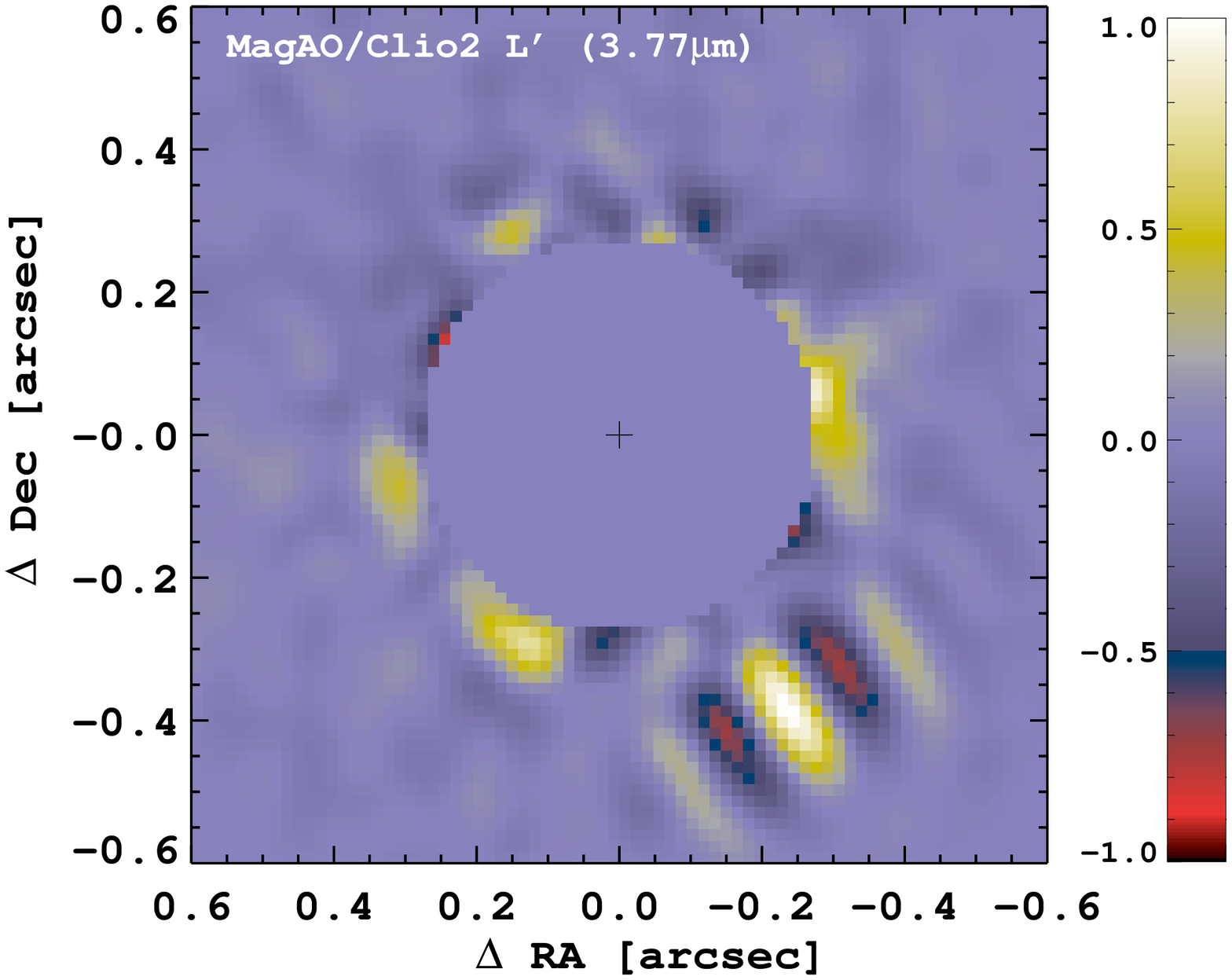}
		\includegraphics[height=0.33\linewidth,angle=90,trim=0.5cm 0.5cm 15.6cm 3.2cm,clip=true]{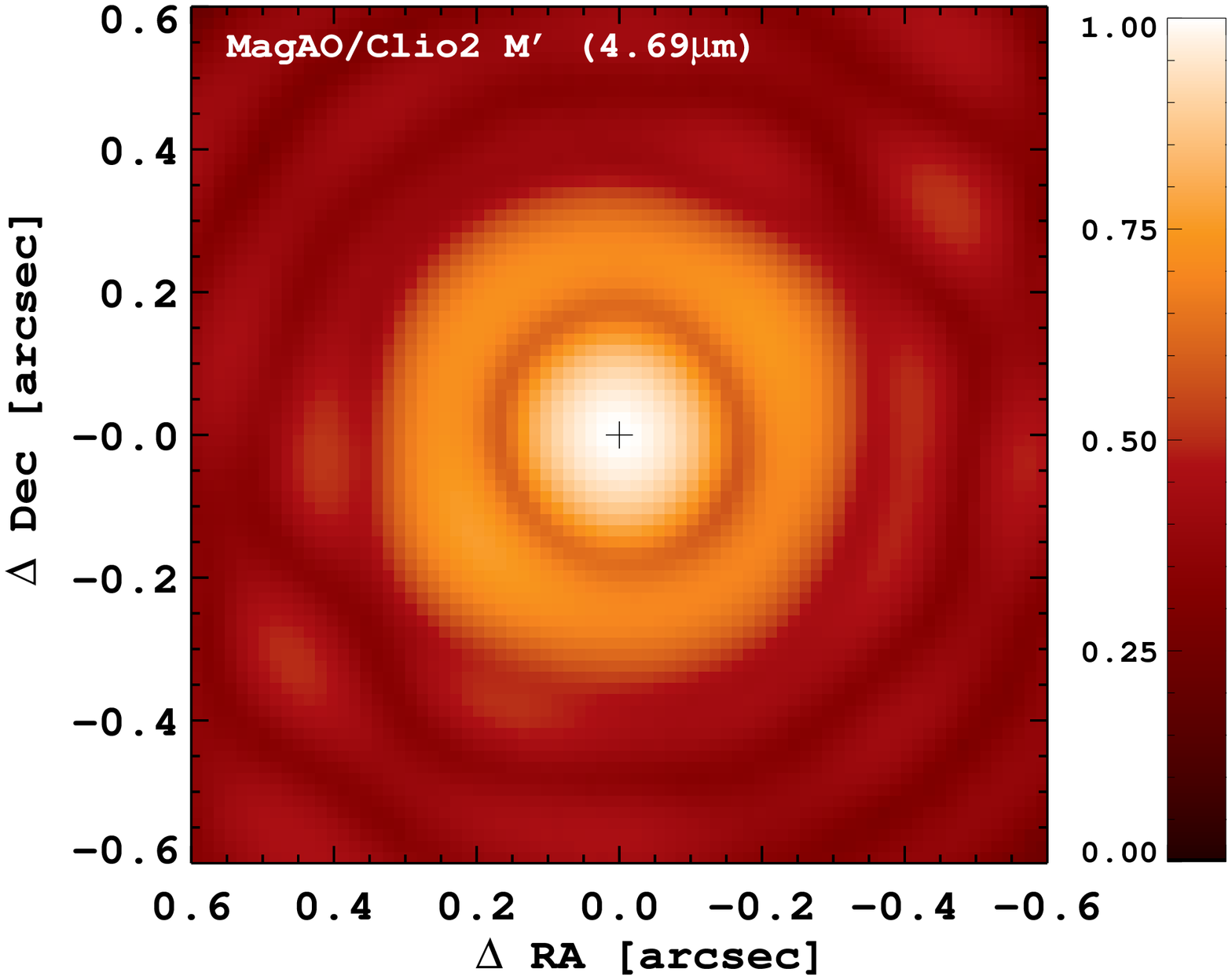}
		\includegraphics[height=0.33\linewidth,angle=90,trim=0.5cm 0.5cm 1.1cm 3.2cm,clip=true]{bpicb_m_image_unsat.eps}
		\includegraphics[height=0.33\linewidth,angle=90,trim=0.5cm 0.5cm 1.1cm 3.2cm,clip=true]{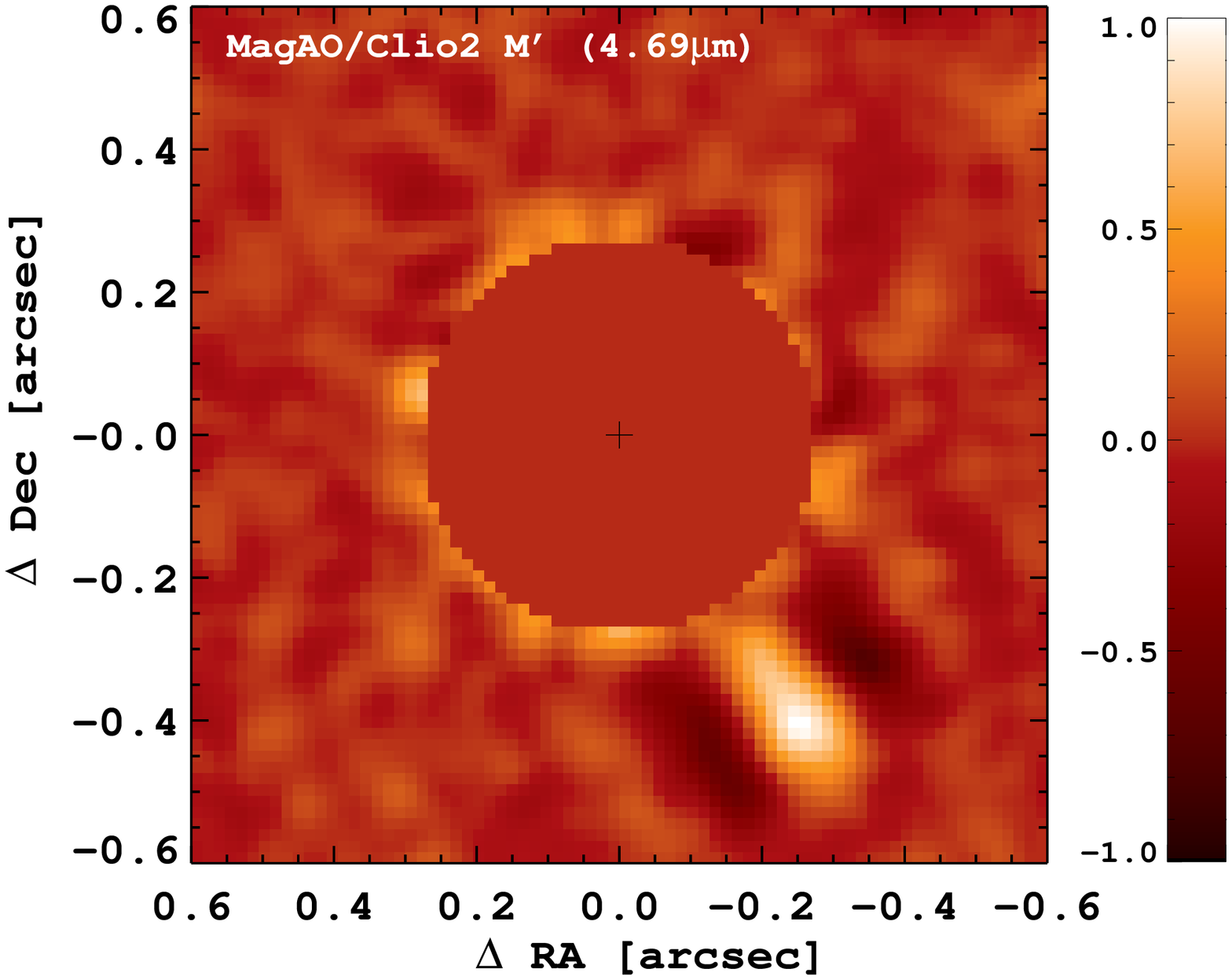}
		\caption{\label{fig:psfsandresids}
			Left: Median saturated PSFs (log scale) --- deep exposures to image the planet;
			Center: Median unsaturated PSFs (log scale) --- short exposures for calibration;
			Right: Final residuals after PSF subtraction, revealing the planet (linear scale).
			From top: $Y_s$, [3.1], [3.3], $L^\prime$, $M^\prime$.
			There is no image at lower left because the $M^\prime$ PSF was never saturated (to avoid sky saturation).
			All images are 1\farcs2 on a side and oriented North up, East left.
			The cross denotes the astrometric center of each image as determined by the algorithm described in Appendix~\ref{sec:app_cent}.
			The unsaturated PSFs were used to simulate the planet in the grid search.
			(VisAO's $Y_s$ coronagraphic images were first presented in Paper I.)
		}
	\end{flushright}
\end{figure*}

\paragraph{$[3.3]$}
On UT 2012 Dec.\ 1 (local night of 2012 Nov.\ 30--Dec.\ 1), we used the 3.3~$\mu$m filter and the Narrow camera (8\arcsec $\times$ 16\arcsec\ field of view).  Unsaturated exposures were 43~ms and taken in the ``stamp'' mode (400 x 200-pixel subarray), and deep exposures were 280~ms taken in full frame mode (1024 x 512 pixels) with no co-adding.  We obtained 1081 frames total, nodding from side to side with the star always on the chip.  To trouble-shoot vibrations, we would occasionally shut off the solid nitrogen pump inside Clio, which was on the Nasmyth platform at the time.  However, these occasions did not affect the detector temperature: it stayed within 1 K of the set point at 55~K. (The solid nitrogen pump is now located in the basement of the telescope, in order to avoid any vibration issues.)
Furthermore, the peak flux of the unsaturated star images (taken four times, about every 45 minutes over the course of the 2.5 hour observation) have a standard deviation of 4\% that is dominated by Strehl/sky/flat variation within a sequence rather than over time or temperature variation.
The point-spread function (PSF) was extremely stable in the core and inner halo.  The final median saturated and unsaturated PSFs are shown in in Fig.~\ref{fig:psfsandresids}.

\paragraph{$L^\prime$}
On UT 2012 Dec.\ 2, we observed in the $L^\prime$ filter and the Narrow camera.  Unsaturated exposures were 43 ms and taken in stamp mode, and deep exposures were 600 ms taken in full frame mode with no co-adding.  We obtained 641 frames total, nodding from side to side with the star always on the chip.  37 of the frames were bad, often at the start of a new nod position when the loop was not yet closed again.  The PSF was extremely stable in the core and inner halo.

\paragraph{$M^\prime$}
On UT 2012 Dec.\ 4, we observed $\beta$ Pic in $M^\prime$.  $M^\prime$ and $Y_s$ were obtained simultaneously (see Paper I).  We nodded $\sim$8\arcsec\ across the chip by hand, taking 4 images per nod position.  This was so that we could manually start an integration when the AO loop re-closed.
All images were obtained in ``strip'' mode (1024 x 300-pixel subarray) at 400 ms integration with no co-adding --- the star was not saturated at this integration time.  We obtained 556 images with the Narrow camera (highly over-sampled, to avoid saturation and to better smooth over bad pixels).  The PSF was extremely stable in the core and inner halo.

\paragraph{$[3.1]$}
On UT 2012 Dec.\ 7 we switched to the Wide camera (14\arcsec $\times$ 28\arcsec\ field of view) to attempt to simultaneously image the disk in the ice line at 3.1~$\mu$m.  Unsaturated exposures were 280~ms and taken in full frame, and deep exposures were 2.5--3~s, also in full frame, all with no co-adding.  We obtained 602 frames total.  However, the seeing was poor and worsening, the PSF was less stable, and the saturation radius is angularly larger in the Wide camera (see the extreme saturation in Fig.~\ref{fig:psfsandresids}), all reducing the quality of these data.  Finally, we discovered later that the Wide camera was out of focus during the first commissioning run, affecting these data.

\subsection{Data}

\subsubsection{Data Reduction}
The data were reduced using custom IDL routines to carry out the following steps: linearity correction; bad pixel correction; sky subtraction via nod pairs; distortion correction; coarse registration via a star-finding algorithm; and fine centroiding via a custom subpixel centering algorithm to $<$1 mas precision for each wavelength (see Tab.~\ref{tab:astrom_err}).
Further details on these steps are given in Appendices \ref{sec:app_photom} (photometric calibration of Clio), \ref{sec:app_astrom} (astrometric calibration of Clio), and \ref{sec:app_cent} (subpixel centroiding of the Clio PSF).
Note that non-linearity was only significant for the $[3.1]$ Wide-camera data, where the linearity correction had the effect of increasing the peak flux of the unsaturated PSF by $\sim$5\%.

\subsubsection{PSF Estimation}
Clio images were taken without a coronagraph.  The PSF is extremely stable at high Strehls and thus excellent high-contrast performance can be obtained by estimating and subtracting out the PSF \citep{morzinski2011}.
We observed with the instrument rotator fixed (maintaining a constant pupil orientation) in order to allow the data to be reduced using angular differential imaging \citetext{ADI, \citealt{marois2006adi}}.
The median unsaturated PSFs are shown in Fig.~\ref{fig:psfsandresids}.

Algorithms for PSF estimation have grown more sophisticated in recent years.
We use the Karhunen-Lo{\`e}ve Image Projection (KLIP) algorithm \citep{soummer2012klip,pueyo2015}, a version of principal-component analysis (PCA) in which the PSF is orthogonalized into modes and the best estimation is found for each mode in annular rings of successively larger radii --- our version is described in \citet{skemer2012,skemer2014}.
This processing determines the most relevant PSF to each mode in each image.
The number of KLIP modes was determined experimentally to reveal the planet with minimal self-subtraction: 20 modes for [3.1], 25 modes for [3.3], 25 modes for $L^\prime$, and 20 modes for $M^\prime$; using more modes had a negligible effect.
The planet is clearly visible in the residual images seen in Fig.~\ref{fig:psfsandresids}.
No masking was employed, and negative lobes are seen at the sides of the planet where it is over-fit in each annulus.
This has no effect on the photometry and astrometry, as described in the following section.

\subsubsection{Photometric and Astrometric Measurement}
The PSF subtraction process described above reveals the planet against a background of residual speckle error as seen in Fig.~\ref{fig:psfsandresids} (right-most column).
To determine the contrast ratio and position of the planet, we conduct a grid search using simulated planets.
We model the planet as the unsaturated PSF, scaled.
For the case of $M^\prime$ where none of the images are saturated, each individual image is its own simulated planet.
For all other cases, the simulated planet is scaled from the mean of the unsaturated PSFs.
We scale the model planet to a particular flux and position it at a particular location;
we then subtract this simulated planet from each reduced frame before running KLIP; finally, we run KLIP to determine the residuals with the model planet subtracted out.
The modeled flux and position of the planet are iterated via a grid search, until we find the minimum KLIP residuals with the best-fit planet subtracted out (described in further detail in Appendix~\ref{sec:app_grid}).

\paragraph{Photometric Error Budget}
The measurement error for the photometry of the planet is given by the signal-to-noise ratio (SNR) of the planet in the residual images.
We calculate the SNR of the planet by determining the
Gaussian-smoothed peak height over the standard deviation in an annulus 2$\times$FWHM wide, centered on the final planet location.
We also verify this via inspection in the grid-search residual images, in that a two-sigma change in modeled flux is enough to make the too-bright or too-faint simulated planet visible against the noise background of the grid-search KLIP residuals.
The next photometric error term is the uncertainty in our measurement of the star's flux, used for calibration of the final contrast ratio.  We attribute this uncertainty to Strehl variation, sky-subtraction errors, telluric transmission, and flatfielding errors, as measured on the unsaturated images which were interspersed throughout the saturated images.  We measure this term as the standard deviation in the peak height of the set of unsaturated PSFs, divided by the square root of the number of unsaturated PSFs.
The number of unsaturated images is N = 81, 65, 36, and 534, for [3.1], [3.3], $L^\prime$, and $M^\prime$, respectively.
Although there is no Strehl variation error in the case of $M^\prime$ (because all the images were unsaturated; therefore, the simulated planet is always an exact copy of the PSF in each image), we include this error for $M^\prime$ to encompass the flat-field variation error (see Appendix~\ref{sec:app_photom}).
The photometry error budget is determined by combining the SNR-based error in the photometry with the calibration error due to Strehl/sky/flat variation.
The measured photometry, given as contrast ratios in each bandpass with respect to the star, is given in Tab.~\ref{tab:phot_err}.

\begin{table}[htb]
	\caption{
		Photometry error budget and final adopted contrast.  All units in magnitudes.
	}
	\label{tab:phot_err}
	\centering
	\begin{tabular}{llllll}
		\hline
		\hline
& Contrast	& Planet	& Star	& Total	& \textbf{Final}	\\
& meas.		& meas.	& cal. & meas.	& \textbf{meas.}	\\
\textbf{Filter} & [$\Delta$mag] & error$^\star$ & error$^*$ & error & \textbf{$\Delta$mag} \\
\hline
$\mathbf{[3.1]}$	& 8.14	& 0.2410	& 0.0104	& 0.2412	& $\mathbf{8.14\pm0.24}$	\\
$\mathbf{[3.3]}$	& 8.34	& 0.1256	& 0.0057	& 0.1257	& $\mathbf{8.34\pm0.13}$	\\
$\mathbf{L^\prime}$	& 8.02	& 0.1116	& 0.0060	& 0.1118	& $\mathbf{8.02\pm0.11}$	\\
$\mathbf{M^\prime}$	& 7.45	& 0.0587	& 0.0035	& 0.0588	& $\mathbf{7.45\pm0.06}$	\\
\hline
\hline
\multicolumn{6}{l}{}	\\ 
\multicolumn{6}{p{0.95\linewidth}}{$^\star$Planet measurement error based on SNR in the KLIP residuals.  Details are given in Tab.~\ref{tab:photmeaserr}.}	\\
\multicolumn{6}{p{0.95\linewidth}}{$^*$Star calibration error due to Strehl/sky/flat error based on standard deviation in peak flux of unsaturated images throughout the observation sequence, divided by the square root of the number of unsaturated images.}	\\
	\end{tabular}
\end{table}

\paragraph{Astrometric Error Budget}
The mean astrometry for the Clio data were summarized in Paper I and \citet{nielsen2014}.
Here we give the details, shown in Tab.~\ref{tab:astrom_err}.
We find the best separation and PA according to the best parabola fits in the grid search (see Appendix~\ref{sec:app_grid}).
We find the errors in separation and PA using the Gaussian fits to the grid search for each filter separately, based on assuming random errors.
We propagate the errors in platescale to errors in astrometry by multiplying astrometric separation
platescale error over the platescale.
Then we combine the 4 filters as follows:
The best-fit separation and PA are the weighted mean of all the measurements, as they are independent measurements.
We find the total measurement error in the star-planet separation by adding the errors in quadrature for each filter and camera.
We calculate the measurement error in finding the platescale (see Appendix~\ref{sec:app_astrom}) for each filter and camera by a simple mean.
Finally, we determine the total astrometric error by combining the star-planet separation errors in a weighted-mean-error, added in quadrature to the platescale errors.
Our final determination of the position of the planet in Dec.\ 2012 is a separation of \textbf{461$\pm$14 mas} and a position angle of \textbf{211.9$\pm$1.2 deg}.

\begin{table*}[htb]
	\footnotesize
	\caption{
		$[3.1]$, $[3.3]$, $L^\prime$, and $M^\prime$ astrometry error budget for UT (average) 2012 December 4.
	}	 \label{tab:astrom_err}
	\centering
	\begin{tabular}{lllllllllllll}
		\hline
		\hline
		&				&				&			& Planet	& Star		& Star-planet			& Platescale			& Platescale			& 	& Planet		& NorthClio			& NorthClio			\\
		&				& UT epoch		& Best		& meas.	& centroid	& total meas.			& meas.				& fiducial			& Best	& meas.		& meas.				& fiducial			\\
		&				&			& sep.$^\star$	& error		& error		& error$^{\star \ddagger}$	& error$^{\dagger1}$	& error$^{\dagger2}$	& PA		& error$^\star$	& error$^{\dagger1}$	& error$^{\dagger2}$	\\
		Camera	& Filter			&		& /mas			& /mas		& /mas		& /mas					& /mas 				& /mas				& /deg.		& /deg.			& /deg.				& /deg.				\\
		\hline
		\hline
		Wide	& $[3.1]$		& 2012 Dec 07	& 466.1			& 74.0		& 0.14		& 74.0001				& 1.20				& 0.80				& 212.62	& 6.26			& 0.16				& 0.3				\\
		Narrow	& $[3.3]$		& 2012 Dec 01	& 463.0			& 19.9		& 0.08		& 19.9002				& 1.26				& 1.37				& 211.92	& 1.65			& 0.16				& 0.3		 		\\
		Narrow	& $L^\prime$	& 2012 Dec 02	& 456.9			& 28.0		& 0.08		& 28.0001				& 1.24				& 1.36				& 211.65	& 2.31			& 0.16				& 0.3				\\
		Narrow	& $M^\prime$	& 2012 Dec 04	& 460.6			& 33.8		& 0.08		& 33.8001				& 1.25				& 1.37				& 212.10	& 2.84			& 0.16				& 0.3				\\
		\hline
		\multicolumn{3}{l}{Weighted mean$^\star$} & 461.1		&			&			&						&					&					& 211.91	&				& 					&					\\
		\multicolumn{3}{l}{Weighted errors$^\star$} &				&			&			& 14.347				&					&					&			& 1.19			& 					&					\\
		\multicolumn{3}{l}{Averaged errors$^{\dagger1,\dagger2}$} & &			&			&						& 1.238				& 1.225				&			&				& 0.16				& 0.3				\\
		\hline
		\multicolumn{3}{l}{\textbf{Combined final$^*$}} & \multicolumn{6}{l}{\textbf{461 $\pm$ 14 mas}}														& \multicolumn{4}{l}{\textbf{211.9 $\pm$ 1.2 deg.}}						\\
		\hline \hline
		\multicolumn{13}{l}{}										\\ 
		\multicolumn{13}{l}{$^\star$Values and errors from measuring the position of the planet for each data set are combined with a weighted mean.} \\
		\multicolumn{13}{l}{$^\ddagger$Planet Gaussian-fit and star centroiding error are added in quadrature to obtain total star-planet measurement error for each data set.} \\
		\multicolumn{13}{l}{$^{\dagger1}$Clio measurement errors of the Trapezium are simply averaged.} \\
		\multicolumn{13}{l}{$^{\dagger2}$Fiducial Trapezium errors from \citet{close2012} are simply averaged.} \\
		\multicolumn{13}{l}{$^*$Errors from previous two lines added in quadrature to get the final astrometry error budget.} \\
		\multicolumn{13}{l}{$^*$Final astrometry measurement also reported in Paper I and \citet{nielsen2014}.} \\
		\multicolumn{13}{l}{NOTE: The final measurement is given in Paper I, their Tab.~2 (p.\ 8), but the Date should have read ``2012 Dec 1--7''.} \\
	\end{tabular}
	\normalsize
\end{table*}

\section{Results}
In the previous section we described how the data were reduced and the flux and position of $\beta$ Pic b were found in our MagAO+Clio images.
In this section, we describe creating a uniform spectral energy distribution (SED) using our new data as well as the literature data.

\subsection{Contrast Measurements in the Literature}
The raw quantity measured in high-contrast photometry is the contrast ratio between the star and planet.
The flux of the planet is derived from the contrast measurement, and must include assumptions about the flux of the star, the instrument system, and the distance.
Different works employ different calibrations for both the apparent magnitude of $\beta$ Pic A and the distance to the system.
Furthermore, there is a lack of consistency in photometric systems when combining different data sets.
Therefore, here we place all available star-planet contrast measurements on a uniform system.

$\beta$ Pic b has been imaged with the following instruments and cameras, as reported in the literature:
VLT UT4/NaCo at $L^\prime$, [4.05], and $M^\prime$ with the L27 camera;
NaCo at $K_s$ with the S27 camera;
NaCo at $J$, $H$, and $K_s$ with the S13 camera;
Gemini South/NICI at CH$_{4s,1\%}$, $H$, $K_s$, and $K_\mathrm{cont}$;
Gemini South/GPI spectra at $J$ and $H$;
Magellan Clay/MagAO+VisAO at $Y_s$;
and MagAO+Clio at [3.1] in the wide camera and [3.3], $L^\prime$, and $M^\prime$ in the narrow camera.
Works with reported photometry of $\beta$ Pic b at the time of this writing are 
\citet{lagrange2009},	
\citet{lagrange2010},	
\citet{quanz2010},	
\citet{bonnefoy2011},	
\citet{currie2011},	
\citet{bonnefoy2013},	
\citet{boccaletti2013},	
\citet{currie2013},	
\citet{absil2013},	
Paper I,	
\citet{bonnefoy2014},	
\citet{chilcote2015},	
and This Work.
We compile the contrast ratios measured in these works in Tab.~\ref{tab:litphotometry}.
For completeness, we note not only the bandpass, but also the telescope, instrument, and camera.

\begin{table*}[htb]
	\small 
	\caption{\label{tab:litphotometry}
		Compilation of all $\beta$ Pic b photometry to date.
		Contrast ratios in parentheses are not included in our SED (see notes).
	}
	\centering
	\begin{tabular}{llllll}
		\hline
		\hline
		Epoch UT [yyyy Mon dd]						& Telescope, Instrument, Camera, Filter		& Delta Mag.	& Duplicate					& Ref.	 & Ref.\ Note \\
		\hline
		2003 Nov 10, 13						& VLT UT4 NaCo L27 $L^\prime$				& 7.7$\pm$0.3		&				& 1		\\
		2009 Oct 25, Nov 24-25, Dec 17, 26, 29	& VLT UT4 NaCo L27 $L^\prime$				& (7.8$\pm$0.3) & $^{\dagger 1}$			& 2		\\
		2010 Apr 03							& VLT UT4 NaCo L27 APP [4.05]				& 7.75$\pm$0.23		&			& 3		\\
		2010 Mar 20, Apr 10						& VLT UT4 NaCo S13, S27 $K_s$				& 9.2$\pm$0.1			&			& 4		\\
		2008 Nov 11							& VLT UT4 NaCo L27 $M^\prime$				& ($\sim$8.02$\pm$0.50)	&			& 5 & 5$^*$	\\
		2009 Dec 29							& VLT UT4 NaCo L27 $L^\prime$				& 7.71$\pm$0.06 & $^{\dagger 1}$		& 5		\\
		2011 Dec 16							& VLT UT4 NaCo S13 $J$						& (10.5$\pm$0.3) & $^{\dagger 2}$		& 6		\\
		2011 Dec 18, 2012 Jan 11					& VLT UT4 NaCo S13 $H$					& (10.0$\pm$0.2) & $^{\dagger 3}$			& 6		\\
		2012 Nov 26							& VLT UT4 NaCo L27 $M^\prime$				& 7.5$\pm$0.2					&	& 6		\\
		2010 Dec 25							& Gemini South NICI	 $K_s$					& ($\sim$8.8$\pm$0.6) & $^{\dagger 4}$	& 7 & 7$^*$	\\
		2011 Dec 16							& VLT UT4 NaCo S13 $J$						& 10.59$\pm$0.21 & $^{\dagger 2}$		& 8 & 8$^*$	\\
		2012 Jan 11					& VLT UT4 NaCo S13 $H$					& 9.83$\pm$0.14 & $^{\dagger 3}$		& 8 & 8$^*$	\\
		2013 Jan 09							& Gemini South NICI	 $H$						& 9.76$\pm$0.18		&			& 8 & 8$^*$	\\
		2013 Jan 09							& Gemini South NICI	 $K_s$					& 9.02$\pm$0.13		&			& 8 & 8$^*$	\\
		2012 Dec 23							& Gemini South NICI	 $[3.09]$					& 8.26$\pm$0.27		&			& 8 & 8$^*$	\\
		2012 Dec 16							& VLT UT4 NaCo L27 $L^\prime$				& 7.79$\pm$0.08		&			& 8 & 8$^*$	\\
		2012 Dec 16							& VLT UT4 NaCo L27 $[4.05]$					& 7.59$\pm$0.08		&			& 8 & 8$^*$	\\
		2012 Dec 15							& VLT UT4 NaCo L27 $M^\prime$				& 7.50$\pm$0.13		&			& 8 & 8$^*$	\\
		2013 Jan 31							& VLT UT4 NaCo L27 AGPM $L^\prime$			& 8.01$\pm$0.16		&	 		& 9		\\
		2012 Dec 04							& Magellan II MagAO VisAO $Y_s$				& 11.97$\bfrac{+0.34}{-0.33}$	&		& 10		\\
		2011 Oct 20							& Gemini South NICI	 $[$CH$_\mathrm{4s,1\%}]$	& 9.65$\pm$0.14		&			& 10		\\
		2010 Dec 25							& Gemini South NICI	 $K_s$					& 8.92$\pm$0.13 & $^{\dagger 4}$		& 10		\\
		2011 Oct 20							& Gemini South NICI	 $K_\mathrm{cont}$			& 8.23$\pm$0.14		&			& 10		\\
		2013 Dec 10							& Gemini South GPI IFS $J$					& Spectrum			&			& 11 & 11$^*$	\\
		2013 Nov 18, Dec 10					& Gemini South GPI IFS $H$					& Spectrum			 &			& 12 & 12$^*$	\\
		2012 Dec 07							& Magellan II MagAO Clio Wide $[3.1]$			& 8.14$\pm$0.24		&			& 13		\\
		2012 Dec 01							& Magellan II MagAO Clio Narrow $[3.3]$		& 8.34$\pm$0.13		&			& 13		\\
		2012 Dec 02							& Magellan II MagAO Clio Narrow $L^\prime$		& 8.02$\pm$0.11		&			& 13		\\
		2012 Dec 04							& Magellan II MagAO Clio Narrow $M^\prime$	& 7.45$\pm$0.06		&			& 13		\\
		\hline
		\hline
		\multicolumn{6}{l}{}	\\ 
		\multicolumn{6}{p{\linewidth}}{
			References:
			[1] \citet{lagrange2009};	
			[2] \citet{lagrange2010};	
			[3] \citet{quanz2010};	
			[4] \citet{bonnefoy2011};	
			[5] \citet{currie2011};	
			[6] \citet{bonnefoy2013};	
			[7] \citet{boccaletti2013};	
			[8] Personal communication (Currie 2014) and \citet{currie2013};	
			[9] \citet{absil2013};	
			[10] Paper I;	
			[11] \citet{bonnefoy2014};	
			[12] \citet{chilcote2015};	
			[13] This work.
		}	\\
		\hline
		\multicolumn{6}{p{\linewidth}}{$^\dagger$Duplications:
		In cases where the same data are re-reduced, we use the more recent result, which in all cases has smaller error bars:
		$^{\dagger 1}$While \citet{lagrange2010} present $L^\prime$ NaCo data from the various epochs listed above, the final quoted value in the text of the paper is that of 2009 Dec 29, which is the data set re-reduced in \citet{currie2011}.
		$^{\dagger 2}$\citet{currie2013} present a re-reduction of the 2011 Dec 16 $J$ NaCo point first presented in \citet{bonnefoy2013}.
		$^{\dagger 3}$While \citet{bonnefoy2013} present $H$ NaCo data from two epochs listed above, the final quoted value in the text of the paper is consistent with the RADI photometry of 2012 Jan 11 only, which is the data set re-reduced in \citet{currie2013}.
		$^{\dagger 4}$Paper I presents an independent reduction of the 2010 Dec 25 $K_s$ NICI point also presented in \citet{boccaletti2013}  (without a quoted value in the text).
		}\\
		\hline
		\multicolumn{6}{p{\linewidth}}{
		$^*$Notes on the References:
		5$^*$ Not used due to lack of unsaturated calibration data.
		7$^*$ The contrast we give here is an average of values in a table in the paper.
		8$^*$ Contrast values were determined via personal communication (Currie, 2014).
		11$^*$--12$^*$ Each author sent us their calibrated spectrum in table format (personal communications, Bonnefoy 2014 and Chilcote 2014).
		In order to use the valuable information in the shape of the spectrum, and to maintain our self-consistent photometric system, we independently normalize the spectra to the broad-band NaCo photomety of \citet{currie2013} for $J$ and $H$ (10.59$\pm$0.21 and 9.83$\pm$0.14, respectively).
		}	\\
	\end{tabular}
	\normalsize
\end{table*}

In more detail, to create Tab.~\ref{tab:litphotometry} we first compile the contrast ratios by inspecting each paper; when multiple values for the contrast ratio are given for a single set of data, we take the value quoted in the text.
However, the following caveats apply:
\begin{enumerate} \itemsep -3pt
\item \citet{currie2011} quote an approximate ($\approx$) contrast ratio in $M^\prime$, but no unsaturated calibration data were obtained concurrent with the observations.  Therefore, we do not use this value.
\item \citet{currie2013} give the apparent magnitudes of b but do not give the contrast ratios they measured; the apparent magnitudes used for A were determined via personal communication (Currie 2014) and used to calculate the contrast ratios.
\item \citet{boccaletti2013} report a number of contrast ratios measured in various trials in a table, but they do not quote any value in the text, as the mask transmission was not known.  Therefore, we do not use their photometry.
\item \citet{bonnefoy2014} and \citet{chilcote2015} report GPI spectra.  The focus of these papers is the shape of the spectrum, and per-channel-contrast-ratios have correlated errors that are difficult to compare directly to the photometric measurements.  Therefore, we use these spectra as a check on our model fits but do not combine them with the photometry when calculating the best-fitting model.
\item In order to enforce independence of the photometric measurements, duplicates are avoided by using the most recent value in all cases of re-reductions of the same data set (see details in Tab.~\ref{tab:litphotometry}).
\end{enumerate}

\subsection{Photometric System}
\begin{figure}[htb]
\centering
	\includegraphics[height=\linewidth,angle=90,trim=0.5cm 0.4cm 0.9cm 0.65cm,clip=true]{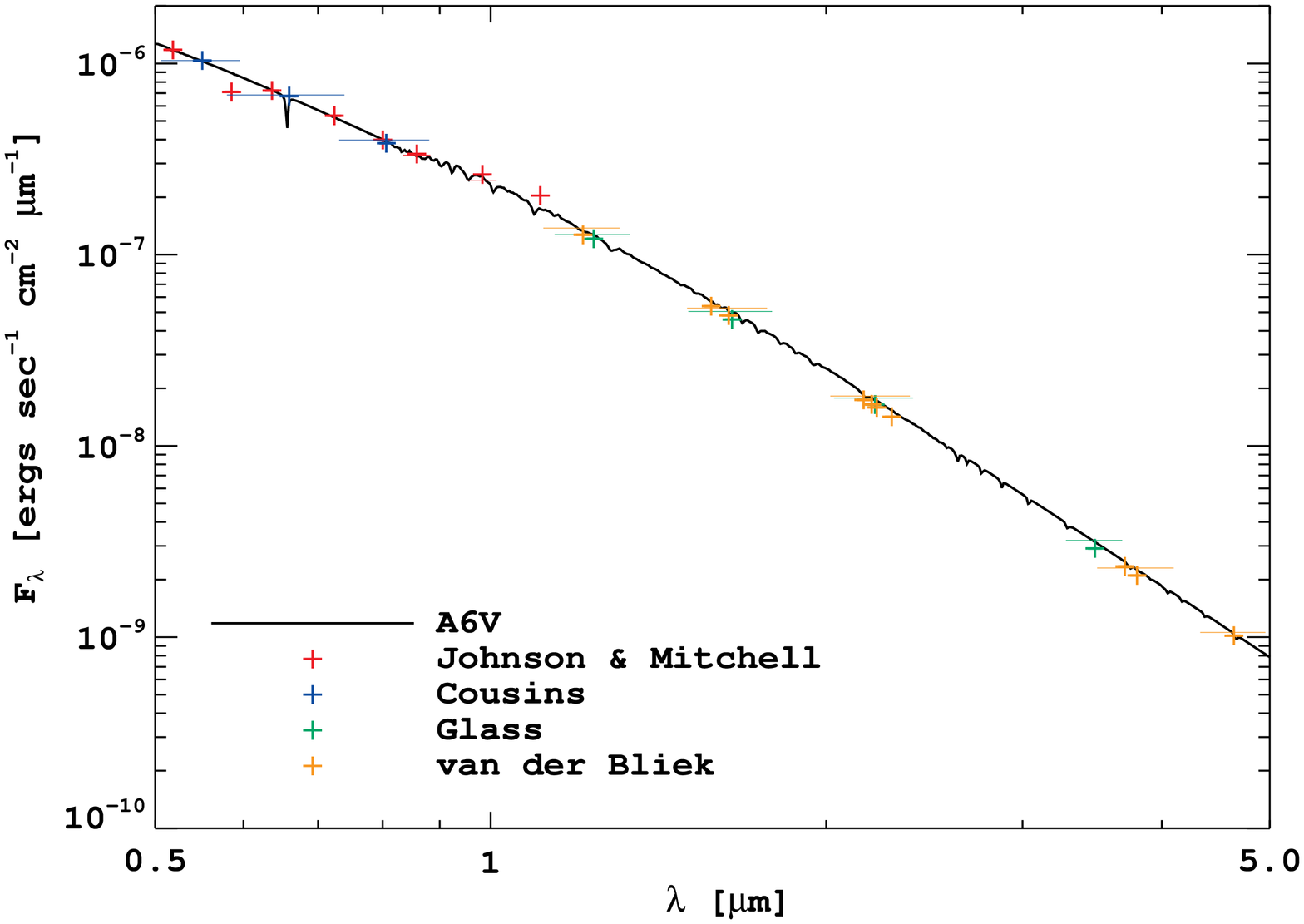}
	\caption{\label{fig:bpica}
	Spectral energy distribution of $\beta$ Pic A.
	The black line is a CalSpec standard A6V spectrum, whereas the points represent photometric measurements of $\beta$ Pic A itself by the following authors: \citet{1975RMxAA...1..299J}, \citet{1980SAAOC...1..234C}, \cite{1974MNSSA..33...53G}, and \cite{1996A&AS..119..547V}.
	}
\end{figure}

We calculate the resultant flux on our uniform photometric system, via a method similar to that used in Paper I, as follows.
First to determine the flux of $\beta$ Pic A (given as an A6V star in \citet{gray2006}), we interpolate an \textit{HST} CalSpec spectrum\footnote{\url{http://www.stsci.edu/hst/observatory/crds/calspec.html}} of the A6V star TYC 4207-219-1 (which is noted to be variable at the $\pm$1\% level \citep{pancino2012}). 
As a check on the applicability of this spectrum we normalized this to the $V$-band photometry from \citet{1980SAAOC...1..166C} and \citet{1980SAAOC...1..234C}, and compared it to the photometry from \cite{1969CoLPL...8....1M}, \citet{1975RMxAA...1..299J}, \cite{1974MNSSA..33...53G}, and \cite{1996A&AS..119..547V}.
This comparison is shown in Fig.~\ref{fig:bpica}.
The A6V spectrum is a good match to the shape of the measured SED to at least 5~$\mu$m.
This demonstrates that, within the 0.5--5~$\mu$m regime, there is no significant attenuation of the starlight of $\beta$ Pic A, neither from interstellar extinction (at a distance of 19~pc) nor from its edge-on disk.
Due to this good match, we also conclude that the 10-mmag-variability of the CalSpec star is negligible, and that $\beta$ Pic A itself lacks significant variability.

The filter curve for each instrument was obtained from its respective website.  We then multiplied these curves by an atmosphere appropriate for the site.  For both MagAO and Gemini/NICI we use the transmission provided by Gemini for Cerro Pachon\footnote{\url{http://www.gemini.edu/?q=node/10789}} based on the ATRAN code \citep{1992NASATM}.  For VLT/NACO we used the ``Paranal-like'' atmosphere available on the ESO website\footnote{\url{http://www.eso.org/sci/facilities/eelt/science/drm/tech\_data/ data/atm\_abs/}}.  We perform all calculations at airmass 1.0 since it has only a minor effect.  The key parameter is precipitable water vapor (PWV), which is not typically documented for these observations.  It is not currently routinely measured at LCO.  In Tab.~\ref{tab:bpicbII_photsys} we list the PWV assumed for each observation.  In the case of Paranal this is the only option provided.  For LCO we used a slightly higher value based on historical measurements \citep{2010SPIE.7733E..4NT} and forecasts made for nearby La Silla Observatory.  As shown in the table, in some filters including the atmosphere in the transmission changes the zero-magnitude flux in that filter by $>$$2\%$, which is a systematic bias.  While this is smaller than the typical precision in any single high-contrast imaging measurement, there are now enough measurements on this object in some filters that combined precision could be approaching this level if we controlled all systematic effects.
Based on this exercise we advocate that observers pay greater attention to atmopheric conditions at the time of such observations, and that capabilities to make coincident PWV measurements should be standard.

\begin{table*}[ht]
	\footnotesize
	\caption{\footnotesize Uniform photometric system and flux of $\beta$ Pic A. \label{tab:bpicbII_photsys}}
	\begin{tabular}{cccccccccccccc}
		\hline
		\hline
		&                &               &        &      & \multicolumn{2}{c}{No Atm.} & \multicolumn{2}{c}{With Atm.} & & & \multicolumn{2}{c}{$\beta$ Pic A}   \\   
		Filter     &  Inst.\    &   Atm.\  &  A.M.\tablenotemark{$\dagger$}  & PWV\tablenotemark{$\ddagger$}  & $\lambda_0$ & $F_\lambda(0)$\tablenotemark{*} & $\lambda_0$ & $F_\lambda(0)$\tablenotemark{*} & Base  & Color   &  mag   & $F_\lambda$\tablenotemark{*}  \\
		&                &               &        & /mm & /$\mu$m & [$\times10^{-8}$] & /$\mu$m & ($\times10^{-8}$) & Filt.\tablenotemark{$\star$}  &&& ($\times10^{-9}$)\\
		\hline
		$Y_s$ & VisAO & C.P. & 1.00 & 4.3 & 0.983 & 679.0 & 0.985 & 674.0 & $99_{13col}$ & 0.000 & $3.561\pm0.035$ & $253.7\pm9.6$ \\
		$J$ & NaCo & Paranal & 1.00 & 2.3 & 1.270 & 292.1 & 1.255 & 302.3 & $J_{ESO}$ & -0.017 & $3.564\pm0.013$ & $113.5\pm1.6$ \\
		$CH_4$ & NICI & C.P. & 1.00 & 2.3 & 1.584 & 127.9 & 1.584 & 127.9 & $H_{ESO}$ & 0.023 & $3.525\pm0.007$ & $49.74\pm0.38$ \\
		$H$ & NaCo & Paranal & 1.00 & 2.3 & 1.665 & 112.5 & 1.655 & 114.5 & $H_{ESO}$ & -0.003 & $3.499\pm0.007$ & $45.62\pm0.34$ \\
		$H$ & NICI & C.P. & 1.00 & 2.3 & 1.665 & 113.9 & 1.658 & 114.6 & $H_{ESO}$ & -0.003 & $3.499\pm0.007$ & $45.69\pm0.35$ \\
		$K_S$ & NaCo & Paranal & 1.00 & 2.3 & 2.150 & 44.59 & 2.159 & 43.81 & $K_{ESO}$ & -0.033 & $3.462\pm0.008$ & $18.07\pm0.16$ \\
		$K_S$ & NICI & C.P. & 1.00 & 2.3 & 2.174 & 42.68 & 2.175 & 42.48 & $K_{ESO}$ & -0.025 & $3.470\pm0.008$ & $17.39\pm0.15$ \\
		$K_{cont}$ & NICI & C.P. & 1.00 & 2.3 & 2.272 & 35.64 & 2.272 & 35.64 & $K_{ESO}$ & 0.068 & $3.563\pm0.008$ & $13.39\pm0.12$ \\
		$3.09 \mu$m & NICI & C.P. & 1.00 & 2.3 & 3.085 & 11.22 & 3.090 & 11.18 & $L'_{ESO}$ & 0.002 & $3.483\pm0.010$ & $4.518\pm0.049$ \\
		$3.1 \mu$m & Clio & C.P. & 1.00 & 4.3 & 3.096 & 11.11 & 3.102 & 11.04 & $L'_{ESO}$ & 0.002 & $3.483\pm0.010$ & $4.464\pm0.048$ \\
		$3.3 \mu$m & Clio & C.P. & 1.00 & 2.3 & 3.329 & 8.543 & 3.345 & 8.406 & $L'_{ESO}$ & 0.003 & $3.484\pm0.010$ & $3.395\pm0.037$ \\
		$L'$ & Clio & C.P. & 1.00 & 4.3 & 3.785 & 5.289 & 3.761 & 5.408 & $L'_{ESO}$ & 0.000 & $3.481\pm0.010$ & $2.190\pm0.024$ \\
		$L'$ & NaCo & Paranal & 1.00 & 2.3 & 3.814 & 5.103 & 3.813 & 5.106 & $L'_{ESO}$ & -0.000 & $3.481\pm0.010$ & $2.069\pm0.022$ \\
		$4.05\mu$m & NaCo & Paranal & 1.00 & 2.3 & 4.056 & 3.884 & 4.056 & 3.883 & $L'_{ESO}$ & -0.008 & $3.473\pm0.010$ & $1.585\pm0.017$ \\
		$M'$ & Clio & C.P. & 1.00 & 7.6 & 4.687 & 2.253 & 4.687 & 2.250 & $M_{ESO}$ & -0.002 & $3.420\pm0.017$ & $0.9643\pm0.0177$ \\
		$M'$ & NaCo & Paranal & 1.00 & 2.3 & 4.789 & 2.105 & 4.818 & 2.048 & $M_{ESO}$ & -0.001 & $3.421\pm0.017$ & $0.8766\pm0.0161$ \\
		\hline
		\hline
		\multicolumn{12}{l}{} \\ 
		\multicolumn{12}{l}{\tablenotemark{$\dagger$} We perform all calculations at airmass 1.0 since AM has only a minor effect compared to PWV.} \\ 
		\multicolumn{12}{l}{\tablenotemark{$\ddagger$} The key parameter is precipitable water vapor (PWV), which is not typically documented for these observations.} \\ 
		\multicolumn{12}{l}{\tablenotemark{*} Flux densities in ergs s$^{-1}$ cm$^{-2}$ $\mu$m$^{-1}$.} \\ 
		\multicolumn{12}{l}{\tablenotemark{$\star$} Measurement to which the template spectrum color was applied.} \\
	\end{tabular}
	\normalsize
\end{table*}

We then use the Vega spectrum of \citet{2007ASPC..364..315B} to determine the zero-magnitude flux density in each filter.  We use the A6V spectrum to calculate corrections from the measured photometry of $\beta$ Pic A to the filters used for imaging the planet, from which we derive flux density of the star.  We show these in Tab.~\ref{tab:bpicbII_photsys}, along with the primary measurement on which each was based.

\subsection{Resultant Flux with Uniform Systematics}
\begin{figure}[htb]
	\centering
	\includegraphics[height=\linewidth,angle=90,trim=0.5cm 1cm 0.9cm 0.65cm,clip=true]{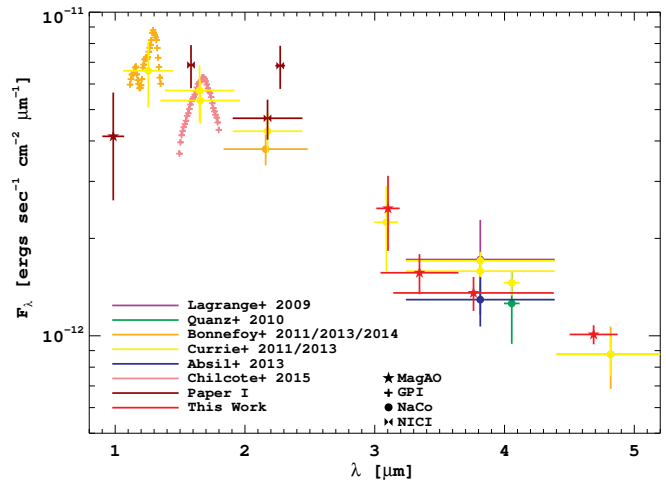}
	\caption{\label{fig:sed}
		Spectral energy distribution of $\beta$ Pic b as of this work.
		The literature data have been re-calibrated to be on a uniform photometric system;
		see Tab.~\ref{tab:litphotometry} for the sources.
		Horizontal error bars are flux-weighted effective filter widths;
		vertical error bars are 1-sigma photometric errors (listed in Tab.~\ref{tab:all_phot}).
	}
\end{figure}

Next we combine the photometric system with the contrast measurements listed in Tab.~\ref{tab:litphotometry} to determine the planet's flux density in each filter.
The absolute magnitudes of $\beta$ Pic b are found by subtracting the contrast ratio, then applying the distance modulus, using the distance 19.44$\pm$0.05 pc \citep{vanleeuwen2007}.
We now have the 0.99 to 4.8 $\mu$m SED of $\beta$ Pic b based on a consistent SED for $\beta$ Pic A.
Tab.~\ref{tab:all_phot} lists the resultant fluxes we determine for the planet.
The final resultant SED of $\beta$ Pic b is shown in Fig.~\ref{fig:sed}.

\begin{table*}[htb]
	\caption{
	Our determination of the bandpasses and flux densities of the photometric measurements of $\beta$ Pic b to date.
	}
	\label{tab:all_phot}
	\centering
	\begin{tabular}{lllll}
		\hline
		\hline
		Instrument \& Filter & Epoch & $\lambda_0$/$\mu$m\tablenotemark{$\star$} & $\Delta\lambda$/$\mu$m\tablenotemark{$*$} & $F_\lambda$ ($\times 10^{-12}$)\tablenotemark{$\dagger$} \\
		\hline
		MagAO/VisAO $Y_s$		& 2012 Dec 04			& 0.985	& 0.084	& $4.1 \pm 1.5$		\\
		NaCo $J$				& 2011 Dec 16			& 1.255	& 0.193	& $6.6 \pm 1.5$		\\
		NICI $CH_4$				& 2011 Oct 20			& 1.584	& 0.017	& $6.9 \pm 1.0$		\\
		NICI $H$				& 2013 Jan 09			& 1.650	& 0.268	& $5.7 \pm 1.1$		\\
		NaCo $H$				& 2012 Jan 01			& 1.655	& 0.307	& $5.33 \pm 0.81$	\\
		NaCo $K_s$				& 2010 Mar 20, Apr 10	& 2.159	& 0.324	& $3.77 \pm 0.41$	\\
		NICI $K_s$				& 2013 Jan 09			& 2.174	& 0.269	& $4.29 \pm 0.61$	\\
		NICI $K_s$				& 2010 Dec 25			& 2.174	& 0.269	& $4.70 \pm 0.66$	\\
		NICI $K_\mathrm{cont}$	& 2011 Oct 20			& 2.272	& 0.037	& $6.8 \pm 1.0$		\\
		NICI $[3.09]$			& 2012 Dec 23			& 3.090	& 0.093	& $2.24 \pm 0.66$	\\
		MagAO/Clio $[3.1]$		& 2012 Dec 07			& 3.102	& 0.090	& $2.48 \pm 0.65$	\\
		MagAO/Clio $[3.3]$		& 2012 Dec 01			& 3.345	& 0.300	& $1.57 \pm 0.22$	\\
		MagAO/Clio $L^\prime$	& 2012 Dec 02			& 3.761	& 0.619	& $1.36 \pm 0.16$	\\
		NaCo $L^\prime$			& 2003 Nov 10, 13		& 3.813	& 0.573	& $1.72 \pm 0.56$	\\
		NaCo $L^\prime$			& 2009 Dec 29			& 3.813	& 0.573	& $1.70 \pm 0.11$	\\
		NaCo $L^\prime$			& 2012 Dec 16			& 3.813	& 0.573	& $1.58 \pm 0.14$	\\
		NaCo $L^\prime$			& 2013 Jan 31			& 3.813	& 0.573	& $1.29 \pm 0.23$	\\
		NaCo $[4.05]$			& 2010 Apr 03			& 4.056	& 0.060	& $1.26 \pm 0.31$	\\
		NaCo $[4.05]$			& 2012 Dec 16			& 4.056	& 0.060	& $1.46 \pm 0.13$	\\
		MagAO/Clio $M^\prime$	& 2012 Dec 04			& 4.687	& 0.185	& $1.010 \pm 0.068$	\\
		NaCo $M^\prime$			& 2012 Nov 26			& 4.818	& 0.416	& $0.88 \pm 0.19$	\\
		NaCo $M^\prime$			& 2012 Dec 15			& 4.818	& 0.416	& $0.88 \pm 0.12$	\\
		\hline
		\hline
		\multicolumn{5}{l}{}	\\ 
		\multicolumn{5}{l}{\tablenotemark{$\star$} Flux-weighted central wavelength.} \\ 
		\multicolumn{5}{l}{\tablenotemark{$*$} Effective flux-weighted filter width.} \\ 
		\multicolumn{5}{l}{\tablenotemark{$\dagger$} Flux densities in ergs s$^{-1}$ cm$^{-2}$ $\mu$m$^{-1}$.} \\ 
	\end{tabular}
\end{table*}

\section{Analysis}
In this section we model the 0.9--5~$\mu$m SED of $\beta$ Pic b using the available spectrophotometry.
We then measure the bolometric luminosity empirically, using a blackbody extension.
Finally, we use evolutionary tracks to determine the planet's mass, radius, and temperature.

\subsection{Exoplanet Atmosphere Models}
For this analysis, we choose to fit publicly-available models to the data.
These are:
Speigel \& Burrows \citep{spiegelburrows2012},
and the PHOENIX models:
AMES ``Cond'' and ``Dusty'' \citep{chabrier2000,allard2001,baraffe2003}
and ``BT Settl'' \citep{allard2012}.
The Spiegel \& Burrows models provide a constraint on formation mode by fitting for initial specific entropy.
The PHOENIX models provide a constraint on current physical properties, such as effective temperature and surface gravity.

We find the best-fit parameters for each model with a chi-squared-minimization process over all the photometry in Tab.~\ref{tab:all_phot}.
The GPI spectra are not independent measurements, as we normalize the $J$- and $H$-band fluxes with broad-band photometry.
Therefore, the GPI spectra are not fit, but are overplotted for visual inspection.
For each model in each grid we calculate the reduced $\chi^2$ statistic, using the photometry and uncertainties.  For the PHOENIX models we scale the overall luminosity, which amounts to choosing the best-fit radius, to minimize $\chi^2$ for any given model. The parameters (gravity, temperature, etc.) of the model with minimum $\chi^2$ are adopted as the best-fit parameters for that grid. 

\subsubsection{Spiegel \& Burrows Models}
We first describe the Spiegel \& Burrows models \citep{spiegelburrows2012}\footnote{Spiegel \& Burrows models downloaded from \url{http://www.astro.princeton.edu/$\sim$burrows/warmstart/}.}.  These models take as input the accretion efficiency during formation.  A range of evolutionary models are calculated with varying initial entropies, including their own version of the ``hot-start'' and ``cold-start'' models of \citet{marley2007}, as well as intermediate ``warm-start'' models and higher-entropy ``hotter-start'' models.

The initial entropies after formation vary because of varying treatments of the energy radiated away during accretion.
Gravitational contraction rather than disk accretion results in a much higher initial entropy, and this is the extreme ``hottest-start'' case considered by \citet{spiegelburrows2012}.
The evolutionary models are then combined with boundary conditions and radiative transfer (COOLTLUSTY) \citep{2003ApJ...594.1011H,2006ApJ...650.1140B} to create synthetic spectra and magnitudes.  For the atmosphere models, they then consider different atmospheric compositions: 1 or 3x Solar metallicity, and hybrid clouds or cloud-free.  Hybrid clouds are a linear combination of cloud-free and cloudy atmospheres to represent patchy or partial clouds \citep{burrows2011}.  We select the 25-Myr-old models as the closest in age to the $\beta$ Pic system, and independently fit each of the four cases for cloud-free or hybrid clouds and 1x or 3x Solar metallicity.

\begin{figure*}[htb]
	\centering
	\includegraphics[height=0.48\linewidth,angle=90,trim=0.5cm 1cm 0.9cm 0.65cm,clip=true]{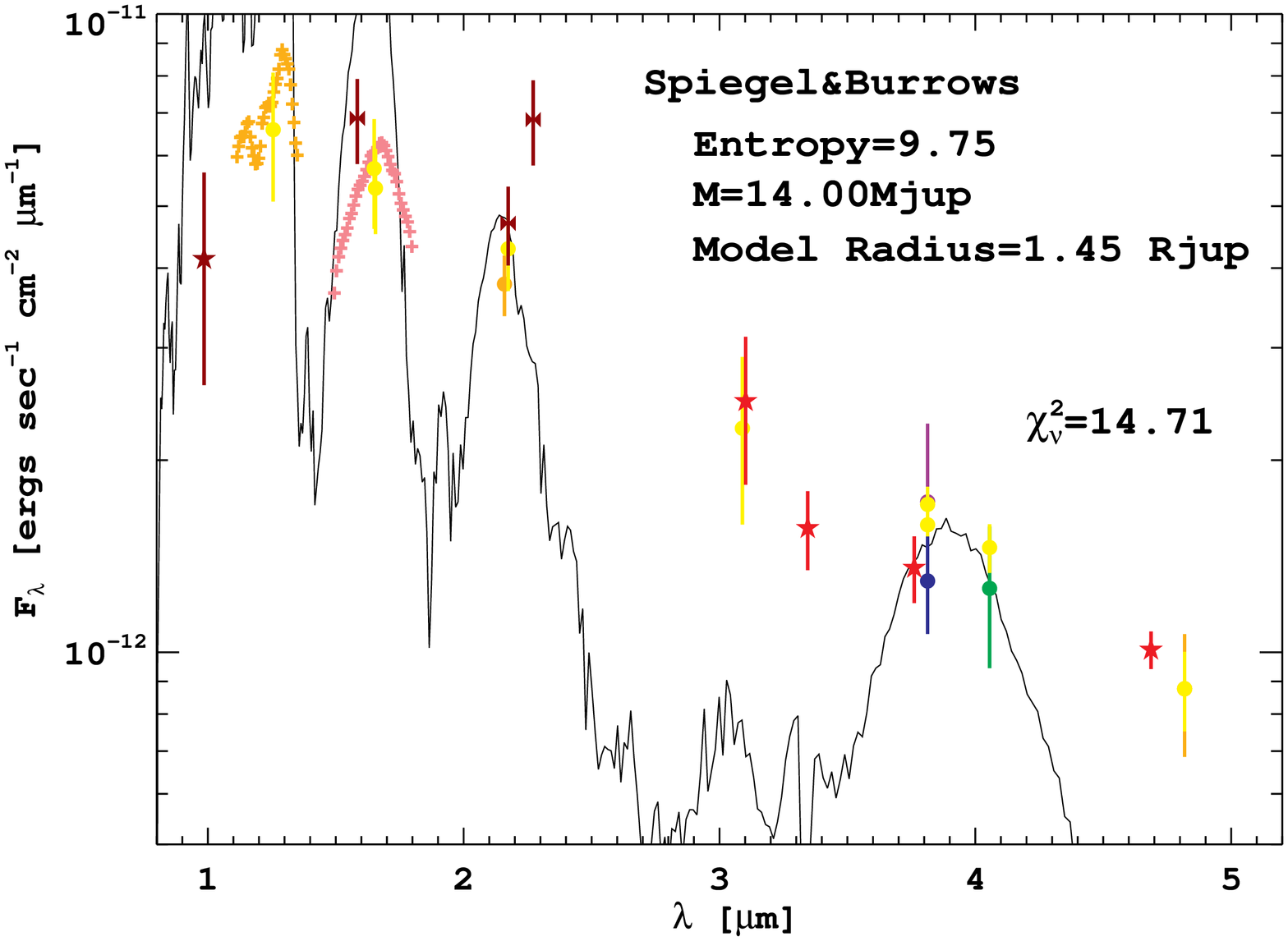}
	\hfill
	\includegraphics[height=0.48\linewidth,angle=90,trim=0.5cm 1cm 0.9cm 0.65cm,clip=true]{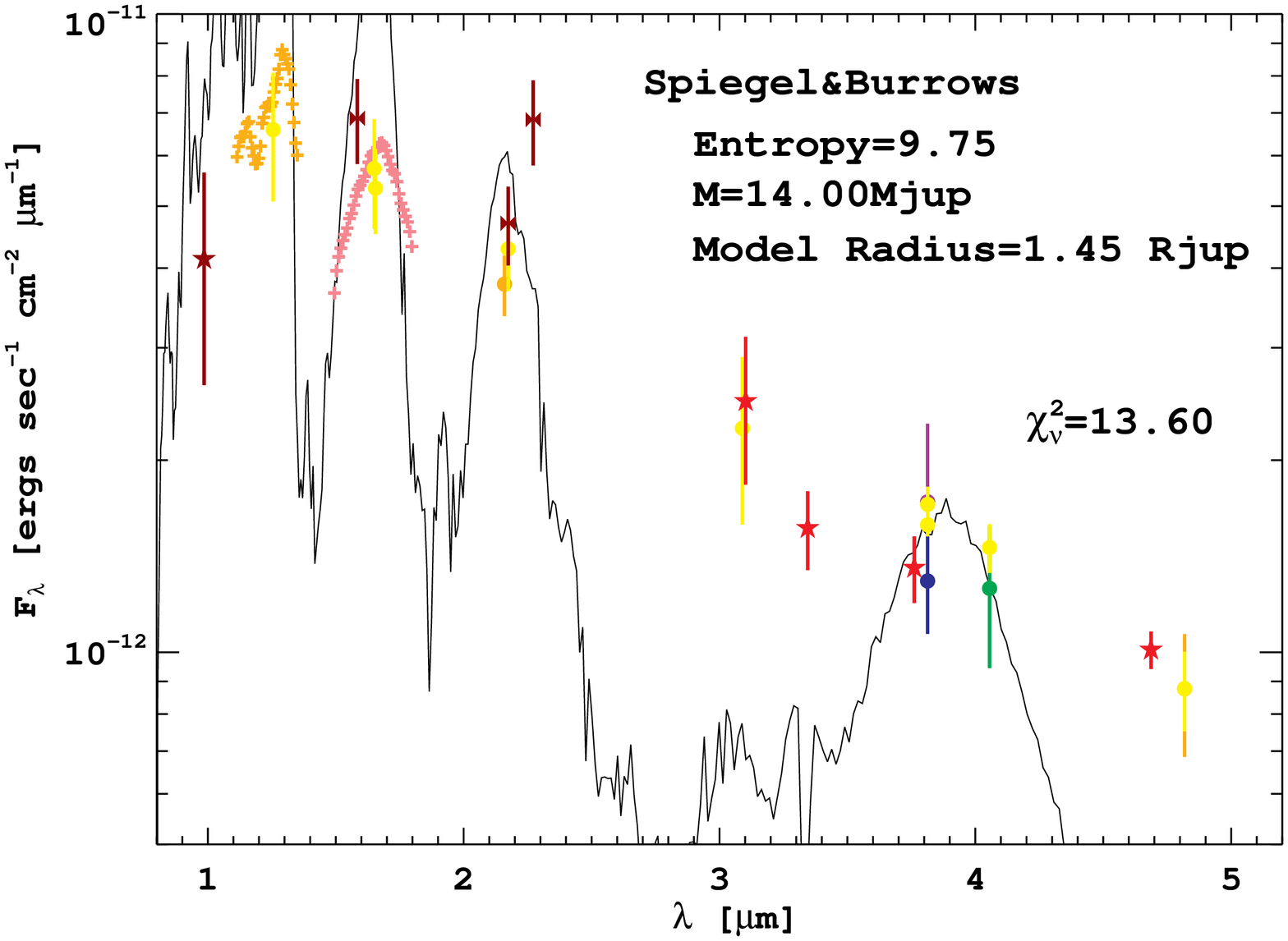}
	\includegraphics[height=0.48\linewidth,angle=90,trim=0.5cm 1cm 0.9cm 0.65cm,clip=true]{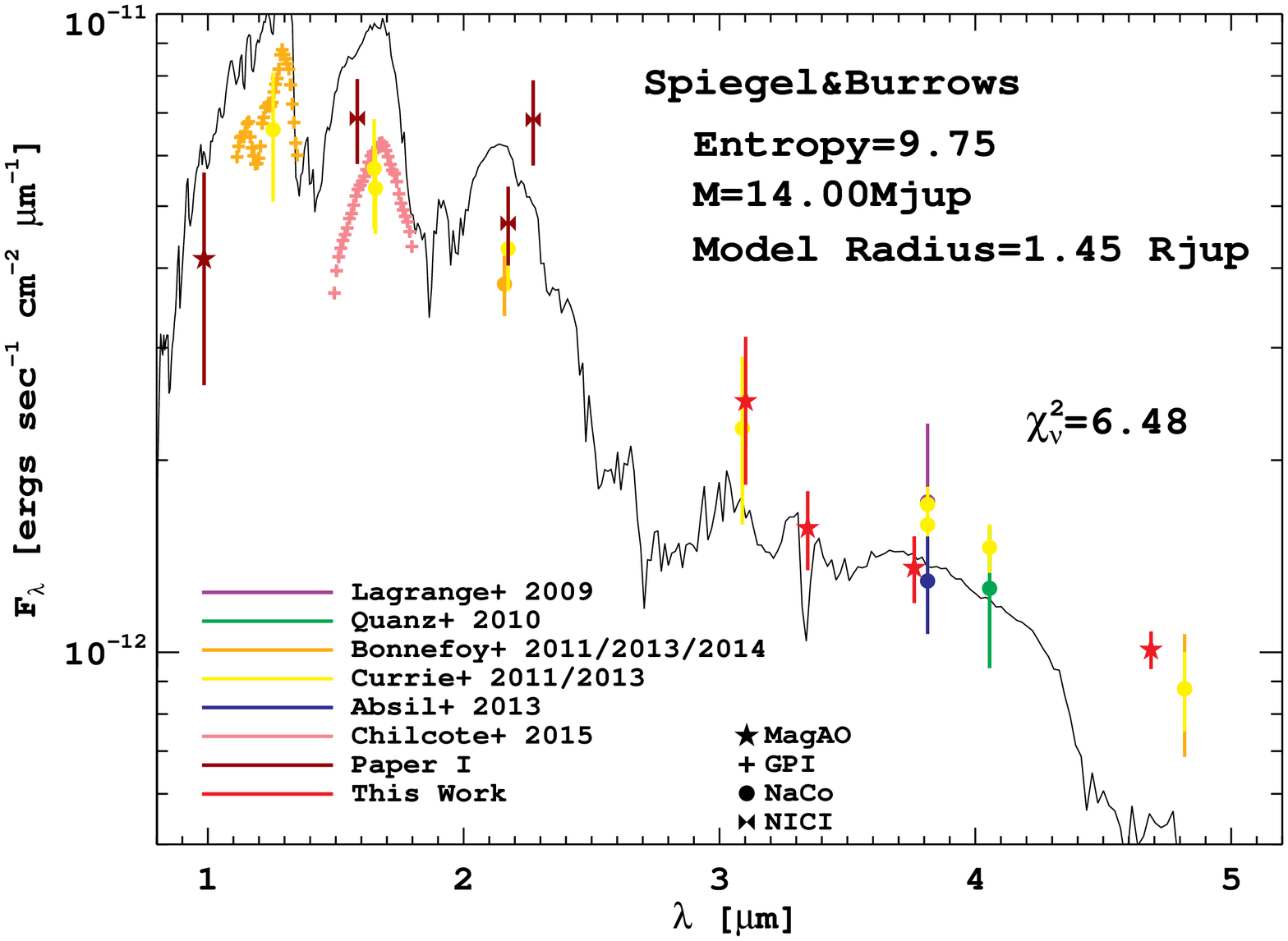}
	\hfill
	\includegraphics[height=0.48\linewidth,angle=90,trim=0.5cm 1cm 0.9cm 0.65cm,clip=true]{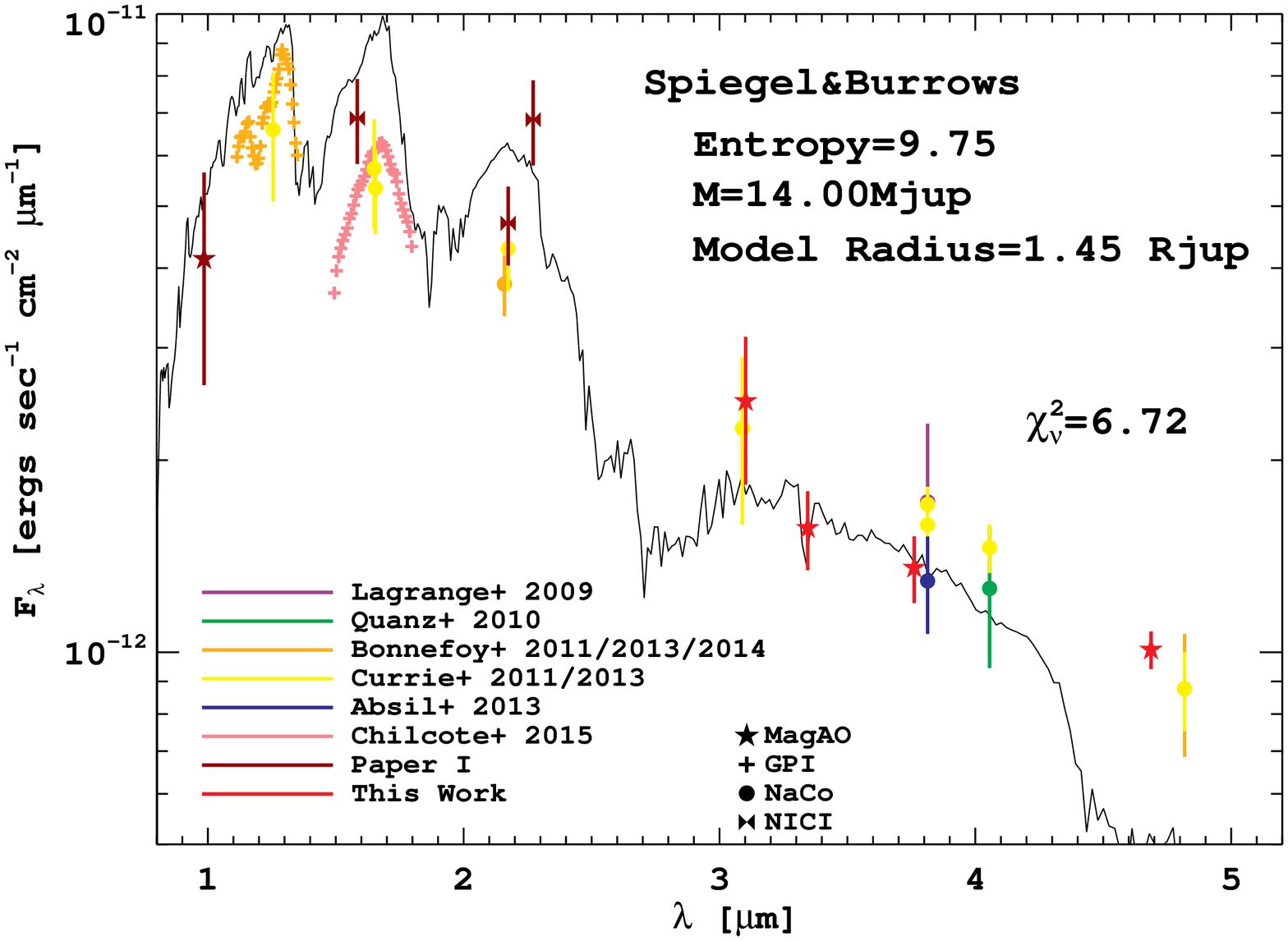}
	\caption{
		\label{fig:spiegelburrows}
		Best fit of the 25-Myr-old \citet{spiegelburrows2012} models.
		The photometric points are fit and the GPI spectra are overlaid for comparison.
		Top row: Cloud-free.
		Bottom row: Hybrid clouds.
		Left: 1x Solar.
		Right: 3x Solar.
		In all cases, the best-fit model has an initial specific entropy of 9.75 $k_B/\mathrm{baryon}$, which in their paradigm shows a warm start to the planet's formation.
	}
\end{figure*}

The model fits are shown in Fig.~\ref{fig:spiegelburrows}, and Tab.~\ref{tab:spiegelburrows} summarizes the best-fit parameters.
The parameters determined in the fit are initial specific entropy, mass, and initial radius.
The best-fit mass (14 \mjup) is high compared to the upper-limit set by the HARPS radial velocity data \citetext{$<$12~\mjup\ for $a$=9~AU, \citealt{lagrange2012rv}}.  The initial radius of 1.45 \rjup\ does not allow room for contraction to the large radii needed to fit the PHOENIX models (below).  The two best-fitting models, at $\chi_\nu^2$=6.5--6.7, are the hybrid cloud models with 1x and 3x Solar metallicity.
Other works have found that $\beta$ Pic b most likely formed with a high initial entropy, as indicated by its brightness \citep{chilcote2015,bonnefoy2014} for a planet-mass object at a young age.
However, all four best-fit models have an initial specific entropy of 9.75 $k_B/\mathrm{baryon}$, which is an intermediate ``warm-start'' value.

\begin{table}[hbt]
	\caption{
		Best fits of 25-Myr-old \citet{spiegelburrows2012} models, plotted in Fig.~\ref{fig:spiegelburrows}.
		The formation is constrained as ``warm-start.''
	}
	\label{tab:spiegelburrows}
	\nopagebreak[4]
	\centering
	\begin{tabular}{lllll}
		\hline
		\hline
								& Init.\ Specific	&			& Init.		& 				\\
								& Entropy			& Mass		& Rad.		& 				\\
		Model					& /$k_B$ baryon$^{-1}$	& /\mjup	& /\rjup	& $\chi_\nu^2$	\\
		\hline
		Cloud-free, 1x Solar	& 9.75				& 14.00		& 1.45		& 14.71			\\
		Cloud-free, 3x Solar	& 9.75				& 14.00		& 1.45		& 13.60			\\
		Hybrid clouds, 1x ''	& 9.75				& 14.00		& 1.45		& 6.48			\\
		Hybrid clouds, 3x ''	& 9.75				& 14.00		& 1.45		& 6.72			\\
	\end{tabular}
\end{table}

\subsubsection{PHOENIX Models}
We also use a suite of publicly-available models based on the PHOENIX atmosphere simulator\footnote{All PHOENIX models were downloaded from \url{http://perso.ens-lyon.fr/france.allard/}.}.
The AMES Cond \citep{allard2001, baraffe2003} and Dusty \citep{chabrier2000, allard2001} models represent limiting cases of dust in the atmosphere:
Cond models do not include dust opacity;
Dusty models treat dust opacity in equilibrium with the gas phase.
Note that the effective temperature of $\beta$ Pic b found by previous studies ($\sim$1600~K) is between the expected valid temperature ranges of the Cond (\teff~$<$~1400~K) and Dusty (\teff~$>$~1700~K) models.
We might expect better results with the BT-Settl \citep{2012IAUS..282..235A} models, which include a cloud model and are valid across the \teff\ range of interest.
We test both the ``CIFIST2011'' and ``CIFIST2011bc'' model grids, which differ in details of how the mixing length is calibrated\footnote{See \url{https://phoenix.ens-lyon.fr/Grids/BT-Settl/README} for additional details.  A new updated grid of BT-Settl models has recently become available \citep{2015A&A...577A..42B}.}.

\begin{figure*}[hbt]
	\centering
	\includegraphics[height=0.48\linewidth,angle=90,trim=0.5cm 1cm 0.9cm 0.65cm,clip=true]{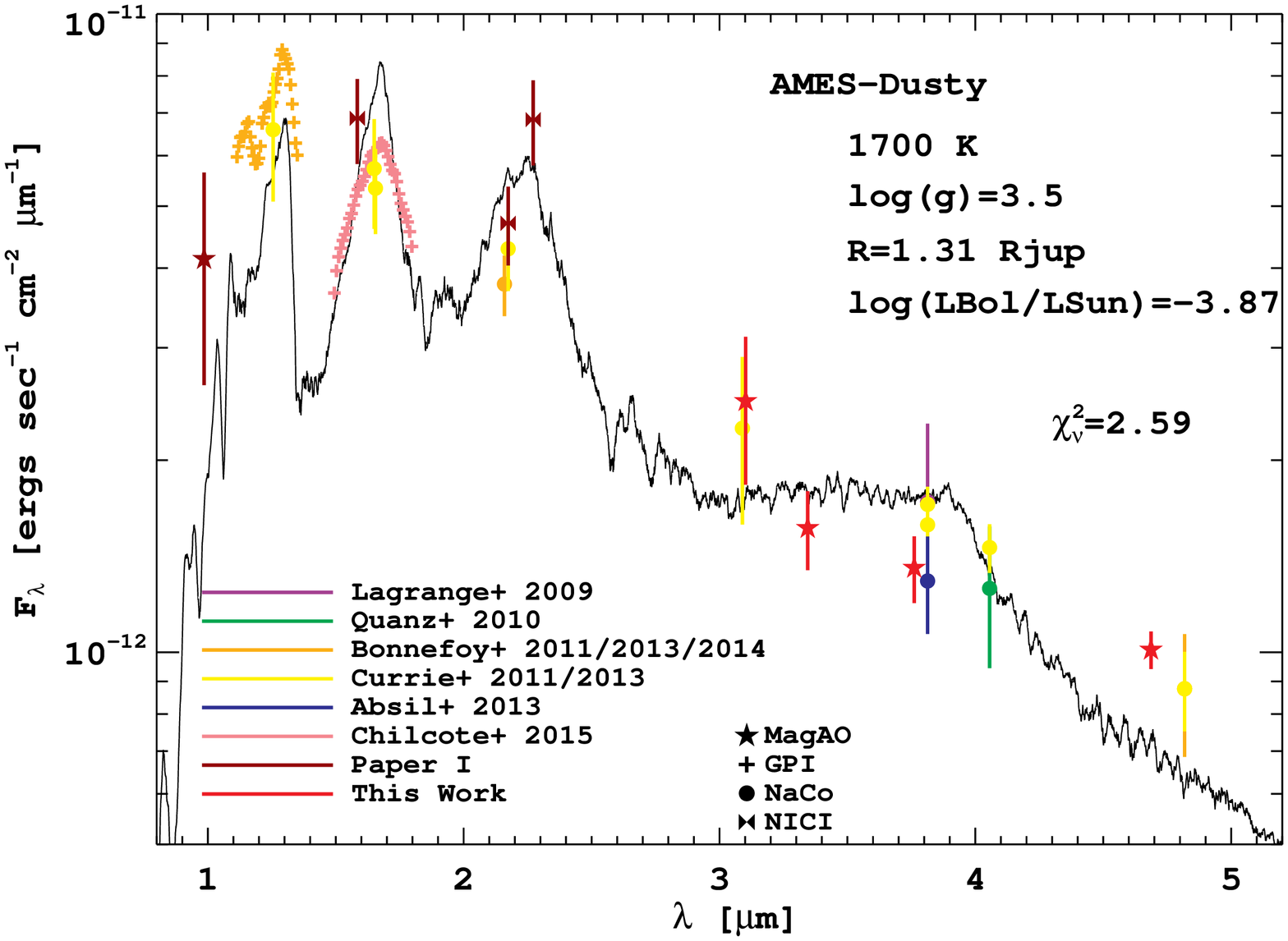}
	\hfill
	\includegraphics[height=0.48\linewidth,angle=90,trim=0.5cm 1cm 0.9cm 0.65cm,clip=true]{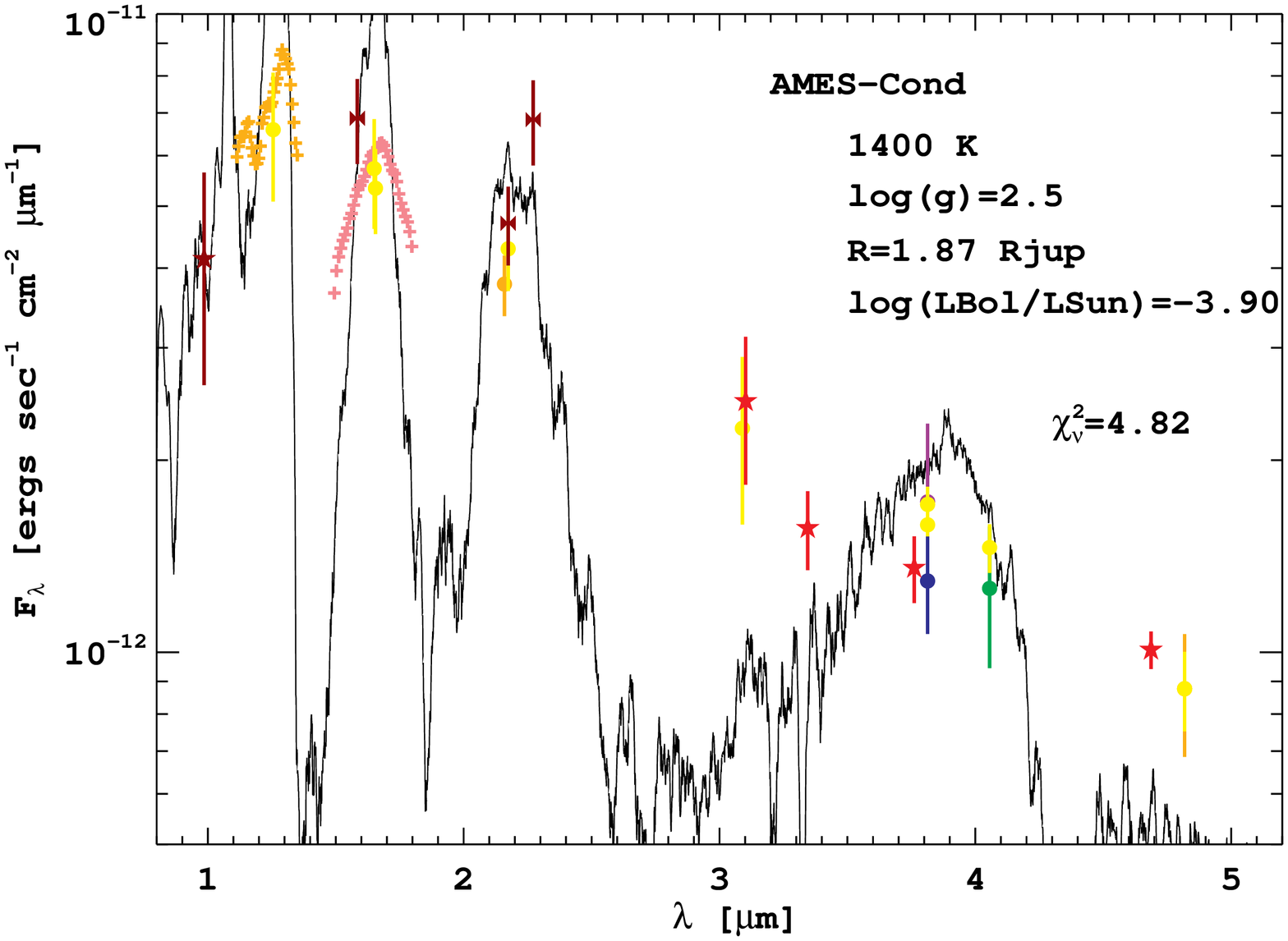}
	\includegraphics[height=0.48\linewidth,angle=90,trim=0.5cm 1cm 0.9cm 0.65cm,clip=true]{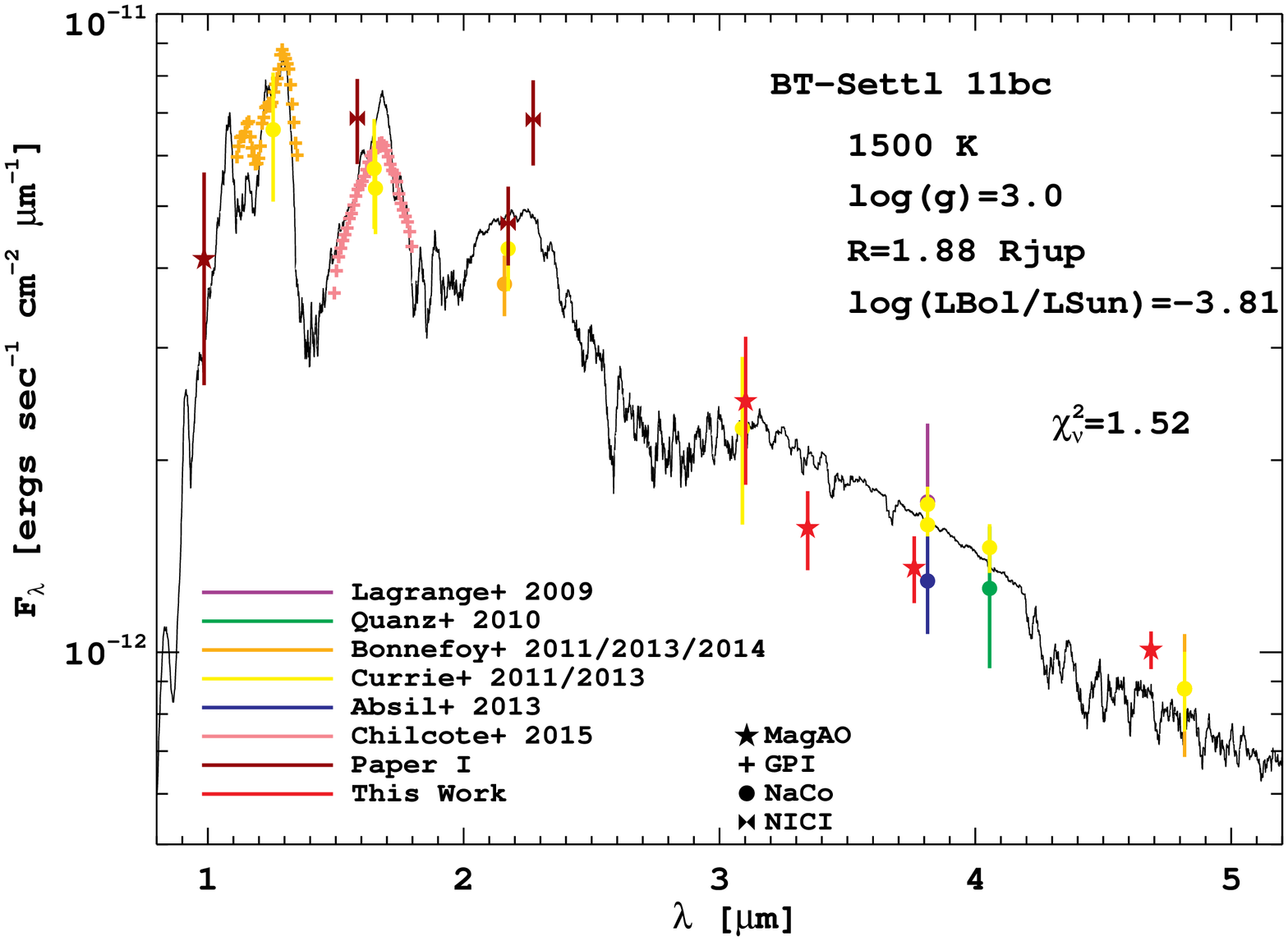}
	\hfill
	\includegraphics[height=0.48\linewidth,angle=90,trim=0.5cm 1cm 0.9cm 0.65cm,clip=true]{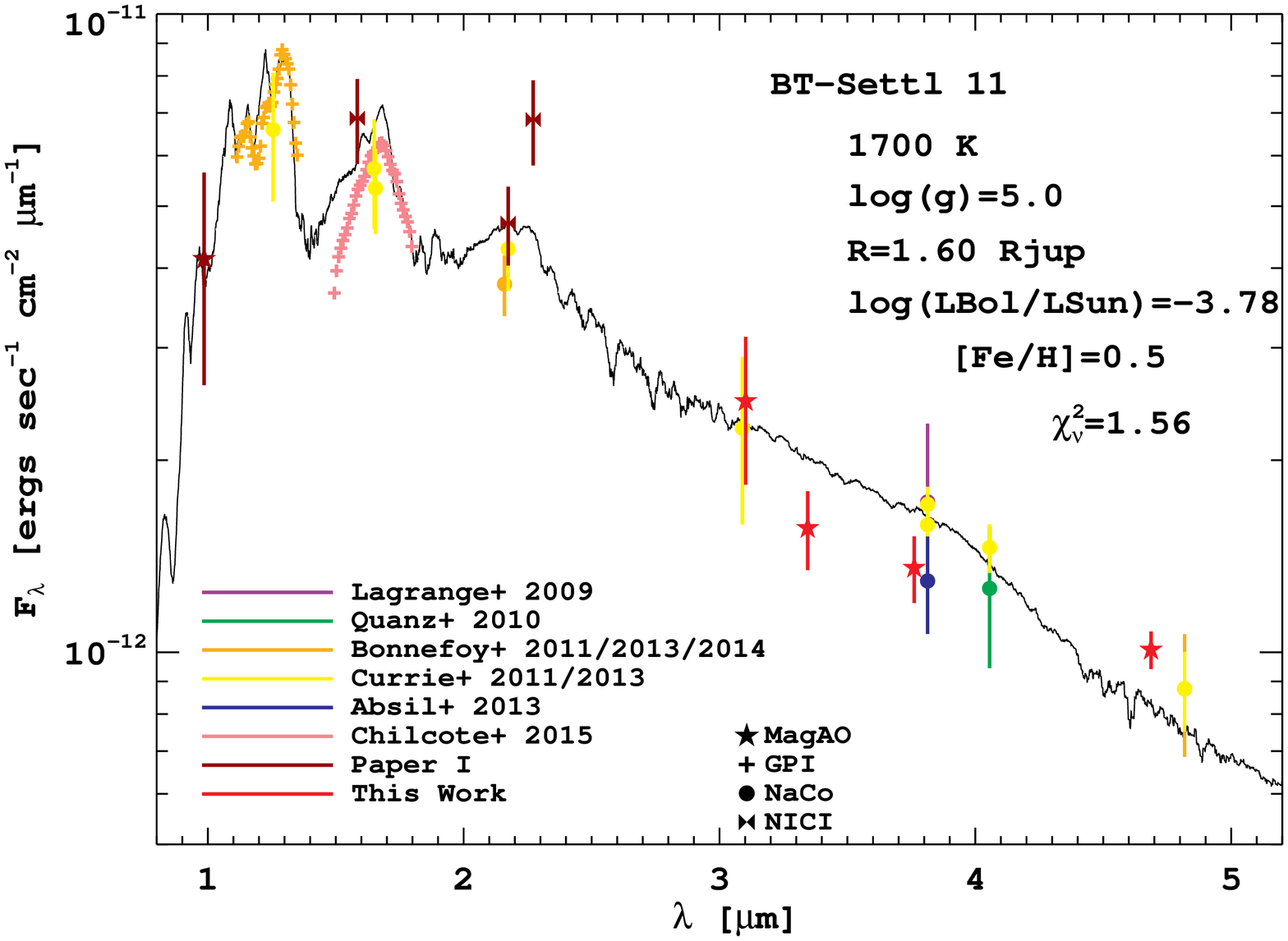}
	\caption{
		\label{fig:phoenix}
		Best-fit of the PHOENIX models.
		The photometric points are fit and the GPI spectra are overlaid for comparison.
		Top row: AMES-Dusty and AMES-Cond.
		Bottom row: BT Settl.
		The best-fit model at lower right has \logg=5 and $R$=1.60~\rjup\, which would correspond to a mass of $\sim$100~\mjup; this mass is ruled out by radial velocity observations of the star \citep{lagrange2012rv,bonnefoy2014}.
		Therefore, the best-fit model at lower-right (BT Settl) is likely unphysical, leaving the model at lower-left (BT Settl bc) as the best fit of the PHOENIX atmosphere models.
	}
\end{figure*}

The fits to the PHOENIX models are shown in Fig.~\ref{fig:phoenix}.
The parameters determined in the fit are \teff, \logg, radius, and sometimes [Fe/H].
Each spectrum in the grid corresponds to a gravity, temperature, and metallicity (which is Solar for all but BT Settl where it varies, and the best-fit metallicity is shown).
We find the scale factor for each spectrum by minimizing $\chi^2$ over the photometry.  This scale factor is (radius/distance) squared, and thus gives the best-fit radius.  We then calculate the bolometric luminosity by integrating the scaled spectrum out to 10~$\mu$m where we had truncated the models to facilitate the minimization algorithm; we then add 2\% for the contribution of flux from 10--100~$\mu$m to determine the total \lbol.
The \teff s, surface gravities, radii, and luminosities determined from the models are listed in Tab.~\ref{tab:phoenix}.

In addition, we calculate what the mass of the planet would be, given the model gravity and best-fit radius. 
We include this quantity in Tab.~\ref{tab:phoenix} to help give intuition for the gravities and radii determined via the models, with the caveat that mass is not an explicit output of the model atmospheres, as they do not include evolutionary tracks.
Nevertheless, it shows that the high gravity of the best-fit BT Settl 11 model is likely unphysical, because the calculated mass of $\sim$100~\mjup\ has been definitively ruled out by radial velocity observations \citep{lagrange2012rv,bonnefoy2014}.

\begin{table}[htb]
	\caption{	\label{tab:phoenix}
	Best fits of the PHOENIX models (shown in Fig.~\ref{fig:phoenix}).
	}
	\nopagebreak
	\centering
	\begin{tabular}{lllllll}
		\hline
		\hline
						&		& log	&			& log(		& Calc.				& 				\\
						& \teff	& $(g)$	& Rad.		& \lbol		& Mass$^\dagger$	& 				\\
		Model			& /K	& [cgs]	& /\rjup	& /\lsun)$^*$	& /\mjup			& $\chi_\nu^2$	\\
		\hline
		AMES-Dusty		& 1700	& 3.5	& 1.31		& $-3.86$		& 2.2				& 2.59			\\
		AMES-Cond		& 1400	& 2.5	& 1.87		& $-3.89$		& 0.4				& 4.82			\\
		BT Settl 11bc	& 1500	& 3.0	& 1.88		& $-3.80$		& 1.4				& 1.52			\\
		BT Settl 11$^\ddagger$		& 1700	& 5.0	& 1.60		& $-3.77$		& 103$^\star$		& 1.56		\\
		\hline
		\hline
		\multicolumn{7}{l}{}	\\ 
		\multicolumn{7}{l}{$^\ddagger$ Best-fit metallicity is [Fe/H]=0.5.} \\
		\multicolumn{7}{l}{$^\dagger$ Calculated mass, using model \logg\ and best-fit radius.} \\
		\multicolumn{7}{l}{$^\star$ Mass $>$20 \mjup\ is ruled out by \citet{lagrange2012rv}.} \\
		\multicolumn{7}{l}{$^*$ Added 2\% for the 10--100 $\mu$m flux region (see text).} \\
	\end{tabular}
\end{table}

\subsection{Spectral Features}
The results of the model fits show some general trends.

\subsubsection{Thick Clouds}
Cloud-free models are poor fits to the broad-band photometry.
AMES-Cond and the cloud-free models of Spiegel \& Burrows are faint at 3 and 5 $\mu$m, where the planet is bright.
Molecular gas opacity modeled in this region includes H$_2$, He, H$_2$O, CO, and CH$_4$ \citep{sharp2007}, which provide some of the absorption features seen in Fig.~\ref{fig:spiegelburrows} (top), but provide too much absorption at 3 $\mu$m and too much flux particularly in the $J$ and $H$ bands.
The hybrid cloud models of Spiegel \& Burrows are somewhat better, but are still too faint at 5~$\mu$m compared to the planet.

The best fits are provided by AMES-Dusty and the BT Settl models.
These include the most dust and cloud opacity, particularly in the 3- and 5-$\mu$m regions.
Therefore, $\beta$ Pic b's atmosphere is dominated by thick clouds.

\begin{figure}[htbp]
	\centering
	\includegraphics[height=\linewidth,angle=90,trim=0.5cm 1cm 0.9cm 0.65cm,clip=true]{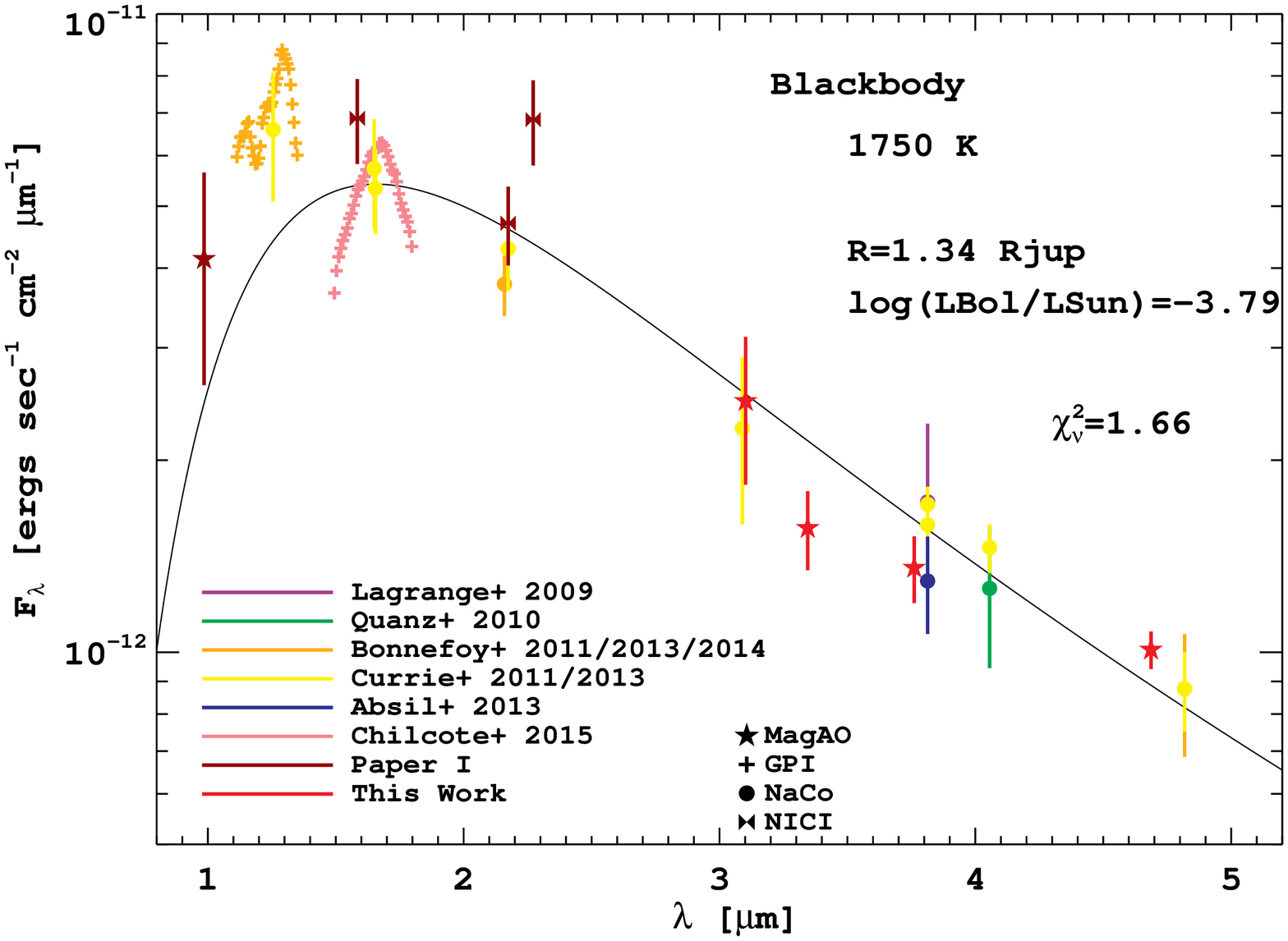}
	\caption{\label{fig:blackbody}
		Best-fit blackbody:
		\teff\ = 1750~K, radius = 1.34~\rjup.
	}
\end{figure}

\paragraph{Blackbody}
An object dominated by thick clouds, as the atmosphere models demonstrate for $\beta$ Pic b, will predominantly exhibit thermal radiation rather than molecular emission/absorption features.
We therefore explore fitting a blackbody curve to the visible through mid-IR photometry.  Figure~\ref{fig:blackbody} shows the best-fit blackbody curve.  Its physical parameters are 
\teff=1750~K and $R$=1.34~\rjup, giving a bolometric luminosity of log(\lbol/\lsun)=$-3.79$.
As with the atmosphere models, the GPI spectra are not included in the $\chi^2$-minimization but are overplotted for visual inspection.

The blackbody is the third-best-performing fit to the photometry, after the two BT Settl models.
While the broad wavelength coverage for $\beta$ Pic b is an asset for determining its overall luminosity, its cloudy atmosphere means that spectral features are flattened as compared to cloud-free atmospheres.
This underscores the importance of spectra and narrow-band photometry, as well as broad wavelength coverage, to fitting atmosphere models.

\subsubsection{Possiblity of Methane}
The $\nu_3$ fundamental bandhead of CH$_4$ is centered at 3.31181~$\mu$m \citep{albert2009} and absorbs strongly from 3.1--3.6~$\mu$m \citetext{3200--2800~cm$^{-1}$, \citealt{burch1962}}.
Clio's 3.3-$\mu$m filter, with an effective width from 3.05--3.65~$\mu$m (Tab.~\ref{tab:all_phot}), well-samples the CH$_4$ fundamental bandhead.
The suppressed flux at 3.3~$\mu$m in the Clio data is thus suggestive of methane absorption.
While \citet{chilcote2015} find no CH$_4$ absorption in the GPI $H$-band spectrum, methane could be detected at 3.3~$\mu$m where CH$_4$ absorption is strongest \citep{noll2000}.
Despite methane-onset in the near-IR being a typical marker of the L--T transition, \citet{noll2000} and \citet{sorahanayamamura2011,sorahanayamamura2012} show spectra of field dwarfs as early as L5 with CH$_4$ absorption at 3.3~$\mu$m.
One possible mechanism is enhanced metal abundance, shown in the \citet{sorahanayamamura2014} models of an 1800-K L dwarf, where increasing metallicity decreases CH$_4$ in the deeper layers but increases CH$_4$ abundance in the atmosphere.
In any case, this is the first detection of $\beta$ Pic b at 3.3~$\mu$m, and the observation will be repeated in a future work to confirm whether CH$_4$ is present. 

\subsubsection{Luminosity, Temperature, and Radius}
The two best-fit BT Settl models, with nearly identical reduced $\chi^2$s, have different temperatures (1500 and 1700 K, respectively) and radii (1.88 and 1.60 \rjup, respectively) but similar luminosities (log(\lbol/\lsun) of $-3.80$ and $-3.77$, respectively).
Taking these two models as the envelope of best fits, the radius is determined to 16\% and the \teff\ to 13\%, yet the \lbol\ is determined to 7\%.
That is, temperature and radius are dually fit by the models, whereas luminosity is well-constrained.

We see this same trend with the best-fit models of other works, listed in Tab.~\ref{tab:bestfit}.
\teff\ and radius are inversely-correlated in all cases, whereas \lbol\ is almost unchanged.
For all best-fit models, \teff\ and radius are constrained to 7\% and 14\%, respectively, while \lbol\ is constrained to within 4\%.
We conclude that the models are best providing a constraint on the luminosity rather than the temperature or radius.
Therefore, in the next section we extend the SED with blackbody curves to understand the fundamentals of the planet's energy budget.

\begin{table}[htb]
	\caption{	\label{tab:bestfit}
		Best-fit modeled parameters for $\beta$ Pic b, presented without error bars for brief inspection of the trends compared to the best model fits in this work.
		(See Tab.~\ref{tab:summary} for a fuller listing, including the best determination from this work.)
		Atmosphere models fit to the photometry are constraining \lbol, not \teff.
	}
	\nopagebreak
	\centering
	\begin{tabular}{lllll}
		\hline
		\hline
		&		& log		&			& log(				\\
		& \teff	& $(g)$		& Rad.		& \lbol				\\
		Work						& /K	& [cgs]		& /\rjup	& /\lsun)		\\
		\hline
		BT Settl 11bc (this work)	& 1500	& 3.0		& 1.88		& $-3.80$		\\
		\citet{baudino2015}			& 1550	& 3.5		& 1.76		& $-3.77$$^*$	\\
		\citet{chilcote2015}		& 1650	& 4.0		& $\cdots$	& $\cdots$		\\
		\citet{bonnefoy2014}		& 1650	& $<$4.7	& 1.5		& $-3.80$$^*$	\\
		BT Settl 11	(this work)		& 1700	& 5.0		& 1.60		& $-3.77$		\\
		Blackbody	(this work)		& 1750	& $\cdots$	& 1.34		& $-3.79$		\\
		\hline
		\hline
		\multicolumn{5}{l}{}	\\ 
		\multicolumn{5}{l}{$^*$ This \lbol\ is calculated from the modeled \teff\ and radius.} \\
	\end{tabular}
\end{table}

\subsection{Empirical Bolometric Luminosity}

Fitting atmospheric models (above) demonstrated that $\beta$ Pic b has a cloudy/dusty atmosphere.  However, because of the correlation between temperature and gravity in fitting the spectra, the excercise of fitting the models points more clearly towards a luminosity than an effective temperature.
Given the simplicity and relative success of the blackbody fit (Fig.~\ref{fig:blackbody}), we next determine the full SED of $\beta$ Pic b by extending the spectrum with a blackbody, and empirically calculating the bolometric luminosity.
See Tab.~\ref{tab:units} for the units we use in this work.
\begin{table}[htb]
	\caption{
		Nominal units and measures used in this work.
	}
	\label{tab:units}
	\centering
	\begin{tabular}{lrll}
		\hline
		\hline
		Quantity$^\dagger$			& Value						& Unit			& Reference	\\
		\hline
		Solar Luminosity \lsun		& $3.827 \times 10^{33}$	& erg s$^{-1}$	& [1,2]$^*$	\\
		Jupiter Radius \rjup		& $7.1492 \times 10^9$		& cm			& [3,4,5]$^\star$	\\
		Jupiter Mass \mjup			& $1.8983 \times 10^{30}$	& g				& [4,5]		\\
		Distance to $\beta$ Pic A	& 19.44$\pm$0.05			& pc			& [6]		\\
		\hline
		\hline
		\multicolumn{4}{l}{}	\\ 
		\multicolumn{4}{l}{[1] \citet{mamajek2012}} \\
		\multicolumn{4}{l}{[2] \citet{mamajeksite}} \\
		\multicolumn{4}{l}{[3] \citet{fortney2007}} \\
		\multicolumn{4}{l}{[4] \citet{jupiterfact}} \\
		\multicolumn{4}{l}{[5] \citet{iau2015b3}} \\
		\multicolumn{4}{l}{[6] \citet{vanleeuwen2007} (Parallax of 51.44$\pm$0.12 mas.)} \\
		\multicolumn{4}{p{0.95\linewidth}}{$^\dagger$Script letters for nominal Solar and Jovian values ($\mathcal{L,M,R}$) are used according to the recommendation given in [5], to emphasize that these quantities are used as constants.} \\
		\multicolumn{4}{p{0.95\linewidth}}{$^*$Updated to $3.828 \times 10^{33}$ erg s$^{-1}$ in \citet{iau2015b2}.}	\\
		\multicolumn{4}{l}{$^\star$Equatorial radius at 1 bar.} \\
	\end{tabular}
\end{table}

\begin{figure}[htbp]
	\centering
	\includegraphics[height=\linewidth,angle=90,trim=0.5cm 0.65cm 0.9cm 0.65cm,clip=true]{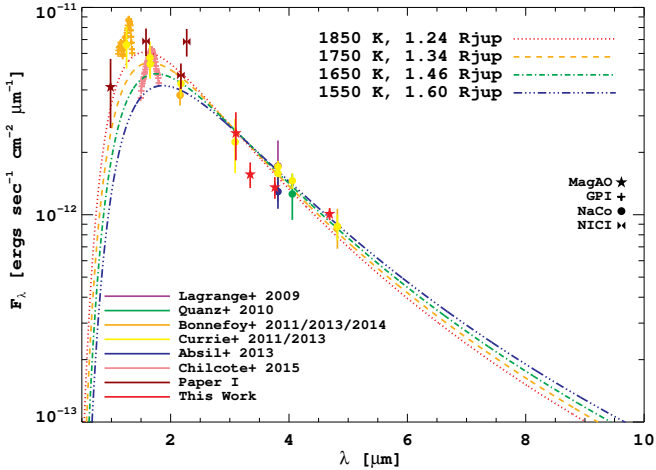}
	\caption{\label{fig:blackbody_lbol_bestfitBBs}
		Best-fit blackbodies for \teff s 1550--1850~K.
	}
\end{figure}

To extend the spectrum with a blackbody, we use the best-fitting blackbody for a selection of \teff s in 100-K increments, listed in Tab.~\ref{tab:bb}.
Fig.~\ref{fig:blackbody_lbol_bestfitBBs} shows the best-fitting blackbodies for temperatures 1550--1850~K.  The long-wavelength data have more measurements and smaller error bars, so these curves are tightly correlated in the Rayleigh-Jeans tail, crossing over near the middle wavelengths of the sample.

\begin{table}[htb]
	\caption{	\label{tab:bb}
		Best-fit blackbodies at each \teff.
	}
	\centering
	\begin{tabular}{rrrr}
		\hline
		\hline
		\teff/K	& Rad./\rjup	& log(\lbol/\lsun)	& $\chi_\nu^2$	\\
		\hline
		1550	& 1.60	& $-3.85$ 	& 2.46		\\
		1650	& 1.46	& $-3.82$	& 1.88		\\
		1750	& 1.34	& $-3.79$	& 1.66		\\
		1850	& 1.24	& $-3.77$	& 1.78		\\
		1950	& 1.15	& $-3.74$	& 2.19		\\
	\end{tabular}
\end{table}

\begin{figure}[ht]
	\centering
	\includegraphics[height=\linewidth,angle=90,trim=0.5cm 0.65cm 0.9cm 0.65cm,clip=true]{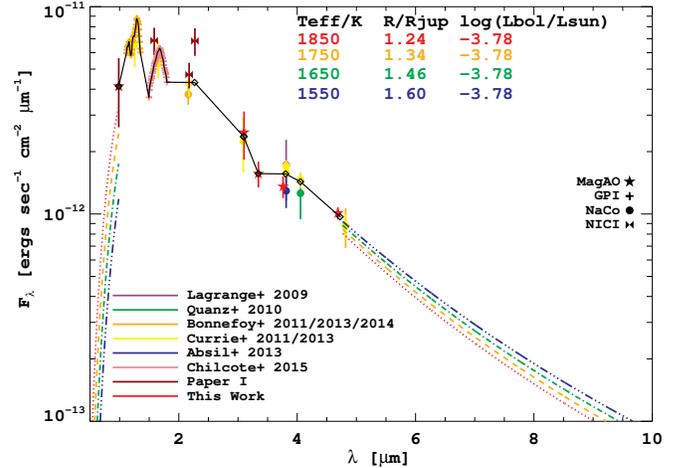}
	\caption{\label{fig:lbol}
		Blackbody extension of the measured SED.
		The black points show the weighted-mean spectrophotometry for each bandpass.
	}
\end{figure}

Next we use these blackbodies to extend the SED.
To combine the spectrophotometry, we find the weighted mean at each broad bandpass.
At the $J$ and $H$ bands we use the GPI spectra which have been normalized by the broadband photometry.
We use the effective filter widths to determine the mean wavelength for the various filters $Y_s$, $K$, 3.1~$\mu$m, 3.3~$\mu$m, $L^\prime$, 4.1~$\mu$m, and $M^\prime$.
The mean photometric points are shown in Fig.~\ref{fig:lbol} and listed in Tab.~\ref{tab:weighted_mean_phot}.
\begin{table}[htb]
	\caption{	\label{tab:weighted_mean_phot}
		Weighted-mean broad-band photometry used to calculate the empirical bolometric luminosity.
	}
	\nopagebreak
	\centering
	\begin{tabular}{rr}
		\hline
		\hline
		$\lambda$/$\mu$m	& $F_\lambda$ ($\times 10^{-12}$)$^\star$	\\
		\hline
		$Y_s$ 0.985			& $4.13\pm1.51$	\\
		$J$-band			& GPI spectrum	\\
		$H$-band			& GPI spectrum	\\
		$K$ 2.27			& $4.31\pm0.29$	\\
		3.10				& $2.35\pm0.46$	\\
		3.34				& $1.57\pm0.22$	\\
		$L^\prime$ 3.80		& $1.56\pm0.07$	\\
		4.10				& $1.43\pm0.12$	\\
		$M^\prime$ 4.72		& $0.97\pm0.06$	\\
		\hline
		\hline
		\multicolumn{2}{l}{}	\\ 
		\multicolumn{2}{l}{$^\star$ Flux densities in ergs s$^{-1}$ cm$^{-2}$ $\mu$m$^{-1}$} \\
	\end{tabular}
\end{table}

To find the bolometric luminosity, we integrate the total SED (data + blackbody extension) from 0.001--100~$\mu$m.
To determine the error bars in the bolometric luminosity, we include three error sources.
Photometric errors from the data lead to error in the trapezoidal integration of the weighted mean photometric points listed in Tab.~\ref{tab:weighted_mean_phot}, contributing 2.0\% to the total \lbol\ error.
Error from the unknown blackbody extension is found by the bounds of the various \teff\ and radius selections in Tab.~\ref{tab:lbol}, contributing 4.7\% (error based on the range).
Finally, the unsampled region around 2.7~$\mu$m could have a deep absorption feature such as that seen in the AMES-Dusty models (Fig.~\ref{fig:phoenix}), so we insert the modeled value here and repeat the integral --- the luminosity drops by 3.9\%, which we take as the error (error also based on the range) due to the unsampled region between $K$ and [3.1].
Summing these errors in quadrature gives a total error in the integrated luminosity of 6.4\%.
\textbf{Therefore, we find log(\lbol/\lsun)=$\mathbf{-3.78\pm0.03}$.}

\begin{table}[htb]
	\caption{	\label{tab:lbol}
		Empirical bolometric luminosity, integrating the data and extending with a blackbody curve.
	}
	\nopagebreak
	\centering
	\begin{tabular}{lrr}
		\hline
		\hline
		SED											& Rad./\rjup	& log($\mathcal{L}$/\lsun)	\\
		\hline
		Data only (0.985--4.77~$\mu$m)				&			& $-3.86$		\\
		Best-fit b.b., 0.985--4.77~$\mu$m only		& 1.34		& $-3.85$		\\
		\hline
		&			& log(\lbol/\lsun)	\\
		\hline
		Data + 1550-K best-fit b.b.	& 1.60	& $-3.78$	\\
		Data + 1650-K best-fit b.b.	& 1.46	& $-3.78$	\\
		Data + 1750-K best-fit b.b.	& 1.34	& $-3.78$	\\
		Data + 1850-K best-fit b.b.	& 1.24	& $-3.78$	\\
	\end{tabular}
\end{table}

\subsection{Physical Parameters of $\beta$ Pic b}
\begin{figure}[htbp]
	\centering
	\includegraphics[height=\linewidth,angle=90,trim=0.3cm 0.4cm 0.9cm 0.4cm,clip=true]{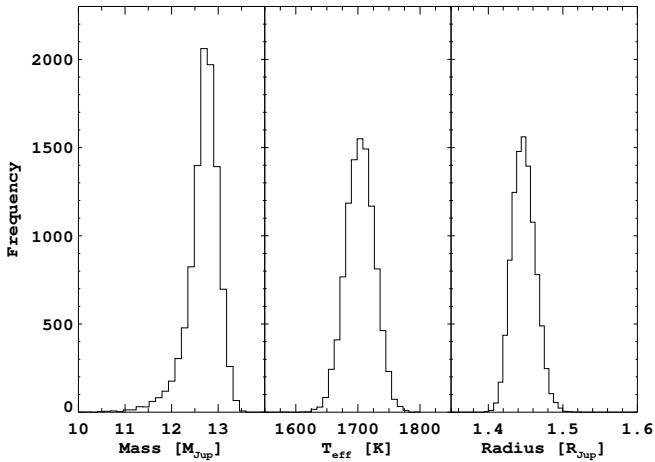}
	\caption{\label{fig:physpar_freqs_kt}
		Physical parameters of $\beta$ Pic b (mass = 12.7$\pm$0.3~\mjup, \teff\ = 1708$\pm$23~K, and radius = 1.45$\pm$0.02~\rjup)
		determined via a Monte Carlo experiment for the input of empirical luminosity log(\lbol/\lsun)=$-3.78\pm0.03$ into hot-start evolutionary models.
	}
\end{figure}

Given the luminosity and age, evolutionary models determine the mass, radius, and temperature as the planet cools and contracts.
We use our empirically-determined \lbol\ and age 23$\pm$3-Myr \citep{mamajek2014}, with the same method as in Paper I.
The results of a 10,000-trial Monte Carlo experiment are shown in Fig.~\ref{fig:physpar_freqs_kt}.
Luminosity and age are assumed to be described by Gaussian statistics with the mean and width specified.  A value was chosen from each distribution, and then the mass, \teff, and radius were found in the AMES-Cond hot-start evolutionary models \citep{chabrier2000,baraffe2003}.
The negative skewness of the mass distribution is because of the extreme luminosity increase for deuterium-burning objects above $\sim$13~\mjup.
\textbf{The median physical parameters from the distribution are mass = 12.7$\pm$0.3~\mjup, \teff = 1708$\pm$23~K, and radius = 1.45$\pm$0.02~\rjup}.
The error bars represent the uncertainty from the Monte Carlo trials, but do not include systematic errors of the evolutionary tracks themselves.

\section{Discussion}

In this work we have analyzed the 0.9--5~$\mu$m spectrophotometry of the young giant exoplanet $\beta$ Pic b.
Atmospheric models demonstrated that the planet is cloudy, and inspired us to measure the bolometric luminosity empirically, finally using evolutionary tracks to determine mass, radius, and temperature.

\subsection{Modeled, Matched, and Measured \lbol}
Table~\ref{tab:summary} compares our results to recent papers on this planet.
We have separated the luminosities determined in each work by method --- \citet{bonnefoy2013,bonnefoy2014} and Paper I used bolometric corrections to determine the luminosity, whereas \citet{currie2013} quoted a value from the models.  We also calculate the \lbol\ from the model-derived \teff\ and radius in cases where it is not presented in the paper.

\begin{table*}[htb]
	\caption{	\label{tab:summary}
		Physical parameters of $\beta$ Pic b.
		Other works use either the luminosity determined from atmosphere models
		or via a bolometric correction applied to the $K$-band photometry.
		Here we determine the luminosity empirically for the first time,
		and use the \lbol\ and age to determine the \teff, radius, and mass.
	}
	\nopagebreak
	\centering
	\begin{tabular}{llllllllll}
		\hline
		\hline
		& \teff	& \logg	& Radius & \multicolumn{3}{c}{log(\lbol/\lsun)} & Mass	& Init.\ Spec.\ Entropy \\ \cline{5-7}
		Reference & /K & [cgs] & /\rjup & Modeled & B.C. & Empirical & /\mjup & /$k_B$ baryon$^{-1}$ \\
		\hline
		\citet{currie2013} & 1575$^*$ & 3.8$\pm$0.2 & 1.65$\pm$0.06 & $-3.80$$\pm$0.02 & $\cdots$ & $\cdots$ & 6.9$^*$ & $\cdots$ \\
		\citet{bonnefoy2013} & 1700$\pm$100	& 4.0$\pm$0.5 & 1.5--1.6$^*$ & $-3.72$$^*$ & $-3.87$$\pm$0.08 & $\cdots$ & 9--10 & $\geq$9.3 \\
		\citet{bonnefoy2014} & 1650$\pm$150	& $<$4.7 & 1.5$\pm$0.2 & $-3.80$$^*$ & $-3.90$$\pm$0.07 & $\cdots$ & $<$20	& $>$10.5 \\
		\citet{chilcote2015} & 1600--1700 & 3.5--4.5 & $\cdots$ & $\cdots$ & $\cdots$ & $\cdots$ & $\cdots$ & $\cdots$ \\
		\citet{baudino2015} & 1550$\pm$150 & 3.5$\pm$1 & 1.76$\pm$0.24 & $-3.77$$^*$ & $\cdots$ & $\cdots$ & 4.0$^*$ & $\cdots$ \\
		Paper I		& 1643$\pm$32	& 4.2$^*$	& 1.43$\pm$0.02	& $\cdots$ & $-3.86$$\pm$0.04 & $\cdots$ & 11.9$\pm$0.7 & $\cdots$ \\
		This Work & 1708$\pm$23 & 4.2$^*$ & 1.45$\pm$0.02 & $\cdots$ & $\cdots$ & $-3.78\pm0.03$ & 12.7$\pm$0.3 & 9.75 \\
		\hline
		\hline
		\multicolumn{9}{l}{}	\\ 
		\multicolumn{9}{l}{$^*$ Value calculated based on other parameters.} \\
	\end{tabular}
\end{table*}

Fascinatingly, the luminosities from modeling are typically around log(\lbol/\lsun)=$-3.8$, while $-3.9$ dex is the typical value for bolometric-correction-determined values.
The empirical value we measured is in agreement with the luminosities determined by modeling.
This result confirms that the atmosphere models we and others have used for $\beta$ Pic b are pointing most strongly to a luminosity but not necessarily a temperature.
The radius is unknown and the thick clouds prevent strong molecular absorption features from dominating the broad SED, so the emergent flux is dominated by blackbody-like opacity, thus giving a good measurement of the overall luminosity yet obscuring the gravity and \teff.

We have integrated the combined spectrophotometry from 0.985--4.7~$\mu$m to determine the energy output that has been empirically measured.
We find that $>$80\% of the total energy has been measured in the data (Tab.~\ref{tab:lbol}).
We have sampled the SED across the peak of the blackbody, with the combination of data points yielding photometric errors as small as 0.06--0.07 ergs s$^{-1}$ cm$^{-2}$ $\mu$m$^{-1}$ at the $L$ and $M$ bands (Tab.~\ref{tab:weighted_mean_phot}).

Because of its brightness and proximity, many different types of observations have led to a wealth of information about this exoplanet.
Astrometric monitoring of its orbit show that its semimajor axis is 9.1$\bfrac{+0.7}{-0.2}$~AU \citep{macintosh2014,nielsen2014,millarblanchaer2015}.
For this orbital separation, radial velocity observations place an upper limit on its mass of $<$15~\mjup\ \citep{lagrange2012rv}.
Comparison with models demonstrate that the planet's SED is dominated by thick clouds (this work).
We have constrained its luminosity to log(\lbol/\lsun)=$-3.78\pm0.03$.
Our method of empirically determining \lbol\ is a unique approach enabled by the broad wavelength coverage of MagAO.

Although $J$ and $H$ spectra are available thanks to GPI, $\beta$ Pic b's atmosphere is not dominated by molecular absorption in the way that a cloud-free atmosphere is.  Instead the overall brightness is constrained by the models.

\subsection{A Self-Luminous Planet}
Many exoplanet atmospheres are characterized by transmission spectroscopy as the planet transits its star.  These exoplanets are necessarily on extremely close orbits (of order 0.1 AU) and are thus so highly-irradiated that their energy budgets are predominantly due to stellar insolation.  Jupiter, on the other hand, is still glowing from its formation, and is thus ``self-luminous''.
To determine whether a planet is self-luminous, the energy input from its star must be compared to its total energy output.

First we consider the energy input from the star.
The spectral type, surface gravity, and effective temperature of $\beta$ Pic A are determined in \citet{gray2006} via moderate-resolution (R$\sim$1300) optical spectroscopy to be A6V, \logg=4.15, and \teff=8052~K, respectively.  The mass of $\beta$ Pic A is determined via astrometric measurements of $\beta$ Pic b to be 1.60$\pm$0.05~\msun \citep{millarblanchaer2015}.  Given this mass and surface gravity, the radius and luminosity can be calculated; these values are summarized in Tab.~\ref{tab:bpica}.

\begin{table}[htb]
	\caption{Empirical measures of $\beta$ Pic A physical parameters.}
	\label{tab:bpica}
	\begin{center}
		\begin{tabular}{lll} \hline \hline
			Parameter			& Value	& Ref								\\
			\hline
			\teff/K				& 8052	& \citet{gray2006}					\\
			\logg [cgs]			& 4.15	& \citet{gray2006}					\\
			Mass/\msun			& 1.60$\pm$0.05	& \citet{millarblanchaer2015}		\\
			Radius/\rsun		& 1.78$\pm$0.24	& \citet{betapicradius}				\\
			log(\lbol/\lsun)	& 1.07	& Calculation using above values	\\
		\end{tabular}
	\end{center}
\end{table}

Next we consider the total energy output.
The ratio of emitted to absorbed energy is
\begin{equation}
E = \frac{4}{1-A} \left( \frac{r}{R_{*}} \right) ^2 \left( \frac{T_p}{T_{*}} \right) ^4
\end{equation}
\citep{ingersoll1978}
where $A$ is the Bond albedo of the planet.
\citet{cahoy2010} find that the Bond albedo can vary from 0.3--0.9 for directly-imaged Jupiter analogs depending on temperature. 
The snow line in the $\beta$ Pic system (the radius where the equlibrium temperature is 170~K) is around $\sim$10 AU, in agreement with \citet{devries2012}.  The ``naive snow line'' \citep{lecar2006} in the Solar System is $\sim$2.7 AU \citetext{\textit{e.g.}, \citealt{hayashi1981}, \citealt{podolak2010}}.
Therefore, we choose the albedo value near 2.7~AU in \citet{cahoy2010} to give a Bond albedo near the snow line of $A=0.7$.
For the size of the orbit of $\beta$ Pic b we use the mean semi-major axis of $9.1\bfrac{+0.7}{-0.2}$~AU
\citep{macintosh2014,chilcote2015,millarblanchaer2015}.
Thus we find that $E\approx10^4$:
$\beta$ Pic b is highly self-luminous.

The equilibrium temperature of a planet is given by
\begin{equation}
T_{eq} = T_{*} \sqrt{\frac{R_{*}}{r}} \left( \frac{1-A}{4} \right) ^{1/4}
\end{equation}
\citetext{\textit{e.g.}, \citealt{ingersoll1975}, \citealt{guillot1996}}.
For $\beta$ Pic b we calculate $T_{eq}$$\sim$130~K, while the planet is much hotter at \teff$\sim$1700~K.
This is in contrast to hot Jupiters which have effective temperatures close to their equilibrium temperatures \citep{cowanagol2011}.
A planet at equilibrium temperature has lost memory of its formation; a self-luminous planet like $\beta$ Pic b, on the other hand, is hot enough to point towards its initial entropy --- which provides the ``warm-start'' constraint on its formation for $\beta$ Pic b.

\section{Conclusions}
In this paper we have presented four images of $\beta$ Pic b from 3--5~$\mu$m obtained with MagAO+Clio.
We then placed 22 independent photometric measurements from 3 telescopes on a uniform photometric system, taking the reported contrast measurements and applying them in a consistent way to the detailed atmospheric and bandpass throughput, the spectrum of $\beta$ Pic A, and the updated distance.
This provided us with a spectral energy distribution (SED) from 0.9--5~$\mu$m with uniform systematics.
Clouds were indicated particularly by the 3- and 5-$\mu$m regions of the spectrum where the planet is bright compared to cloud-free models.
The flux we measured at 3.3~$\mu$m is fainter than the modeled flux in this region, hinting at a possibility of CH$_4$ absorption.

We extended the SED with a blackbody, where the Wien's tail is constrained by VisAO's $Y_s$ point and the Rayleigh-Jeans tail is constrained by Clio's $M^\prime$ point, which were measured simultaneously.
Photometry and spectra to date have sampled the optical through thermal-IR region containing $>$80\% of the planet's energy, so our blackbody extension to the SED only accounts for $<$20\% of the energy.
Integrating the extended SED gave an empirical measure of log(\lbol/\lsun)=$-3.78\pm0.03$.

Evolutionary tracks for this luminosity gave a mass of 12.7$\pm$0.3~\mjup, a \teff\ of 1708$\pm$23~K, and a radius of 1.45$\pm$0.02~\rjup, excluding model-dependent errors.
Our empirically-determined luminosity is in agreement with values from atmospheric models (typically $-3.8$ dex), but brighter than values from the field-dwarf-calibrated bolometric correction (typically $-3.9$ dex), illustrating the limitation of comparing young self-luminous exoplanets to old brown dwarfs.
The success of a blackbody fit to the photometry and the degeneracy of radius and gravity in the atmosphere models underscores the importance of spectra to constraining detailed atmospheric properties.
However, photometry with broad spectral coverage is useful for constraining \lbol\ in a wholly empirical fashion.


\section*{Acknowledgments}
We thank the Magellan and Las Campanas Observatory staff for making this well-engineered, smoothly-operated telescope and site possible.
We would especially like to thank Povilas Palunas for help over the entire MagAO commissioning run.
Juan Gallardo, Patricio Jones, Emilio Cerda, Felipe Sanchez, Gabriel Martin, Maurico Navarrete, Jorge Bravo, Victor Merino, Patricio Pinto, Gabriel Prieto, Mauricio Martinez, Alberto Pasten, Jorge Araya, Hugo Rivera, and the whole team of technical experts helped perform many exacting tasks in a very professional manner.
Glenn Eychaner, David Osip, and Frank Perez all gave expert support which was fantastic.
The entire logistics, dining, housekeeping, and hospitality staff provide for an excellent, healthy environment that ensured the wellness of our team throughout the commissioning runs.
It is a privilege to be able to commission an AO system with such a fine staff and site.

The MagAO system was developed with support from the NSF, MRI and TSIP programs.
The VisAO camera was developed with help from the NSF ATI program.
K.M.M. and J.R.M. were supported under contract with the California Institute of Technology, funded by NASA through the Sagan Fellowship Program.
J.R.M. is grateful for the generous support of the Phoenix ARCS Foundation.
L.M.C.'s and Y.-L.W.’s research were supported by NSF AAG and NASA Origins of Solar Systems grants.
V.B. was supported in part by the NSF Graduate Research Fellowship Program (DGE-1143953).

We thank the anonymous referee for a careful, timely review that significantly improved the manuscript.

\facility{Magellan:Clay (MagAO+Clio)}

\appendix

This work involved first-light data from MagAO/Clio, and thus necessitated a detailed calibration of the camera.
In the Appendix we present calibrations done during commissioning, as well as give more details on the photometric and astrometric methodology.
We describe: (A) the photometric calibrations; (B) the astrometric calibrations; (C) the rotational centering algorithm on the saturated and unsaturated $\beta$ Pic A PSFs; and (D) the grid search method to measure the high-contrast astrometry and photometry of $\beta$ Pic b.

\vspace{10pt}

\section{Photometric Calibration of Clio}\label{sec:app_photom}
The Clio instrument was moved from the MMT to Magellan for the MagAO project in 2012.
Here we describe calibrating Clio for MagAO at commissioning.

The calibrations that affect photometry are linearity, bad pixels, flat fielding, crosstalk, and persistence.
Here we determine the linearity correction and the bad pixel map.
We also show the attempted ``flat field'' images to explain why flat fielding was not performed on this data set.

\subsection{Linearity Calibration}
\begin{figure}[ht]
	\centering
	\includegraphics[width=\linewidth]{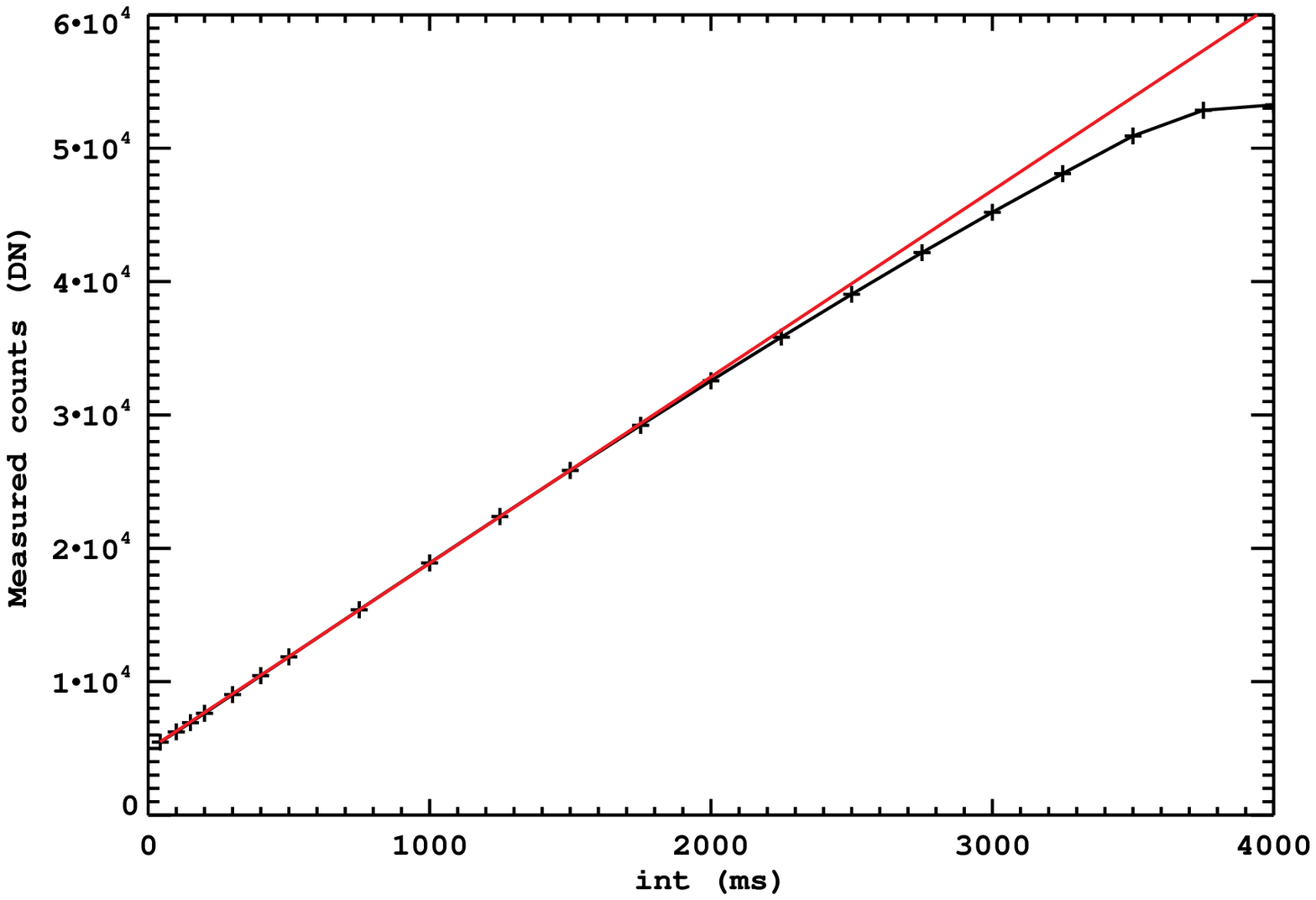}
	\caption{Best-fit line (red) to the raw linearity data (black).
		The detector is only strictly linear up to $\sim$27,000--30,000 DN.
		A linearity correction must be applied above this threshold.
		}
	\label{fig:line}
\end{figure}

We calibrated the detector linearity during the second commissioning run (2013 March 30th, UT), with the same detector temperature to within a milli-Kelvin ($\beta$ Pic data: 54.994 K; Linearity data: 54.999 K) and identical bias voltage (VOS voltage = 2.5 V in both cases) to the first commissioning run.  We obtained data in the 3.4-$\mu$m filter with the rectangle field stop in the wide camera, illuminating a large fraction of the detector while keeping a small fraction un-illuminated for reference.  We stepped up the integration time and recorded the median counts within the illuminated rectangle, for 5 images taken at each integration time.  We average the count rate for each integration time, and the standard deviation is found to be $<$0.5\%.  Figure~\ref{fig:line} plots the measured counts rate in data numbers (DN) against the integration time in ms.  The data are not linear, and need to be corrected.

\begin{figure}[ht]
	\centering
	\includegraphics[width=\linewidth]{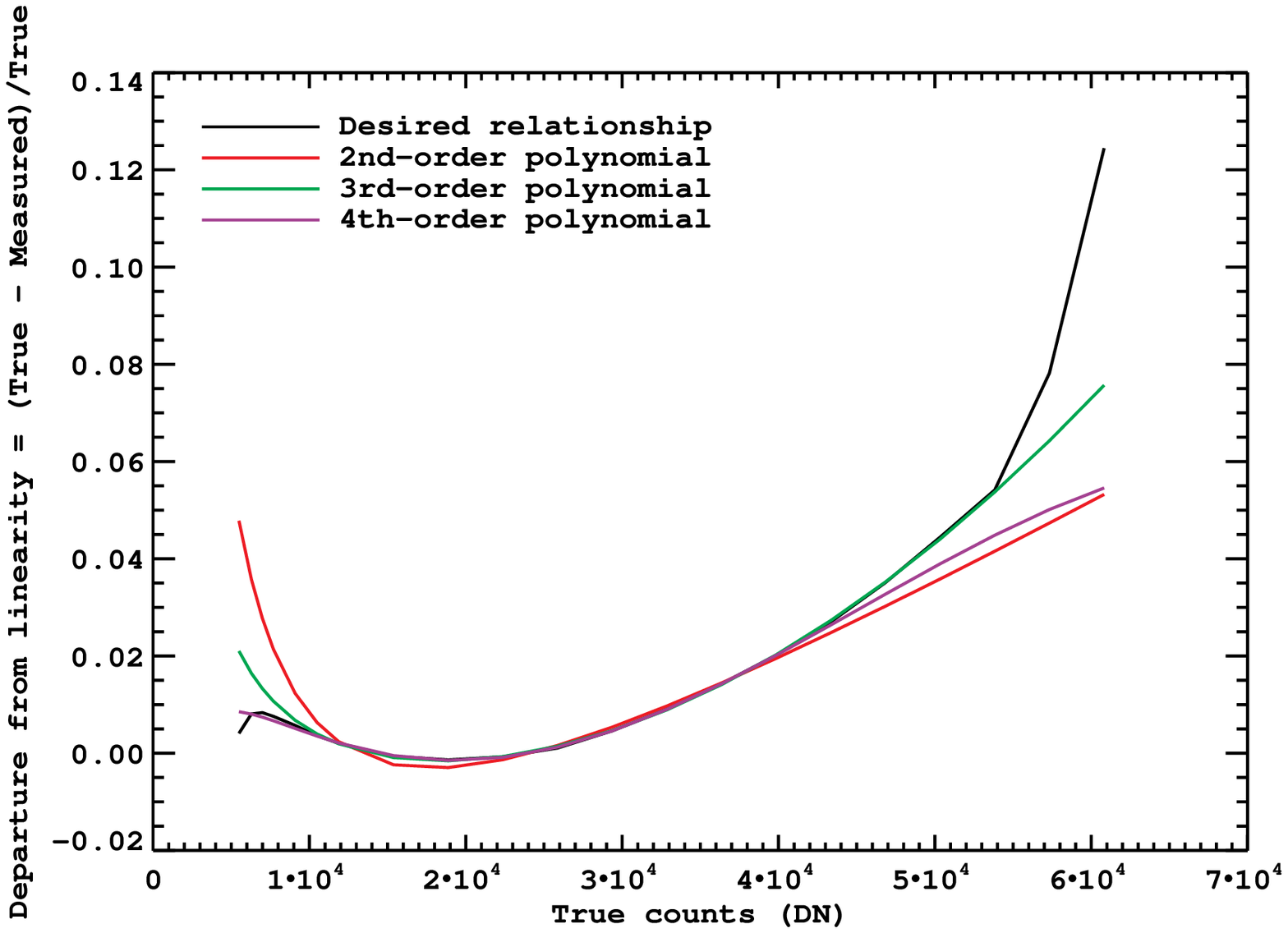}
	\caption{Best fit functions of 2nd (red), 3rd (green), and 4th (purple) order to the linearity data (black).  The y-axis is a metric for linearity which is equal to the difference of true counts and measured counts, divided by true counts.  The best fit is the 3rd-order polynomial (green), which can be applied to raw images for pixels above a count rate of 27,000 counts.}
	\label{fig:fit}
\end{figure}

``True'' counts are the count rate that give a linear relationship with integration time.  Departure from linearity is determined by subtracting true from measured counts, normalized by true counts --- plotted in Fig.~\ref{fig:fit}.  A series of polynomials are fit to this line, from 2nd order to 4th order, and the best fit is determined to be a 3rd-order polynomial above 27,000 counts.

\begin{figure}[ht]
\centering
	\includegraphics[width=\linewidth]{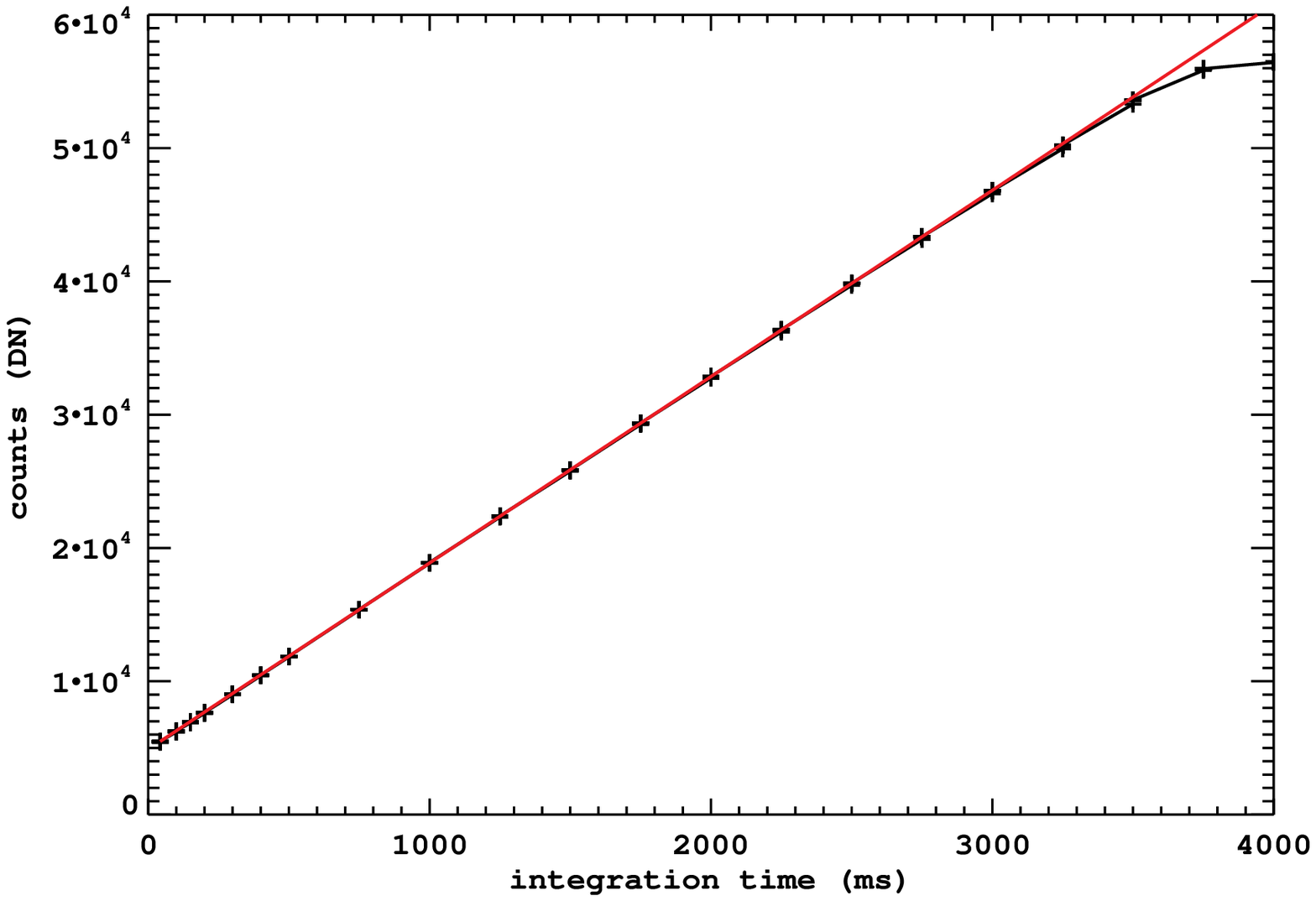}
		\caption{Linearity-corrected data using Eqtn.~\ref{eq:linearize} applied to pixels above 27,000 raw DN.  Counts up to $\sim$45,000 DN in the raw frames (which is $\sim$46,000 DN, corrected) can be corrected for linearity; avoid raw counts above this threshold.}
\label{fig:linresult}
\end{figure}

Based on this fit, the linearity correction formula is:
\begin{equation}\label{eq:linearize}
y = A + B x + C x^2 + D x^3
\end{equation}
where y is the true counts and x is the measured counts,
where the coefficients are:
A = 112.575; B = 1.00273; C = -1.40776e-06; and D = 4.59015e-11,
and where x must be $>$ 27,000 counts.
FITS files with coadded data must first be divided by the number of coadds before applying the linearity correction.
Figure~\ref{fig:linresult} shows the result of the linearity correction.  The data have been linearized up to $\sim$46,000 DN.

\subsection{Bad Pixels}
\begin{figure}[ht]
	\centering
	\includegraphics[width=\linewidth]{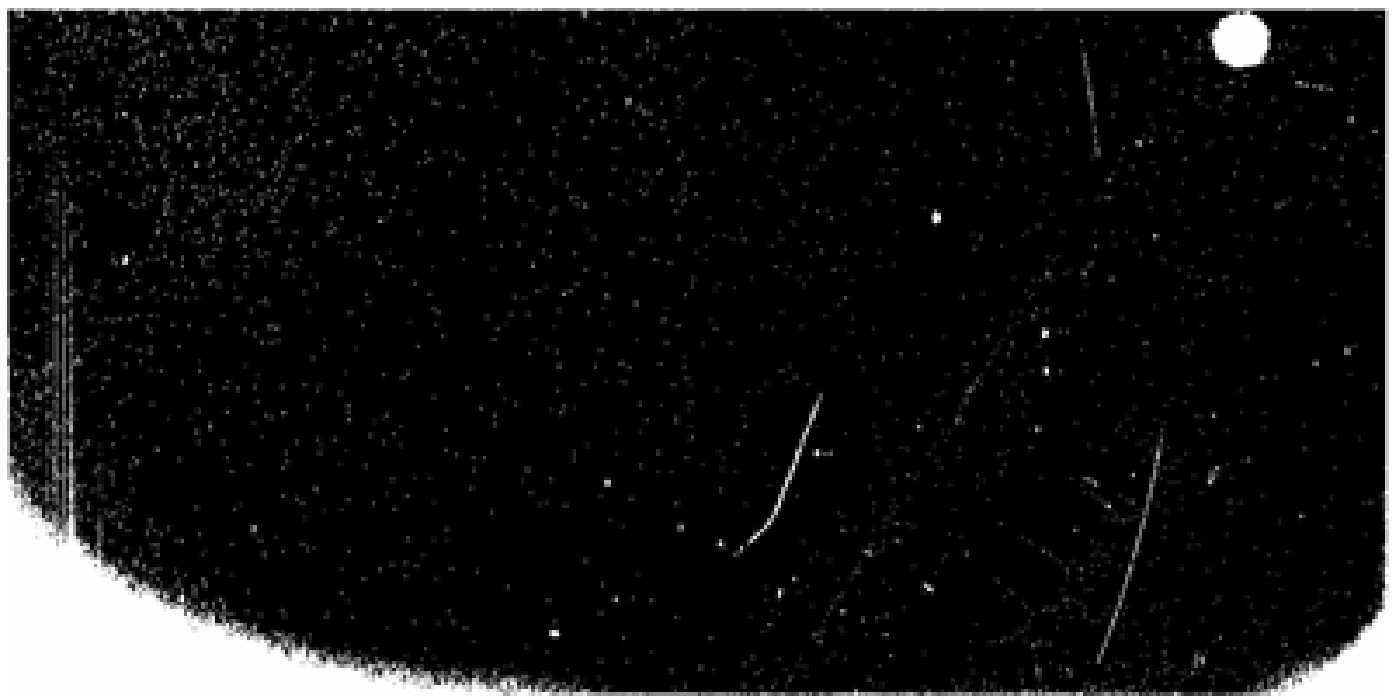}
	\caption{  \label{fig:badpix}
		Bad pixel map for Clio.  Black = good pixels.  White = bad pixels.
	}
\end{figure}

The Clio2 detector is a Hawaii-I HgCdTe with a deep well depth ($\sim$56,000 DN), and is thus ideal for high-contrast imaging.
The physical array has 1024 pixels on a side, but only 2 readout amplifiers are usable, bringing the array size to $1024 \times 512$.
Although the chip has poor cosmetics (see Fig.~\ref{fig:badpix}), we compensate in high-contrast imaging by greatly oversampling the PSF in the Narrow camera, allowing for better averaging over the bad pixels while also increasing the dynamic range.

We create the bad pixel map as follows.
Several short-exposure darks are averaged; pixels that are bright in these images are flagged as hot pixels.
Several long-exposure flats are averaged; pixels that are dark in these images are flagged as dead pixels.
The flagging threshold is a sigma-clipping chosen by inspection.
The hot and dead pixel maps are combined to make the bad pixel map, with a resultant 7\% of the pixels being flagged as bad.
Note that a high fraction of those are in two lower corners, which we avoid --- see Fig.~\ref{fig:badpix}.
The bad pixel map is available as a FITS file on the web in the Clio manual\footnote{\url{http://zero.as.arizona.edu/groups/clio2usermanual/wiki/6d927/ Calibration\_Data.html}}, in full frame and subarray versions.

\subsection{Other Photometric Calibrations}

\begin{figure}[ht]
	\centering
	\includegraphics[width=\linewidth]{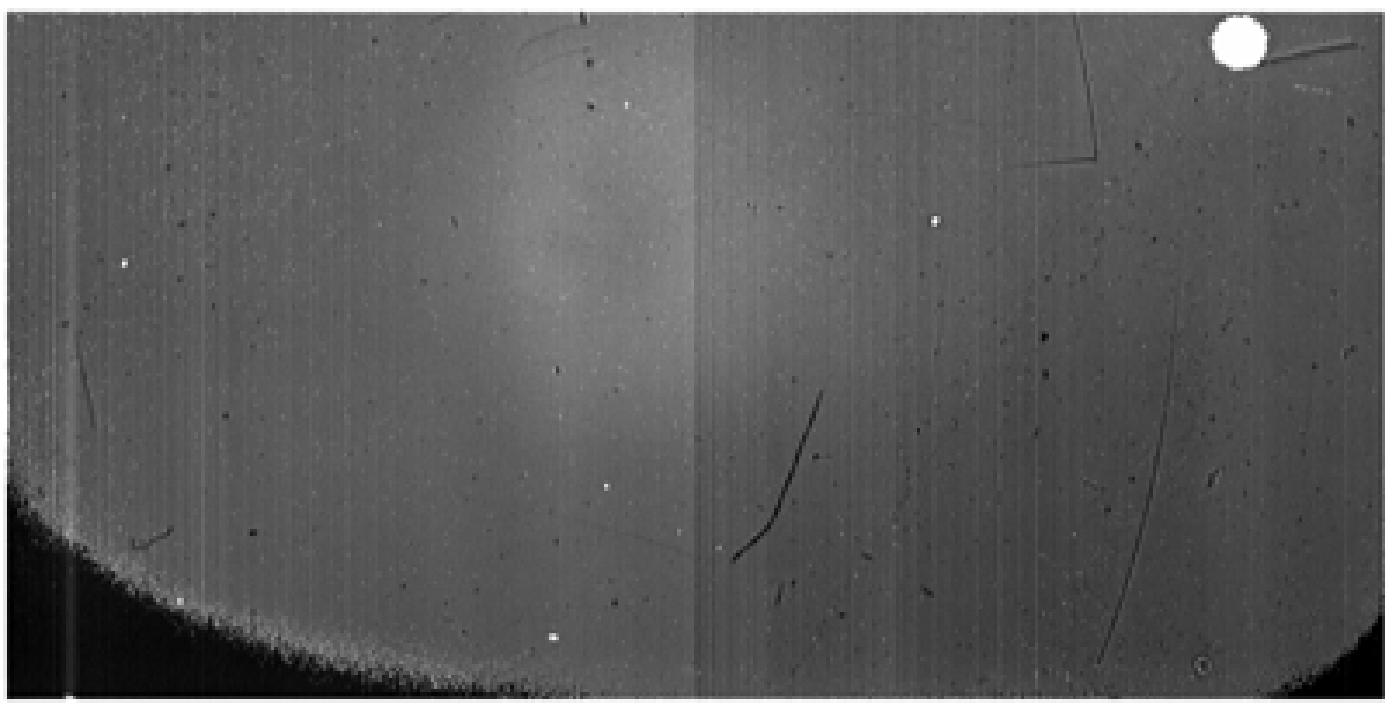}
	\includegraphics[width=\linewidth]{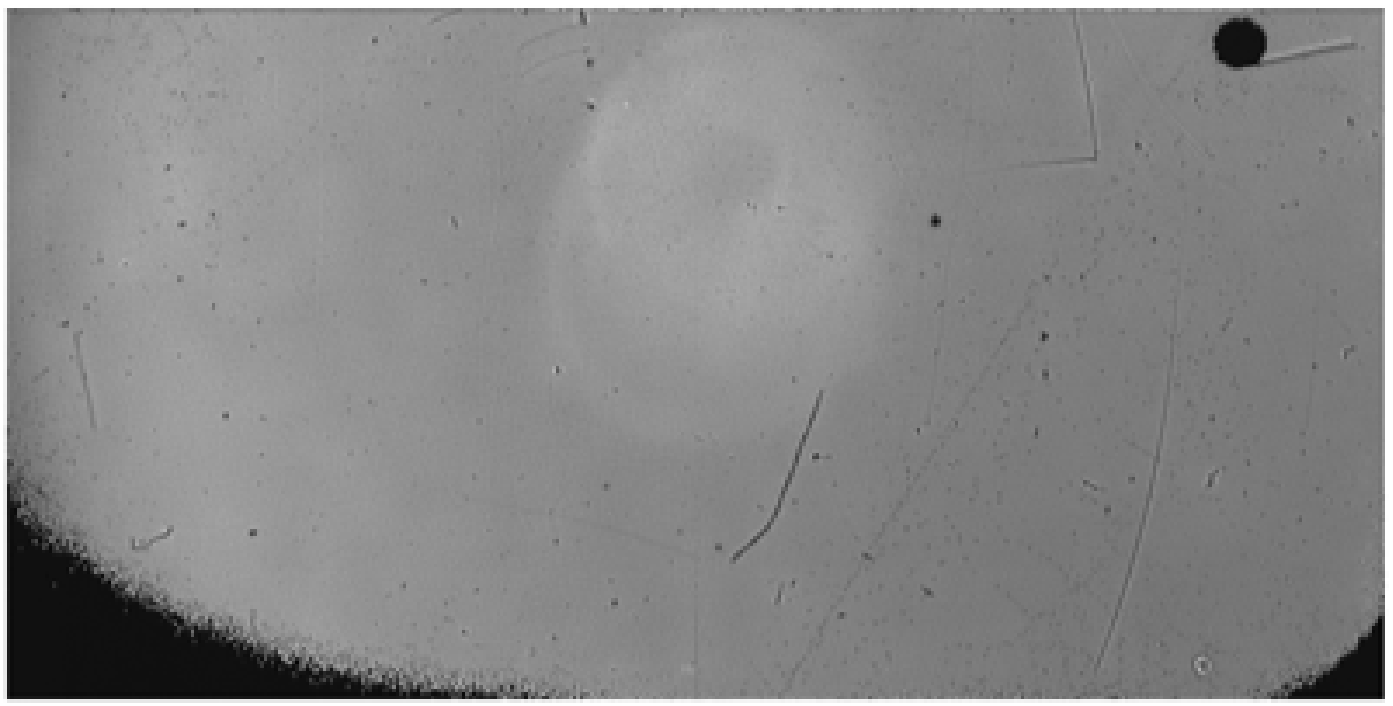}
	\caption{
		Best attempt at creating a flat, both displayed in linear scale.
		Top: Narrow-camera at 3.3~$\mu$m.
		Bottom: Wide-camera at $H$-band.
		\label{fig:flats}}
\end{figure}

\paragraph{Flat Fielding}
We have put significant effort into attempting to acquire a flat illumination of the Clio detector.  There is no flat-field screen in the dome that works with MagAO, so we took twilight flats, either at dusk or dawn.  We took flats with a constant integration time as the sun was setting or rising, to vary the illumination level.  We tried different integration times.  We tried subtracting darks at the same integration time, as well as flats with less solar illumination at the same integration time.  We tried all the different filters and both cameras.  In all cases, we are not able to remove the ``glow'' in the top center of the images (see Fig.~\ref{fig:flats}).
This photometric variation is on the order of 10\% across the field for the wide camera. It looks like an out-of-focus image of the pupil, and we posit that it is either caused by a pinhole of light leaking through, or an imperfect anti-reflective coating allowing the last concave optic to focus a glow onto the detector.

In any case, this image is not usable as a flat, because the pixels that are bright in the pupil-glow portion of the ``flat'' are \emph{not} more sensitive to light, and therefore the raw data should \emph{not} be divided by this image.
Therefore, rather than flat-fielding our data, we depend on the sky subtraction to remove pixel-to-pixel variation in the sky images.
Furthermore, the [3.1], [3.3], and $L^\prime$ images were all saturated, so the variation in the unsaturated PSF (after nodding and dithering around the detector) folds in the flat-fielding errors.
However, we must add a ``flat-fielding'' error term for the $M^\prime$ images, because all the images were unsaturated; thus, the simulated planet is unique to each image and has no variation that could include a flat-fielding error.  We find this by the variation in the peak of the nodded and dithered PSFs throughout the data set.

\paragraph{Crosstalk}
There is electronic cross-talk between the two amplifiers, that manifests as a negative image of the star, located exactly 512 pixels away, with a flux of 0.7\% that of the original image.  We avoid this crosstalk with our nod pattern; because $\beta$ Pic b is only $<$50 pixels from the star, it was not a problem.

\paragraph{Persistence}
Persistence was observed in dark images taken at the end of a long night including hours of $\beta$ Pic observations, an extremely bright star that was saturated for most of the observations.  However, we do not typically record or correct for this effect, which was negligibly small.

\section{Astrometric Calibration of Clio}\label{sec:app_astrom}

$\beta$ Pic b is much fainter in the optical as compared to the infrared. Therefore, to improve the confidence of the detection and photometry on VisAO (which has 8-mas pixels) in Paper I, we must report with confidence the subpixel-position of the planet on Clio, as seen during simultaneous observations. Our goal is to measure the position of $\beta$ Pic b relative to A, at the epoch of observation, with absolute astrometric accuracy at the mas level. Toward this end, we carried out a set of astrometric calibrations for the Clio cameras.

\subsection{Trapezium Observations}
We observed the Trapezium Cluster in the Orion Nebula as a reference for our platescale, instrument angle, and distortion.
Over the nights of 2012 Dec.\ 3rd, 4th, 6th, and 8th, we observed various Trapezium fields.
We locked the AO loop on either $\theta ^1$ Ori B or C.  Stars $\theta ^1$ Ori A and E were also in the fields, as were many fainter Trapezium stars, as listed in Tab.~\ref{tab:trapobs}.

\begin{table*}[ht]
\footnotesize
	\caption{Trapezium observations, December 2012}
	\label{tab:trapobs}
	\begin{center}
	\leavevmode
\begin{tabular}{lllll} \hline \hline
Filter		& Camera	& Epoch	& Frame \#s	& $\theta ^1$ Ori stars in field$^{\dagger, \ddagger}$ \\
\hline
$J$		& Narrow	& 2012-12-03	& 472--491	& C only (100-pixel stamp subarray)
\\
		&	 	&			& 492--596	& A1, E, 43, 48, 59, 63, (A2) \\
$K_s$	& Wide	& 2012-12-03	& 1--140		& A1, B1, C, E, 43, 48, 49, 59, 63, 70, 73, (A2, B2, B3, B4, 72) \\
		& 		& 2012-12-06	& 1--20		& B1, E, 48, 49, 78, 81, (B2, B3, B4) \\
		& 		& 2012-12-08	& 1--10		& A1, C, E, (A2, \ldots) \\
		& Narrow	& 2012-12-03	& 141--471	& A1, E, 43, 48, 59, 63, (A2) \\
		& 		& 2012-12-04	& 1--40		& B1, 49, (B2, B3, B4, 35) \\
		& 		& 2012-12-08	& 91--124		& A1, E, 38, 43, 59, 63, (A2) \\
		& 		& 2012-12-08	& B 1--10		& B1, E, (B2, B3, B4, \ldots) \\
$[3.1]$	& Wide	& 2012-12-08	& 1--20		& A1, C, 39, 46, 59, 63, 70, 73, (A2, 72) \\
$[3.3]$	& Narrow	& 2012-12-08	& 21--40		& A1, E, 43, 48, 59, 63, (A2) \\
$L'$		& Narrow	& 2012-12-08	& 41--60		& A1, E, 43, 48, 59, 63, (A2) \\
$M'$		& Narrow	& 2012-12-08	& 61--90		& A1, E, 63, (A2) \\
\hline
\hline
\multicolumn{5}{l}{}											\\ 
\multicolumn{5}{l}{$^\dagger$ See \citet{trapezium_astrometry} for star identification (A1=45, B1=60, B2=56, C=68, and E=40).} \\
\multicolumn{5}{p{\linewidth}}{$^\ddagger$ C is not resolved in the Clio images and is listed singly; A1-A2 and B1-B2-B3-B4 are resolved as listed.  Stars in parantheses were not used as part of the astrometric solution, due to either having a known high proper motion (72), being too faint to centroid well on (35), being a non-round proplyd (74), or being a close companion with contamination and/or orbital motion creating centroiding and positional uncertainty (A2, B2, B3, B4).} \\
\end{tabular}
	\end{center}
\normalsize
\end{table*}

\subsection{Fiducial Trapezium Positions}
\citet{trapezium_astrometry} produced a comprehensive work on absolute astrometry of the Trapezium cluster, including tying the infrared observations to their radio counterparts, and resulting in an absolute precision of 30 mas rms.  This work is often used as a fiducial for referencing the positions of Trapezium stars, but it is now two decades out of date.  Nevertheless, we attempt an astrometric calibration referenced to \citet{trapezium_astrometry}, as a cross-check to another method.

We start with the $K_s$ Wide-camera images of 2012-12-03 (with the most stars in a single frame) and measure the separations of each pair.  
After reducing the data in the standard way, we find the centroids for each of the stars listed excluding A2, B2, B3, and B4, giving a total of 12 stars. (We exclude the companions to A1 and B1 because, while they are well-resolved, they show the most orbital motion and have the least certain positions.)
Comparing these measurements to those of \citet{trapezium_astrometry} gives a platescale and instrument angle estimate for each pair of stars.

We also use the fiducial separations in a more recent work: that of \citet{close2012}, who measured Trapezium positions with near-IR first-light LBT AO with the Pisces camera in Nov.\ 2011.
We find discrepancies in separations at the 10-mas level over 2\arcsec\ between these two fiducials. To determine whether the discrepancy is due to proper motion or to distortion in Clio, we take the 12 stars as tie points to de-warp the images using IDL's \textit{warp\_tri}. Repeating the process on the undistorted images still results in separation discrepancies of up to 5 mas over 2\arcsec.
We attribute the difference to true astrophysical variation such as orbital and proper motion in the intervening two decades, and find that the \citet{trapezium_astrometry} positions are now out-of-date for mas-level astrometry.
Therefore, we use the \citet{close2012} positions as fiducials for the distortion correction that follows, in which we use all our Trapezium observations of Dec.\ 2012.

\subsection{Determining the Distortion}
\citet{close2012} used a pinhole grid to calibrate distortion on \textit{LBT/Pisces}, and used \textit{HST} images of the Trapezium \citep{ricci_trapezium} for calibrating the instrument angle and platescale.  Therefore, tying our distortion correction to \cite{close2012} is one step away from tying it to a pinhole calibration.

\begin{figure}[htb]
\centering
	\includegraphics[width=\linewidth,trim=0.5cm 0.4cm 0.5cm 3.8cm,clip=true]{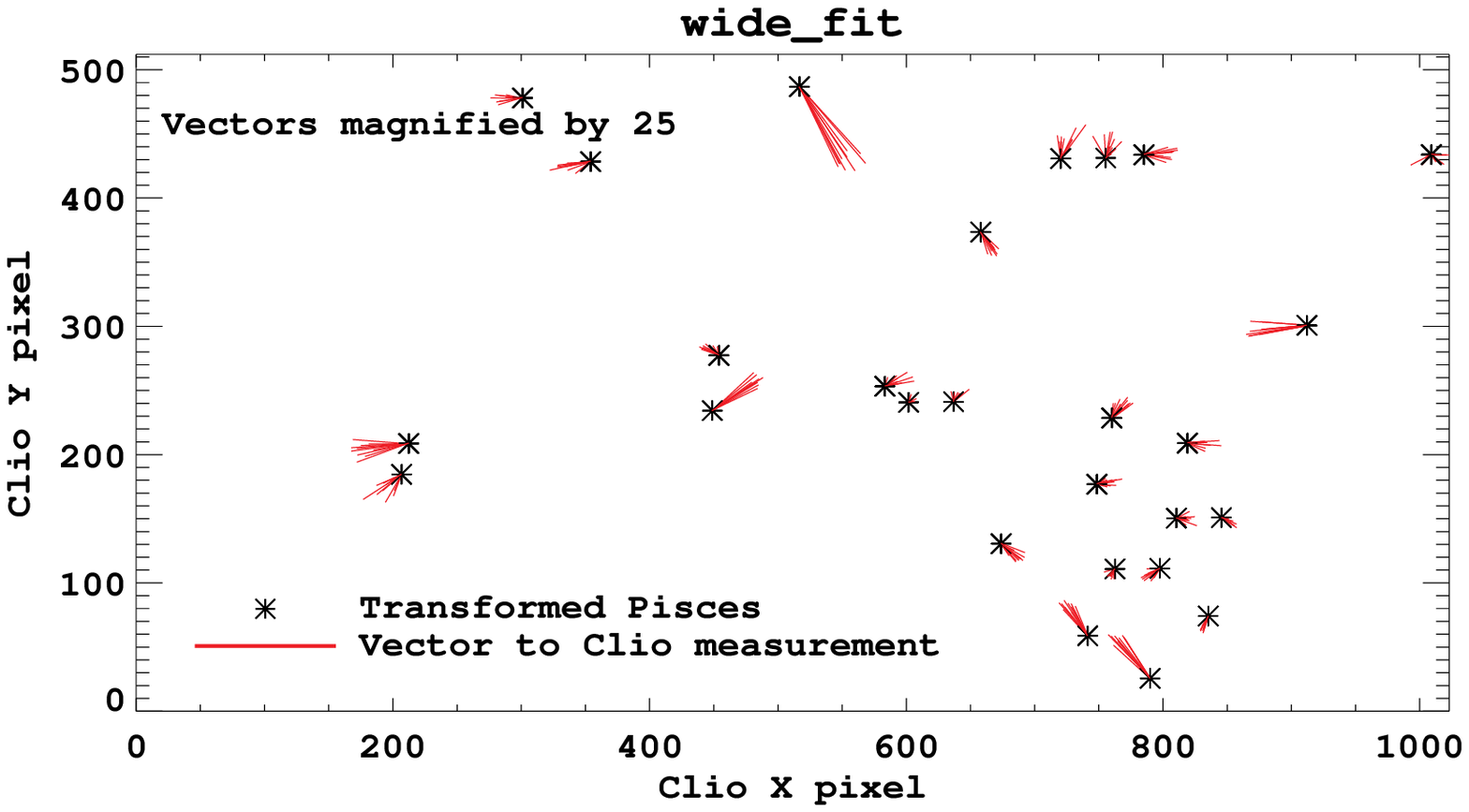}
	\includegraphics[width=\linewidth,trim=0.5cm 0.4cm 0.5cm 3.8cm,clip=true]{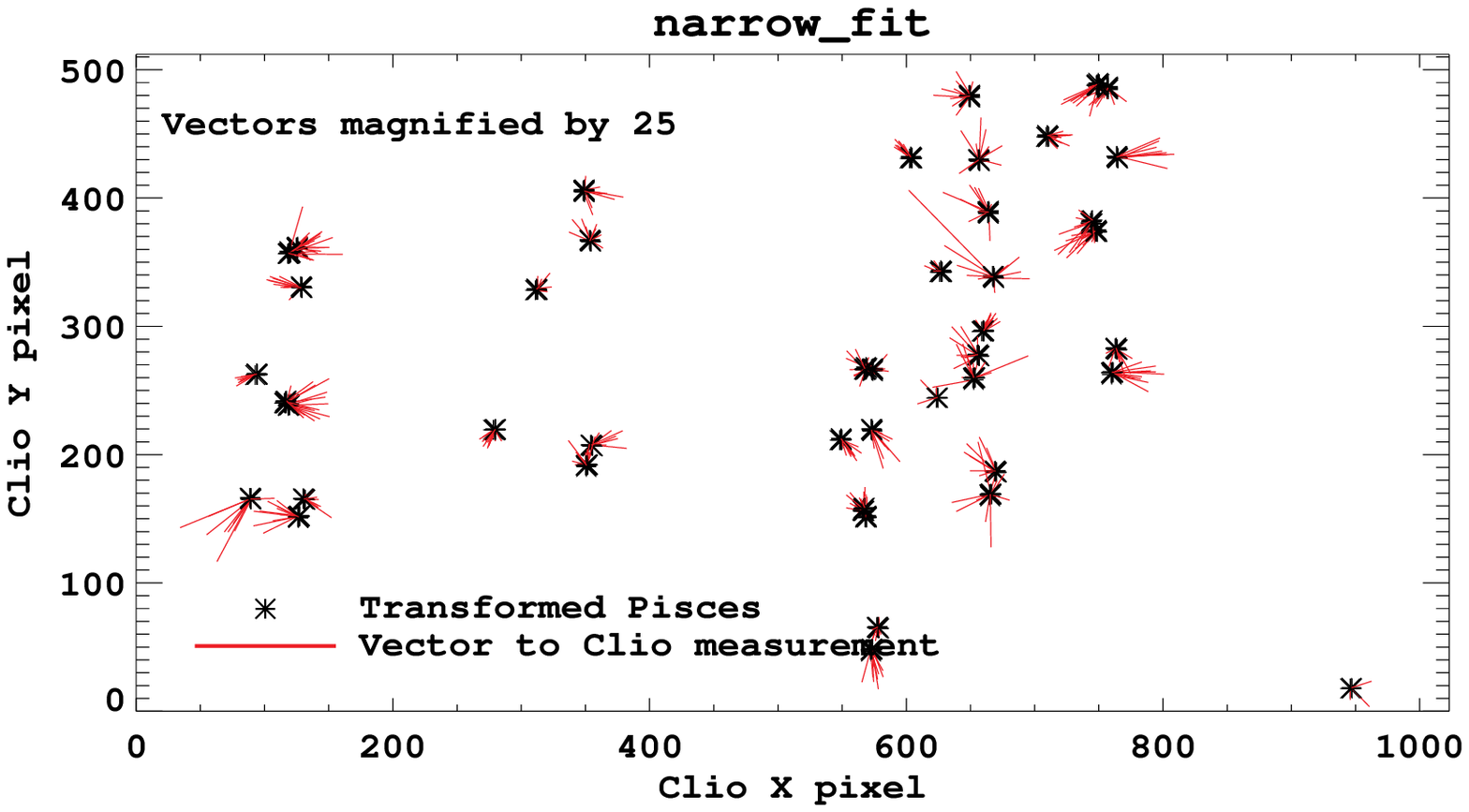}
	\caption{
		Vectors of measured to true positions for Trapezium stars, in the Wide (top) and Narrow (bottom) cameras.
		\label{fig:distortion}}
\end{figure}

To compute the distortion solution, we compare our Trapezium observations to the first-light LBTAO/Picses observations \citep{close2012}.
We group our data sets by camera and filter (see Tab.~\ref{tab:trapobs}).
For each set of [camera, filter] we have a series of frames dithered around the Trapezium cluster.  A single frame does not have sufficient numbers of stars to get a good astrometric calibration.  Therefore, we solve an inverse problem using \textit{MPFIT} in IDL \citep{markwardt} 
to determine the shifts, scales, and rotations that transform the Pisces coordinates to Clio coordinates for each image.
The output are $x$ and $y$ shifts $[dx, dy]$ for each position on the array that imaged a star.
While we perform the fit for each set of camera and filter, the distortion is not significantly different for the various filters, so in the end we combine the data into one group for each camera.
The distortions are shown as red vectors in Fig.~\ref{fig:distortion}.

\begin{figure}
	\centering
	\leavevmode\epsfxsize=\linewidth\epsfbox{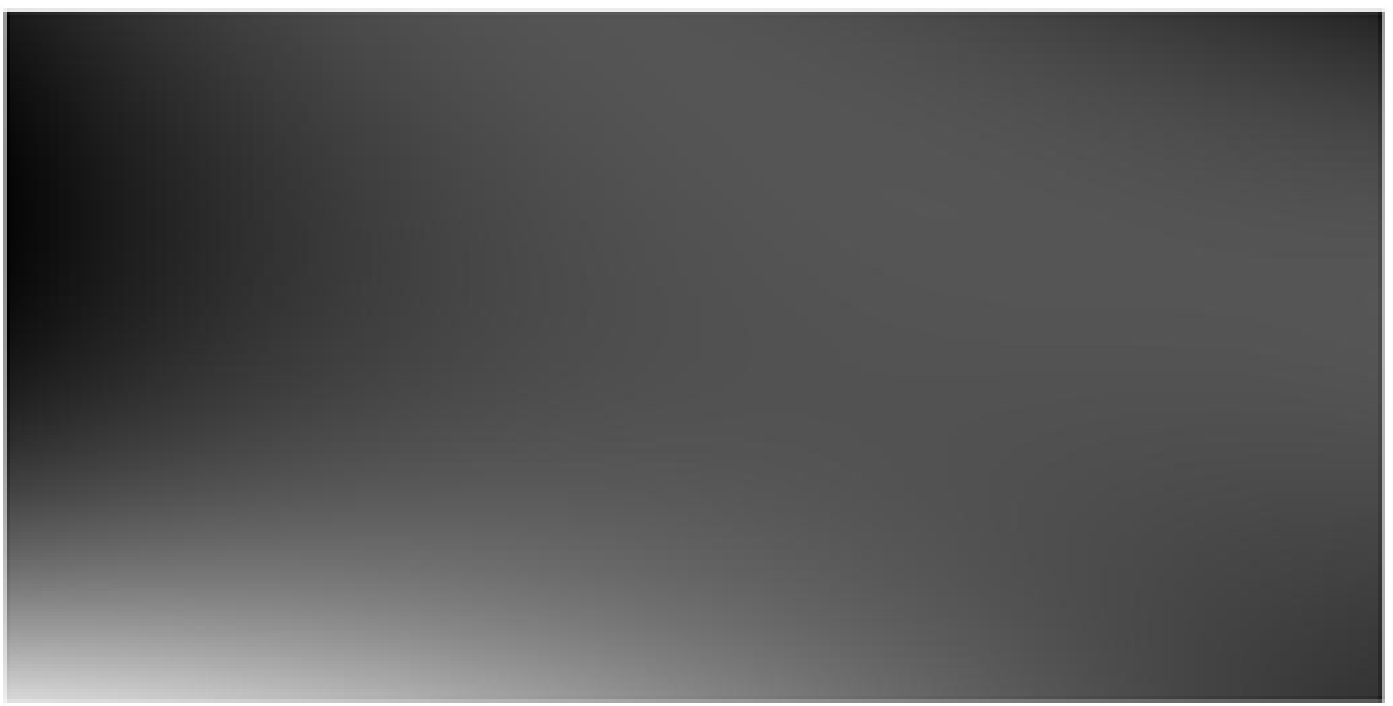}
	\leavevmode\epsfxsize=\linewidth\epsfbox{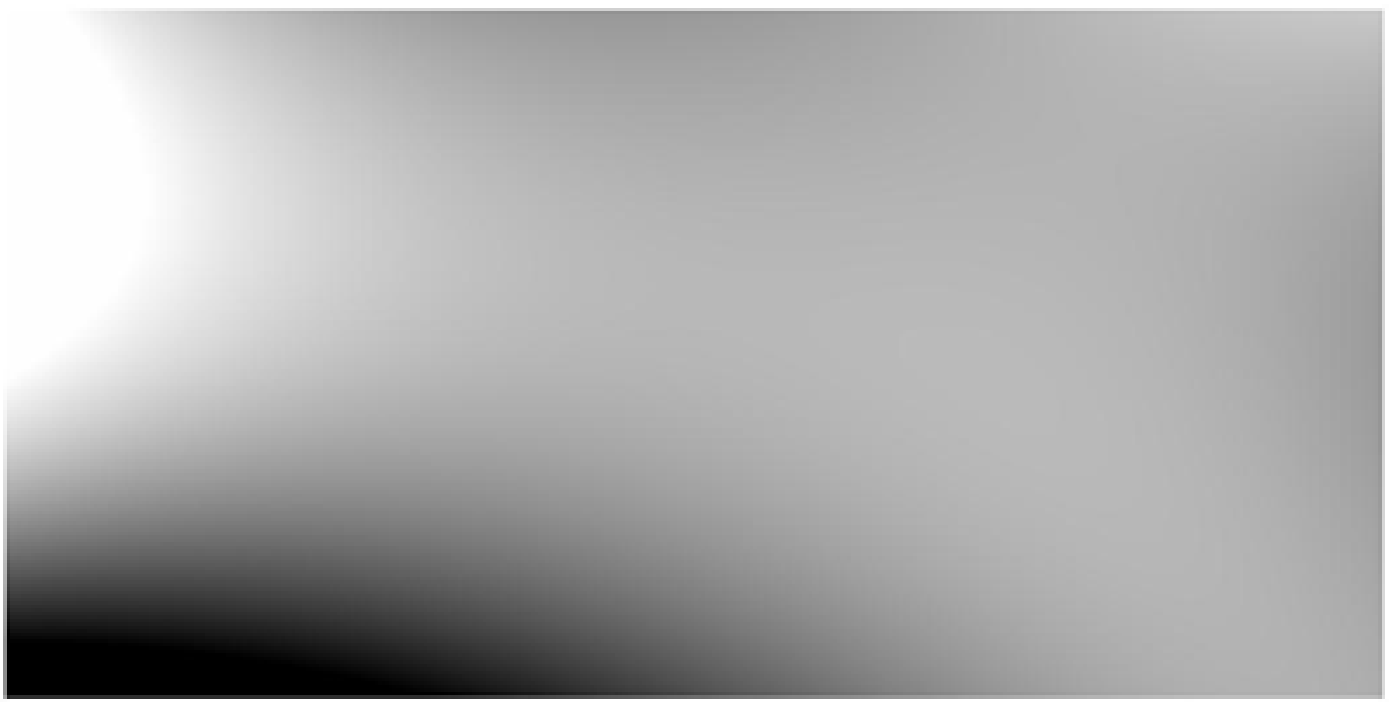}
	\caption{
		dx (top) and dy (bottom) that correct distortion for the Wide camera.
		\label{fig:wide}}
\end{figure}

\begin{figure}
	\centering
	\leavevmode\epsfxsize=\linewidth\epsfbox{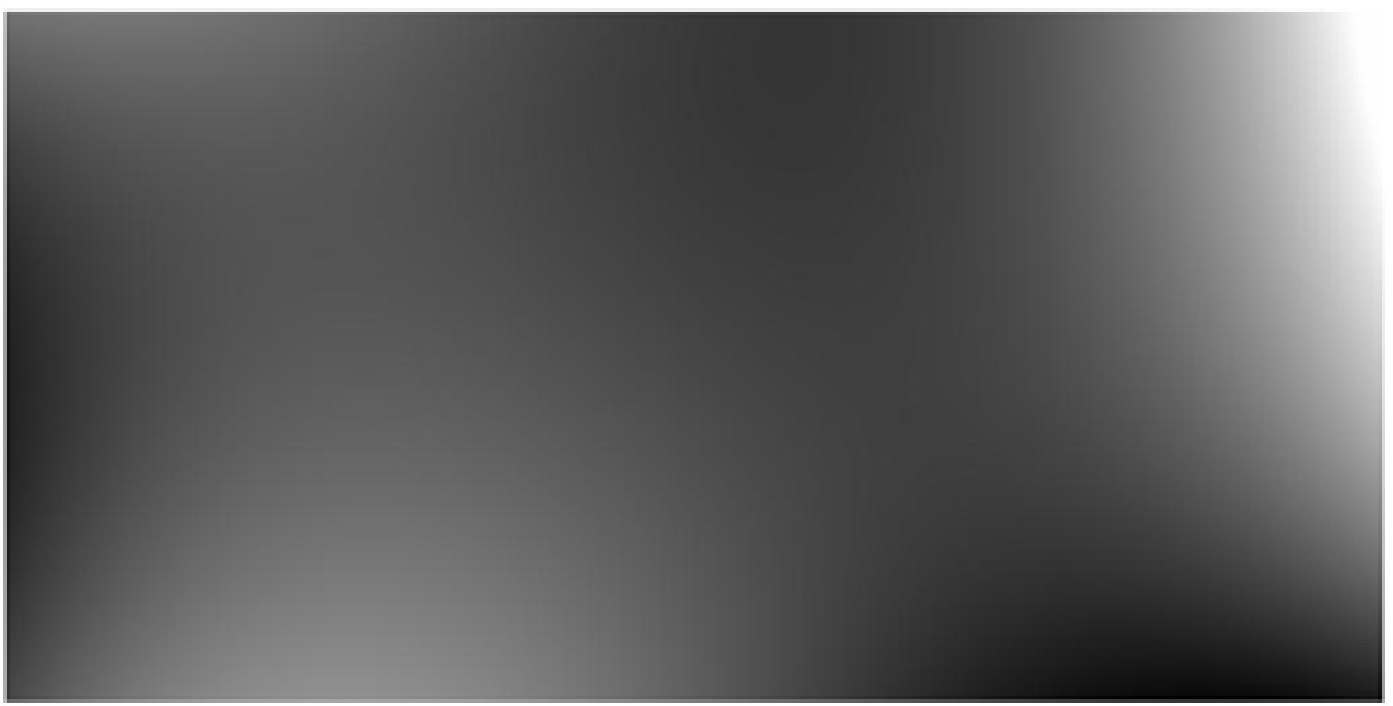}
	\leavevmode\epsfxsize=\linewidth\epsfbox{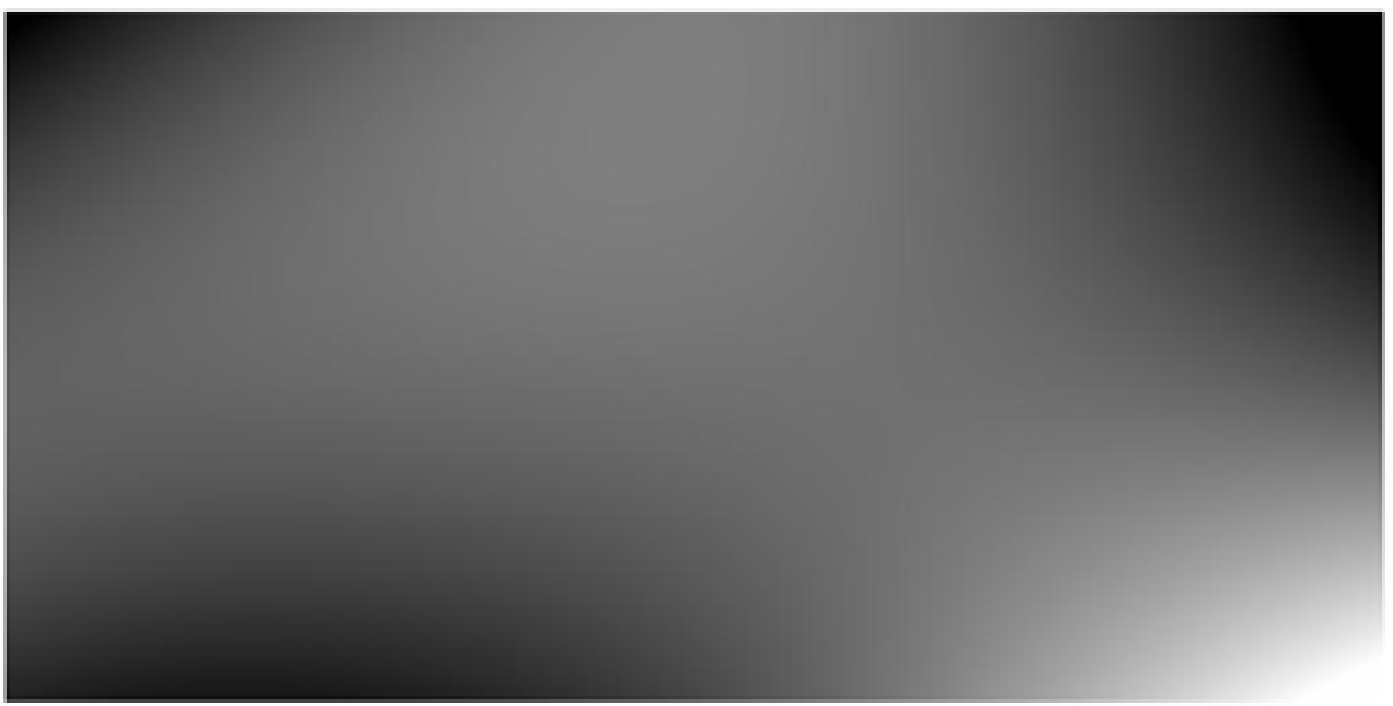}
	\caption{
		dx (top) and dy (bottom) that correct distortion for the Narrow camera.
		\label{fig:narrow}}
\end{figure}

These shifts $[dx, dy]$ are fit to a 3-dimensional polynomial to create a smooth surface using \textit{MPFIT2DFUN} \citep{markwardt}.
These are shown in Fig.~\ref{fig:wide}--\ref{fig:narrow}.
These images are saved as FITS files for input into the IRAF \textit{Drizzle} package \citep{drizzle}.

\subsection{Undistorted Platescale}
Finally, we use the distortion correction to get a best value for the platescale and instrument angle.
We correct the distortion in the images and determine the platescale and instrument angle by comparing pairs of stars to the fiducial separations and position angles.
After correcting the Trapezium data for distortion, the resulting platescale and instrument angle are reported in Tab.~\ref{tab:platescale}.

\begin{table}[htb]
	\scriptsize
	\caption{Astrometric calibrations, December 2012}
	\label{tab:platescale}
	\begin{center}
		\begin{tabular}{llllll} \hline \hline
			&	&		& Meas.	& Fid.	& Total \\
			Quantity & Value & Unit	 & Error & Error & Error \\
			\hline
			Platescale, Narrow					& 15.846	& mas/pix.		& 0.043				& 0.047		& $\pm$ 0.064	\\
			Platescale, Wide						& 27.477	& mas/pix.		& 0.071				& 0.047		& $\pm$ 0.085 \\
			$\text{NORTH}_\text{Clio}$$^\dagger$	& -1.797	& \degr		& 0.159				& 0.3			& $\pm$ 0.34 \\
			\hline
			\hline
			\multicolumn{6}{l}{}											\\ 
			\multicolumn{6}{l}{$^\dagger$Derotation Angle: 
				$\text{DEROT}_\text{Clio} = \text{ROTOFF} - 180 + \text{NORTH}_\text{Clio}$
			} \\
			\multicolumn{6}{l}{A positive angle is counterclockwise to get North up \& East left.}
		\end{tabular}
	\end{center}
	\normalsize
\end{table}

Measurement error is the sample standard deviation of our 10 to 16 frames with 6 to 12 stars per frame. Fiducial error is the uncertainty in the reference positions, as given in \citet{close2012}. Errors are added in quadrature with measurement error to obtain the total astrometric calibration error.
The instrument angle ``NORTH Clio'' is used to derotate the images counter-clockwise as given in the table.

\section{Subpixel Centroiding of the Clio PSF}	\label{sec:app_cent}

Now that we have calibrated the photometry and astrometry of Clio, we turn to measuring photometry and astrometry of $\beta$ Pic b.  For measuring the flux and position of the planet, centroiding on the star is critical:
our PSF-subtraction method is based on rotation and thus depends on all PSFs being registered to sub-pixel accuracy in order to ensure the same center of rotation.
Furthermore, separation and position angle of the planet are sensitive to centroiding because they are referenced to the center of the star.
Yet the broad saturated PSF means centroiding is challenging at the sub-pixel level.

We use the IDL function \textit{rot} for all rotations, so therefore we choose to work at its preferred center as our point of origin.  For an image of dimensions $nx \times ny$, the central point about which \textit{rot} rotates is located at $[cx,cy]$, where $cx=(nx-1)/2$ and $cy=(ny-1)/2$, and where the pixel in the lower-left corner of the array has the coordinate $[0,0]$, at the lower-left corner of the pixel.  All of the arrays used here have even-numbered dimensions.

\begin{figure}[ht]
\centering
	\includegraphics[width=0.49\linewidth]{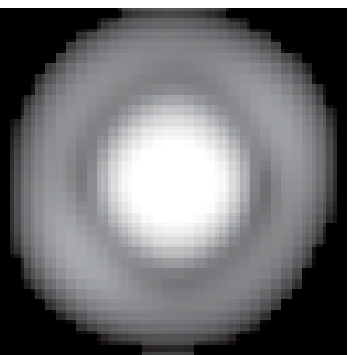}
	\includegraphics[width=0.49\linewidth]{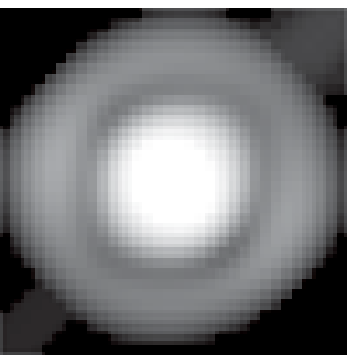}
	\caption{
	Result of sub-pixel rotational centering.
	Left: Centered PSF.
	Right: Rotated by 45 degrees using IDL \textit{ROT}, to illustrate rotational symmetry.
	Log scale.
	\label{fig:centroid_result}
	}
\end{figure}

Our procedure is to find the point $[cx,cy]$ that minimizes the residuals in the difference images when subtracting images rotated by all angles from 0\degr\ to 360\degr.  In practice, we choose a center of rotation, subtract images rotated in 10-degree increments from 5\degr\ to 355\degr\ from the upright image, and iterate on the center of rotation until the residuals are minimized.
We repeat the procedure with smaller pixel step sizes for the center of rotation, iterating until we have centration to 0.005 pixel accuracy.
Figure~\ref{fig:centroid_result} shows the result of sub-pixel rotational centration for a saturated PSF at 3.3~$\mu$m.

\section{Grid Search Method to Determine the Photometry and Astrometry of $\beta$ Pic \lowercase{b}} \label{sec:app_grid}

Finally, once the images are centered to 0.005 pixel accuracy, we subtract modeled planets to determine the photometry and astrometry of the true planet.
Here we describe the grid search and the metric for best guess used to determine the photometry and astrometry of $\beta$ Pic b.

We choose a value for the $[x,y]$ position and flux ratio $f$.
We create a model for the planet at this chosen value, by scaling the unsaturated PSF in flux $f$ and in translation to position $[x,y]$.  The modeled planet $[x,y,f]$ is subtracted from each reduced image.  All of the images with the chosen planet subtracted are then run through our KLIP pipeline.
To determine how well the planet was subtracted out, we calculate the residuals in a fixed location in all trials (``uniform'' mask), and in a location just around the planet's modeled position (``local'' mask).  
The metric we use to compare trials is the standard deviation of the residuals.
The modeled planet $[x,y,f]$ that minimizes the residuals is the best fit, giving the flux and position of the planet.

\begin{figure*}[ht]
\centering
	\includegraphics[width=0.33\linewidth]{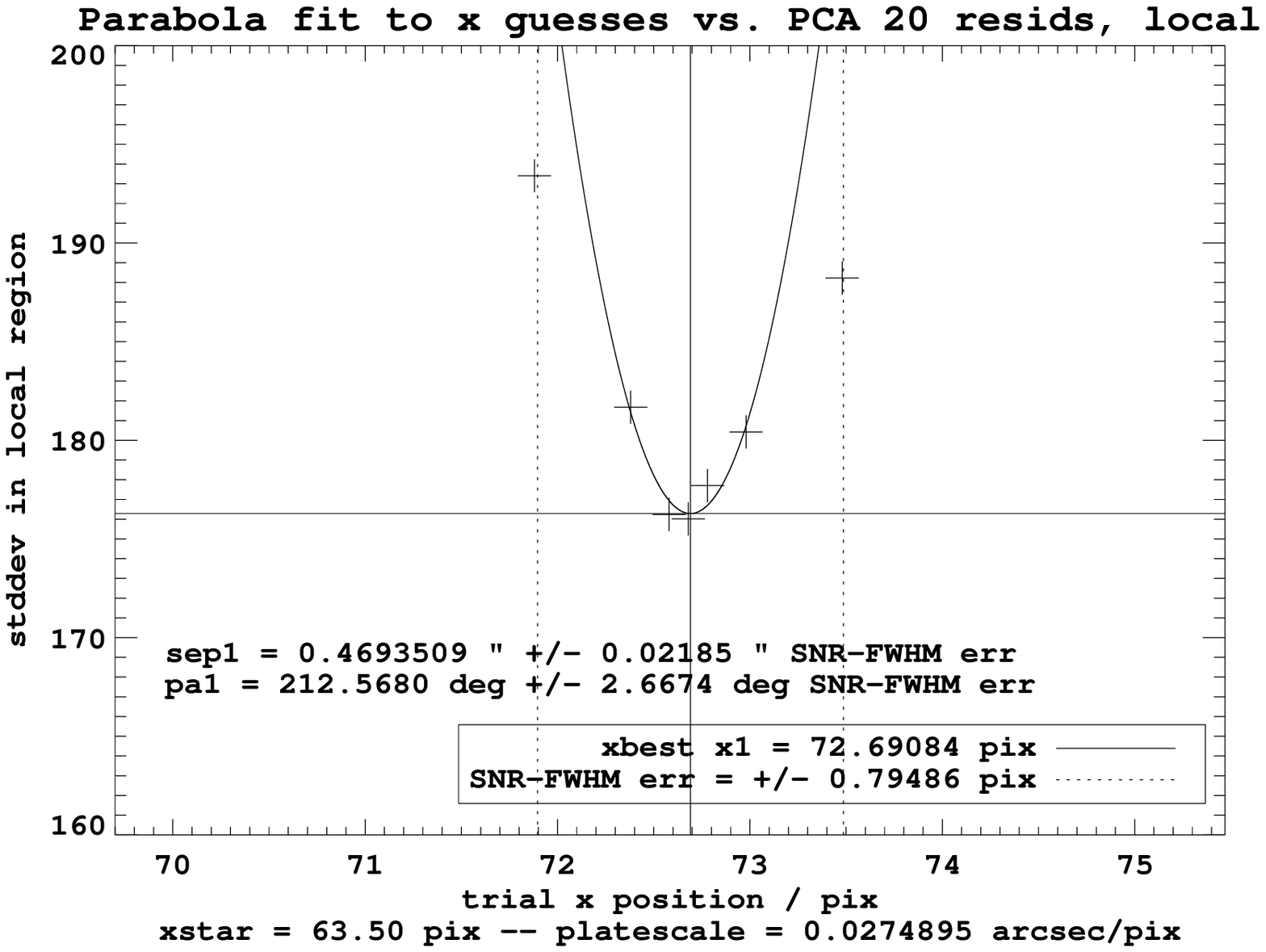}
	\includegraphics[width=0.33\linewidth]{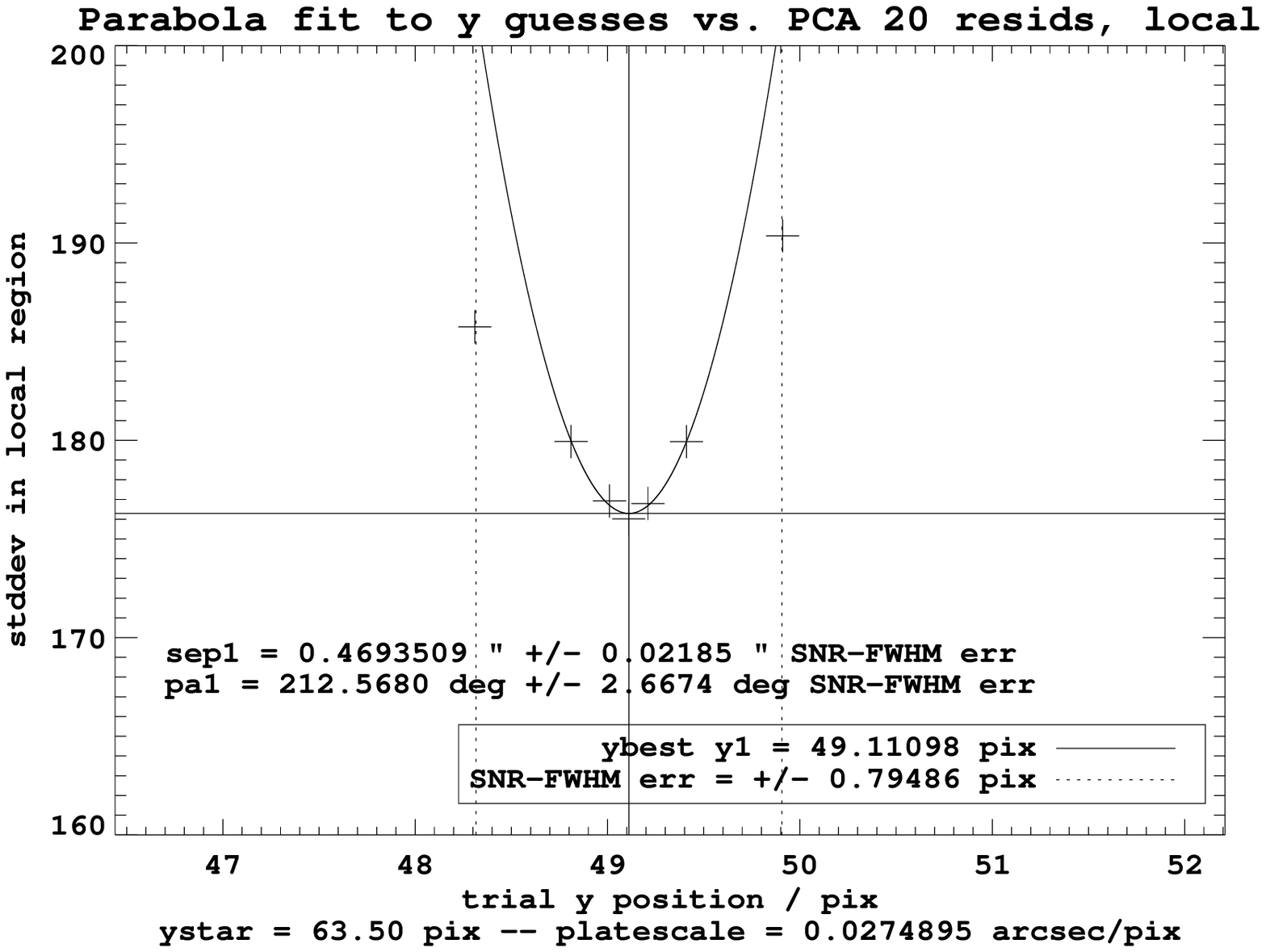}
	\includegraphics[width=0.33\linewidth]{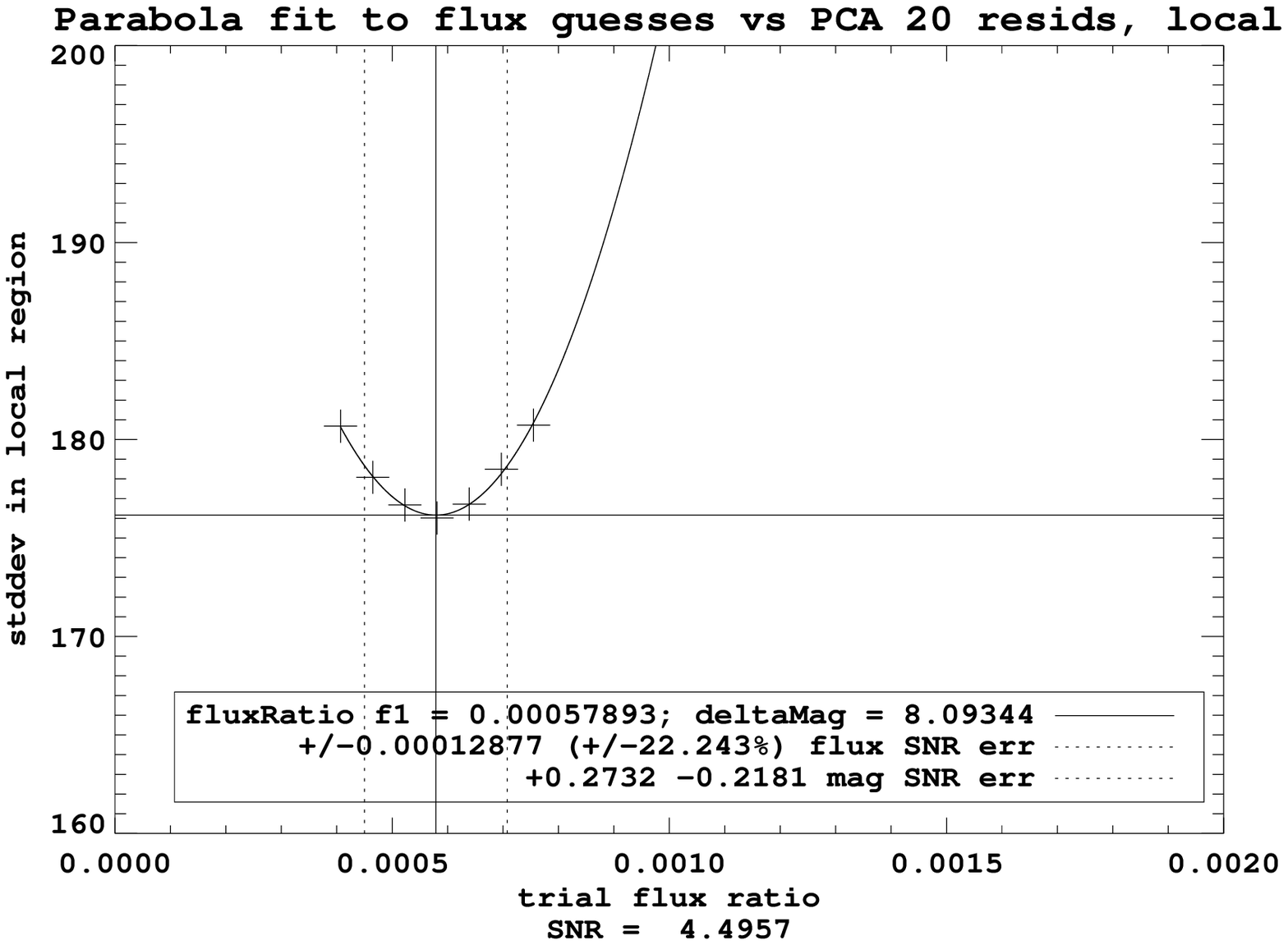}
	\includegraphics[width=0.33\linewidth]{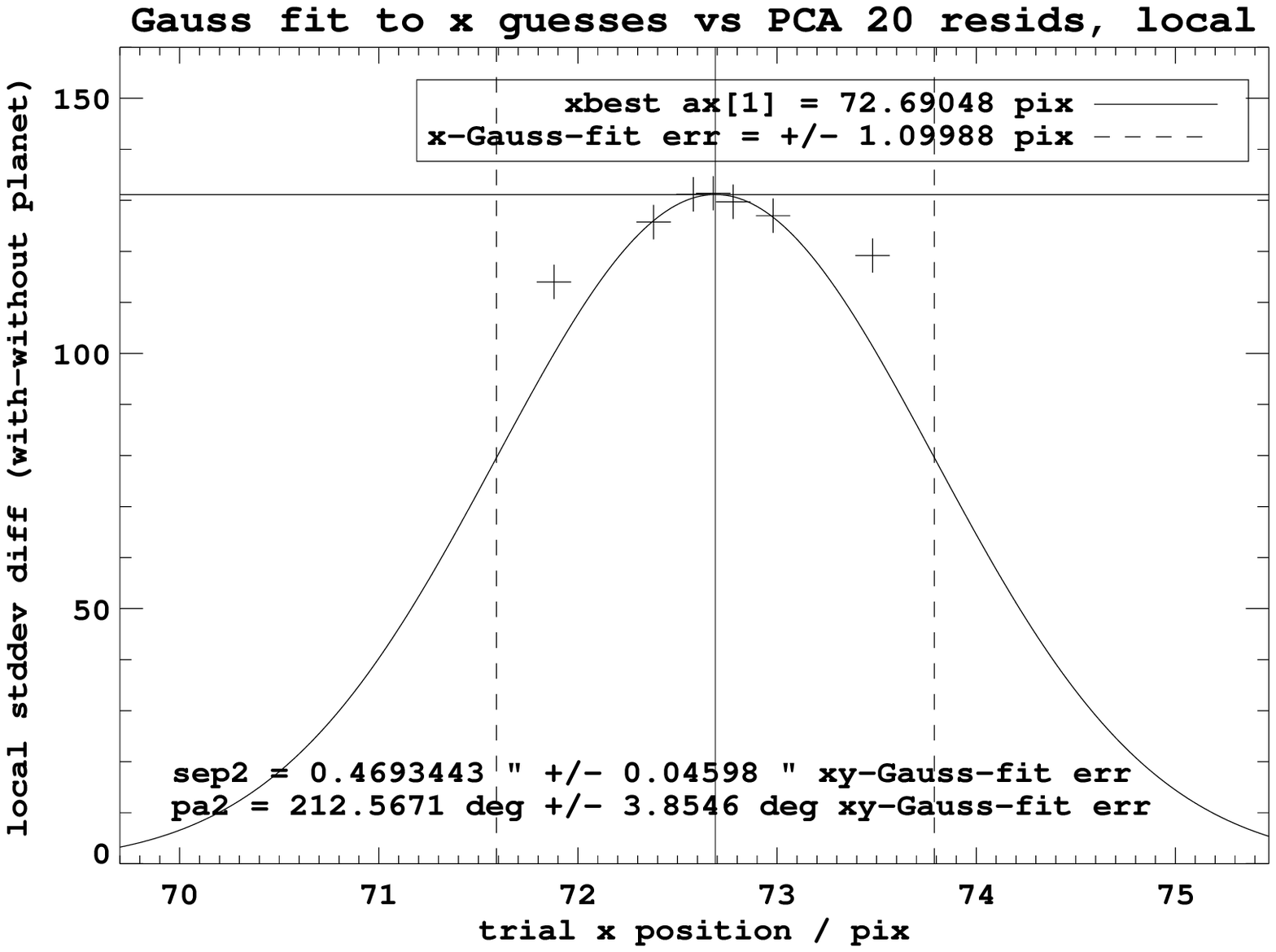}
	\includegraphics[width=0.33\linewidth]{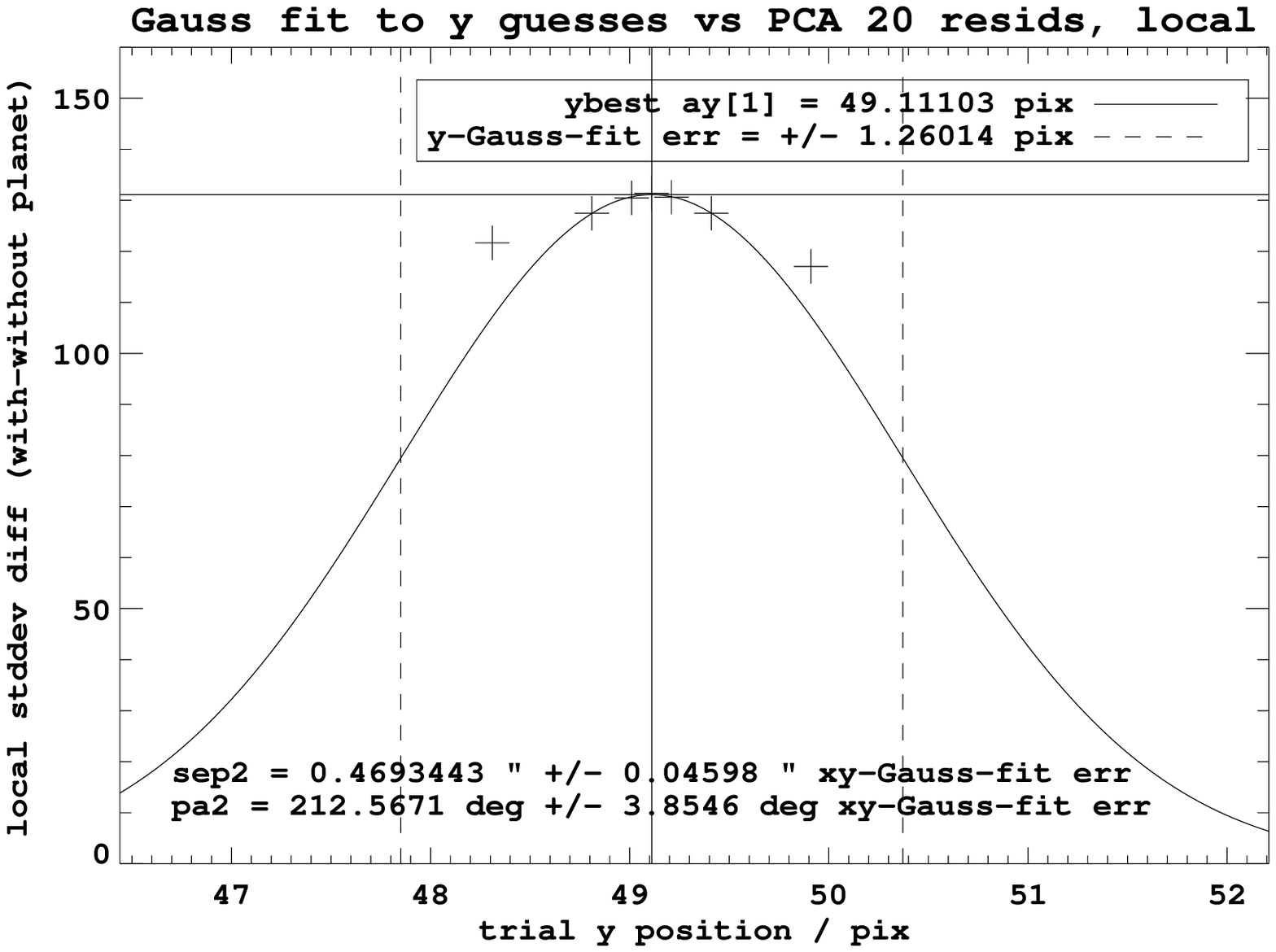}
	\includegraphics[width=0.33\linewidth]{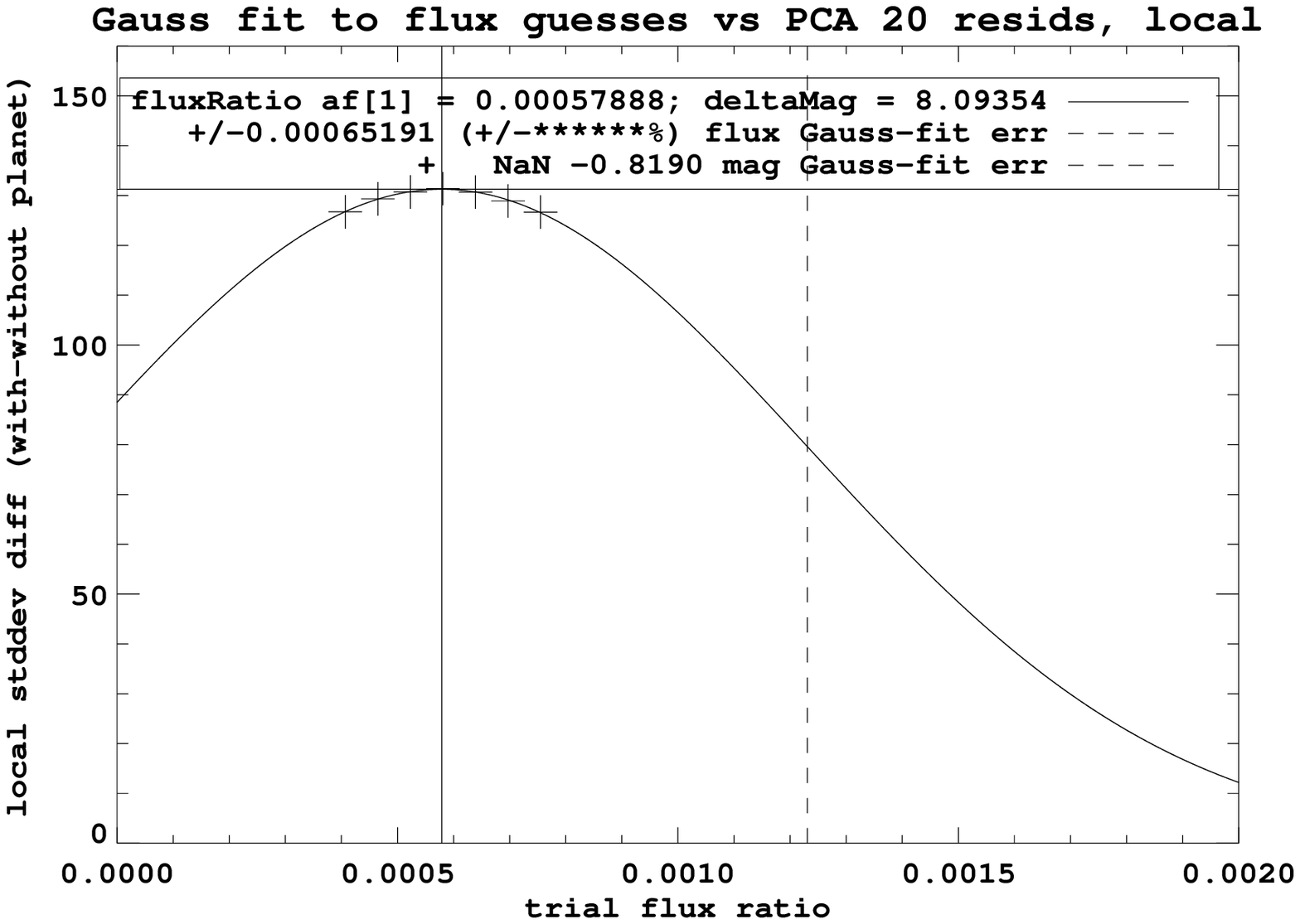}
	\includegraphics[width=0.33\linewidth]{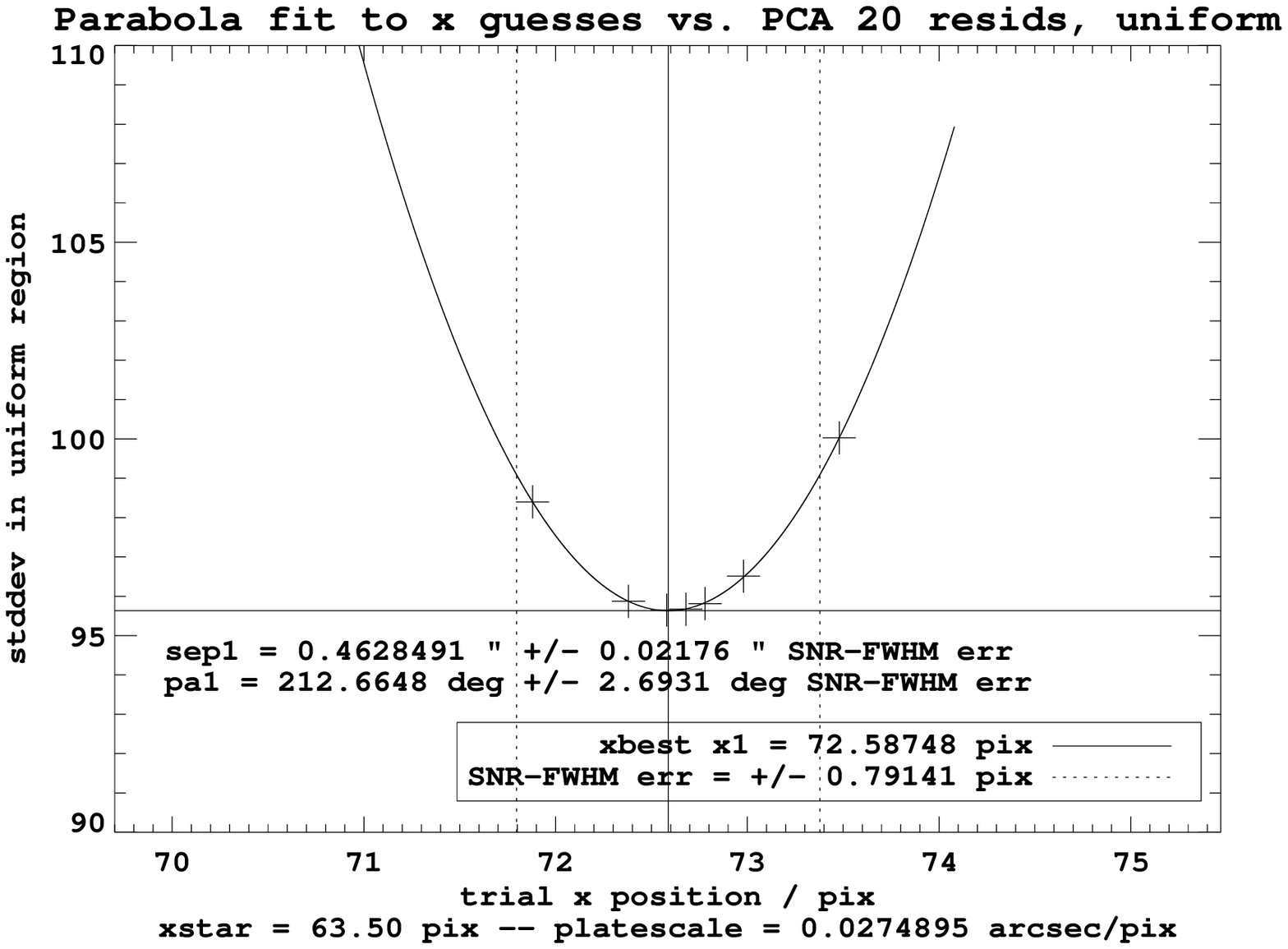}
	\includegraphics[width=0.33\linewidth]{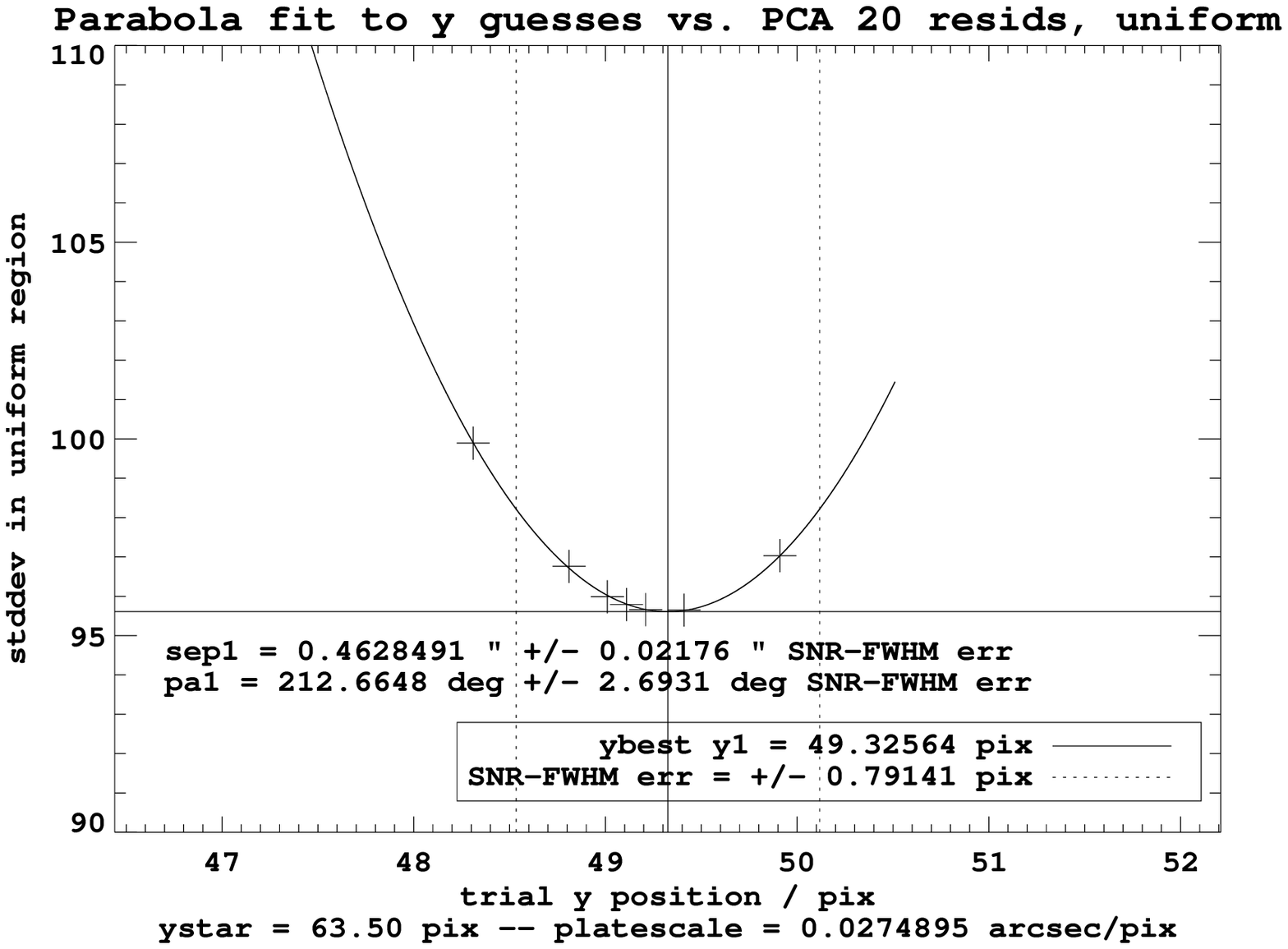}
	\includegraphics[width=0.33\linewidth]{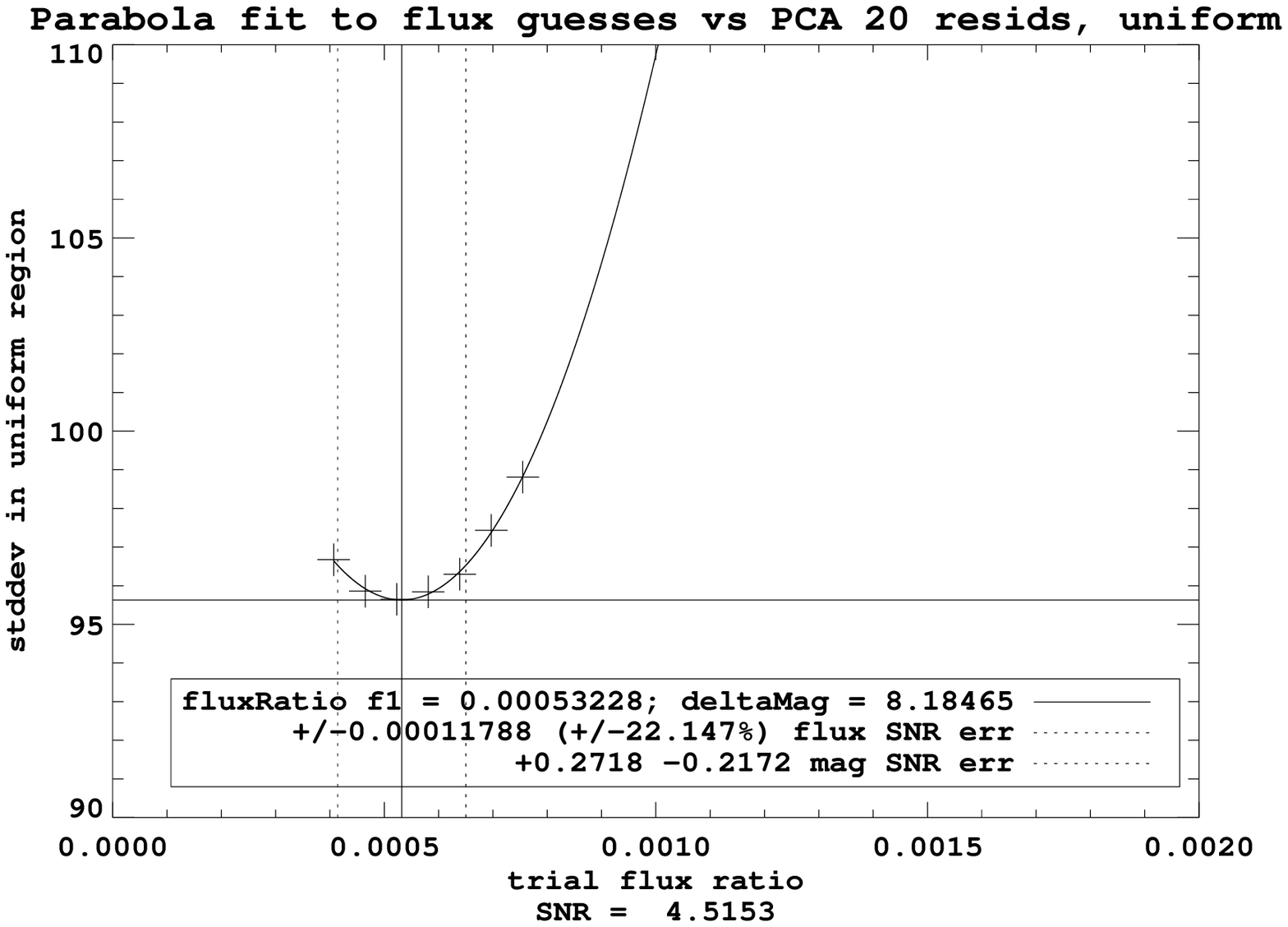}
	\includegraphics[width=0.33\linewidth]{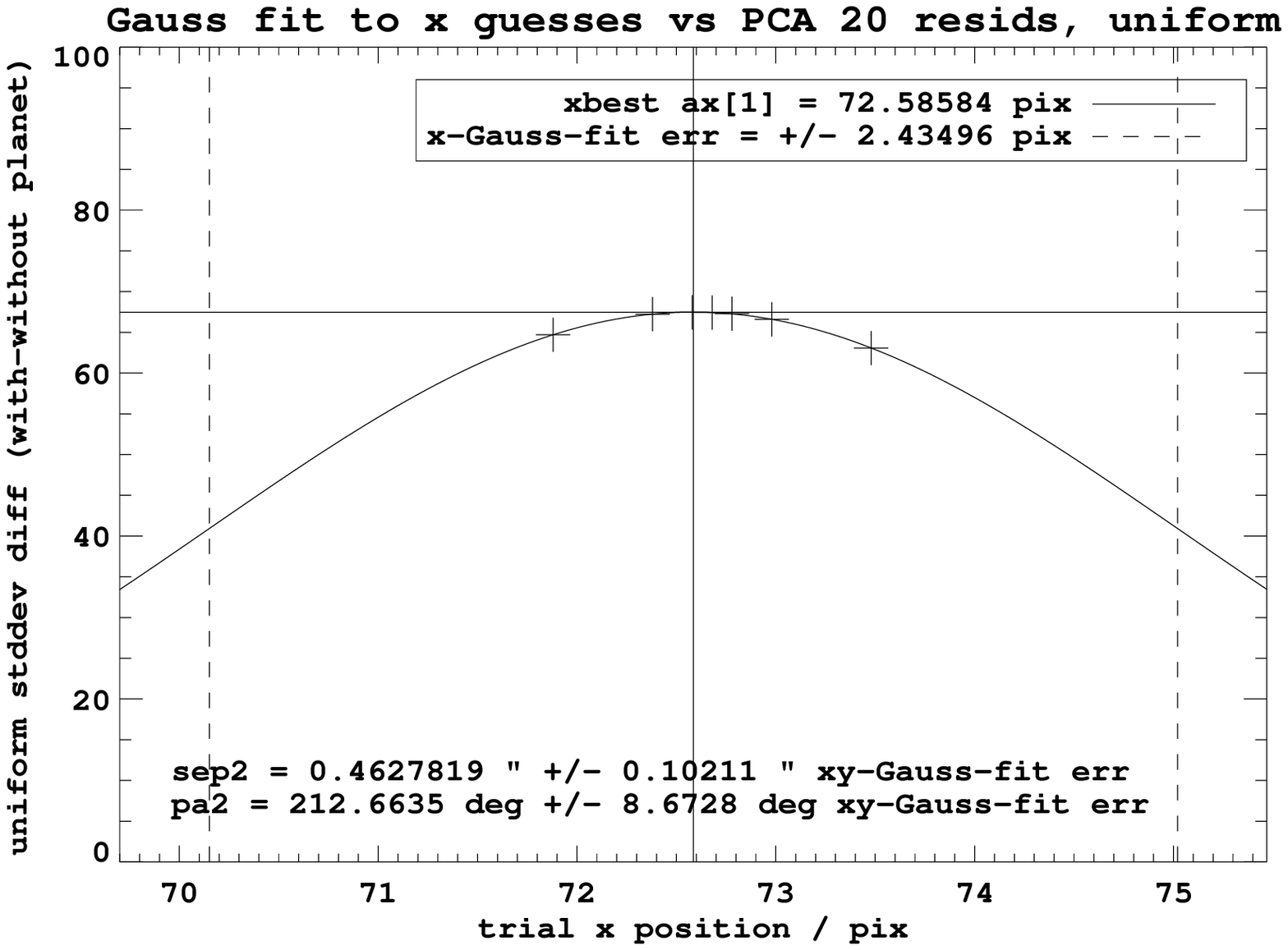}
	\includegraphics[width=0.33\linewidth]{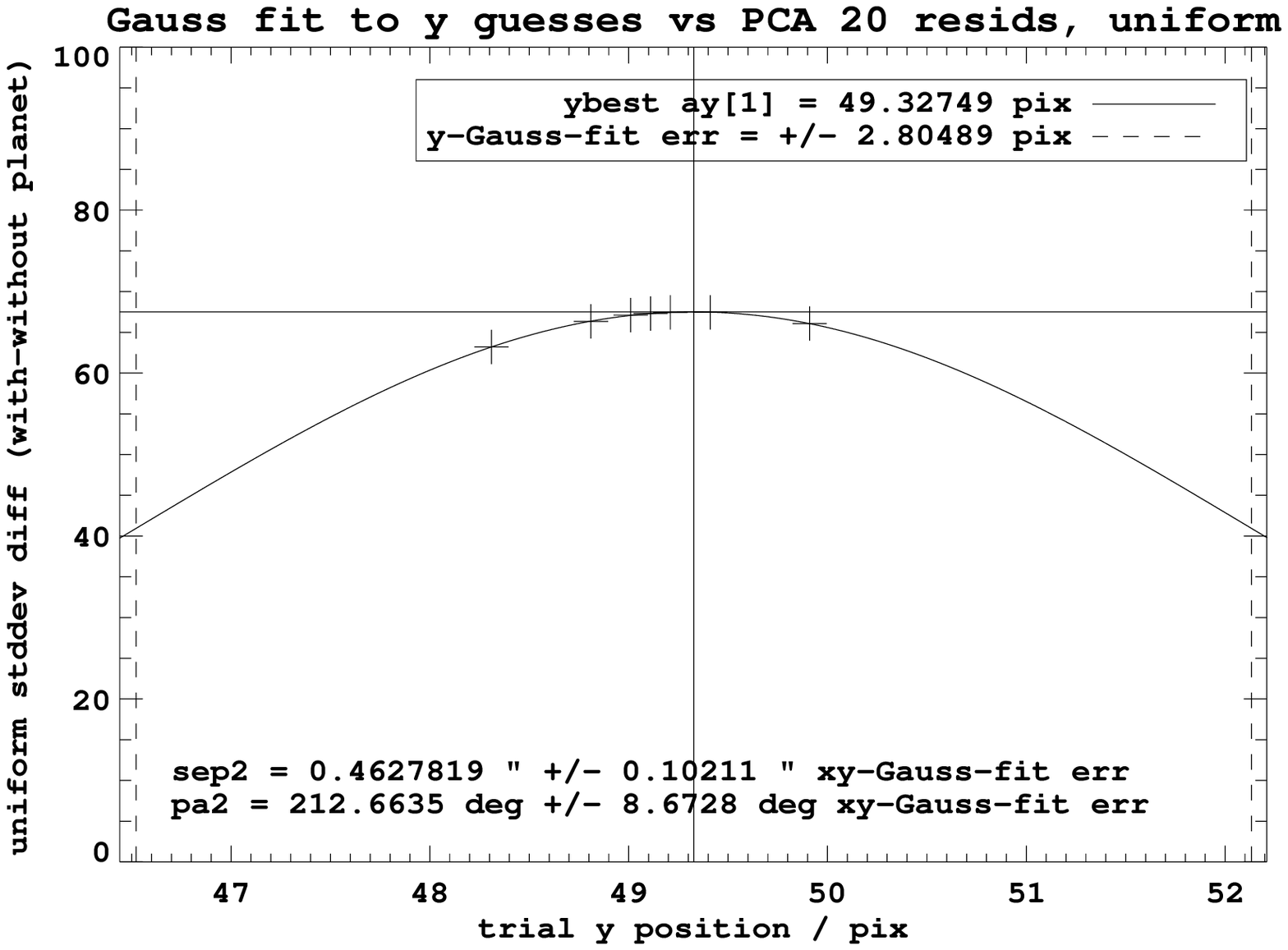}
	\includegraphics[width=0.33\linewidth]{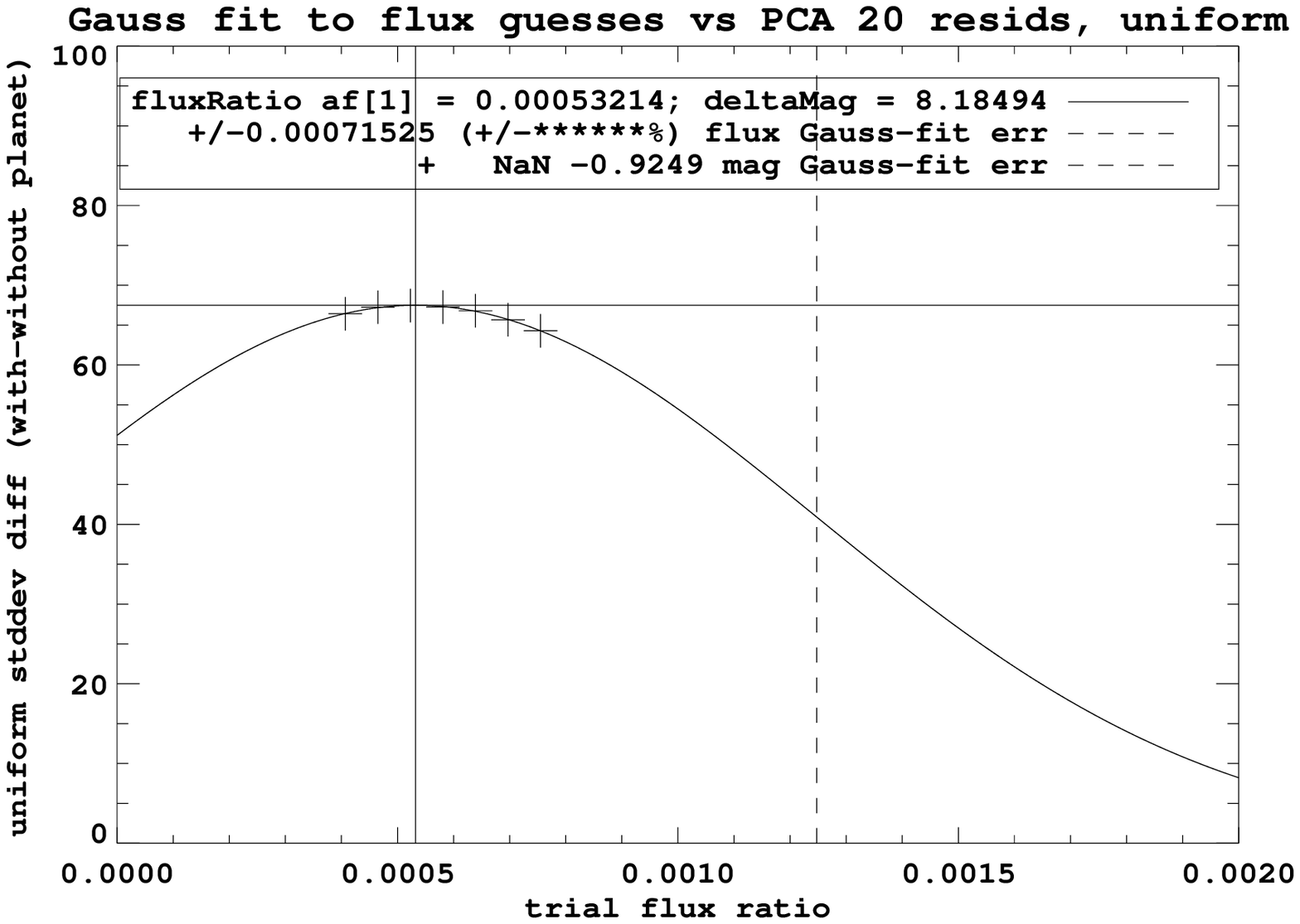}
	\caption{
		Grid search in the $[3.1]$ images for the best-fit photometry and astrometry of the planet,
		using PCA with 20 modes.
		Left column: x position (detector coordinates);
		Center column: y position (detector coordinates); and
		Right column: flux ratio.
		Top row: Parabola fit, local regions;
		Second row: Gaussian fit, local regions;
		Third row: Parabola fit, uniform regions; and
		Bottom row: Gaussian fit, uniform regions.}
	\label{fig:gridsearch31}
\end{figure*}

\begin{figure*}[ht]
\centering
	\includegraphics[width=0.33\linewidth]{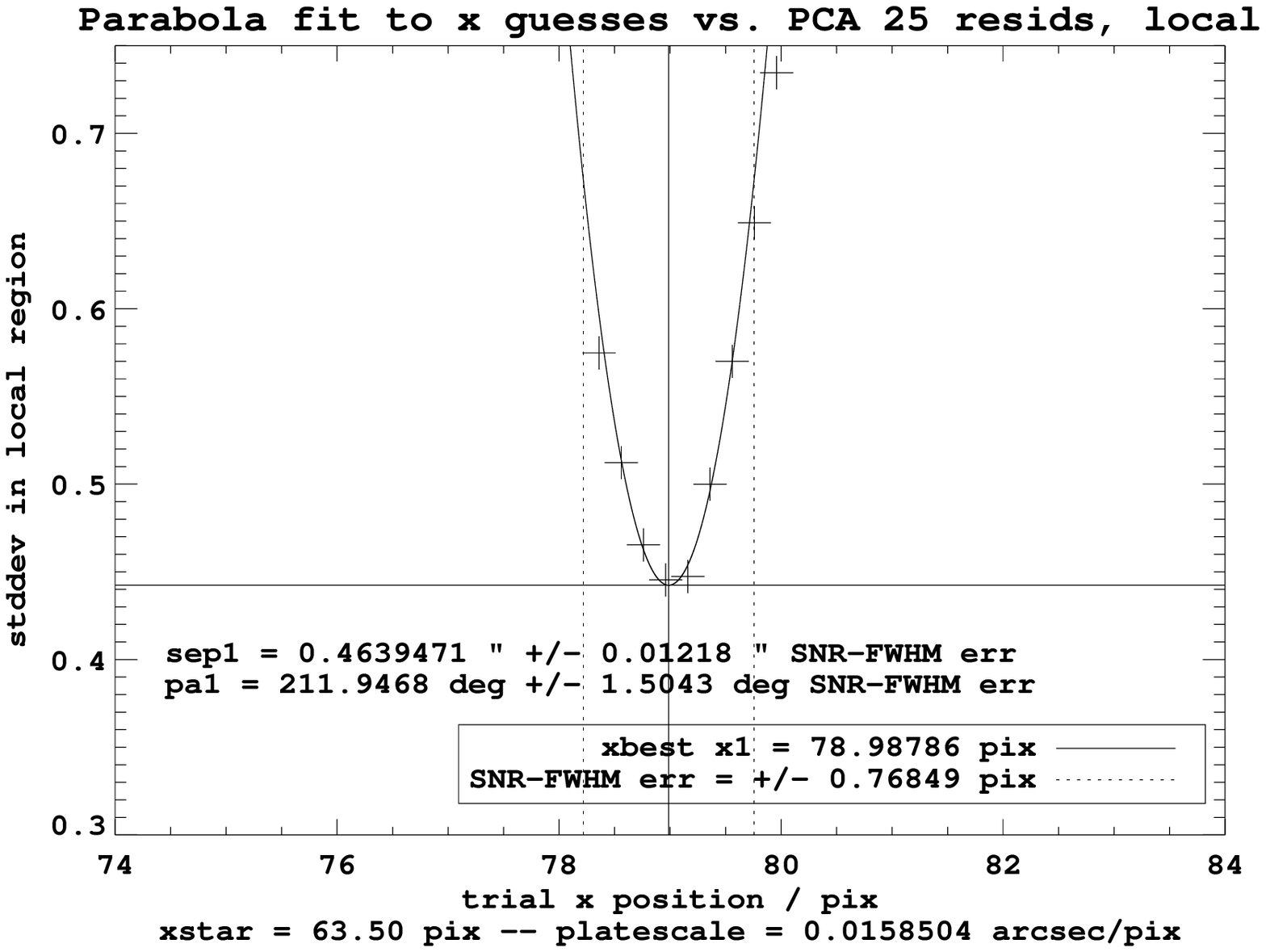}
	\includegraphics[width=0.33\linewidth]{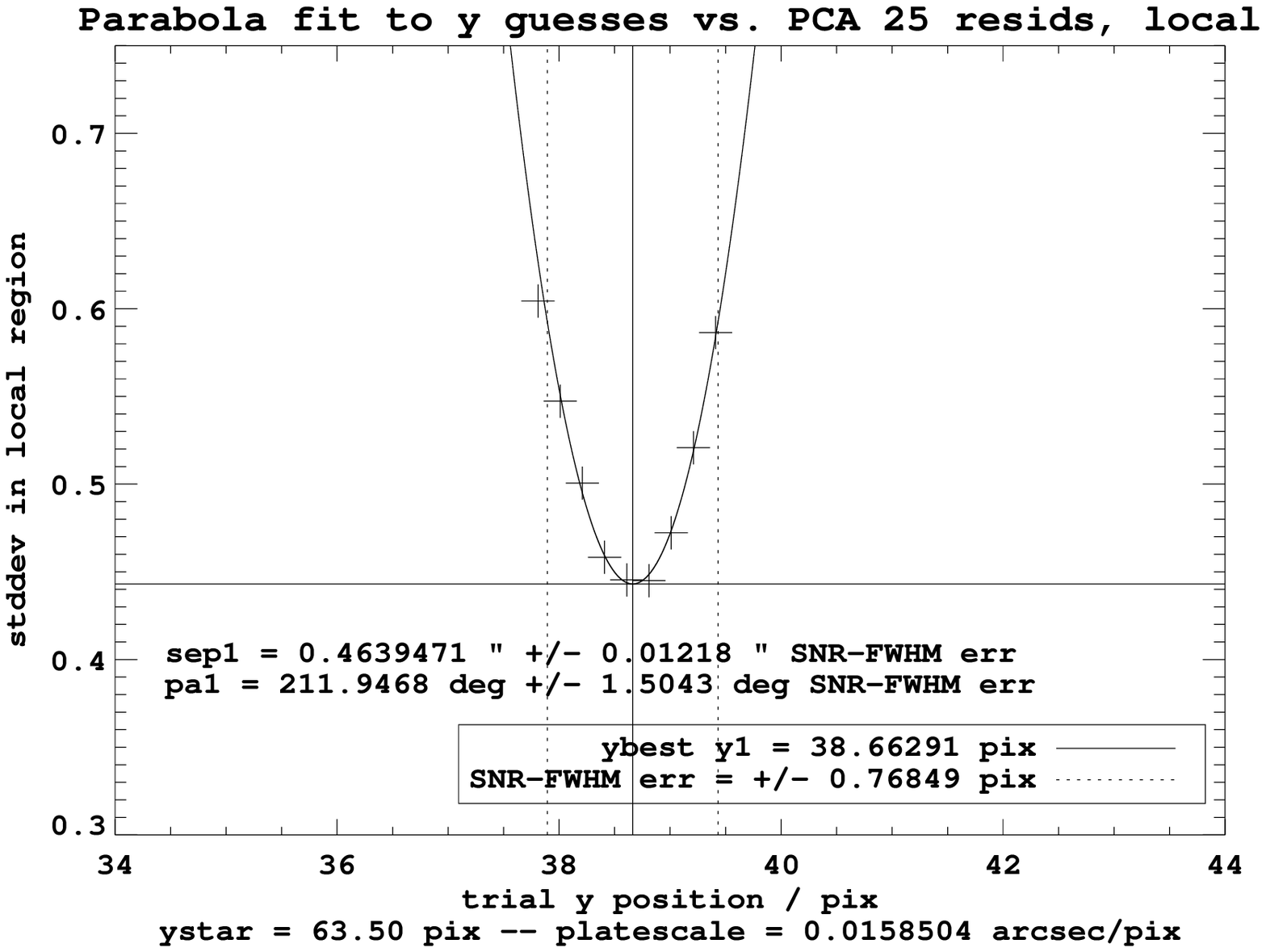}
	\includegraphics[width=0.33\linewidth]{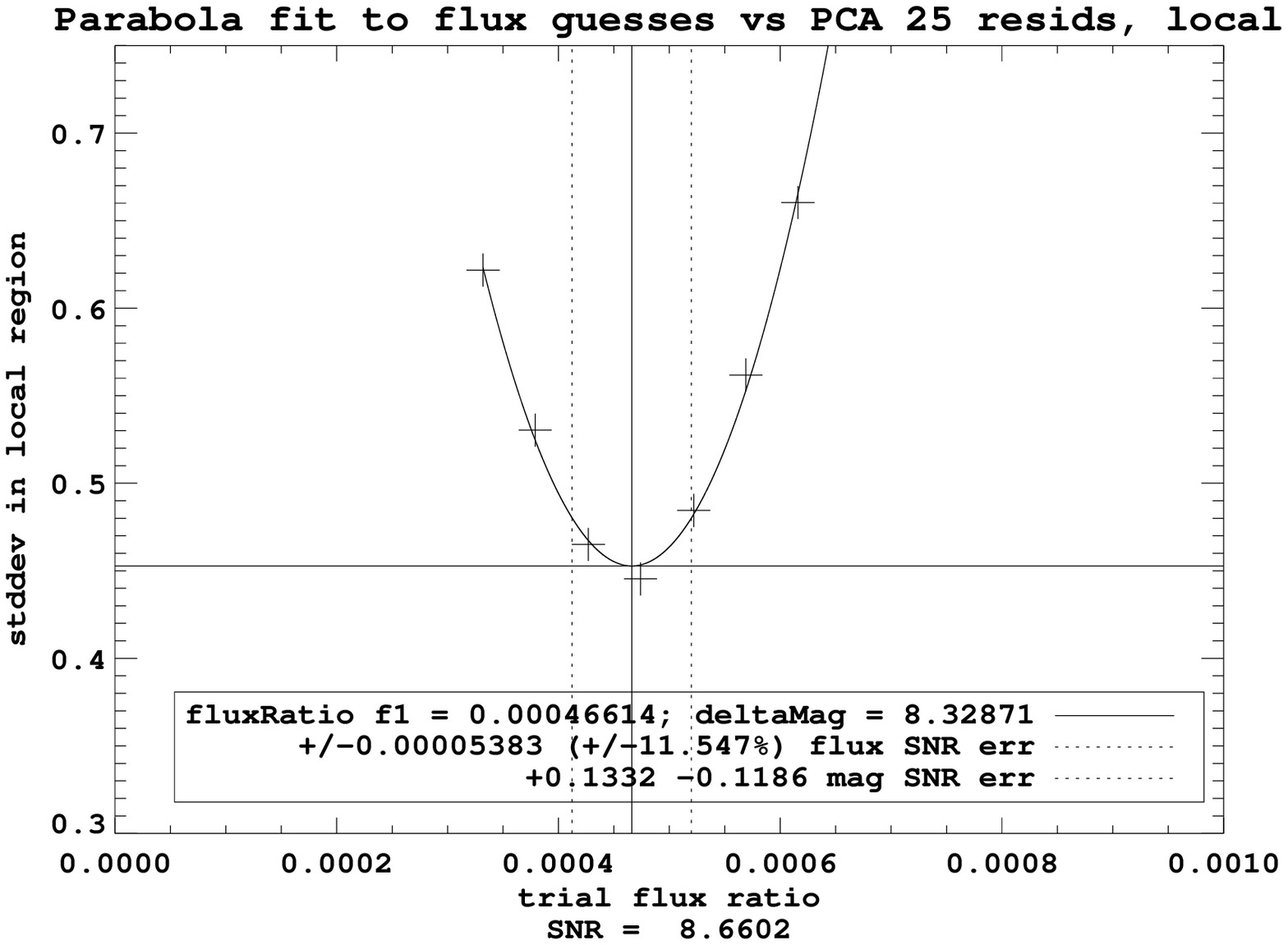}
	\includegraphics[width=0.33\linewidth]{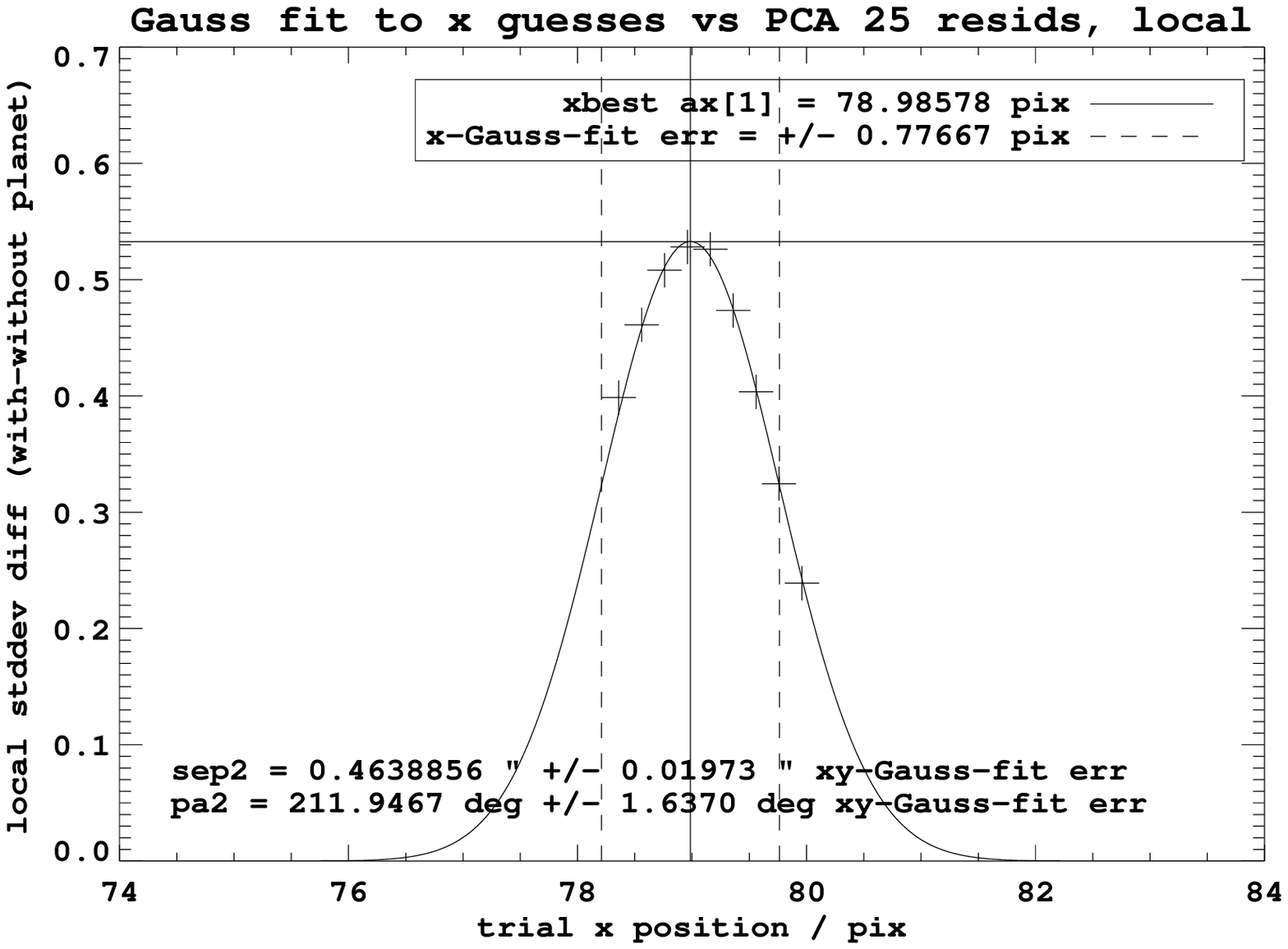}
	\includegraphics[width=0.33\linewidth]{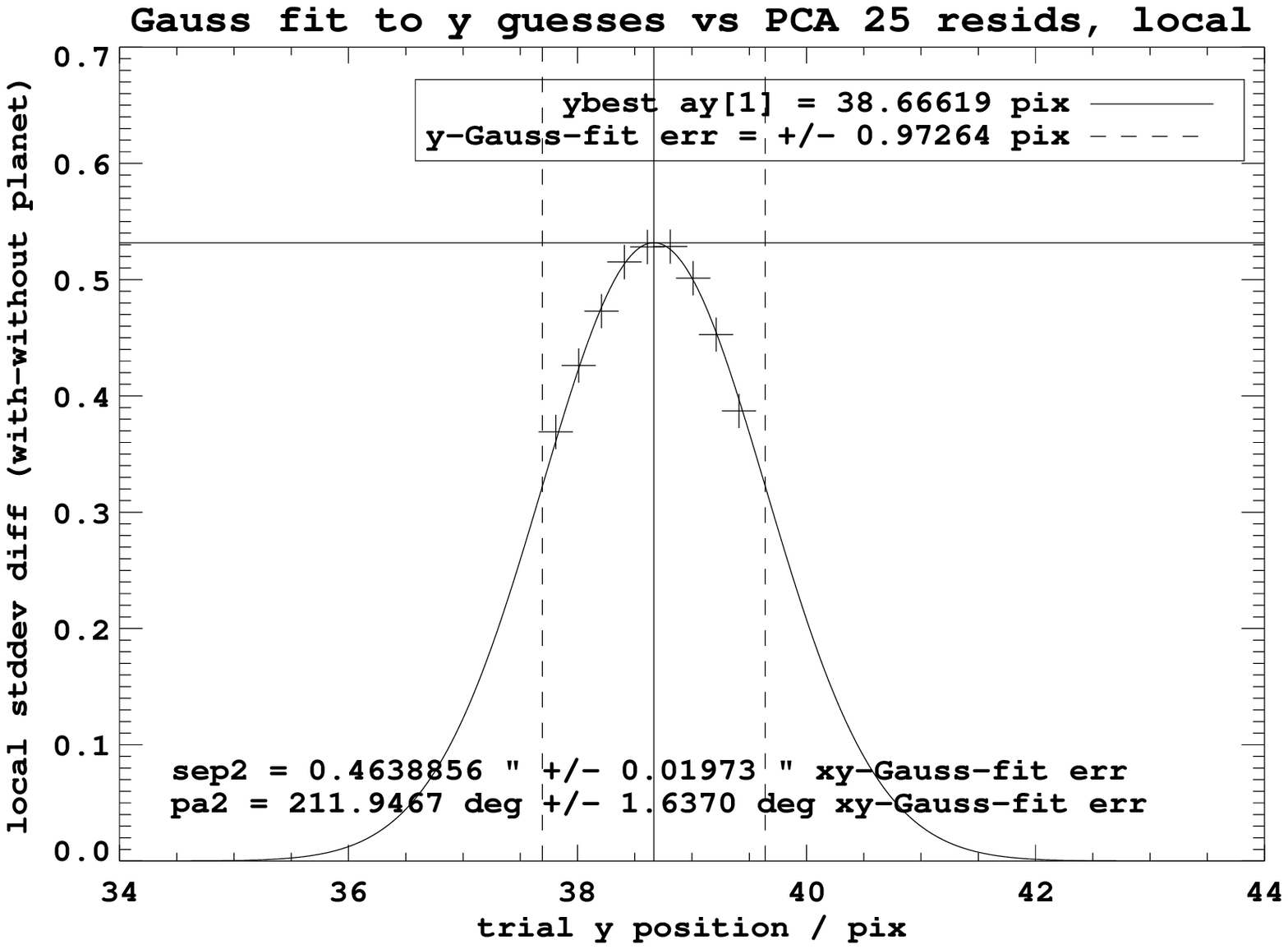}
	\includegraphics[width=0.33\linewidth]{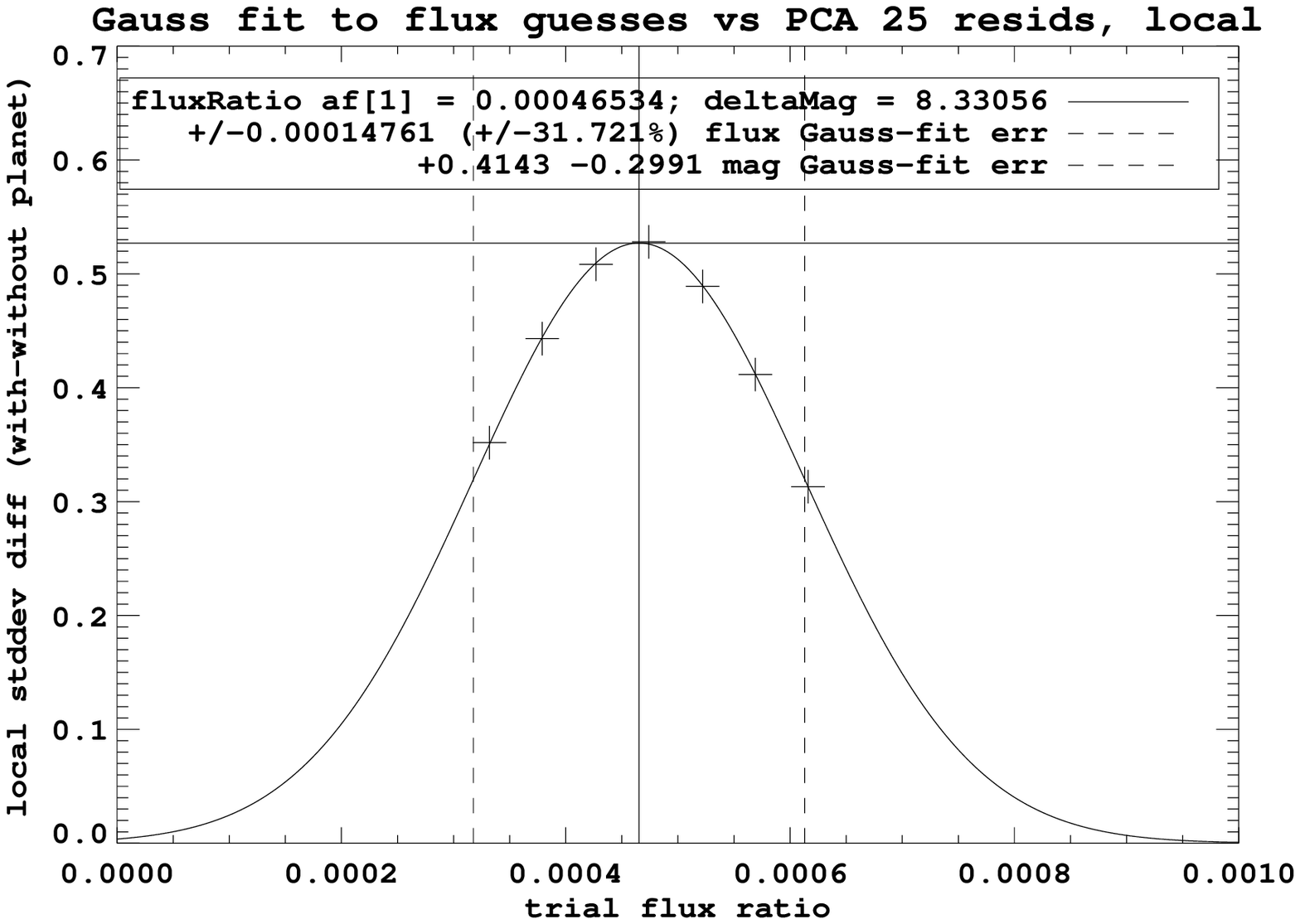}
	\label{}
	\includegraphics[width=0.33\linewidth]{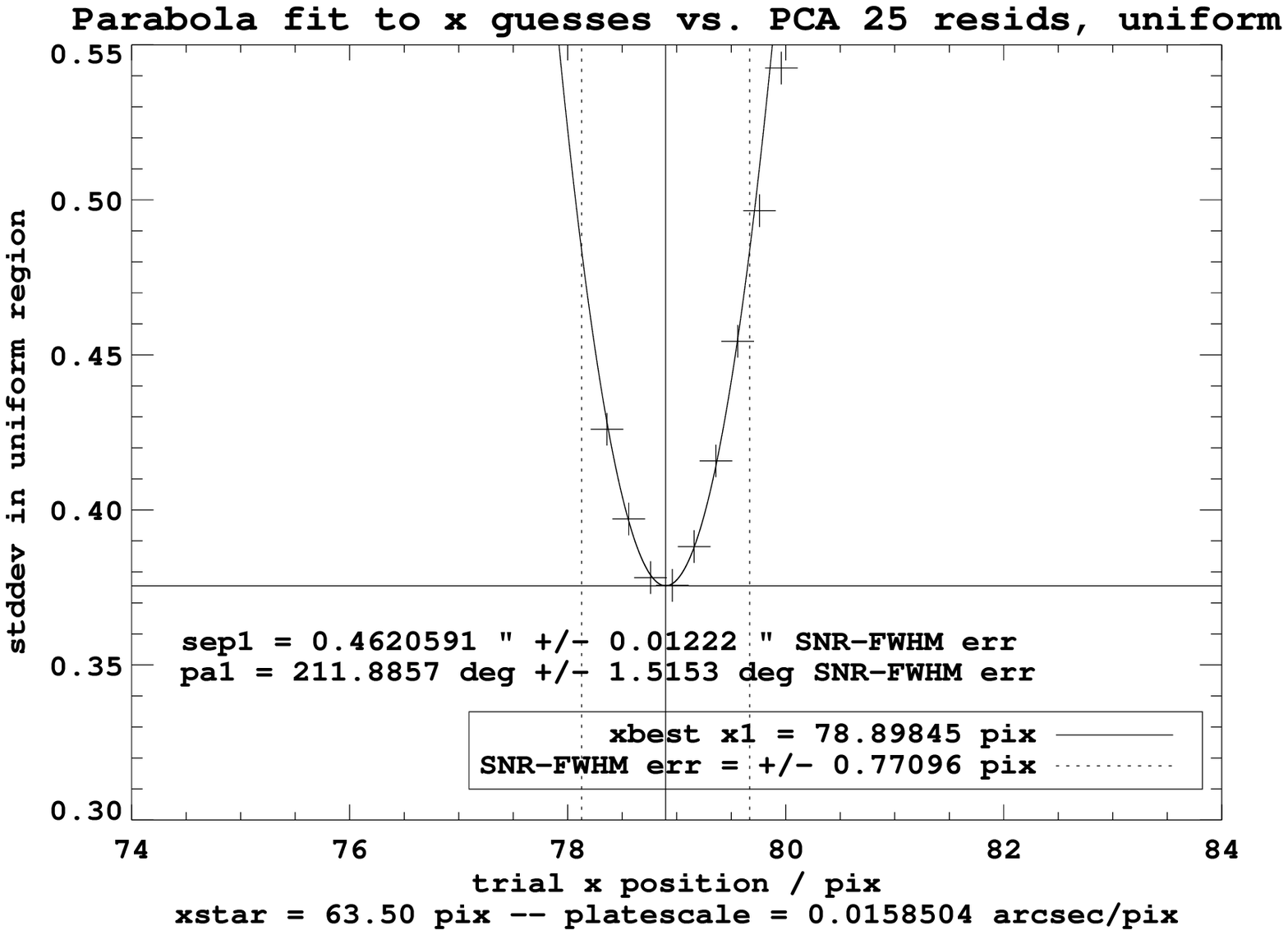}
	\includegraphics[width=0.33\linewidth]{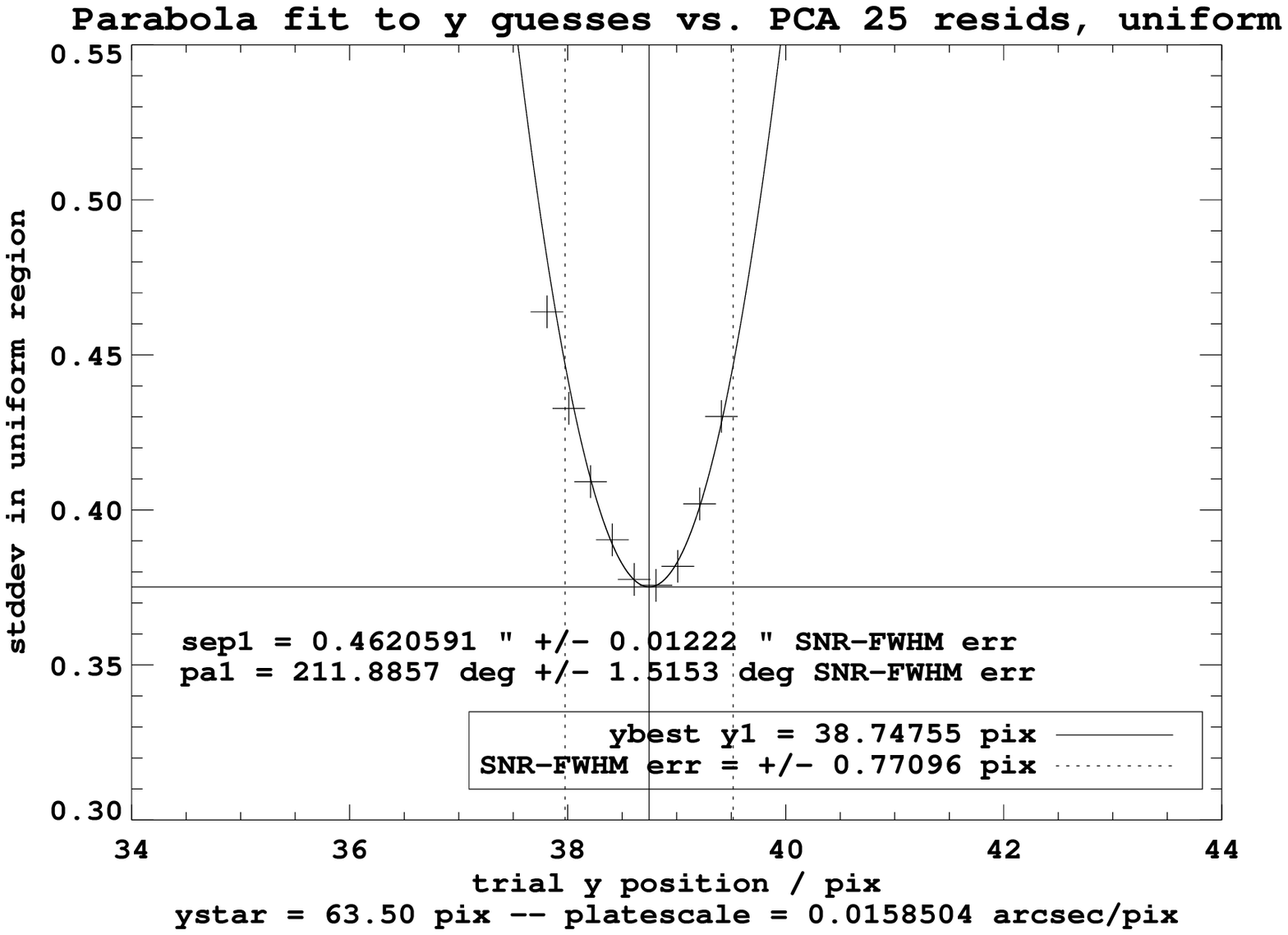}
	\includegraphics[width=0.33\linewidth]{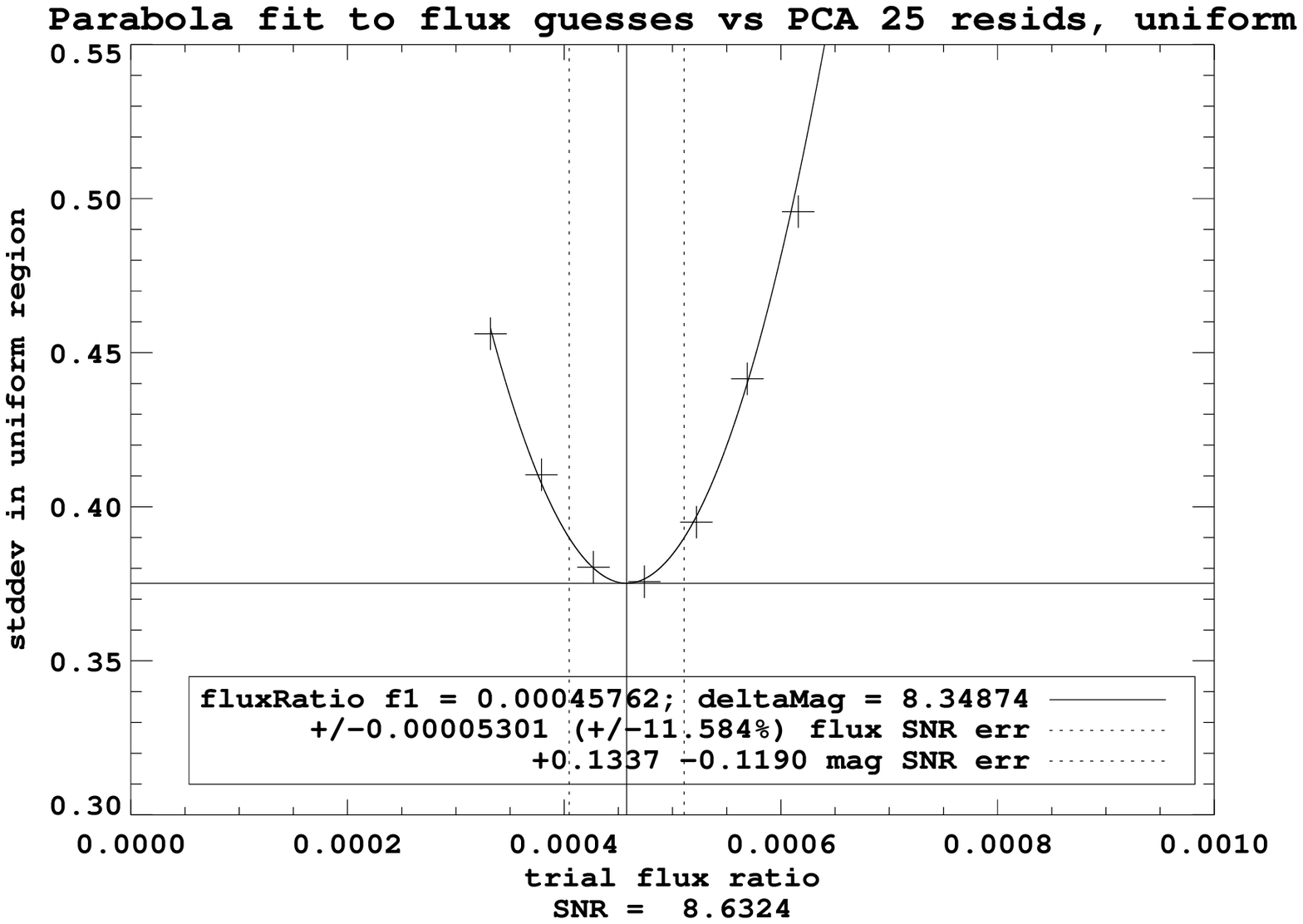}
	\includegraphics[width=0.33\linewidth]{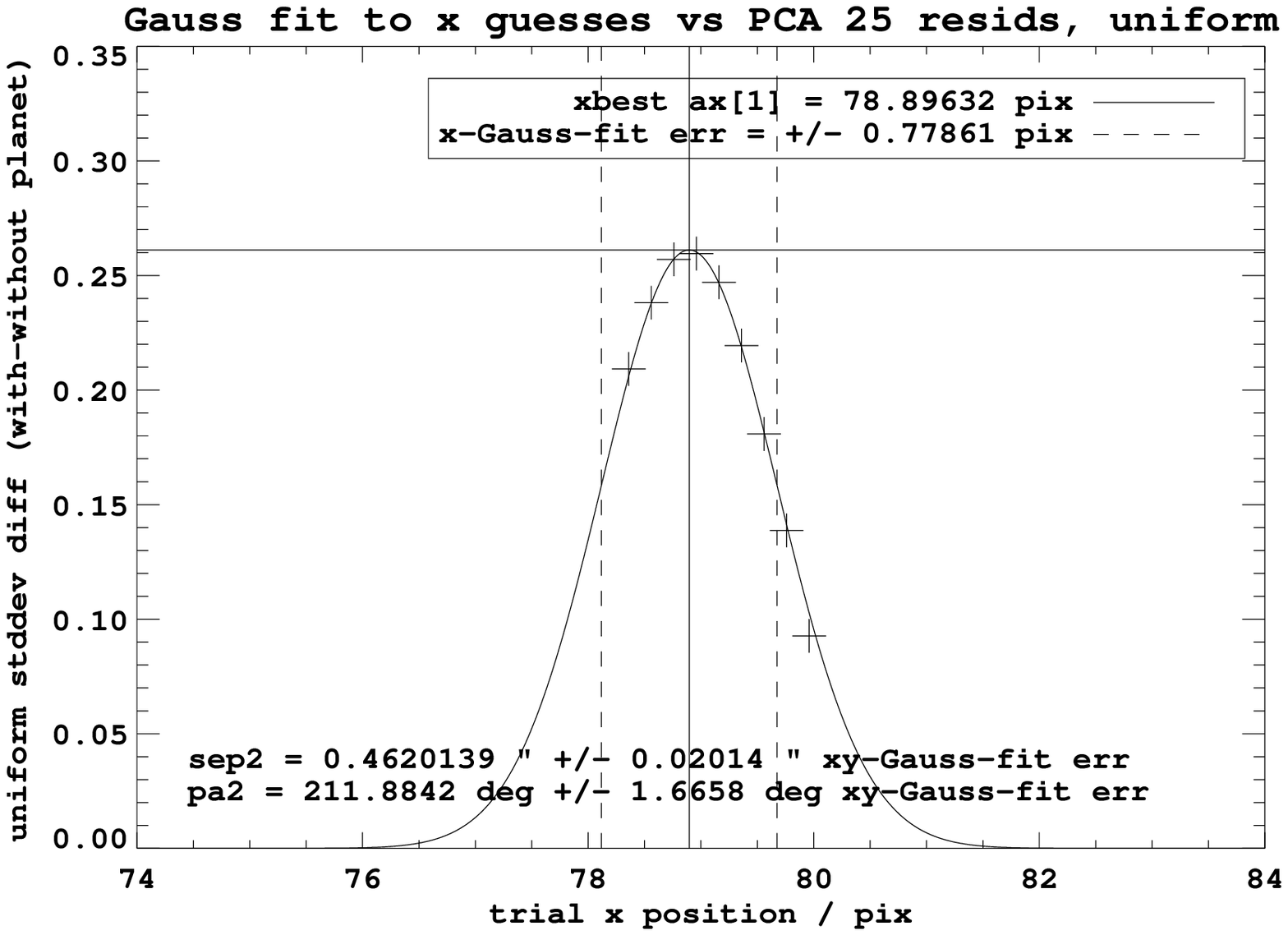}
	\includegraphics[width=0.33\linewidth]{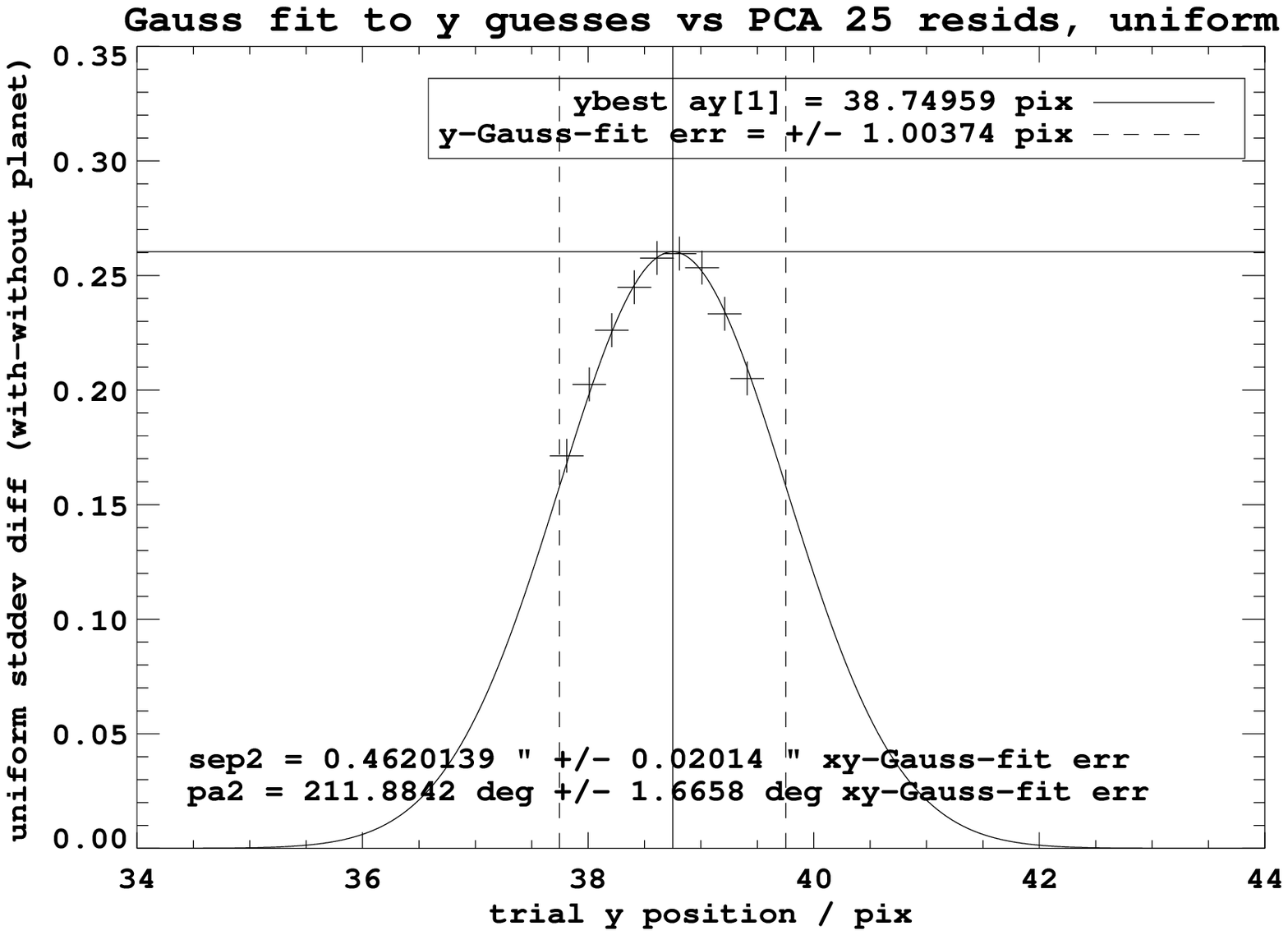}
	\includegraphics[width=0.33\linewidth]{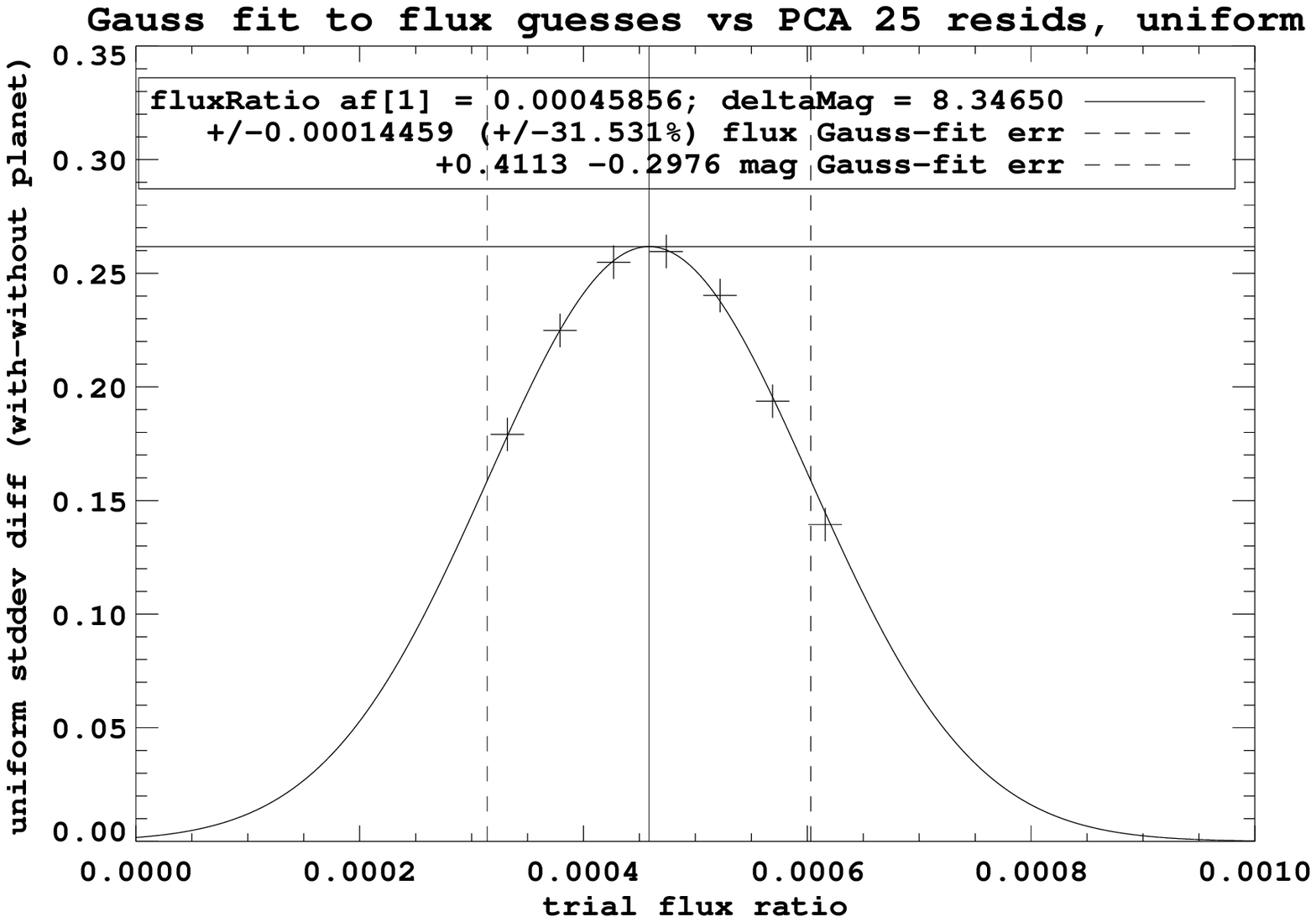}
	\caption{
		Grid search in the $[3.3]$ images for the best-fit photometry and astrometry of the planet,
		using PCA with 25 modes.
		Left column: x position (detector coordinates);
		Center column: y position (detector coordinates); and
		Right column: flux ratio.
		Top row: Parabola fit, local regions;
		Second row: Gaussian fit, local regions;
		Third row: Parabola fit, uniform regions; and
		Bottom row: Gaussian fit, uniform regions.}
	\label{fig:gridsearch33}
\end{figure*}

\begin{figure*}[ht]
\centering
	\includegraphics[width=0.33\linewidth]{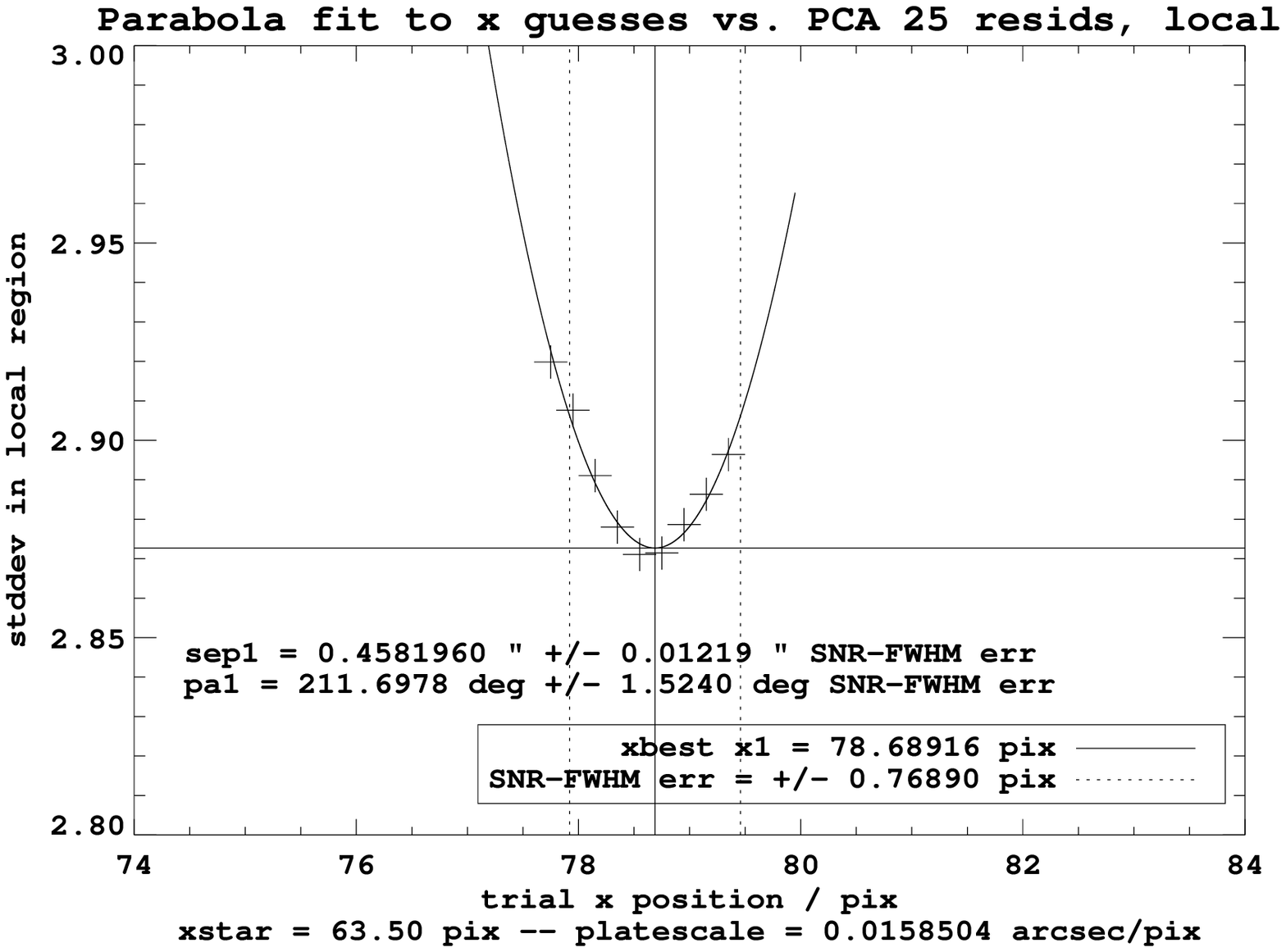}
	\includegraphics[width=0.33\linewidth]{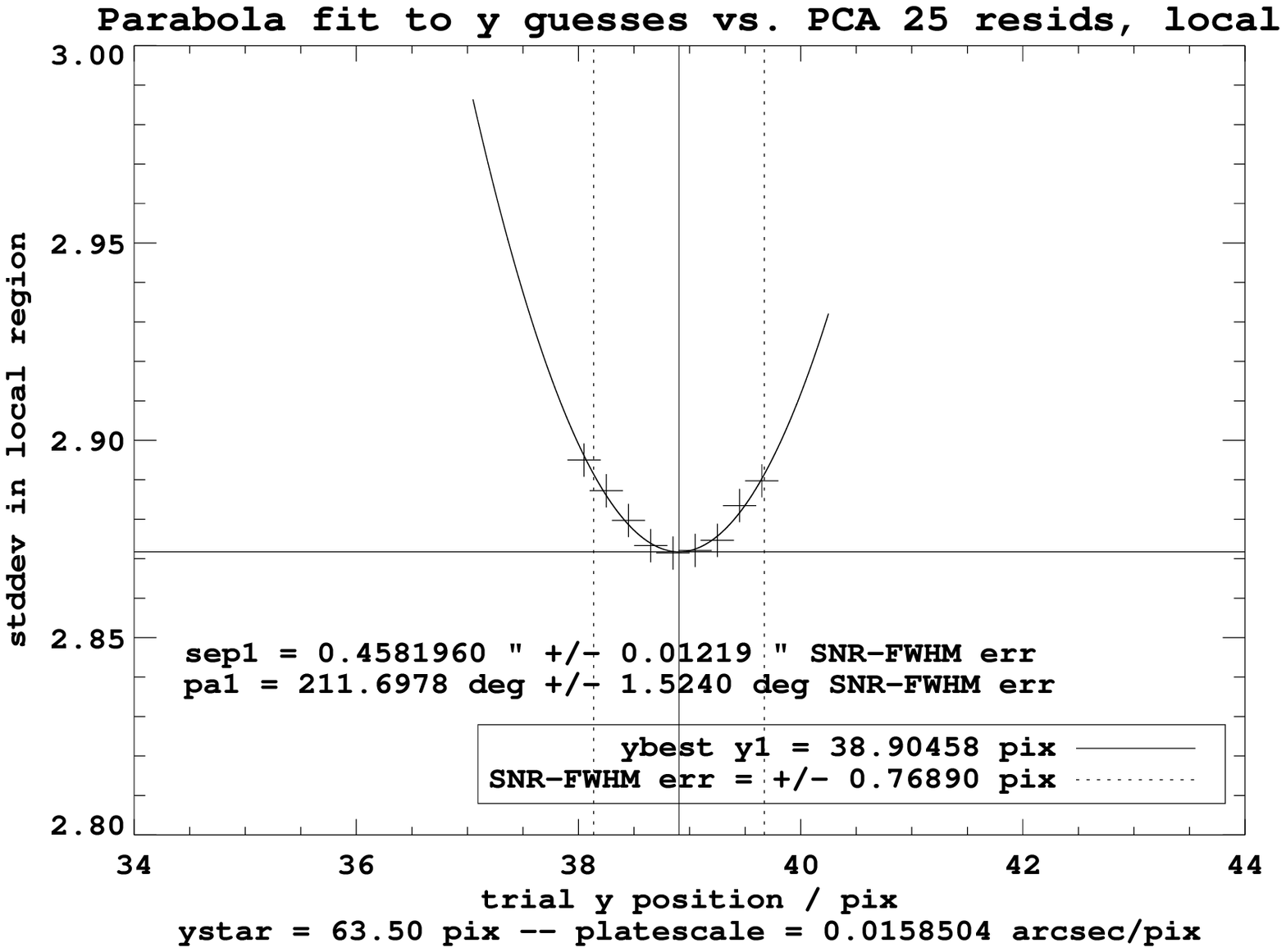}
	\includegraphics[width=0.33\linewidth]{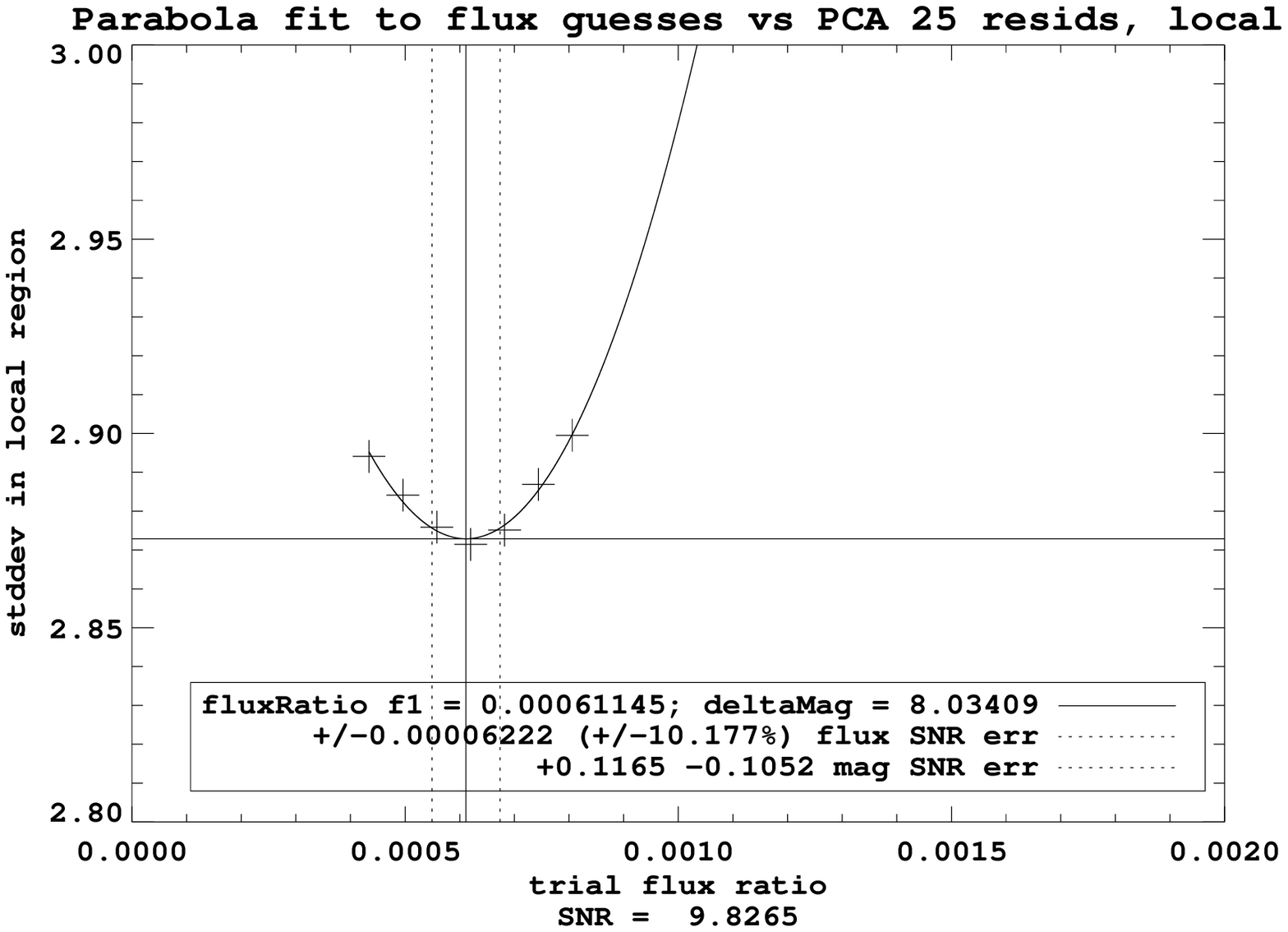}
	\includegraphics[width=0.33\linewidth]{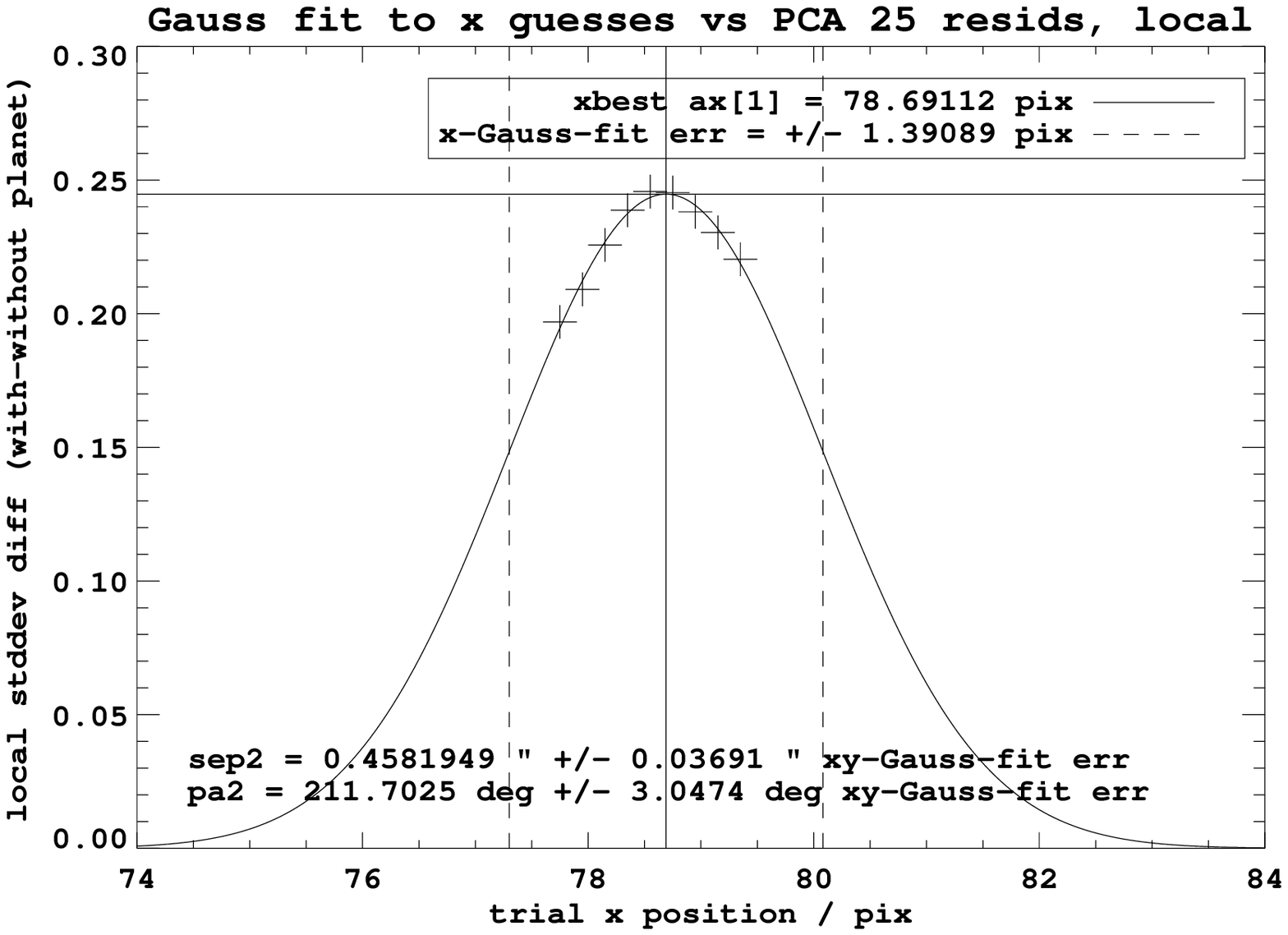}
	\includegraphics[width=0.33\linewidth]{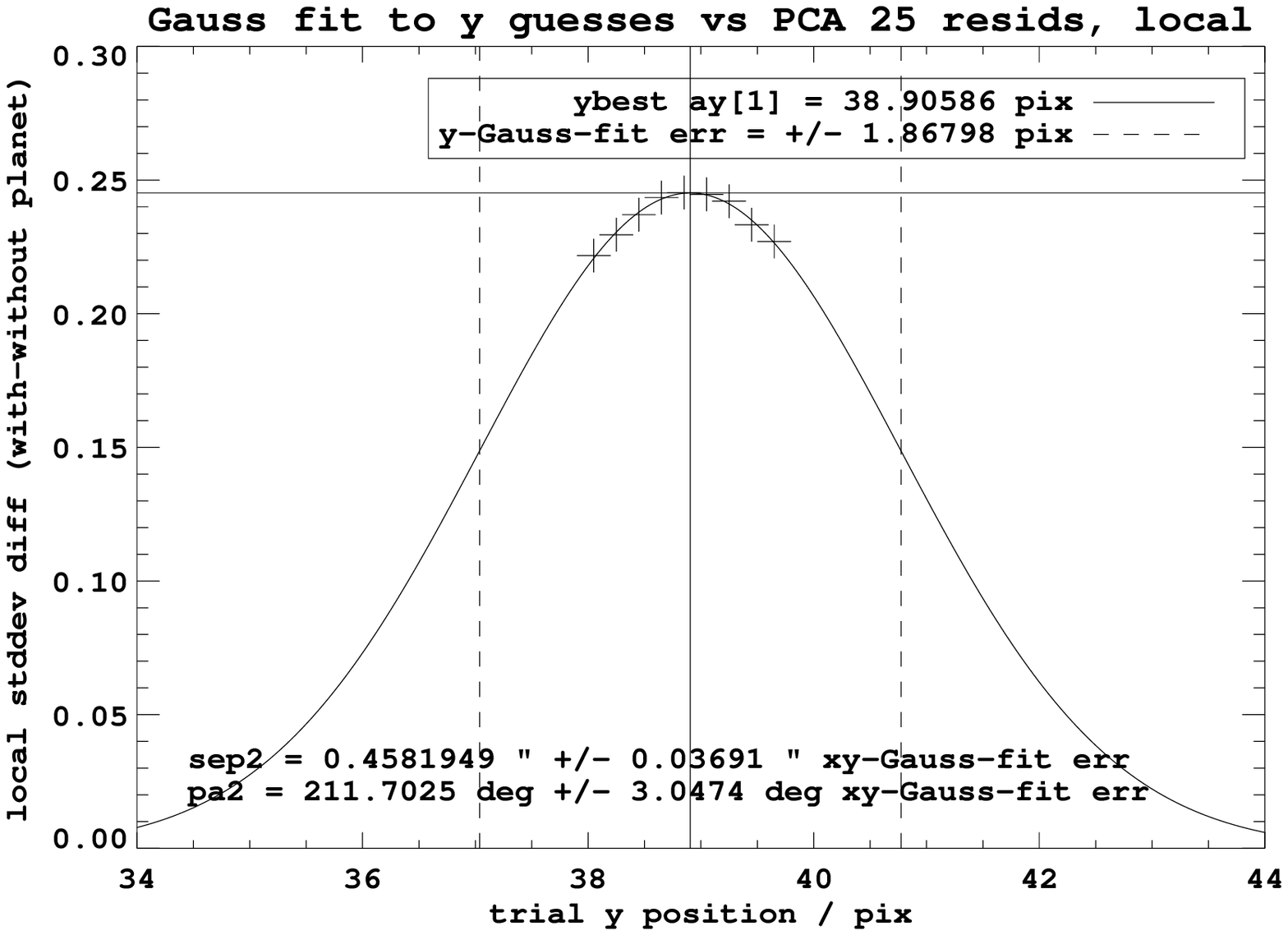}
	\includegraphics[width=0.33\linewidth]{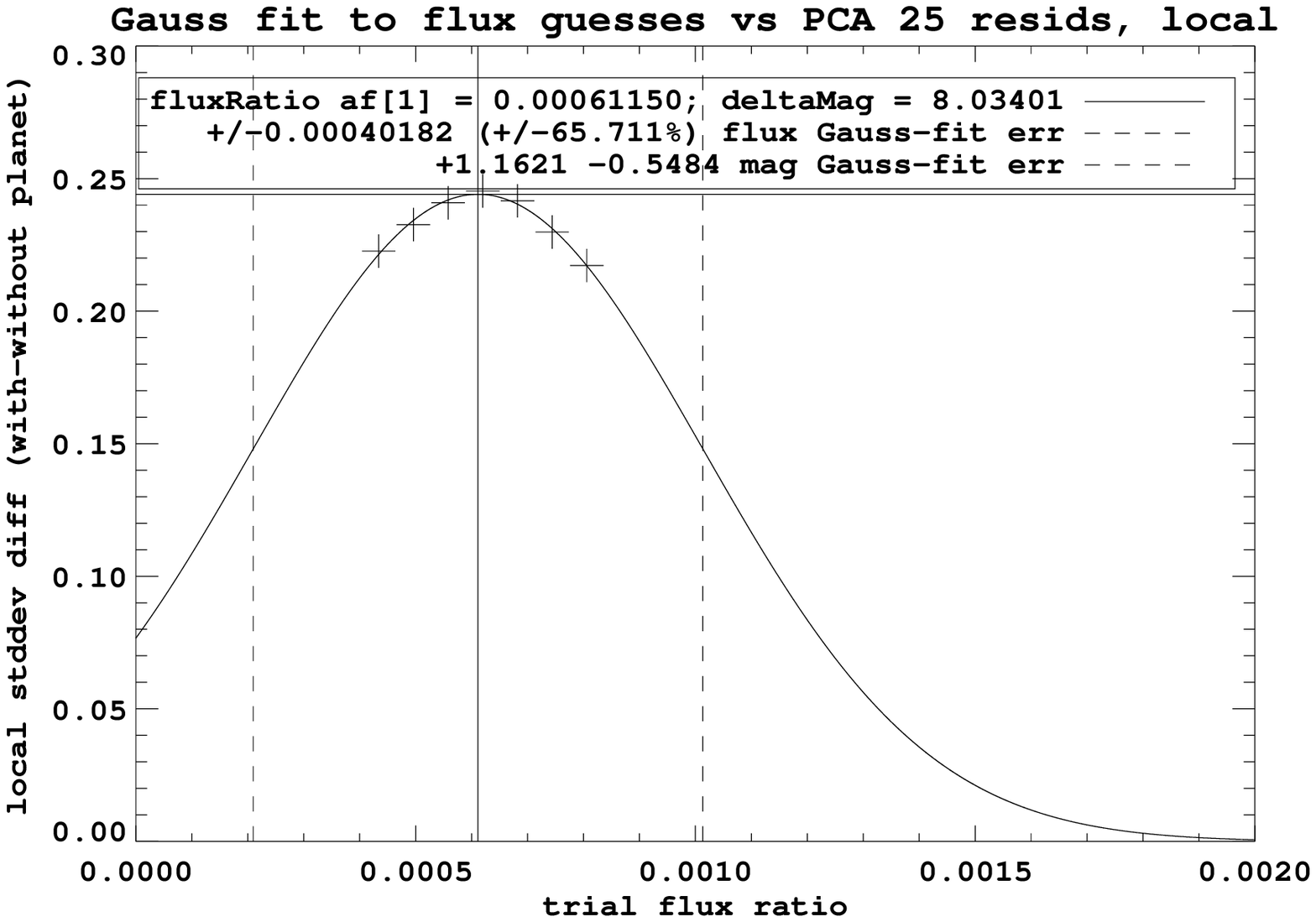}
	\includegraphics[width=0.33\linewidth]{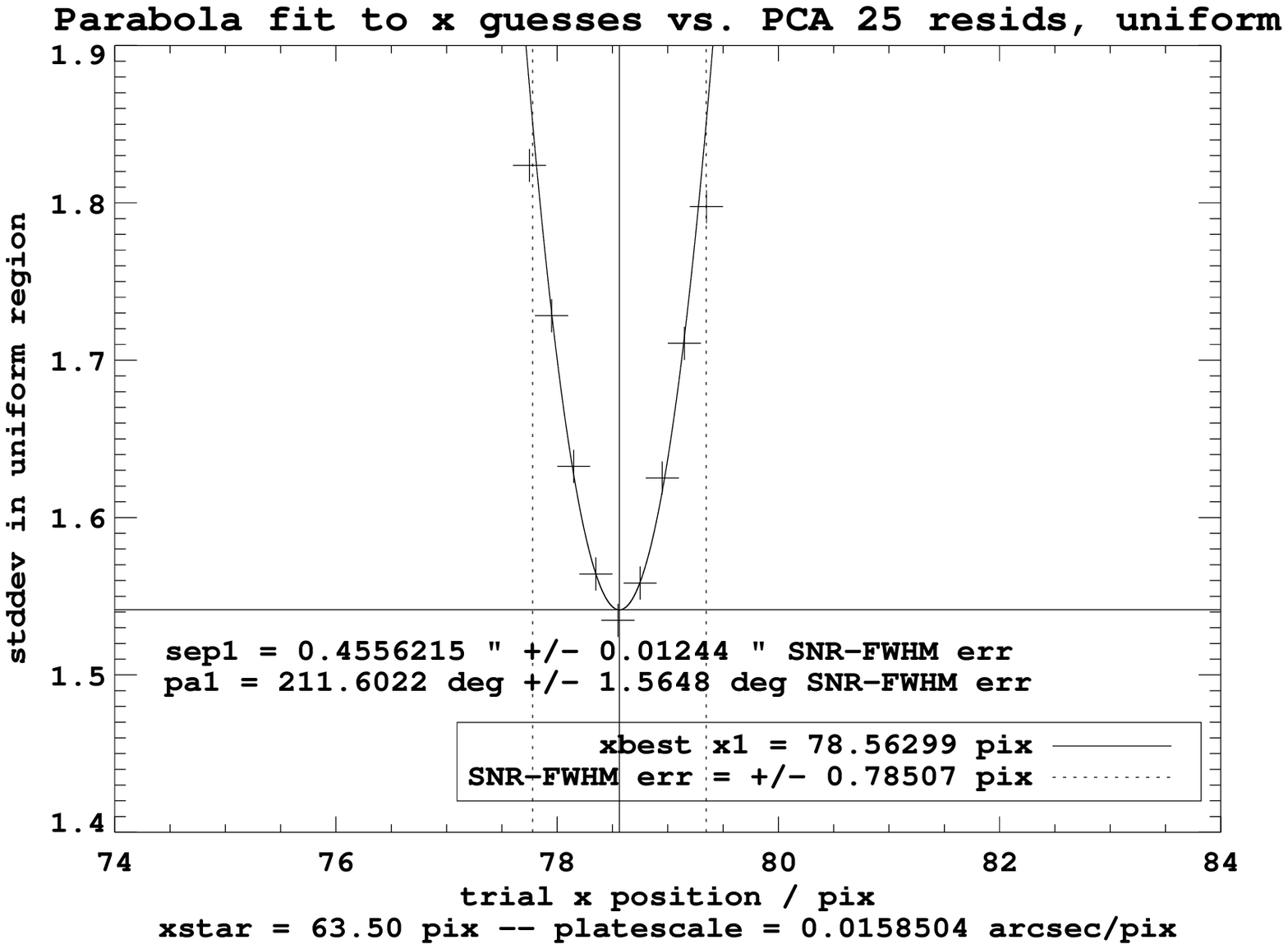}
	\includegraphics[width=0.33\linewidth]{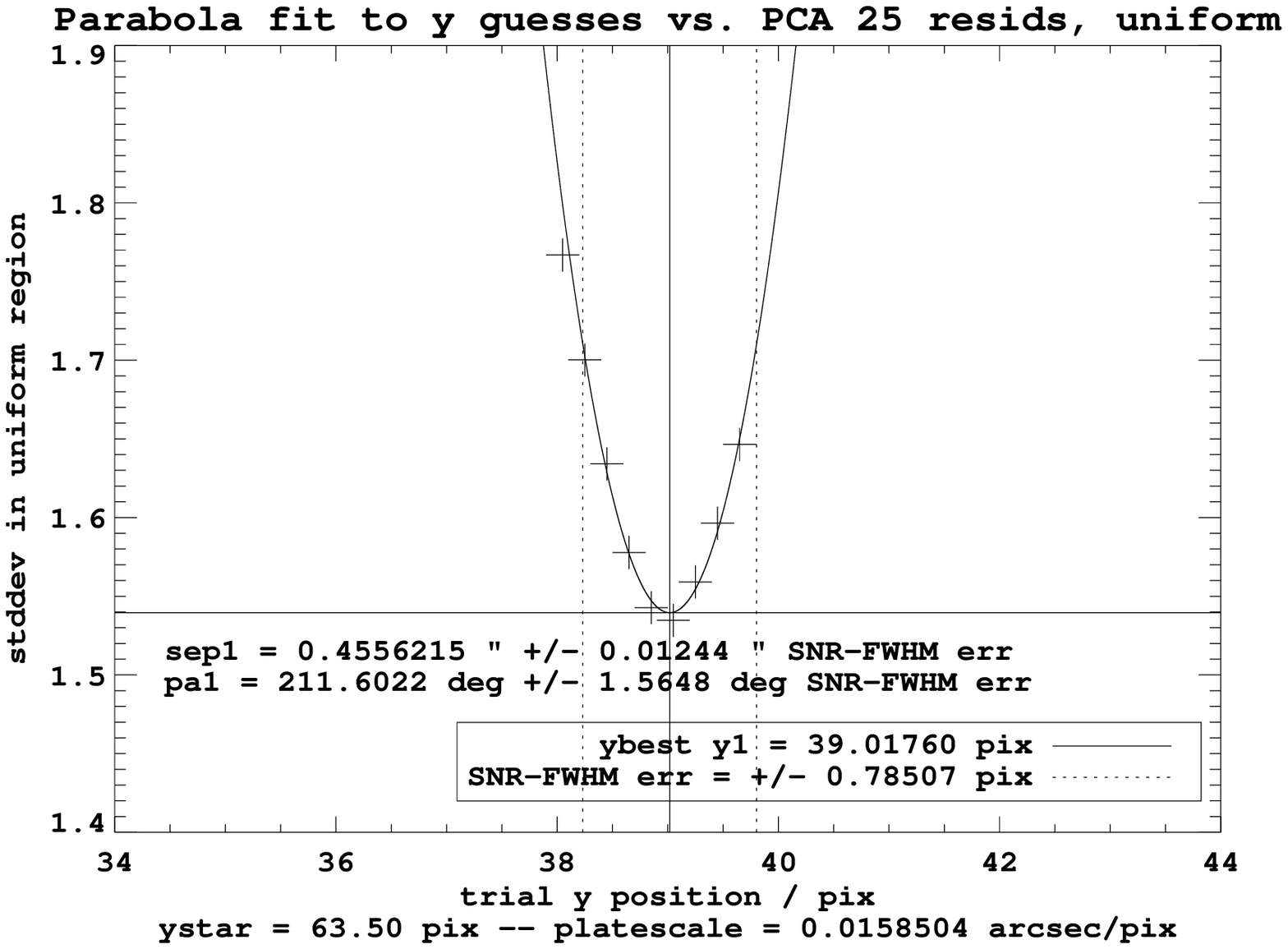}
	\includegraphics[width=0.33\linewidth]{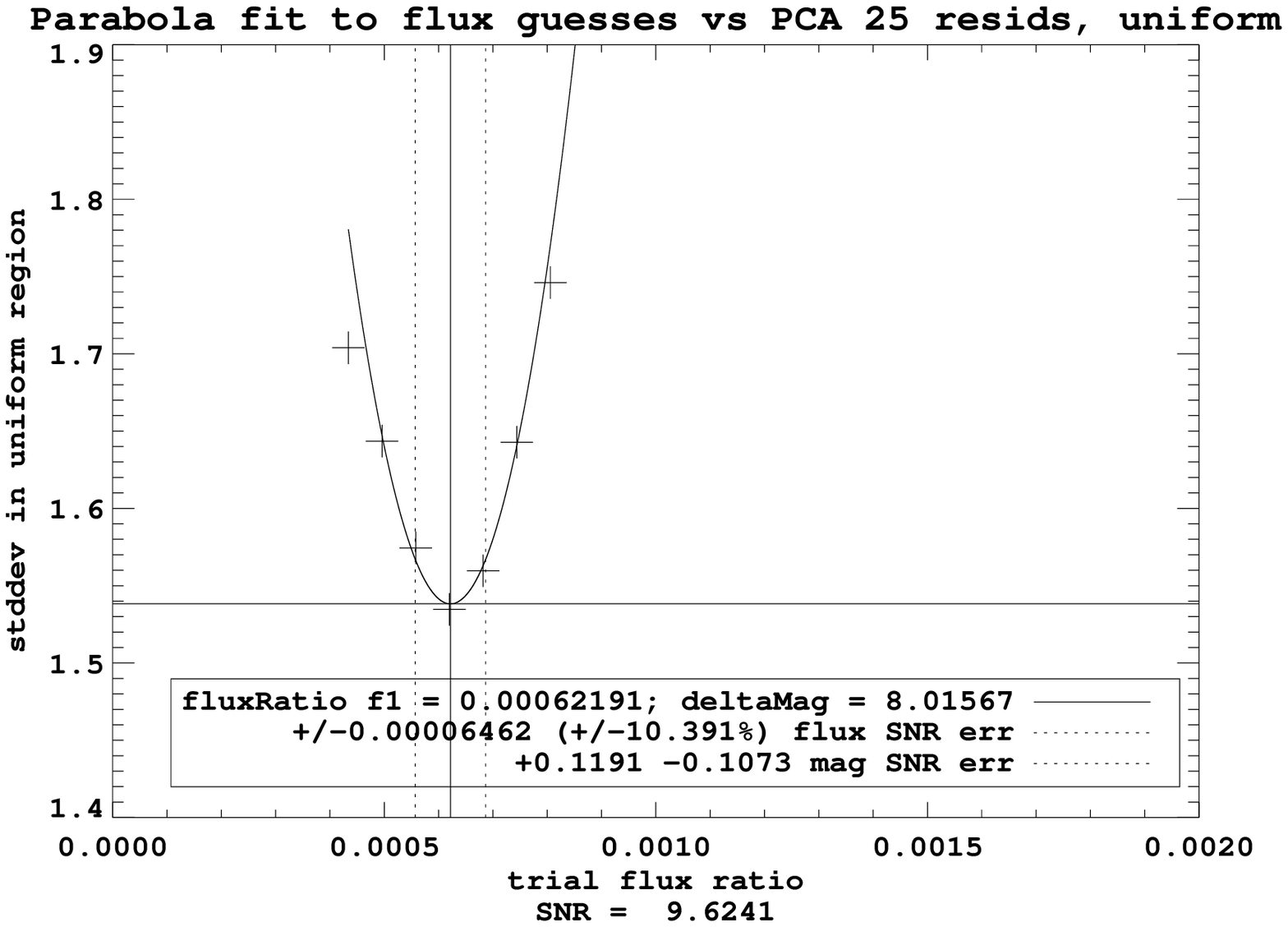}
	\includegraphics[width=0.33\linewidth]{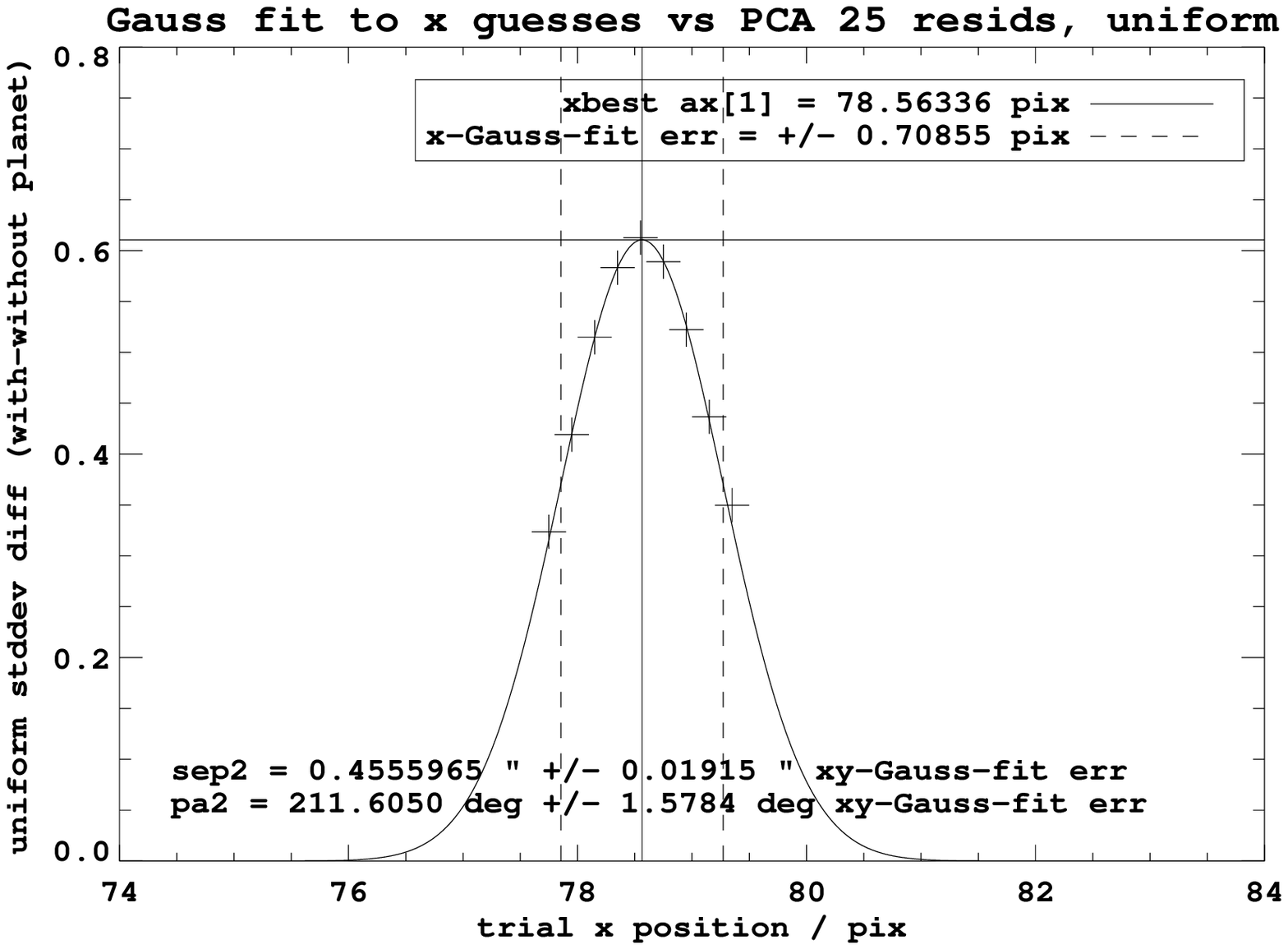}
	\includegraphics[width=0.33\linewidth]{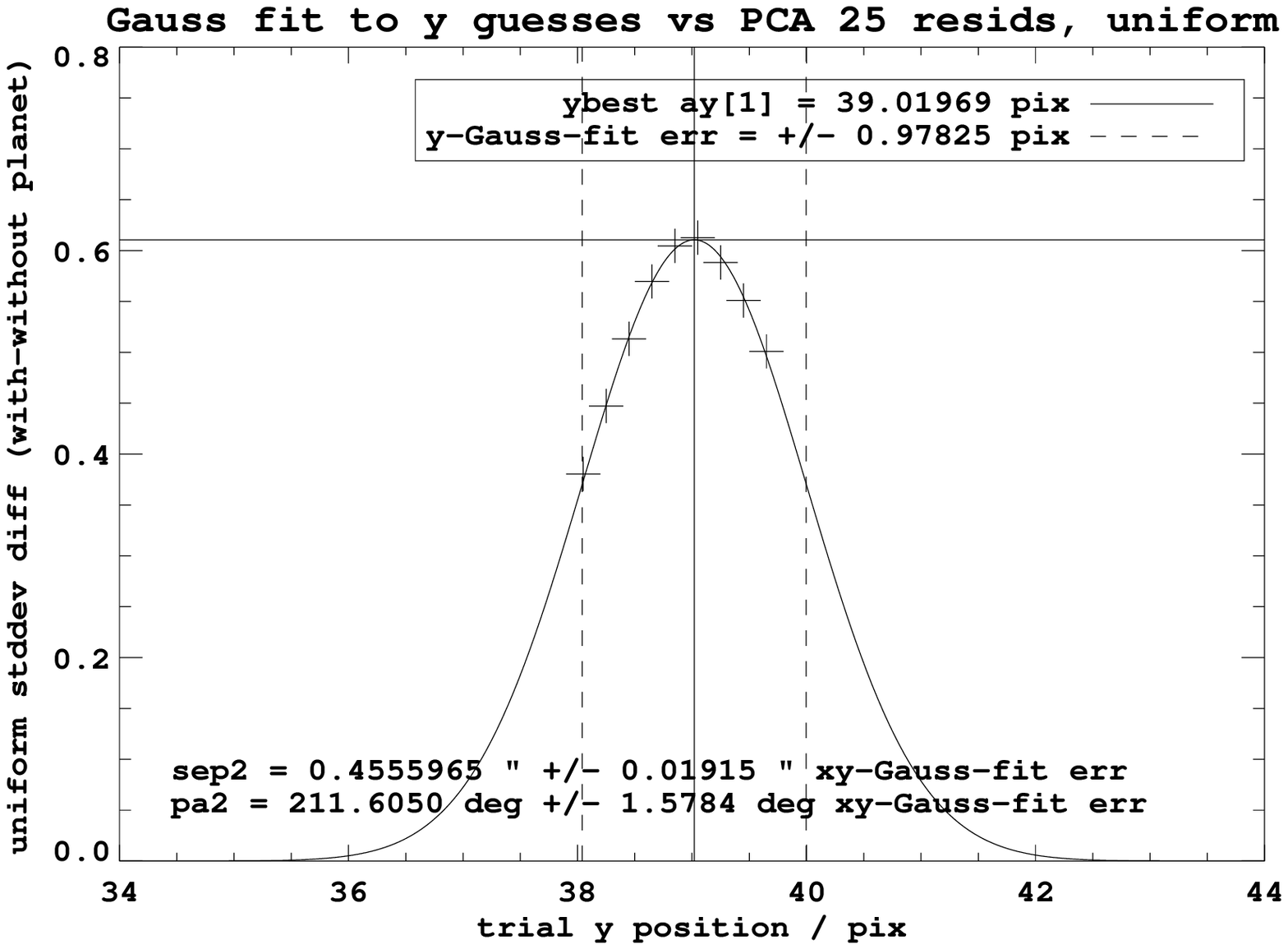}
	\includegraphics[width=0.33\linewidth]{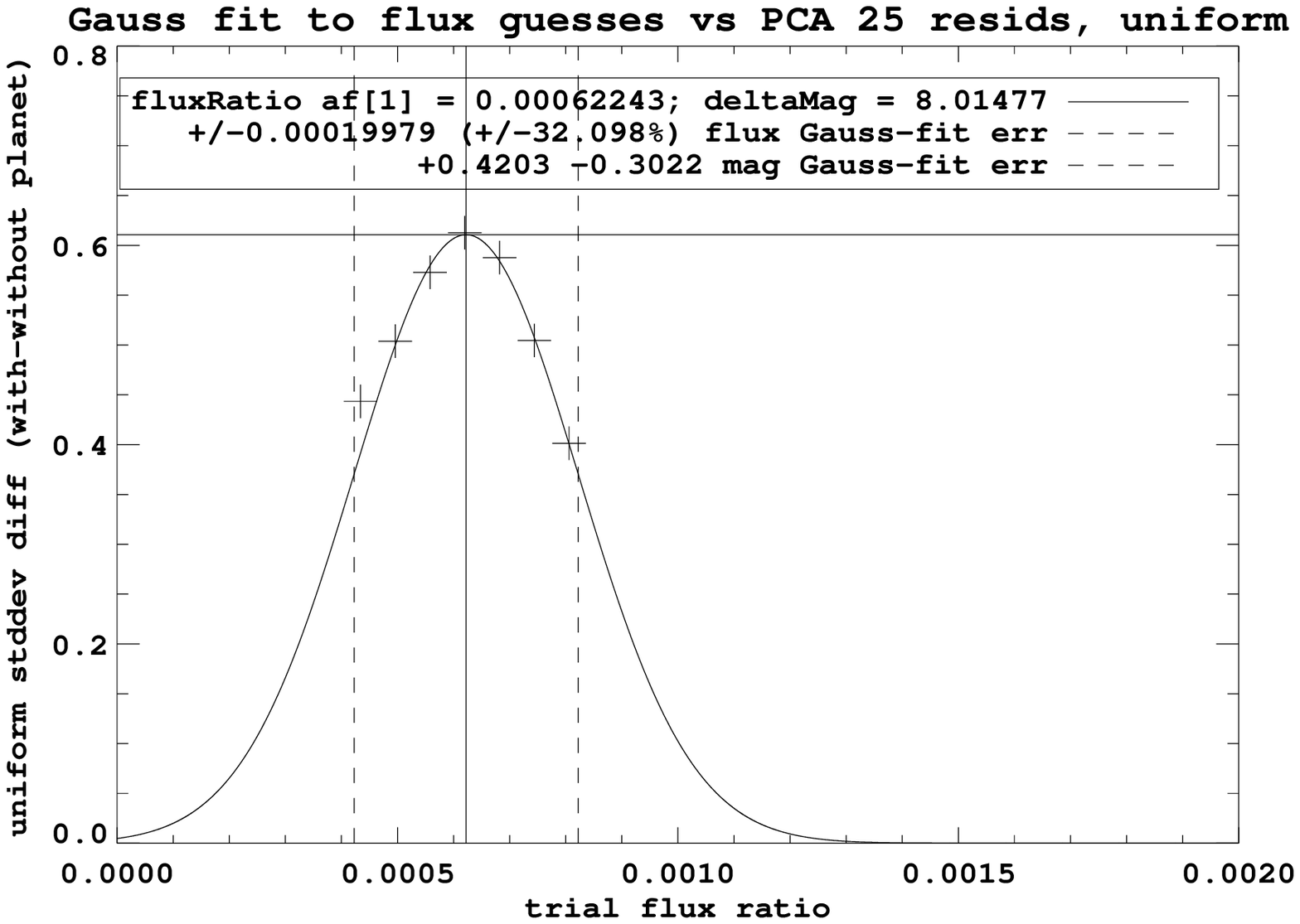}
	\caption{
		Grid search in the $L^\prime$ images for the best-fit photometry and astrometry of the planet,
		using PCA with 25 modes.
		Left column: x position (detector coordinates);
		Center column: y position (detector coordinates); and
		Right column: flux ratio.
		Top row: Parabola fit, local regions;
		Second row: Gaussian fit, local regions;
		Third row: Parabola fit, uniform regions; and
		Bottom row: Gaussian fit, uniform regions.}
	\label{fig:gridsearchlprime}
\end{figure*}

\begin{figure*}[!h]
\centering
	\includegraphics[width=0.33\linewidth]{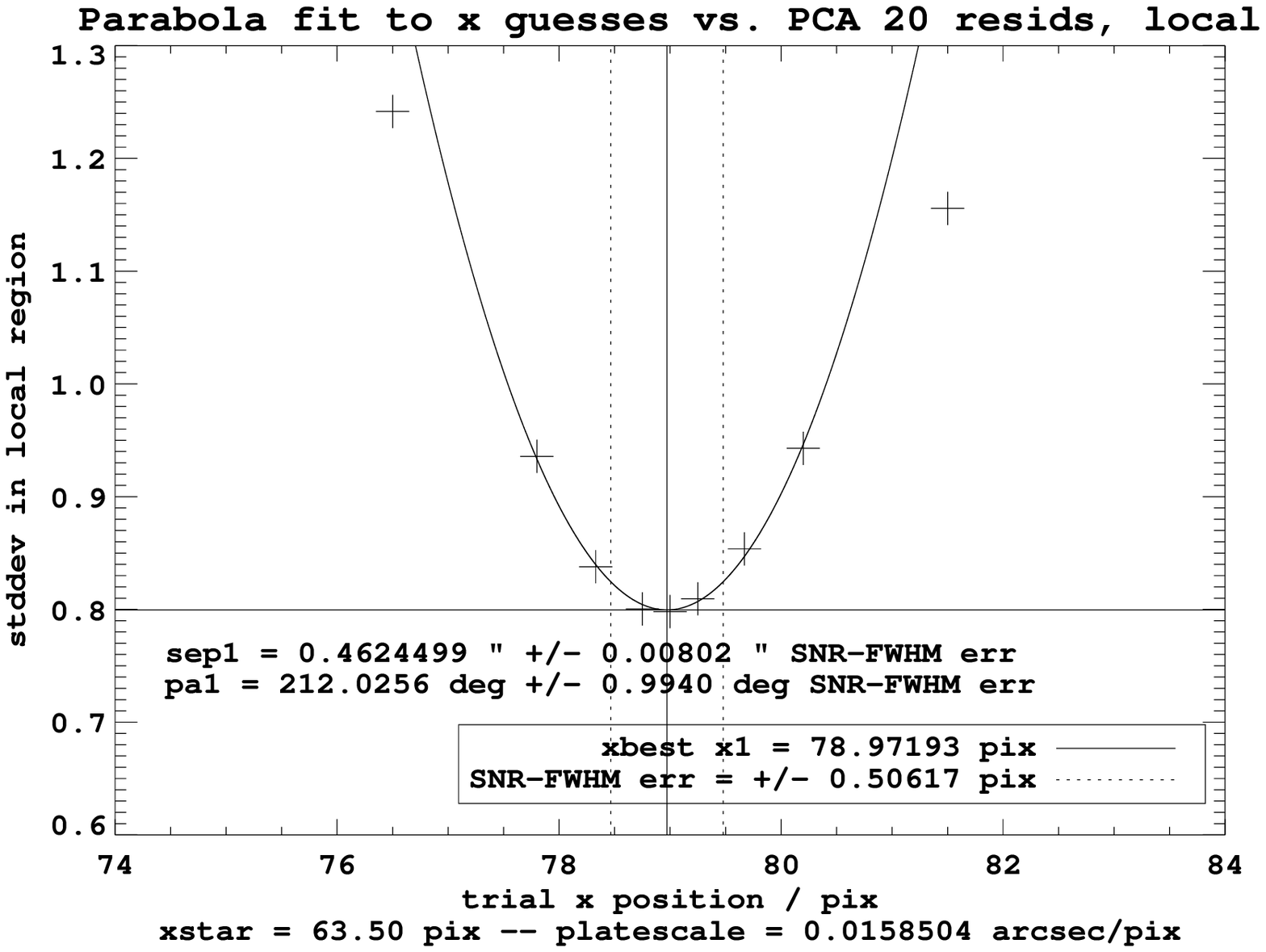}
	\includegraphics[width=0.33\linewidth]{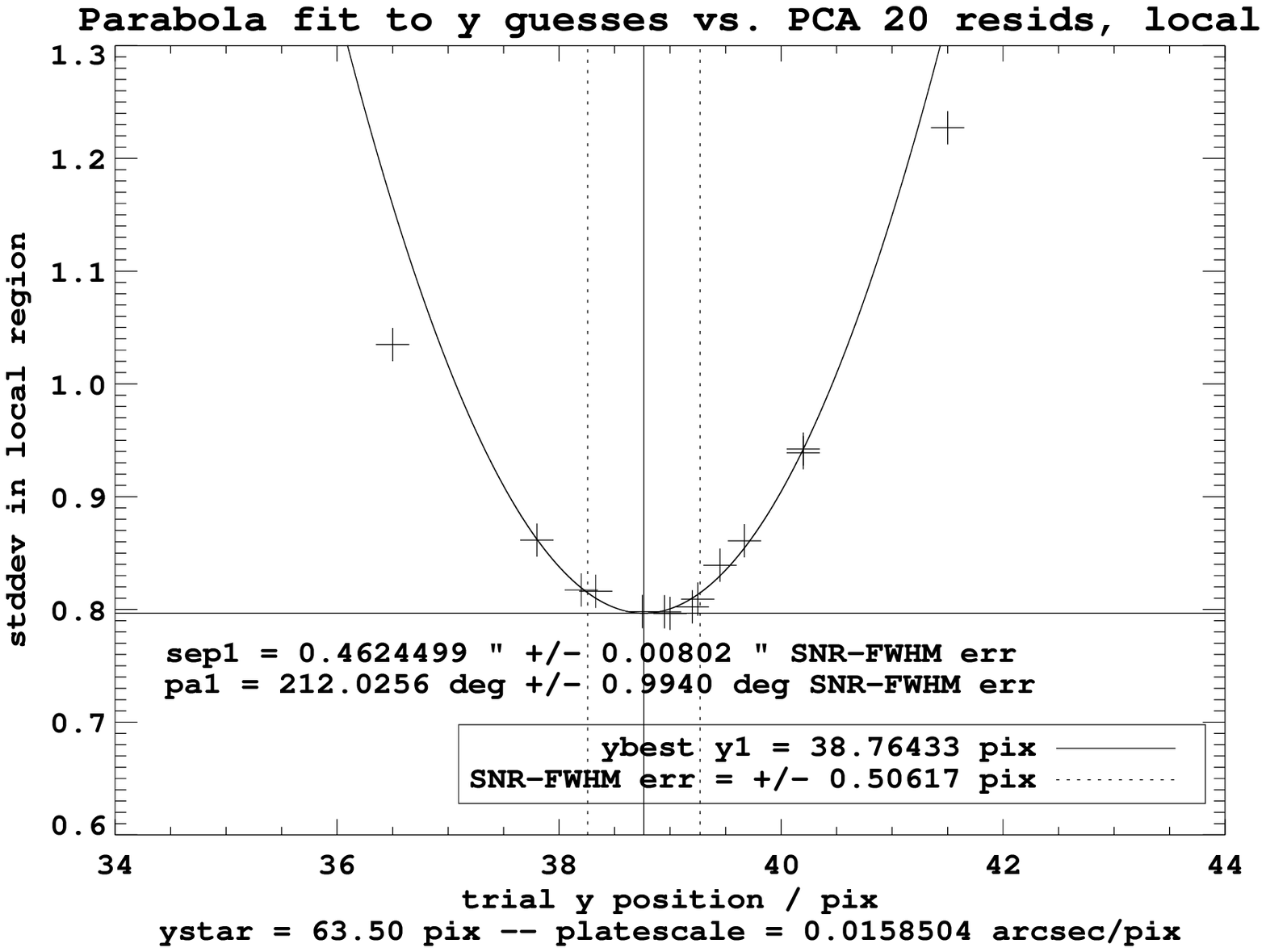}
	\includegraphics[width=0.33\linewidth]{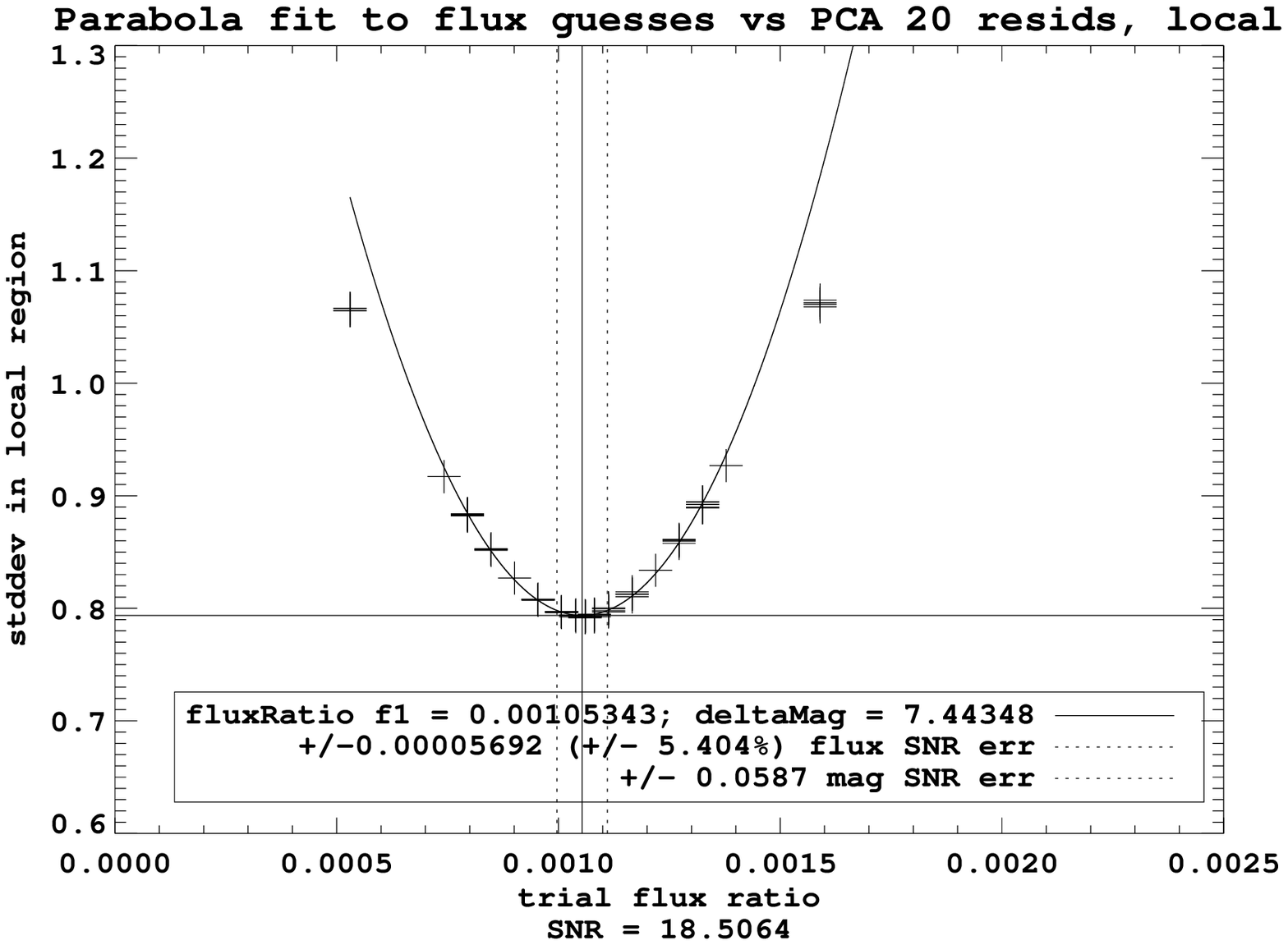}
	\includegraphics[width=0.33\linewidth]{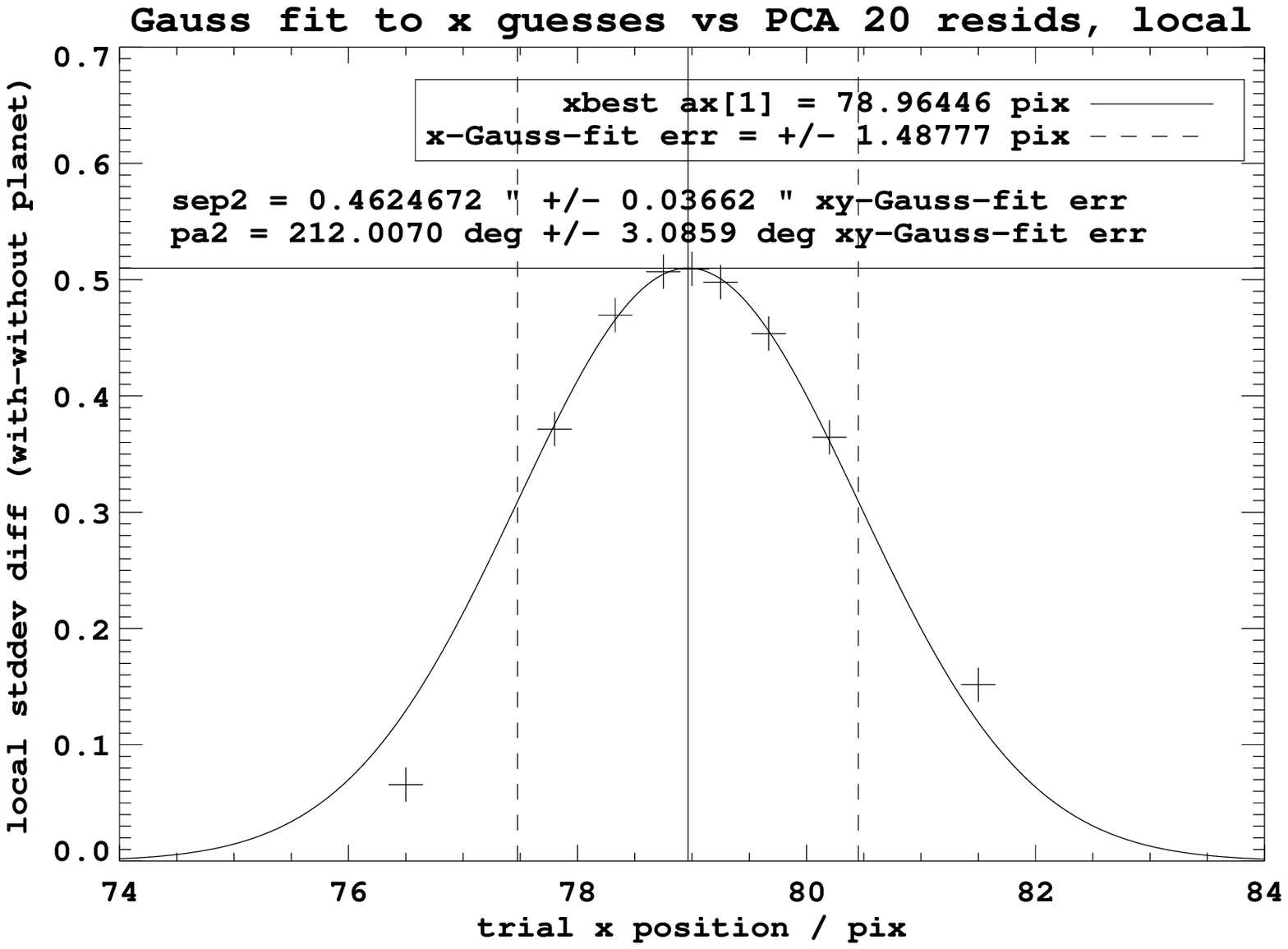}
	\includegraphics[width=0.33\linewidth]{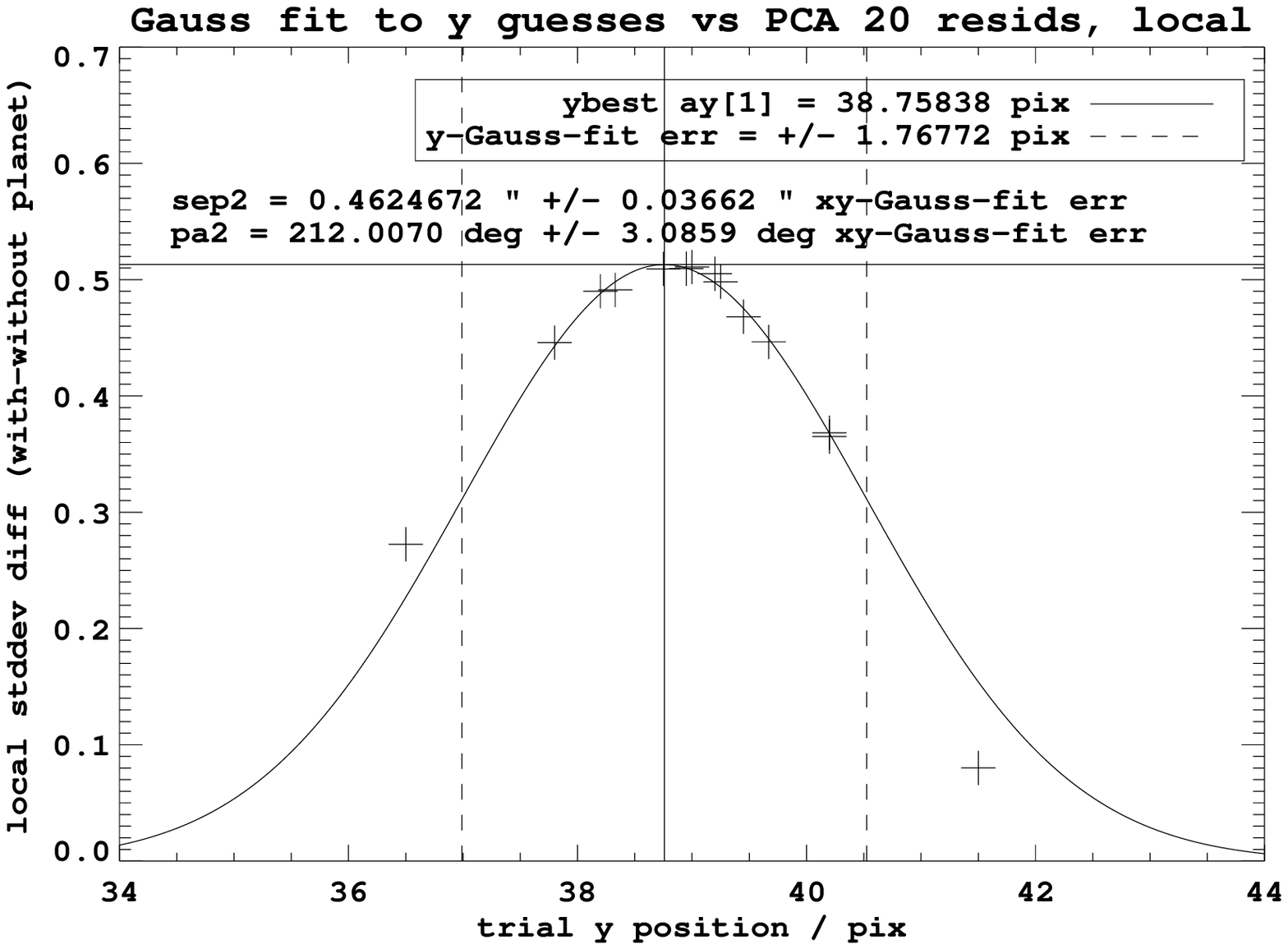}
	\includegraphics[width=0.33\linewidth]{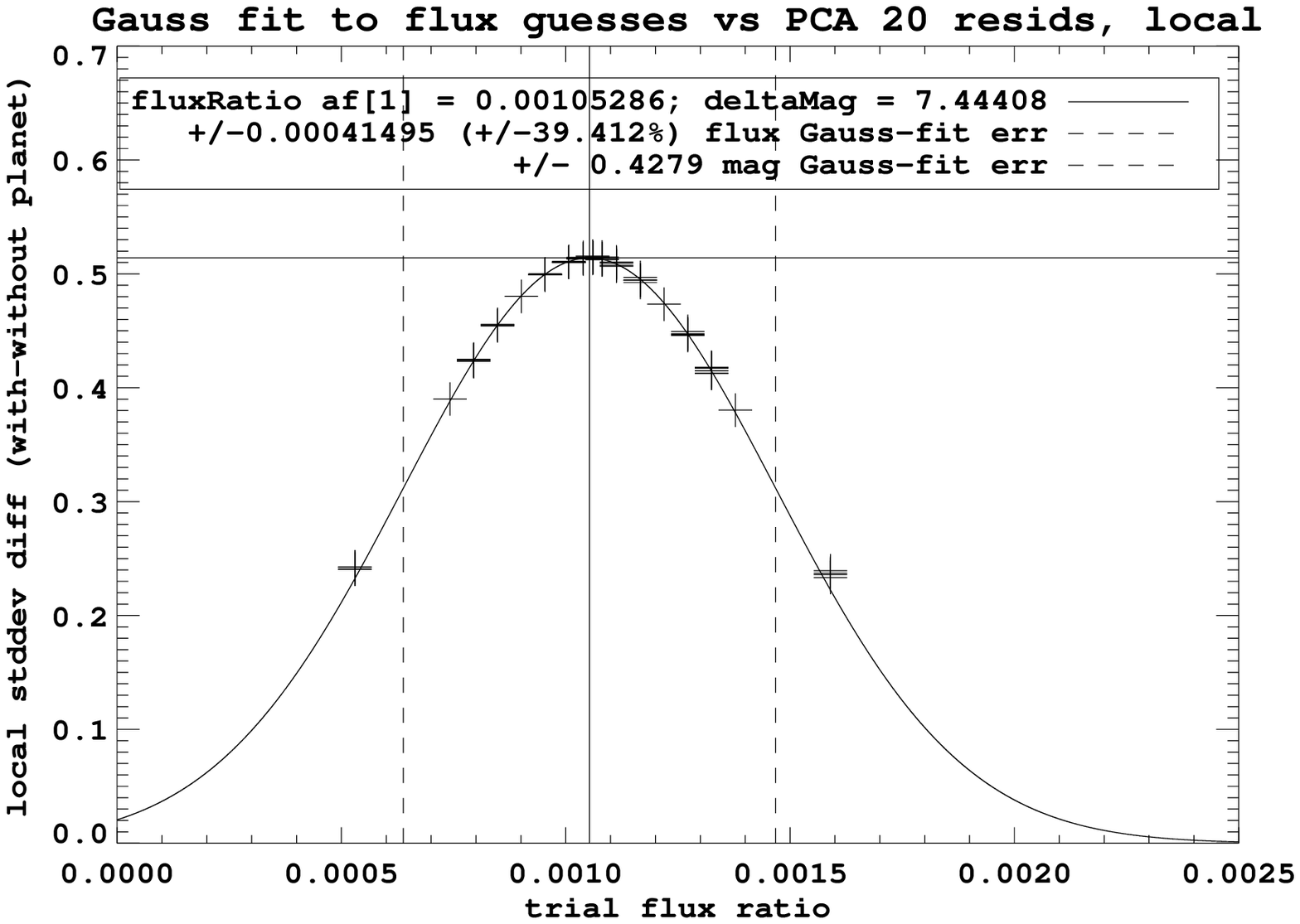}
	\includegraphics[width=0.33\linewidth]{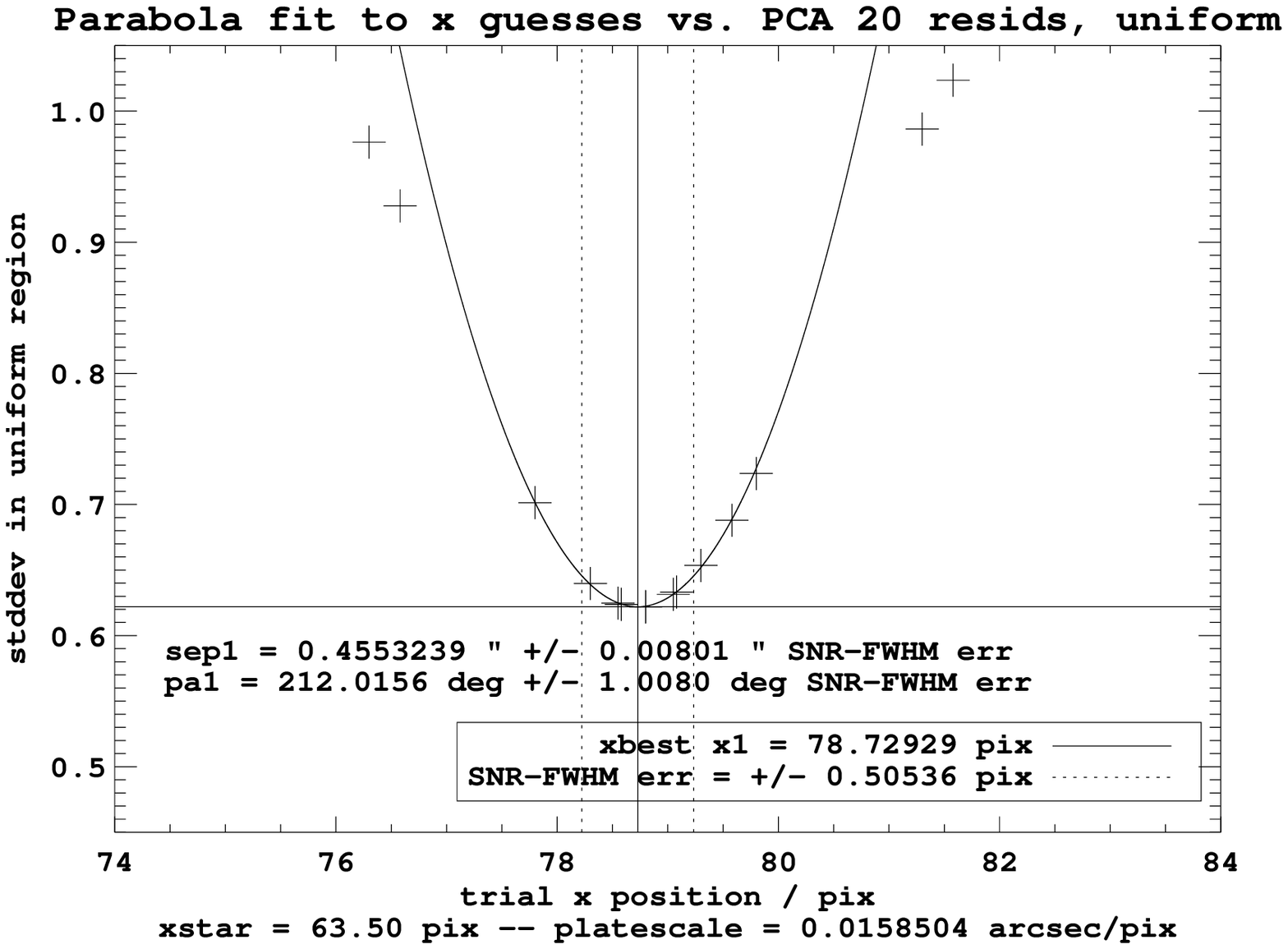}
	\includegraphics[width=0.33\linewidth]{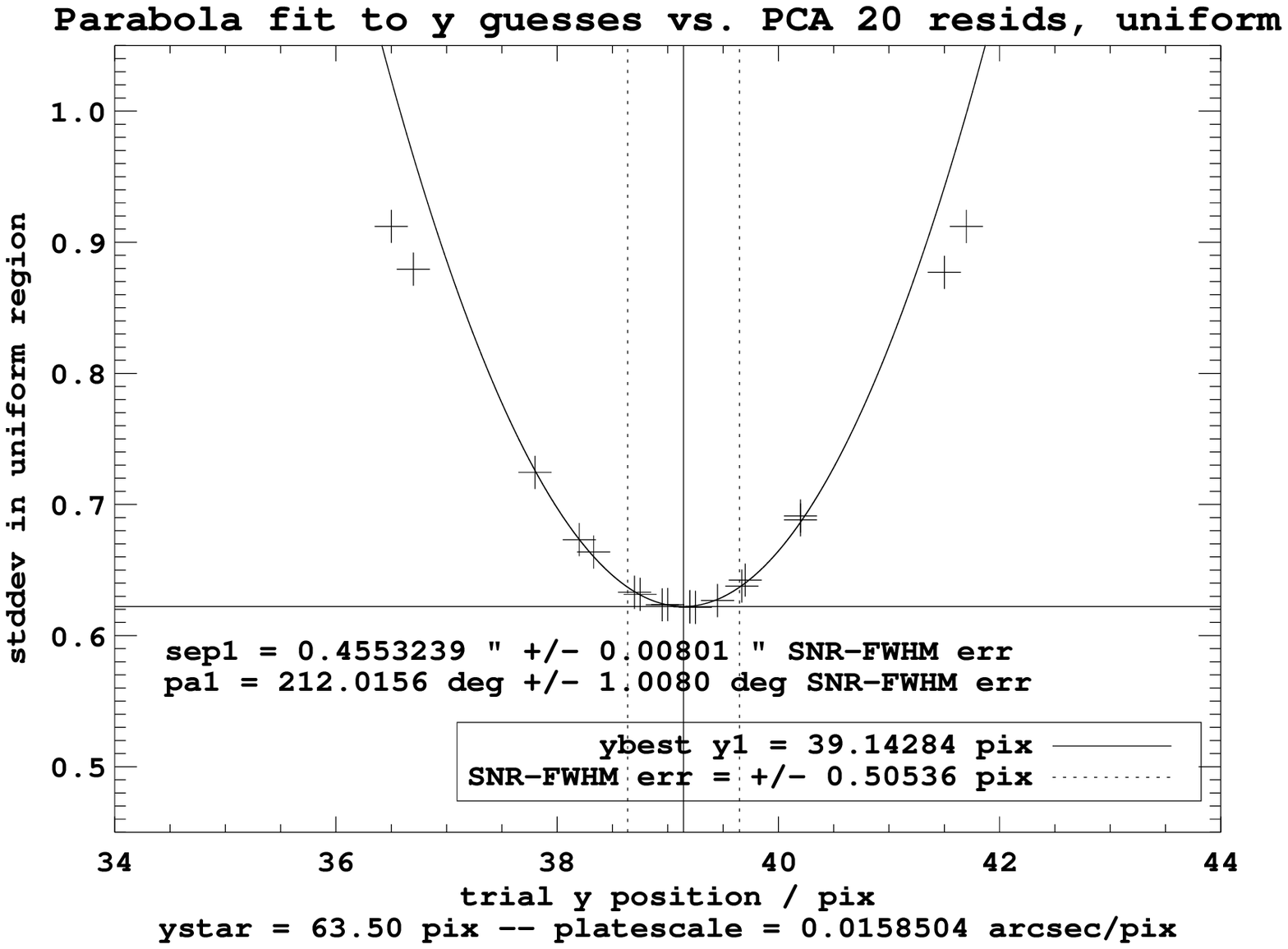}
	\includegraphics[width=0.33\linewidth]{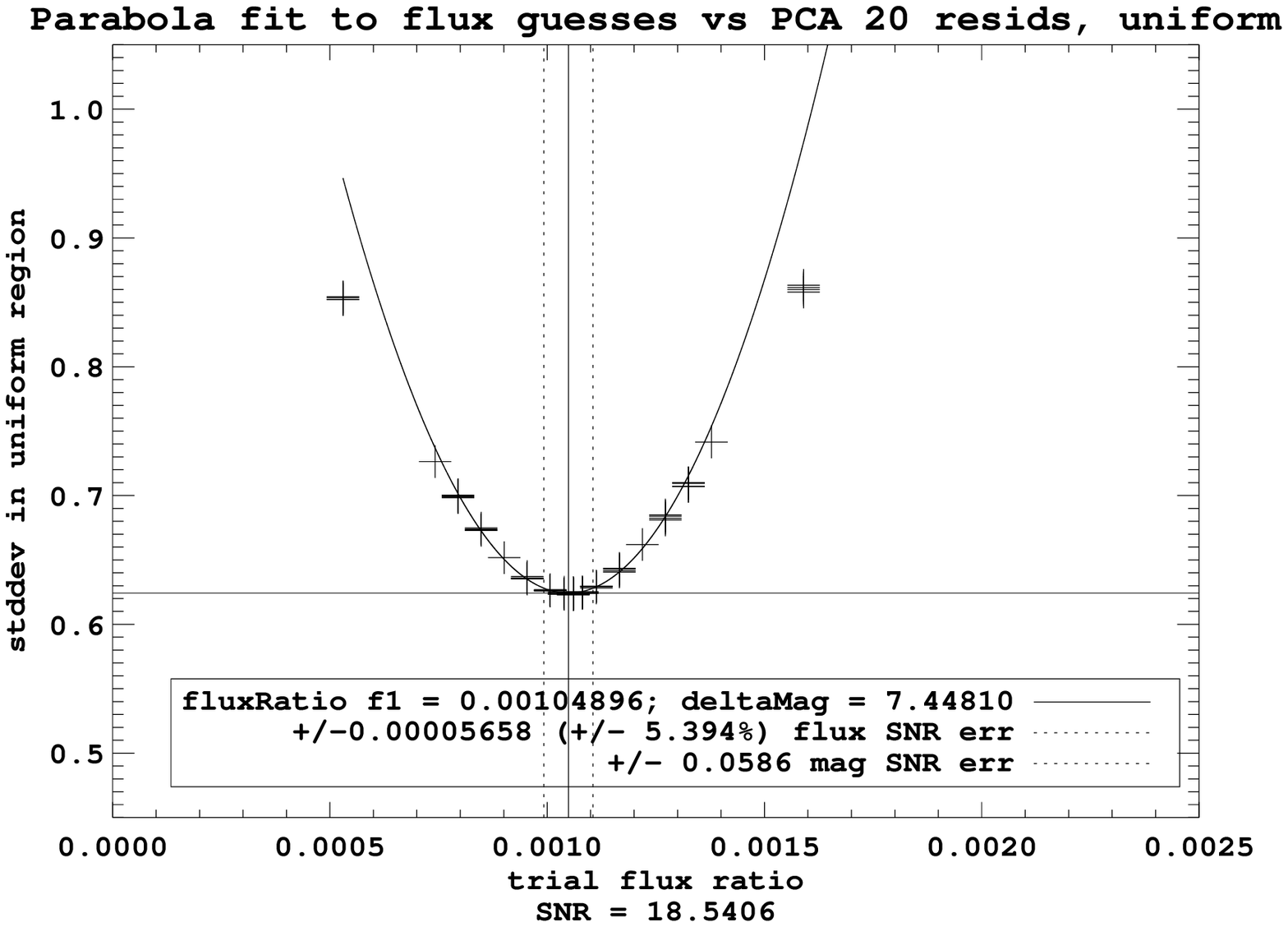}
	\includegraphics[width=0.33\linewidth]{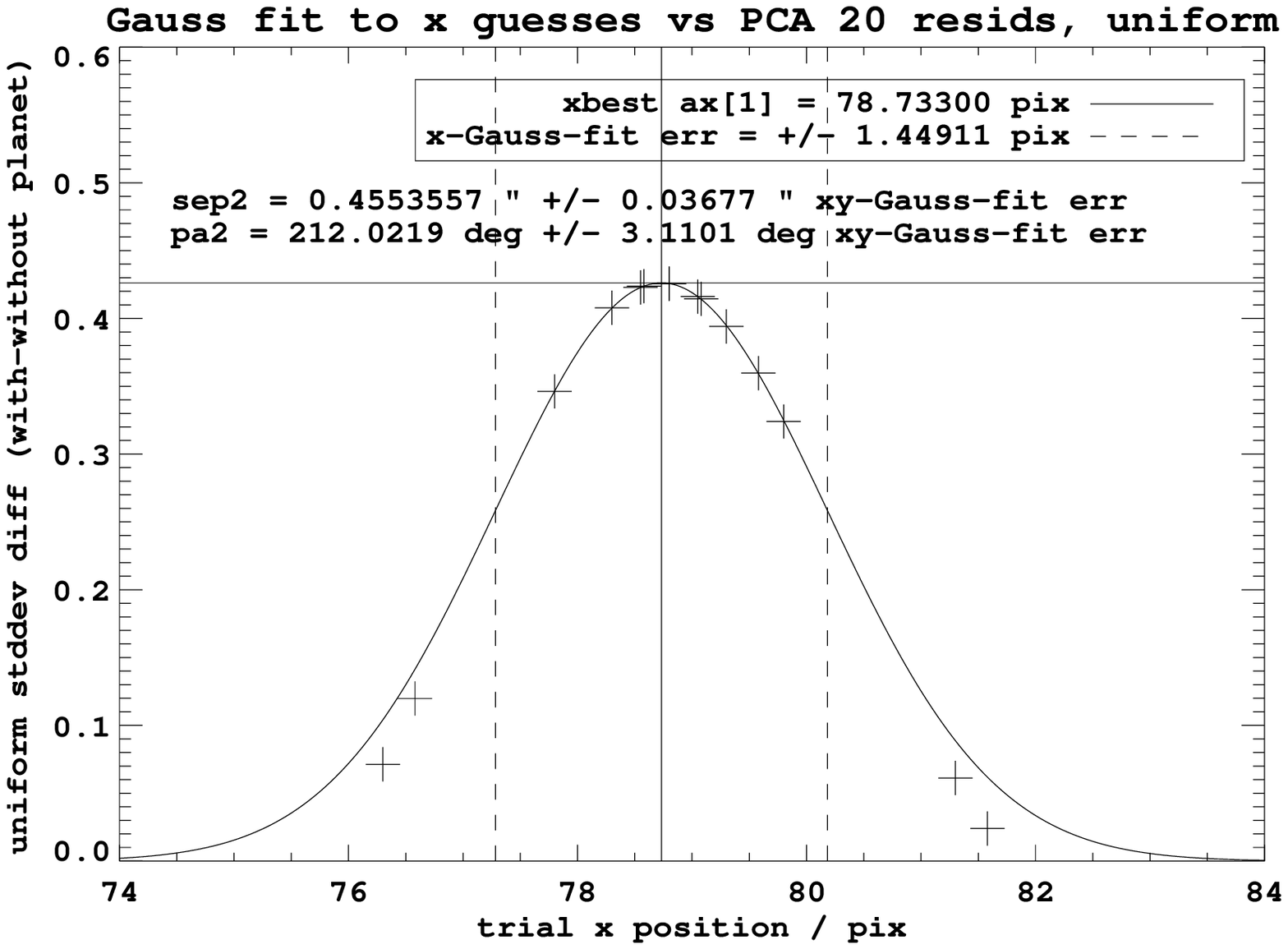}
	\includegraphics[width=0.33\linewidth]{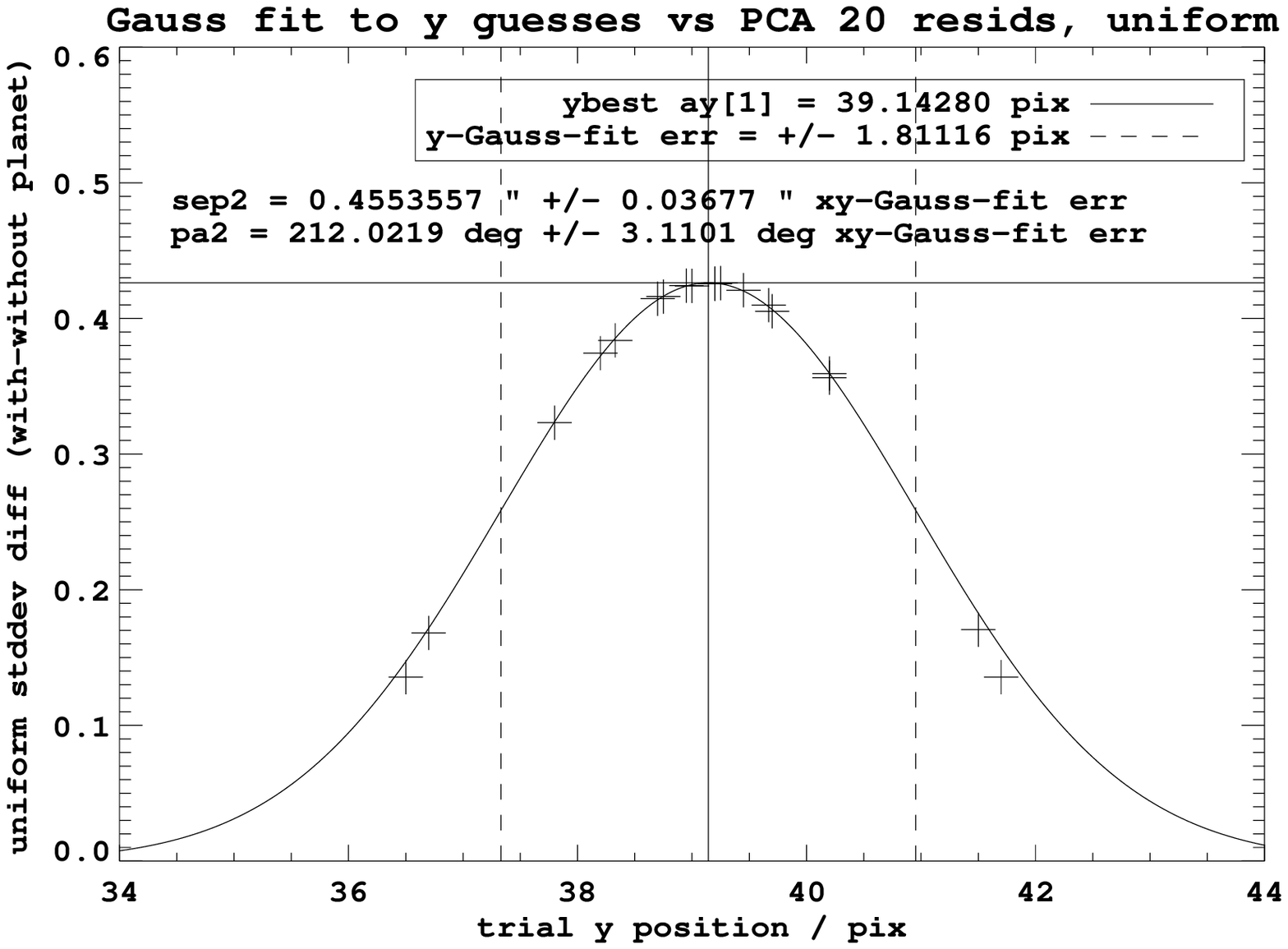}
	\includegraphics[width=0.33\linewidth]{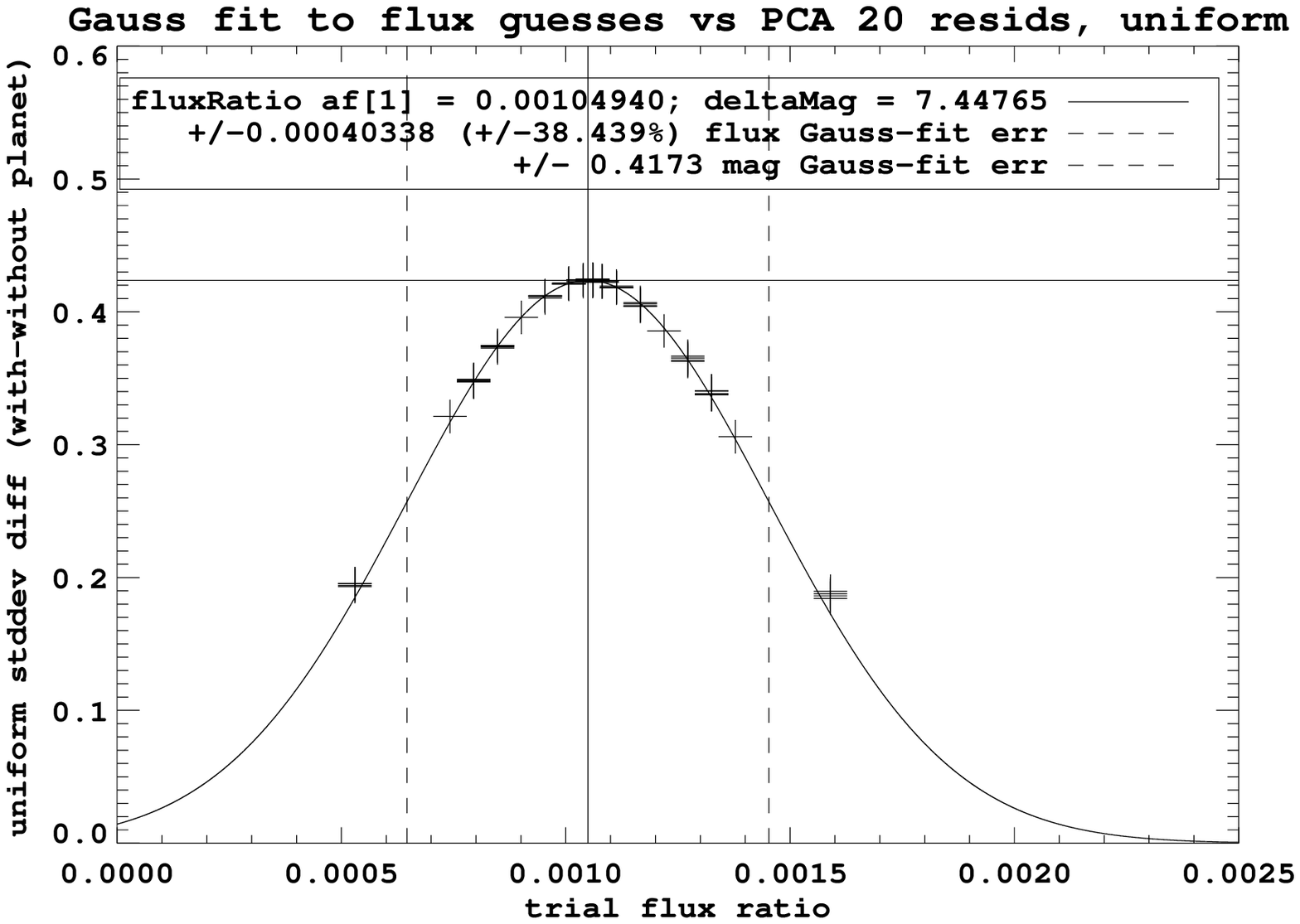}
	\caption{
		Grid search in the $M^\prime$ images for the best-fit photometry and astrometry of the planet,
		using PCA with 20 modes.
		Left column: x position (detector coordinates);
		Center column: y position (detector coordinates); and
		Right column: flux ratio.
		Top row: Parabola fit, local regions;
		Second row: Gaussian fit, local regions;
		Third row: Parabola fit, uniform regions; and
		Bottom row: Gaussian fit, uniform regions.}
	\label{fig:gridsearchmprime}
\end{figure*}

We fit two different types of curves to the standard-deviation of the KLIP residuals: we find the minimum of a parabola-fit to determine the best modeled value; and we use a Gaussian-fit in order to get the standard errors.
We fit a parabola or a Gaussian to the standard deviation (y-axis) as a function of the guessed parameter ($x$, $y$, or $f$, on the x-axis).
We kept the other parameter ($y$ and $f$ if fitting $x$) fixed at its best value --- thus this was an iterative process.
To create the Gaussian-fits, we take the difference of the KLIP residuals with the planet subtracted out and the KLIP residuals with the planet still in, and find the standard deviation of this difference image.
Fig.~\ref{fig:gridsearch31} shows the $[3.1]$ results,
Fig.~\ref{fig:gridsearch33} shows the $[3.3]$ results,
Fig.~\ref{fig:gridsearchlprime} shows the $L^\prime$ results, and
Fig.~\ref{fig:gridsearchmprime} shows the $M^\prime$ results.
Table~\ref{tab:photmeaserr} shows the resulting photometry errors.

\begin{table*}[h!]
	\small
	\caption{
		Measurement errors in flux ratio of planet in $[3.1]$, $[3.3]$, $L'$, and $M'$.
	}	\label{tab:photmeaserr}
	\centering
	\begin{tabular}{lccccccccc}
		\hline
		\hline
		&			&			&					& \multicolumn{2}{c}{\sc Best:}		& \multicolumn{2}{c}{\sc Error:}					& \multicolumn{2}{c}{\sc Error:}					\\
		& \# PCA	& Region	&					& \multicolumn{2}{c}{Parabola-fit}		& \multicolumn{2}{c}{SNR-limit}					& \multicolumn{2}{c}{Gaussian-fit}				\\
		Filt.		& modes	& of metric	& SNR$^\dagger$	& flux ratio		& $\Delta$ mag		& flux$^*$ $\sigma$	& $\Delta$ mag $\sigma$	& flux$^*$ $\sigma$	& $\Delta$ mag $\sigma$	\\
		\hline
		[3.1]	& 20		& local		& 4.50				& 0.00057893	& 8.09344			& 0.00012877		& 0.241498					& 0.00065191		& 1.22260					\\
		&			& uniform	& 4.52				& 0.00053228	& 8.18465			& 0.00011788		& 0.240450					& 0.00071525		& 1.45896					\\
		\cline{5-10}
		\multicolumn{4}{r}{}									& Mean			& 8.13809			& \multicolumn{2}{r}{$\pm$0.240974}				& \multicolumn{2}{r}{$\pm$1.34078}				\\
		\multicolumn{4}{r}{\textbf{Combined measurement}}	& \multicolumn{6}{l}{\textbf{8.14 $\pm$ 0.24 mag}}																						\\
		\multicolumn{10}{l}{}																																										\\ 
		\hline
		[3.3]	& 25		& local		& 8.66				& 0.00046614	& 8.32871			& 0.00005383		& 0.125381					& 0.00014761		& 0.343814					\\
		&			& uniform	& 8.63				& 0.00045762	& 8.34874			& 0.00005301		& 0.125770					& 0.00014459		& 0.343050					\\
		\cline{5-10}
		\multicolumn{4}{r}{}									& Mean			& 8.33868			& \multicolumn{2}{r}{$\pm$0.125575}				& \multicolumn{2}{r}{$\pm$0.343432}				\\
		\multicolumn{4}{r}{\textbf{Combined measurement}}	& \multicolumn{6}{l}{\textbf{8.34 $\pm$ 0.13 mag}}																						\\ 
		\multicolumn{10}{l}{}																																										\\ 
		\hline
		$L'$		& 25		& local		& 9.83				& 0.00061145	& 8.03409			& 0.00006222		& 0.110482					& 0.00040182		& 0.713502					\\
		&			& uniform	& 9.62				& 0.00062191	& 8.01567			& 0.00006462		& 0.112814					& 0.00019979		& 0.348795					\\
		\cline{5-10}
		\multicolumn{4}{r}{}									& Mean			& 8.02484			& \multicolumn{2}{r}{$\pm$0.111648}				& \multicolumn{2}{r}{$\pm$0.531148}				\\
		\multicolumn{4}{r}{\textbf{Combined measurement}}	& \multicolumn{6}{l}{\textbf{8.02 $\pm$ 0.11 mag}}																						\\ 
		\multicolumn{10}{l}{}																																										\\ 
		\hline
		$M'$	& 20		& local		& 18.51				& 0.00105343	& 7.44348			& 0.00005692		& 0.058666					& 0.00041495		& 0.427676					\\
		&			& uniform	& 18.54				& 0.00104896	& 7.44810			& 0.00005658		& 0.058564					& 0.00040338		& 0.417522					\\
		\cline{5-10}
		\multicolumn{4}{r}{}									& Mean			& 7.44579			& \multicolumn{2}{r}{$\pm$0.058615}				& \multicolumn{2}{r}{$\pm$0.422610}				\\ 
		\multicolumn{4}{r}{\textbf{Combined measurement}}	& \multicolumn{6}{l}{\textbf{7.45 $\pm$ 0.06 mag}}																						\\ 
		\multicolumn{10}{l}{}																																										\\ 
		\multicolumn{10}{l}{$^*$ In units of star A brightness.}																																		\\
		\multicolumn{10}{l}{$^\dagger$ Local and Uniform SNR vary (within noise) due to different positions of planet where SNR is calculated.}															\\
		\multicolumn{10}{l}{Magnitude errors calculated as follows: $\sigma_\mathrm{mag} = \frac{2.5}{ln_e(10)} \times \frac{\sigma_f}{f}$.}																\\
		\multicolumn{10}{l}{We have chosen to use the SNR-limit errors for the flux ratio because the speckle noise is not Gaussian.}
	\end{tabular}
	\normalsize
\end{table*}

\clearpage
\bibliography{ktmorz_bib}

\end{document}